Università degli Studi di Torino
Scuola di Dottorato

Dottorato in Fisica - XXXVIII ciclo

# Towards next-gen parton distribution and fragmentation functions

Tanishq Sharma

Università degli Studi di Torino
Scuola di Dottorato

---

Dottorato in Fisica

# Towards next-gen parton distribution and fragmentation functions

Tanishq Sharma

Supervisor: Prof. Emanuele Roberto Nocera

Defended on the 15th of September, 2025

# Abstract

The interest into parton distribution functions (PDFs) and fragmentation functions (FFs) in current high energy physics research is twofold. On the one hand, they are fundamental objects to conduct precision phenomenology studies, e.g. at the Large Hadron Collider (LHC), to determine the Standard Model parameters and search for new physics. On the other hand, they are also a means to understand the inner structure and dynamics of hadrons, e.g. in regards to the proton spin decomposition at the future Electron Ion Collider (EIC). In this thesis, I present several advancements in the determination of PDFs and FFs that will allow for the release of their next versions by the NNPDF collaboration. These include some crucial components that will enter these new PDF and FF sets. Concerning PDFs, I consider three aspects. First, I study the impact of the commonly used K-factor approximation against the use of exact next-to-next-to-leading order (NNLO) computations. Second, I study the compatibility of new data with existing PDFs, taking into account all sources of experimental and theoretical uncertainties (including those coming from PDFs, missing higher orders, and $\alpha_s$). Third, I study the impact of incremental inclusion of new data in PDF fits. Specifically, I focus on data sets that are relevant for the determination of the gluon PDF, namely single-inclusive jet and di-jet production data in proton-proton collisions and in deep-inelastic scattering, and top quark pair production data in proton-proton collisions. Concerning FFs, I extend the NNPDF computational framework in three respects, each one corresponding to three different pieces of software. First, I implement time-like evolution in EKO. Second, I extend PineAPPL to handle multiple convolutions of PDFs and FFs, including the corresponding factorization scales and the polarization of the initial and final states. Third, I develop a new software, called vhf, specifically devised to compute single-inclusive annihilation (SIA) and semi-inclusive deep-inelastic scattering (SIDIS) cross sections. All of these developments are crucial to prepare the release of next-gen PDF and FF sets that will allow us to take advantage of the forthcoming LHC and EIC physics programs as much as possible.

In loving memory of my Papa

and

dedicated to my Mummy

# Acknowledgements

I would like to begin by expressing my heartfelt gratitude to my supervisor Emanuele, for everything, his guidance, support, mentoring, advice and more. I could write a whole chapter on my amazing experience doing my PhD with him, but to summarize it in one sentence, I could not have asked for a better supervisor myself. From the bottom of my heart, Thanks!

I am also grateful to Juan Rojo for allowing me the opportunity to spend the second year of my PhD within his group at Nikhef. I sincerely cherished the pleasing environment during my time as a visitor there.

I am also thankful to all the members of the NNPDF collaboration for the amazing time I have had working and collaborating with all of them, and I hope to continue to do so for a very long time. I would also like to thank Andrea, Christopher, Felix, Giacomo, Juan Cruz and Tanjona for having provided me with valuable guidance at some point(s) of my PhD.

This PhD was supported by a doctoral scholarship by the University of Torino, which was funded by Emanuele's Rita Levi Montalcini grant, awarded by the Italian Ministry of Education, University and Research (MIUR).

"The physics is theoretical, but the fun is real."

~ Sheldon Cooper (B.S., M.S., M.A., Ph.D., Sc.D.), circa 2009

# Contents

# List of Abbreviations

(a)N$^3$LO: (approximate) next-to-next-to-next-to-leading order
(a)NNLO: (approximate) next-to-next-to-leading order
BSM: Beyond the Standard Model
DIS: deep inelastic scattering
DGLAP: Dokshitzer-Gribov-Lipatov-Altarelli-Parisi
EIC: Electron Ion Collider
EFT: effective field theory
FCC: Future Circular Collider
FF: fragmentation function
HERA: Hadron-Electron Ring Accelerator
(HL-)LHC: (High Luminosity-) Large Hadron Collider
LEP: Large Electron-Positron Collider
LO: leading order
MC: Monte Carlo
MHO(U): missing higher order (uncertainties)
NLO: next-to-leading order
PDF: parton distribution function
QCD: Quantum Chromodynamics
QED: Quantum Electrodynamics
QFT: quantum field theory
RGE: renormalization group equation
SIA: single-inclusive annihilation
SIDIS: semi-inclusive deep inelastic scattering
SM: Standard Model of Particle Physics

# Introduction

A very fundamental method of improving our understanding of the fundamental constituents of matter and the forces that govern them is by smashing, or colliding particles and observing the resulting interactions. At the LHC, these experiments typically involve studies that fit in two broad categories: precision physics and new physics searches. The former is concerned with improving the accuracy with which we know the parameters of the SM, while the latter is concerned with searching for new particles or interactions beyond the SM. However, precision physics is also a precursor to many new physics searches, as these often tend to require a very accurate and precise knowledge of the SM parameters such as to be able to identify any deviations from the theoretical predictions. Consequently, the progress in precision physics is necessary for both, improving the accuracy with which we know the SM parameters, and for new physics searches.

In the context of these collider experiments, it is crucial to understand the initial state particles involved in the interactions. These particles can be fundamental particles, such as at the LEP, composite particles, such as at the LHC, or a combination of both, such as at the HERA. The theoretical predictions involving fundamental particles, such as quarks and leptons, are straightforward, as these are computed order by order in perturbation theory. The theoretical predictions involving composite particles, such as protons, are less straightforward, as dealing with bound states involves modelling strong interactions within the framework of QCD in non-perturbative regimes. This is where density functions come into play. Density functions model the probability of encountering a child particle carrying a given fractional momentum of its parent particle, at a given energy scale. These functions can model such behavior in both the transverse and longitudinal directions relative to the momentum of the parent particle. The functions that model it in the longitudinal direction are called collinear distributions. Furthermore, the functions can model this for initial state composite particles, in which case they are called collinear PDFs, or final state composite particles, in which case they are called collinear FFs. In particular, collinear PDFs model the probability of finding a given parton with a given fractional momentum of its parent hadron, when probed at a given energy scale. Collinear FFs model the probability of a given parton hadronizing into a given hadron carrying a given fraction momentum of the initial parton, at a given energy scale. These collinear distributions are the focus of this thesis.

Advancements in PDFs are particularly important in the context of precision phenomenology at high energy hadron colliders, such as the LHC. The extraction of many SM parameters requires a combination of experimental data and theoretical predictions. The ever improving statistics and a finer control over systematics has led to significant improvements in the precision of the experimental data at the LHC. This has led, in some cases, to PDFs becoming the dominant source of uncertainty in the SM parameters' determination, as is discussed in detail in Sec. 3.1. It is therefore vital to improve the accuracy and precision of the PDFs to allow for successful precision phenomenology at the LHC. In particular, we aim to achieve a percent level accuracy in the determination of PDFs. In addition, PDFs also play an important role in improving our understanding of the internal structure and dynamics of hadrons. In this aspect, they are heavily complemented by FFs.

The work done in the culmination of this thesis is towards the next generation determination of both, PDFs and FFs, by the NNPDF collaboration. Concerning PDFs, this involves conducting studies involving incremental improvements to assess the impact of each new update or upgrade. This would allow for a thorough and systematic path towards the new and improved PDF set, where every difference is well understood. In the context of FFs, this involves extending the NNPDF framework to be able to perform FF determinations using the robust and flexible NNPDF methodology. This would allow for a new and improved FF determination.

The structure of the thesis is as follows:

- Chapter 1 provides an overview of the SM. In this chapter, I discuss QED and QCD in the context of quantum field theory, followed by a discussion on renormalization and the running of the coupling constants. This is followed by a discussion on the QCD factorization, which forms the basis for PDFs and FFs.

- Chapter 2 provides an overview of the NNPDF methodology and framework. In this chapter, I discuss technical details involving the NNPDF framework that allow for a robust setup to perform state-of-the-art PDF determinations.

- Chapter 3 discusses the studies I carried out that will lead towards the next generation PDF determination by the NNPDF collaboration. These studies touch upon some of the most pressing aspects currently in the field of PDFs. One such study concerns the impact of K-factor approximation against the use of exact NNLO corrections. The second study concerns the uncertainty quantification in PDF determinations which is heavily impacted by the underlying methodology. I present a study where we assess various PDF sets by their ability to produce sensible theoretical predictions for experimental data, which was not part of the fitting procedure of the respective PDF sets. The third study concerns the utilization of new high precision data in the process of PDF determination which requires systematically evaluating the impact new experimental data might have on the PDFs when included in the fit, such as to be able to identify exactly what difference a given set of data makes. I present a study where I perform inclusion of data corresponding to gluon sensitive processes into the PDF fits to assess the impact on the PDF and its phenomenology. This chapter is fully based on original work carried out as part of my PhD.

- Chapter 4 discusses the extension of the NNPDF framework to be able to perform FF deter-

minations. In this chapter, I provide an overview of how various tools in the NNPDF framework had to be extended to accommodate FFs, by looking at both, the physics behind it and the details of the implementation. This includes extending EKO, a tool that solves DGLAP evolution, to allow for time-like evolution, which is needed to evolve FFs from a given energy scale to another. It also includes extending PineAPPL, an interpolation grid library to make it more flexible such that it can support FFs and the energy scale associated to FFs. I also discuss the development of a new tool, vhf, that allows for computation of theoretical predictions for processes that go into the determination of FFs. This chapter is also fully based on original work done as part of my PhD.

- Conclusion provides a concise summary of the work done in this thesis, and discusses the future outlook.

- Appendices A and B provide additional material that supports the study in Sec. 3.3.

# Chapter 1

# Crash course on particle physics

## 1.1 Introduction

The Standard Model of Particle Physics is, currently, our main theory to explain the fundamental particles and their interactions. However, the SM is not able to account for a number of concepts such as gravity, neutrino masses, dark matter and dark energy (if they exist), just to name a few, and hence would at some point in time be replaced by some other theory that is more general. This makes the SM an effective theory, that is, a theory that can explain some observable phenomena at particular scales at which it is deemed applicable. Within its range of applicability, the SM has so far stood the test of time when confronted with experimental data coming from particle physics experiments. Therefore, the SM is currently our go-to theory, especially for the continuing advancements in the field of precision high energy physics. In this chapter, the SM will be discussed at a surface level, to pave the path towards the discussion of the work carried out in the context of this thesis.

The SM is a quantum field theory, i.e. a culmination of quantum mechanics and special relativity. To achieve this, it is insufficient to find a relativistic generalization of quantum mechanics. At relativistic energies, particles are created and annihilated, and a wave equation cannot account for processes with a variable particle number. To overcome this, instead of quantizing a particle in a classical potential, one associates particles with the modes of a field and quantizes the field itself, promoting it to a quantum field.

The discussion on SM will proceed as follows: first, there will be an example of a toy scalar QFT to introduce the concepts of particles and interactions within a QFT. This will be followed by an overview of electromagnetic interactions and of strong interactions in the SM. Next, there will be a discussion on renormalization, the running of couplings, and how they lead to bound states in strong interactions. Then, QCD factorization will be discussed to explain the method of dealing with bound states. This will also include a discussion on PDFs and FFs. Throughout this chapter, results will be provided as is, without any derivations, as the focus is on a quick review of the topic before delving into the actual research. A reader interested in a more thorough and mathematical outlook may benefit from reading the topic in question in any of the standard graduate textbooks such as Ref. [1–4].

## 1.2 Quantum Field Theories

### 1.2.1 A toy scalar QFT

To begin, let us start with a free scalar field. It can be denoted using the following Lagrangian density $\mathscr{L}$:

$$\mathscr{L} = \frac{1}{2}\left(\partial_\mu \varphi \partial^\mu \varphi - m^2 \varphi^2\right) \tag{1.1}$$

where $\varphi$ is a field, and it depends on $x^\mu$ coordinates. $m$ is the mass of the quanta of the field. $\partial_\mu \varphi \partial^\mu \varphi$ denotes a kinetic term and $m^2\varphi^2$ denotes a mass term. In this theory, the scalar particles have masses and can freely propagate but cannot interact. The internal propagator for this field is simply the inverse coefficient of $\varphi^2$, which in momentum space is given as

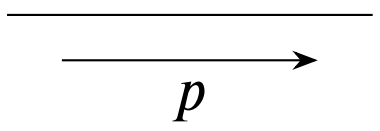

$$\frac{i}{p^2 - m^2}$$

To allow for interactions, the above Lagrangian density could be modified as follows:

$$\mathscr{L} = \frac{1}{2}\left(\partial_\mu \varphi \partial^\mu \varphi - m^2 \varphi^2\right) - \frac{\lambda}{4!}\varphi^4 \tag{1.2}$$

where $-\lambda\varphi^4/4!$ denotes an interaction term such that at an interaction vertex, four internal propagators coincide. Here $\lambda$ is the coupling constant, and $4!$ is the normalization factor. This Lagrangian density corresponds to a well known toy model called the $\varphi^4$ theory (see chapter 4 in Ref. [2]). In this model, an external particle propagator is represented by 1, an internal particle propagator is represented by exactly as above and an interaction vertex is represented by $-i\lambda^4$.

An important consistency check of a given QFT is that it should transform under the Poincaré group such that the action remains invariant. The transformations of the Poincaré group include spacetime translations and Lorentz transformations. Lorentz transformations include boosts and rotations. In the case of $\varphi^4$ theory, the kinetic term, the mass term and the interaction term, are all individually invariant under these transformations. With this warm up example out of the way, the discussion will now shift to more physical theories.

### 1.2.2 Quantum Electrodynamics

The SM explains three types of forces in nature: electromagnetic, weak and strong. In this subsection, we will look at QED, which explains the electromagnetic interactions, in isolation, for simplicity. In general, one should consider the full electroweak theory, which combines electromagnetic and weak interactions. This is important as two of the three massive vector bosons ($W^+$ and $W^-$), that mediate the weak force, also carry an electromagnetic charge.

In QED (see part 1 of Ref. [2]), we have two types of fields, the fermion field and the photon field.

The Lagrangian density is given as:

$$\mathscr{L}_{QED} = \mathscr{L}_{photon} + \mathscr{L}_{fermion} + \mathscr{L}_{int} \tag{1.3}$$

$$\mathscr{L}_{photon} = -\frac{1}{4} F^{\mu\nu} F_{\mu\nu} \tag{1.4}$$

$$\mathscr{L}_{fermion} = \sum_f \bar{\psi}_f \left( i\gamma^\mu \partial_\mu - m_f \right) \psi_f \tag{1.5}$$

$$\mathscr{L}_{int} = \sum_f -eQ_f \bar{\psi}_f \gamma^\mu \psi_f A_\mu \tag{1.6}$$

where $\mathscr{L}_{photon}$ is the free photon field term, $\mathscr{L}_{fermion}$ is the free fermion field term and $\mathscr{L}_{int}$ is the interaction term, where two fermions and one photon interact. The index $f$ corresponds to the different fermion flavors. For simplicity, in the following equations, the dependence of particle fields on $x_\mu$ is implicit.

A fermion is represented by the spinor field $\psi$ (incoming), and its adjoint $\bar{\psi}$ (outgoing). The mass of the fermion is represented by $m$. Each fermion particle has an anti-particle counterpart. $\gamma^\mu$ are the Dirac matrices. All the Feynman rules presented in this thesis are done so in the Feynman gauge. The external propagators for a fermion are denoted as $u$ and $\bar{u}$ for an incoming and an outgoing particle respectively, and as $\bar{v}$ and $v$ for an incoming and an outgoing anti-particle respectively. An internal fermion propagator is given as

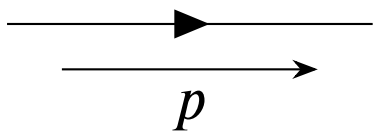

$$\frac{i}{\gamma^\mu p_\mu - m_f}$$

$F_{\mu\nu}$ is the electromagnetic field strength tensor. It is given as

$$F_{\mu\nu} = \partial_\mu A_\nu - \partial_\nu A_\mu$$

where $A_\mu$ is the photon field. The external propagators for a photon are $\epsilon_\mu$ and $\epsilon^*_\mu$ for an incoming and an outgoing particle respectively, where the $\epsilon_\mu$ is the photon's polarization vector. An internal propagator for a photon is given as

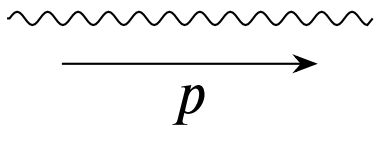

$$-\frac{i\eta_{\mu\nu}}{p^2}$$

where $\eta_{\mu\nu}$ is the metric tensor for the Minkowski space.

The interaction vertex, which couples an incoming fermion, an outgoing fermion and a photon is given as

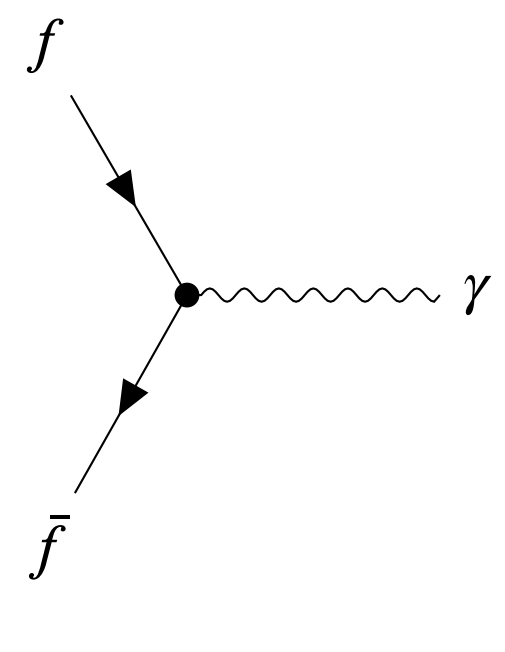

$$-ieQ_f\gamma^\mu$$

where $Q_f$ is the electric charge of the fermion. It is worth noting that the electromagnetic coupling is given as

$$\alpha_{em} = \frac{e^2}{4\pi}.$$

Cross sections are customarily stated in terms of the electromagnetic coupling, $\alpha_{em}$ and not the elementary electric charge, $e$.

With this non-trivial example considered, we now proceed to an even more complicated QFT, that is QCD, which is the foundational theory that forms the bedrock upon which lies the entire research conducted as part of this thesis.

### 1.2.3 Quantum Chromodynamics

QCD is the theory of strong interactions (see chapter 1 of Ref. [4]). Just as QED has its own electric charge, so does QCD, which has its own color charges, namely red, green and blue. A distinct characteristic of QCD, unlike QED, is that in nature and up to about 2 TeraKelvin, one does not find colored particles, but rather colorless bound states that contain colored particles. This will be discussed at length in section 1.3. In this subsection, we proceed as before by taking a look at the Lagrangian density of QCD followed by its Feynman rules.

$$\mathscr{L}_{QCD} = \mathscr{L}_{gluon} + \mathscr{L}_{quark} + \mathscr{L}_{int} \tag{1.7}$$

$$\mathscr{L}_{gluon} = -\frac{1}{4} G^a_{\mu\nu} G^{a\mu\nu} \tag{1.8}$$

$$\mathscr{L}_{quark} = \sum_q \bar{\psi}^i_q \left(i\gamma^\mu \partial_\mu - m_q\right) \psi^i_q \tag{1.9}$$

$$\mathscr{L}_{int} = -g_s \sum_q \bar{\psi}^i_q \gamma^\mu (T^a)_{ij} A^a_\mu \psi^j_q \tag{1.10}$$

Here, the gluon term, $\mathscr{L}_{gluon}$, represents the gluon kinetic term and the gluon self-interaction term. Unlike electroweak vector bosons, vector bosons of QCD can interact amongst themselves, due to the QCD group structure being non-abelian. Gluon self-interactions can happen through a triple gluon vertex or a quartic gluon vertex. $\mathscr{L}_{quark}$ represents the free quark field term. The index $q$ runs over all the quark flavors. The index $i$ (and $j$) are color indices. $\mathscr{L}_{int}$ represents the interaction term where two quarks and one gluon interact.

The $A^a_\mu$ term represents the gluon field, where the index a runs over 1 to 8, corresponding to the 8 generators of the SU(3) group, which is the symmetry group of QCD. $G^a_{\mu\nu}$ is the gluon field strength tensor, given as:

$$G^a_{\mu\nu} = \partial_\mu A^a_\nu - \partial_\nu A^a_\mu + g_s f^{abc} A^b_\mu A^c_\nu,$$

where $f^{abc}$ are the structure constants of the SU(3) group defined as follows:

$$\left[T^a, T^b\right] = i f^{abc} T^c,$$

where $T^a$ are the generators of the SU(3) group and $T^a = \lambda^a/2$ where $\lambda_a$ are the Gell-Mann matrices. The Gell-Mann matrices are an explicit representation of the generators of the SU(3) group. They are given as

$$\lambda^1 = \begin{pmatrix} 0 & 1 & 0 \\ 1 & 0 & 0 \\ 0 & 0 & 0 \end{pmatrix}, \quad \lambda^2 = \begin{pmatrix} 0 & -i & 0 \\ i & 0 & 0 \\ 0 & 0 & 0 \end{pmatrix},$$

$$\lambda^3 = \begin{pmatrix} 1 & 0 & 0 \\ 0 & -1 & 0 \\ 0 & 0 & 0 \end{pmatrix}, \quad \lambda^4 = \begin{pmatrix} 0 & 0 & 1 \\ 0 & 0 & 0 \\ 1 & 0 & 0 \end{pmatrix},$$

$$\lambda^5 = \begin{pmatrix} 0 & 0 & -i \\ 0 & 0 & 0 \\ i & 0 & 0 \end{pmatrix}, \quad \lambda^6 = \begin{pmatrix} 0 & 0 & 0 \\ 0 & 0 & 1 \\ 0 & 1 & 0 \end{pmatrix},$$

$$\lambda^7 = \begin{pmatrix} 0 & 0 & 0 \\ 0 & 0 & -i \\ 0 & i & 0 \end{pmatrix}, \quad \lambda^8 = \frac{1}{\sqrt{3}} \begin{pmatrix} 1 & 0 & 0 \\ 0 & 1 & 0 \\ 0 & 0 & -2 \end{pmatrix}.$$

$g_s$ is the strong coupling constant. The remaining terms in the Lagrangian density such as $\psi$ or $\gamma^\mu$ are analogous to their counterparts in QED.

The Feynman rules for the quarks in QCD are similar to the Feynman rules for the fermions in QED, such that the external propagators are given by $u$ and $\bar{u}$ for an incoming and an outgoing quark respectively, and as $\bar{v}$ and $v$ for an incoming and an outgoing anti-quark respectively. The internal propagator for a quark is given as

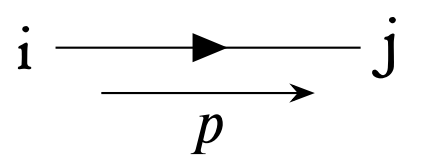

$$\frac{i\delta_{ij}}{\gamma^\mu p_\mu - m_q}.$$

The same holds true for gluons, where an external gluon propagator is given by $\epsilon^a_\mu$ and $\epsilon^{a*}_\mu$ for an incoming and an outgoing gluon respectively, where $\epsilon^a_\mu$ is the gluon's polarization vector. An internal gluon propagator is given as

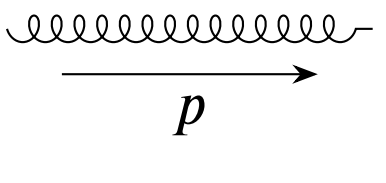

$$\frac{-i\eta_{\mu\nu}\delta^{ab}}{p^2},$$

where $\eta_{\mu\nu}$ is the metric tensor for the Minkowski space.

For the Feynman rules of the interaction vertices, three possible vertices need to be considered: 2 quarks and 1 gluon, 3 gluons and 4 gluons. The 2 quarks and 1 gluon vertex is given as

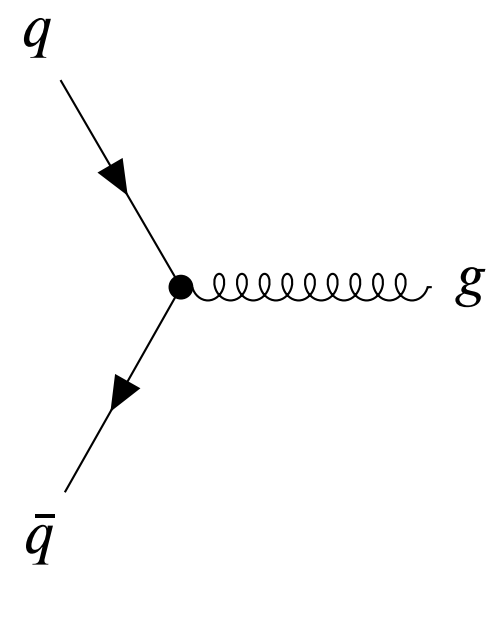

$$-ig_s\gamma^\mu(T^a)_{ij}.$$

The 3 gluon vertex is given as

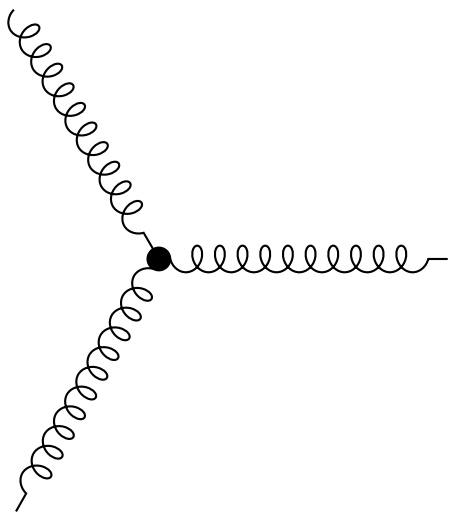

$$-ig_s f^{abc}\begin{bmatrix}\eta^{\mu\nu}(p_1-p_2)^\rho+\\ \eta^{\nu\rho}(p_2-p_3)^\mu+\\ \eta^{\rho\mu}(p_3-p_1)^\nu\end{bmatrix},$$

where $p_i$ are the momenta of the $i^{th}$ gluon. The 4 gluon vertex is given as

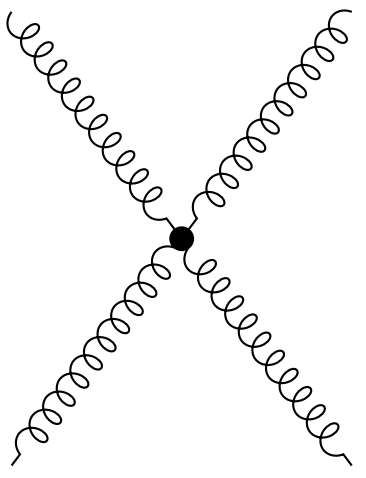

$$-ig_s^2\begin{bmatrix}f^{abe}f^{cde}(\eta^{\mu\rho}\eta^{\nu\sigma}-\eta^{\mu\sigma}\eta^{\nu\rho})+\\ f^{ace}f^{bde}(\eta^{\mu\nu}\eta^{\rho\sigma}-\eta^{\mu\sigma}\eta^{\nu\rho})+\\ f^{ade}f^{bce}(\eta^{\mu\nu}\eta^{\rho\sigma}-\eta^{\mu\rho}\eta^{\nu\sigma})\end{bmatrix}.$$

With this, we have the bare minimum tools that are required for tree-level computations in perturbative QCD.

## 1.3 Couplings and renormalization

### 1.3.1 Divergences

The Feynman rules we looked at in the last section allow us to compute cross sections at the tree level, which means, at the leading order in perturbation theory. This is done by drawing the possible Feynman diagrams for a given interaction, expressing the Feynman diagrams with the Feynman rules of the constituent elements, summing up and squaring the expressions and integrating them over the phase space. To obtain theoretical predictions at an ever increasing precision, one needs to perform calculations at higher and higher orders in order to move towards perturbative convergence. However, as we move to higher orders, we encounter divergences in the calculations. Some of the common divergences encountered include collinear divergences, infrared (IR) divergences and ultraviolet (UV) divergences. Collinear divergences occur when a massless particle is emitted or absorbed by a fermion, parallel to the direction of the fermion's momentum. IR divergences occur when a massless particle's momentum goes to zero. UV divergences occur when the momentum of a loop integral goes to infinity.

In this section, we will only discuss UV divergences, as they require the process of renormalization, to be dealt with. Let us consider, for simplicity, the UV divergent diagrams in QED.

A key rule of scattering amplitude computations is the conservation of momentum at each and every vertex. This means that the sum of all incoming momenta at a vertex should exactly equal the sum of all outgoing momenta at that vertex. Consider the diagram in Fig. 1.1a, where at the first vertex, the incoming momentum is $p$ and the outgoing momentum is $k+(p-k)$, which is $p$. There is a key difference between $p$ and $k$, in that $p$ is an external momentum, which has a physical value as it corresponds to a real particle, whereas $k$ is an internal momentum, corresponding to a virtual particle that can be off-shell. As such, $k$ can take any value. To properly account for loops, one needs to consider all possible field configurations, which requires the integration over all possible momenta. Therefore, the amplitude computations require an integral for every loop momentum as shown below:

$$\int \frac{d^4k}{(2\pi)^4}.$$

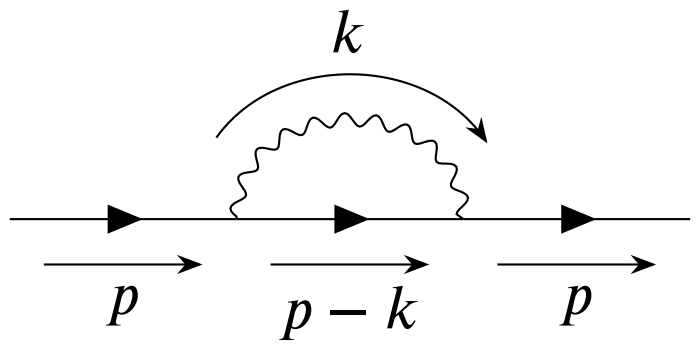

(a) fermion self-energy

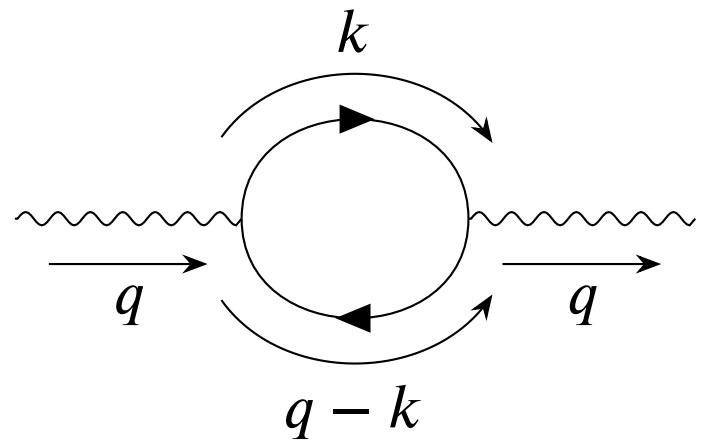

(b) photon self-energy

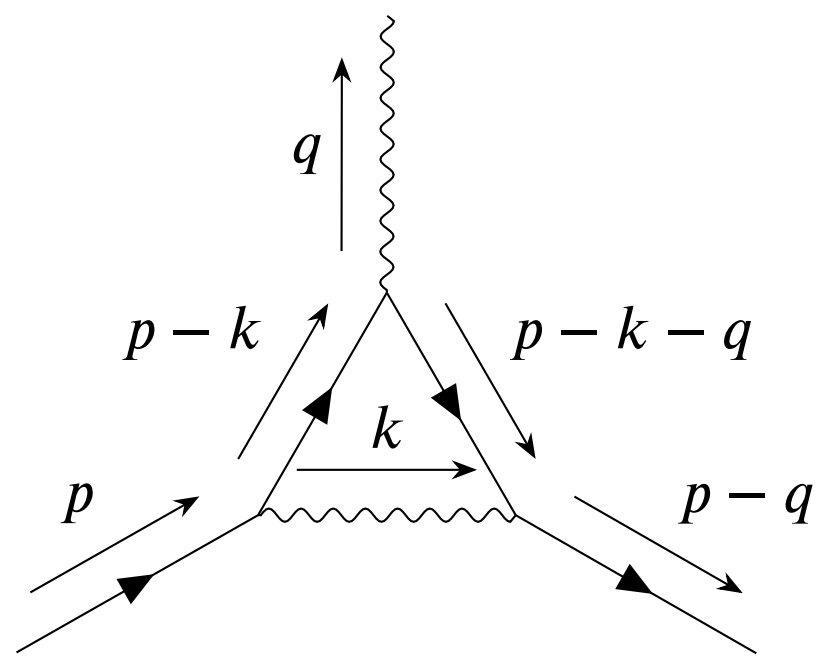

(c) vertex correction

Figure 1.1: One-particle irreducible (1PI) one-loop diagrams in QED

Consider the generic expression:

$$\int \frac{d^4k}{(2\pi)^4} \frac{N(k)}{M(k)}.$$

To determine the UV divergence, one needs to study its behavior as $k \to \infty$. This can be done through power counting. Let's define D = 4 + degree of N(k) - degree of M(k). Then, the following is the behavior of the integral:

$$D > 0 \rightarrow \text{linear or higher divergent}$$
$$D = 0 \rightarrow \text{logarithmically divergent}$$
$$D < 0 \rightarrow \text{convergent}$$

The process of dealing with UV divergences is two-fold. First, one needs to regularize the divergent integrals to make them well defined. Regularization can be achieved by regularization schemes such as dimensional regularization. The second step is to renormalize the theory. This involves redefining the parameters of the theory such that the divergences are absorbed into the redefined pa-rameters. This is done by adding counter-terms to the Lagrangian density. The counter-terms are chosen such that they cancel the divergences. The renormalized parameters are then expressed in terms of the physical parameters. Renormalization can be achieved by renormalization schemes such as the modified minimal subtraction $\left(\overline{\mathrm{MS}}\right)$ scheme. The next section gives a brief overview of this procedure.

### 1.3.2 Regularization & renormalization

While a detailed discussion on regularization and renormalization is beyond the scope of this section, a general sketch of the procedure is provided here. An interested reader is encouraged to refer to chapter 6 and 7 in Ref. [2] for a more thorough discussion. To renormalize a theory, one needs to make a distinction between 'bare' and 'physical' parameters. The parameters that we have looked at so far in the various Lagrangian densities are all bare parameters. They are not directly observed or probed in an experiment. Conventionally, the bare parameters are denoted with a subscript '0', which is what we will do henceforth. The parameters without the subscript will now be used for the physical parameters. Let's consider, for simplicity, the case of QED. We first define the renormalization factors $Z_i$ and their corresponding counter-terms $\delta_i$.

$$
\begin{aligned}
Z_1 &= 1 + \delta_1, \\
Z_2 &= 1 + \delta_2, \\
Z_3 &= 1 + \delta_3, \\
Z_m &= 1 + \delta_m.
\end{aligned}
$$

The field renormalizations are given as:

$$
\begin{aligned}
\psi_0 &= Z_2^{1/2} \psi, \\
A_{\mu,0} &= Z_3^{1/2} A_\mu.
\end{aligned}
$$

The coupling renormalization is given as:

$$
e Z_1 = e_0 Z_2 Z_3^{1/2}.
$$

The mass renormalization is given as:

$$
m = m_0 Z_m.
$$

These redefinitions are inserted into the Lagrangian density, and they allow the Feynman rules to be kept in the same form as before. The Lagrangian density now contains terms with bare parameters and counter-terms. Consider the case of a fermion propagator, based on bare fermion fields. Diagrammatically, it is shown in Fig. 1.2.

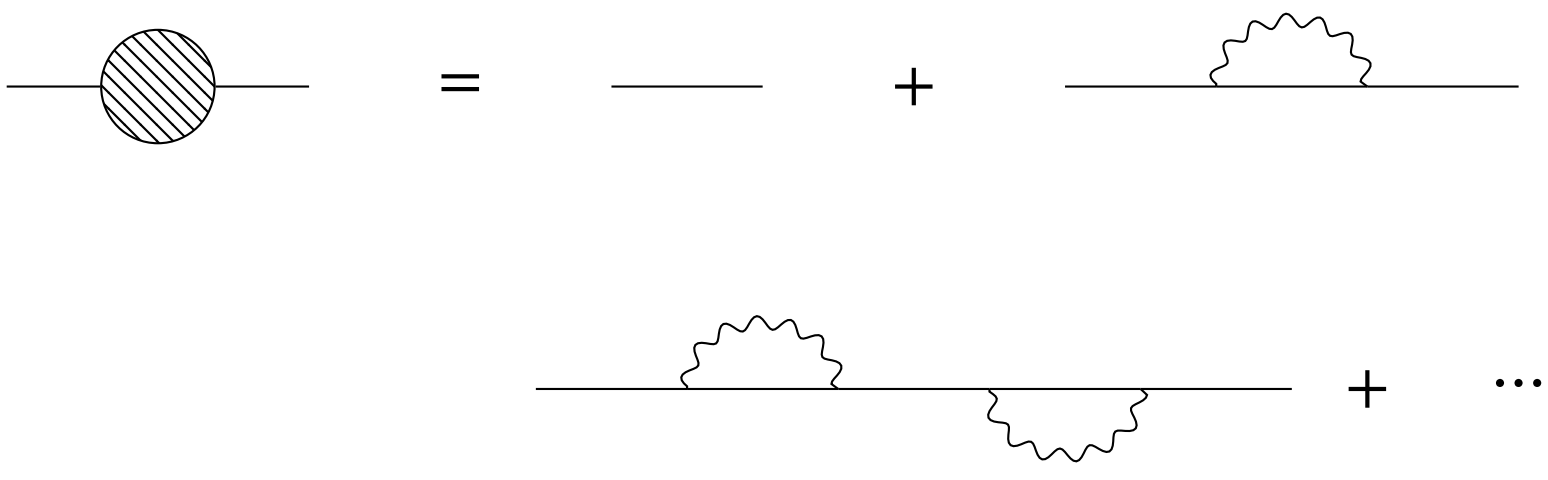

Figure 1.2: Full fermion propagator as a sum of 1PI diagrams

Naively amending the fermion propagator to account for these loops would simply lead to the divergences of Fig. 1.1a. This is solved by the field redefinitions and the counter-terms introduced above. In particular, the introduction of the counter-terms in the Lagrangian density produces another set of Feynman rules, based on the counter terms, that cancel the bare divergences. The QED Feynman rule for the fermion counter-term is given as:

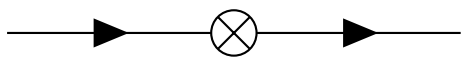

$$-i\left(\delta_2^f \gamma^\mu p_\mu - \delta_m^f m_f\right).$$

The QED Feynman rule for the photon counter-term is given as:

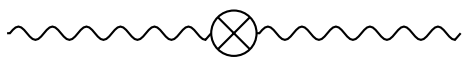

$$-i\delta_3\left(\eta_{\mu\nu} q^2 - q_\mu q_\nu\right).$$

The QED Feynman rule for the interaction counter-term is given as:

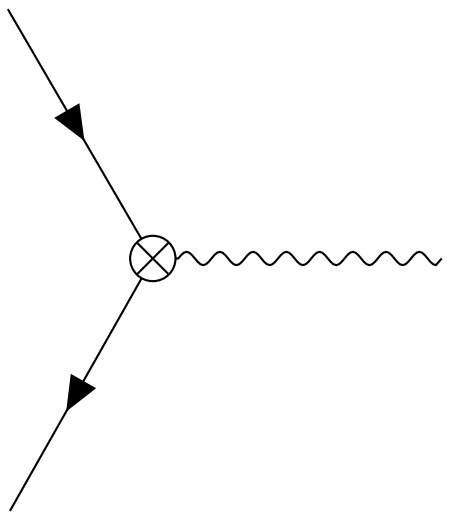

$$-ieQ_f \delta_1^f \gamma^\mu.$$

As is the case for bare terms, which need to be considered at all orders in perturbation theory, the counter-terms also need to be considered at all orders. Diagrammatically, for the fermion propagator, this is shown in Fig. 1.3.

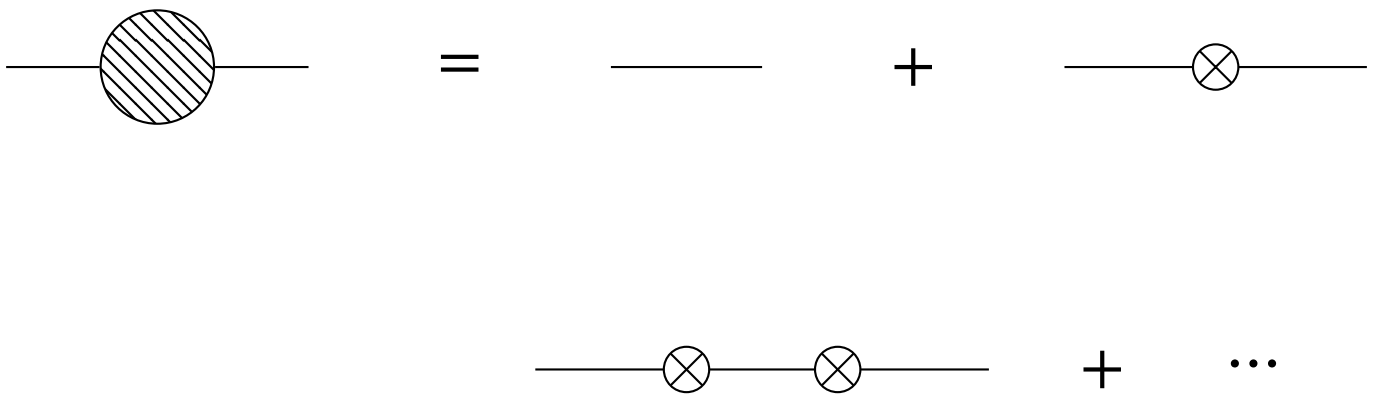

Figure 1.3: Full counter-term fermion propagator summed over all orders

At each order, the counter-terms are chosen such that they cancel the divergences of the bare terms. The specific computation of the counter-terms is done by considering the renormalization conditions, which further depend upon the renormalization scheme chosen. The discussion on renormalization schemes is beyond the scope of this section, however, the key point to note is that this modifies the Feynman rules. The final, renormalized Feynman rules (with fermion indices made implicit, for simplicity) are as follows. The renormalized fermion propagator is given as:

$$S(p) = \frac{i}{\gamma^\mu p_\mu - m - \Sigma(p)},$$

where $\Sigma(p)$ is the fermion self-energy correction. The renormalized photon propagator is given as:

$$D_{\mu\nu}(q) = -\frac{i\eta_{\mu\nu}}{q^2\left(1 - \Pi(q)\right)},$$

where $\Pi(q)$ is the photon self-energy correction. The renormalized interaction vertex is given as:

$$\Gamma^\mu(p, q) = -ieQ_f\left[\gamma^\mu + \Lambda^\mu(p, q)\right],$$

where $\Lambda^\mu(p, q)$ is the vertex correction.

A similar, albeit more complicated, procedure is followed for QCD. The procedure of renormalization ensures the correct treatment of UV divergences, however, it is not sufficient for the computation of the desired result. One still needs to be able to integrate over the Feynman diagrams before the cancellation between the divergences of the bare terms and the counter-terms kicks in. To be able to perform this integration, one needs to regularize the integral, which is to make it manageable and well defined, until renormalization is performed. A common regularization scheme is the cut-off regularization, where a cut-off $\Lambda$ is introduced, such that the integrals are evaluated up to the cut-off, which acts as the upper limit of the integral. This introduces a new energy scale in the theory, $\Lambda$. Another common regularization scheme is dimensional regularization, where the number of space-time dimensions is taken to be $D = 4 - \epsilon$. This also leads to an introduction of a new energy scale in the theory, $\mu^\epsilon$. The important point to note is that regularization imposes a new energy scale in the theory, which is not physical. Of course, a non-physical scale should not affect the physical results, and this is indeed the case, if one performs the computation at all orders in perturbation theory. However, in reality, the computations are truncated to a finite order, and this invariably leads to a dependence on this scale, which is called the renormalization scale. As one moves to higher orders, the dependence on the renormalization scale decreases. To faithfully account for the dependence on the renormalization scale, one needs to take into account an uncertainty associated to this scale in the final result. This will be touched upon in Sec. 2.3.1.

### 1.3.3 Running of couplings

For simplicity, the running of the couplings will be sketched out in the context of QED, and the results of QCD will be stated thereafter. Consider the loop diagrams in Fig. 1.1 at one loop. In an interaction vertex, these loops could appear in four distinct places. There could be a photon self-energy loop in the photon propagator, a vertex correction loop in the interaction vertex, and a fermion self-energy loop in either of the two fermion propagators. The vertex correction loop in the interaction vertex exactly cancels the fermion self-energy loop of the two fermion propagators. This is known as the Ward identity. Hence, it is only the photon self-energy loop that affects the interaction vertex, and hence the coupling. The photon self-energy loop, or the vacuum polarization, modifies the interaction in a manner which also depends on $q^2$, the square of the physical energy scale of the process. However, the renormalized coupling also has a dependence on the renormalization scale, $\mu_r$. The running of the coupling is described by the RGE. Given its importance, let us look at the origin of the RGE (in the context of QED for simplicity) before stating it. Starting with the relation between the bare and renormalized coupling together with dimensional regularization, one gets:

$$e_0 = \mu_r^{\epsilon/2} Z_3^{1/2}(\mu_r)\, e(\mu_r).$$

Note that the bare coupling is independent of the 'renormalization scale' $\mu_r$. With this, the differentiation of the bare coupling with respect to $\mu_r$ yields:

$$\mu_r \frac{d}{d\mu_r} e_0 = \mu_r \frac{d}{d\mu_r} \left( \mu_r^{\epsilon/2} Z_3^{1/2}(\mu_r) e(\mu_r) \right) = 0.$$

After some algebraic manipulation, this simplifies to:

$$\mu_r \frac{de}{d\mu_r} = -\epsilon e - \frac{1}{2} e \cdot \frac{\mu_r}{Z_3} \frac{dZ_3}{d\mu_r}.$$

Using $\alpha = e^2/4\pi$, the expression becomes:

$$\mu_r \frac{d\alpha}{d\mu_r} = -2\alpha\epsilon - \alpha \cdot \frac{\mu_r}{Z_3} \frac{dZ_3}{d\mu_r}.$$

In the physical limit where $\epsilon \to 0$, the Beta Function is given as:

$$\beta(\alpha) = -\alpha \cdot \frac{\mu_r}{Z_3} \frac{dZ_3}{d\mu_r},$$

which leads to the RGE:

$$\mu_r \frac{d\alpha}{d\mu_r} = \beta(\alpha) \tag{1.11}$$

$Z_3$ can be computed order by order using vacuum polarization diagrams and hence the beta function can also be computed order by order in perturbation theory, where the beta function, $\beta(\alpha)$, is given as:

$$\beta(\alpha) = \sum_{n=0}^{\infty} \beta_n \alpha^{n+2}. \tag{1.12}$$

The one-loop beta function for QED is given as:

$$\beta(\alpha) = \frac{2\alpha^2}{3\pi} \tag{1.13}$$

and hence the solution to the RGE at one-loop is given as:

$$\alpha(\mu_r) = \frac{\alpha(\mu)}{1 - \frac{2\alpha(\mu)}{3\pi} \ln\left(\frac{\mu_r}{\mu}\right)}. \tag{1.14}$$

Based on this, given the value of the coupling at a certain energy scale, one can compute the value of the coupling at any other energy scale. To determine the value of the coupling at a certain energy scale, one needs to measure it experimentally and at that point, the physical energy scale $q^2$ is generally taken as the initial energy scale. A standard way of quoting the value of the coupling is to quote it at the $Z$ boson mass scale, $M_Z$. A key point to note is that $\beta_n$ is a scheme-dependent quantity, that needs to be computed at each order in perturbation theory. Invariably, the beta function has a dependence on the renormalization scheme and the renormalization scale.

With this discussion on the running of the coupling in QED, we now proceed to the running of the coupling in QCD. The beta function for QCD is given as:

$$\beta(\alpha_s) = -\sum_{n=0}^{\infty} \beta_n \left(\frac{\alpha_s}{4\pi}\right)^{n+2}. \tag{1.15}$$

The one-loop beta function for QCD is given as:

$$\beta(\alpha_s) = \frac{11}{3} C_A - \frac{4}{3} T_R n_f, \tag{1.16}$$

where $C_A = N_c = 3, T_R = 1/2$ and $n_f$ is the number of active quark flavors. One can similarly solve the RGE for QCD, and determine the value of the coupling at any energy scale given its value at an initial energy scale.

An interesting distinction between the QED and QCD beta functions is the sign of the one-loop beta coefficient. The sign of the beta coefficient has an important physical implication. A positive beta coefficient implies that the coupling increases as the energy scale increases, while a negative beta coefficient implies that the coupling decreases as the energy scale increases. Given the negative beta coefficient of QCD, the coupling decreases as the energy scale increases. This leads to two distinct phases for QCD colored particles, confinement and asymptotic freedom. The energy scale $\Lambda_{\text{QCD}}$ separates these two phases. At energies below $\Lambda_{\text{QCD}}$, the coupling is so large that quarks and gluons form bound states, and perturbation theory, that relies on the smallness of the coupling, breaks down. This is the confinement phase. At energies above $\Lambda_{\text{QCD}}$, the coupling is small, and quarks and gluons can be treated as free particles. This is the asymptotic freedom phase. In this phase, the coupling is small enough to allow for perturbative convergence and QCD computations can be performed in perturbation theory.

## 1.4 QCD Collinear Factorization

The nature of QCD presents some fundamental challenges in the computation of theoretical predictions. In a collider experiment, the initial state hadrons are always color singlet bound states, due to the confinement of quarks and gluons. However, the high energy scattering allows for the constituent quarks and gluons to interact with one another. While we can compute the scattering amplitudes of the interactions amongst the constituent quarks and gluons with other fermions and bosons, we are not able to properly characterize the dynamics of the constituent partons within the hadrons, due to the breakdown of perturbation theory at low energy scales. It is here where the collinear factorization theorem plays a crucial role. It allows for a systematic treatment to separate the short distance and perturbative dynamics from the long distance and non-perturbative dynamics. In this section, an overview of this theorem, collinear factorization, is provided.

To see collinear factorization in action, consider the scattering of an electron and a proton, where the electron is relativistic. These scattering processes can fall in three energy regimes: elastic scattering, inelastic scattering and deep inelastic scattering. In elastic scattering, the electron scatters off the proton as a whole, and this process allows for the determination of the proton's form factor, which parametrizes the spatial distribution of the proton's charge and current. In inelastic scattering, the electron scatters off the proton, with a high enough energy to push the proton into an excited state, such as the $\Delta$ baryon, which is a baryon resonance that decays back into a nucleon and a pion. In deep inelastic scattering, the electron scatters off the proton with a high enough energy to resolve the constituent quarks and gluons inside the proton. This causes the proton to break apart. It is this process, which is of interest to understand the QCD collinear factorization at work.

### 1.4.1 Deep inelastic scattering

The process of deep inelastic scattering involves the scattering of a lepton off a hadron. It can be denoted as:

$$l\left(k_\mu\right) + H\left(p_\mu\right) \rightarrow l'\left(k'_\mu\right) + X\left(p_\mu + q_\mu\right)$$

where $l$ is the incoming lepton with momentum $k_\mu$, $H$ is the incoming hadron with momentum $p_\mu$, $l'$ is the outgoing lepton with momentum $k'_\mu$, $X$ is the final state hadronic system and $q_\mu = k_\mu - k'_\mu$ is the momentum transfer by the vector boson. The intermediate vector boson can be either a photon, $Z$ boson or a $W$ boson. Neutral current interactions involve the exchange of a photon or a $Z$ boson, while charged current interactions involve the exchange of a $W$ boson. Neutral current interactions at an energy scale $Q^2 > m_Z^2$ can involve the exchange of either a photon or a $Z$ boson and lead to an electroweak interaction while neutral current interactions at an energy scale $Q^2 < m_Z^2$ have suppressed $Z$ boson interactions and thus can be approximated with the exchange of only a photon and lead to an electromagnetic interaction. The nature of these interactions is a bit different in case of neutrinos which do not couple to the photon, and hence the discussion here is geared towards charged leptons like electrons. For simplicity, we will focus on electromagnetic interactions where an electron scatters off a proton, shown in Fig. 1.4.

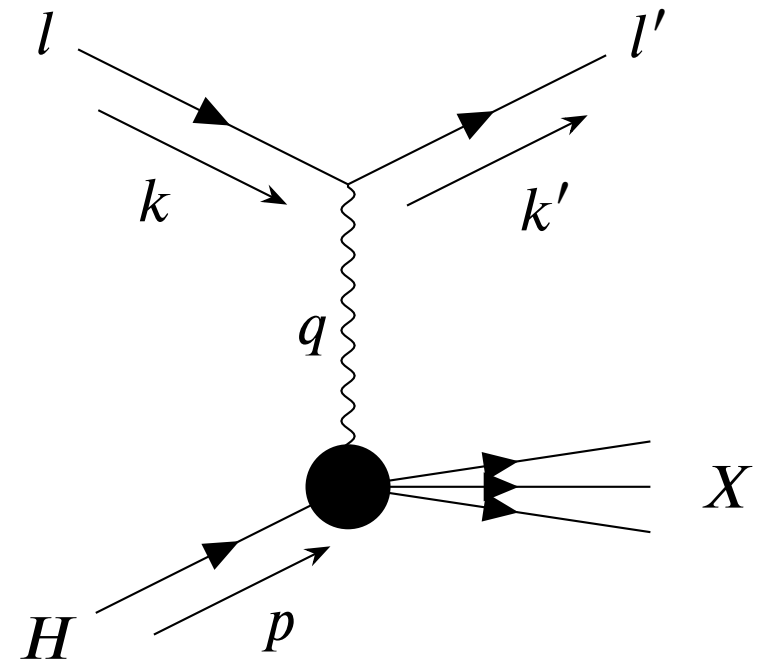

Figure 1.4: Deep inelastic scattering: $l + H \to l' + X$, where $X$ denotes inclusive hadronic final state.

To study the process of deep inelastic scattering, it is useful to define a few standard kinematical variables:

$$\begin{aligned} s &= (p+k)^2, \\ W^2 &= (p+q)^2, \\ Q^2 &= -q^2, \\ x &= \frac{Q^2}{2p\cdot q}, \\ y &= \frac{p\cdot q}{p\cdot k} = \frac{Q^2}{xs}, \end{aligned} \tag{1.17}$$

where $s$ is the center of mass energy squared, $W^2$ is the invariant mass squared of the final state hadronic system, $Q^2$ is the momentum transfer squared, $x$ is the Bjorken scaling variable and $y$ is the inelasticity variable. The amplitude for the process is given as:

$$|\mathscr{M}|^2 \sim L_{\mu\nu} W^{\mu\nu}, \tag{1.18}$$

where $L_{\mu\nu}$ is the leptonic tensor and $W^{\mu\nu}$ is the hadronic tensor. The leptonic tensor is given as:

$$L_{\mu\nu} = k^\mu k'^\nu + k'^\mu k^\nu - \eta_{\mu\nu} k\cdot k' \tag{1.19}$$

The hadronic tensor is given as:

$$W^{\mu\nu} = -\left(\eta^{\mu\nu} - \frac{q^\mu q^\nu}{q^2}\right) F_1(x,Q^2) + \frac{1}{p\cdot q}\left(p^\mu - \frac{p\cdot q}{q^2} q^\mu\right)\left(p^\nu - \frac{p\cdot q}{q^2} q^\nu\right) F_2(x,Q^2) \tag{1.20}$$

where $F_1(x,Q^2)$ and $F_2(x,Q^2)$ are the structure functions. There also exists a third structure function, $F_3(x,Q^2)$, which is only relevant for electroweak interactions, when $Z$ or $W^\pm$ bosons are ex-changed. Since this discussion is limited to electromagnetic interactions, $F_3(x, Q^2)$ is not relevant here. The structure functions $F_1$ and $F_2$ can be recast as longitudinal and transverse structure functions, $F_L$ and $F_T$, respectively.

$$\begin{aligned} F_L &= F_2 - 2xF_1 \\ F_T &= 2F_1 \end{aligned} \tag{1.21}$$

With this, a differential cross section for DIS can be written as:

$$\frac{d^2\sigma}{dxdQ^2} = \frac{2\pi\alpha_{em}^2}{Q^4}\left((1+(1-y)^2)F_T(x,Q^2) + \frac{2(1-y)}{x}F_L(x,Q^2)\right) \tag{1.22}$$

where $\alpha_{em} = \frac{e^2}{4\pi}$.

It is at this stage where collinear factorization comes into play. While a review of the mathematical construction of factorization is beyond the scope of this section, and an interested reader is encouraged to refer to Ref. [5] for details, the key results for the purposes of our discussion are as follows. The collinear factorization theorem allows for the structure functions to be decomposed as follows:

$$F_i(x,Q^2) = \sum_p f_p\left(x,\mu_f^2\right) \otimes C_p^i\left(x,Q^2,\mu_f^2,\mu_r^2\right),\ i = T, L \tag{1.23}$$

where $f_p$ are the PDFs and $C_p^i$ are the coefficient functions. Index $p$ runs over the parton species and $\mu_f$ and $\mu_r$ are the factorization scale and the renormalization scale respectively. A consequence of the collinear factorization procedure is the introduction of a factorization scale, $\mu_f$. The $\otimes$ symbol denotes a convolution, which is defined as:

$$(f \otimes g)(x) = \int_x^1 \frac{dy}{y} f\left(\frac{x}{y}\right) g(y). \tag{1.24}$$

The PDFs are universal functions that describe the distribution of partons inside a hadron. The coefficient functions are process-dependent functions that encode the hard scattering dynamics of the process, and are calculated order by order in perturbation theory as a series expansion in $\alpha_s$ as follows:

$$C_p^i = C_p^{i,(0)} + \frac{\alpha_s(\mu_r^2)}{4\pi}C_p^{i,(1)} + \left(\frac{\alpha_s(\mu_r^2)}{4\pi}\right)^2 C_p^{i,(2)} + \mathcal{O}\left(\alpha_s^3\right) \tag{1.25}$$

The structure functions $F_i$ are measured experimentally, while the coefficient functions $C_p^i$ are computed theoretically, and with these two objects, one can perform a global fit to extract the PDFs $f_p$, as will be discussed in 2. An individual structure function generally constrains a specific combination of PDF(s), and therefore one needs multiple structure functions that constrain different combinations of PDFs to achieve proper flavor separation. The PDFs have to be explicitly extracted from the data as there is no direct way to compute them from first principles [1], however, once they are extracted at

[1]In principle, PDFs can be computed from first principles using lattice QCD, however, these determinations are

any given scale, they can be evolved to any other scale using the DGLAP evolution equations.

### 1.4.2 DGLAP evolution

The process of collinear factorization introduces the factorization scale, $\mu_f$, which is a scale upon which the PDFs depend. Just as is the case for strong and electromagnetic couplings, the energy scale dependence of the PDFs can be described by a set of evolution equations, known as the DGLAP evolution equations. The origin and structure of DGLAP evolution equations is discussed in [6-8]. They are a set of $2n_f + 1$ coupled integro-differential equations, where $n_f$ is the number of active quark flavors. They are given as:

$$\frac{d}{d\ln\mu_f^2} f_p(x,\mu_f^2) = \sum_{p'} \int_x^1 \frac{dy}{y} P_{p'p}(y)\, f_{p'}(\frac{x}{y},\mu_f^2), \tag{1.26}$$

where $P_{p'p}$ are the DGLAP splitting functions, which describe the probability of a parton of type $p$ splitting into a parton of type $p'$ and a parton of type $p$. The splitting functions are universal, and they are computed order by order in perturbation theory as a series expansion in $\alpha_s$ as follows:

$$P_{p'p} = \frac{\alpha_s(\mu_r^2)}{4\pi} P^{(0)}_{p'p} + \left(\frac{\alpha_s(\mu_r^2)}{4\pi}\right)^2 P^{(1)}_{p'p} + \left(\frac{\alpha_s(\mu_r^2)}{4\pi}\right)^3 P^{(2)}_{p'p} + \mathcal{O}\left(\alpha_s^4\right). \tag{1.27}$$

Given that these equations are coupled, it helps to rotate from the flavor basis to an 'evolution basis', where the equations are maximally decoupled. This basis allows for $2n_f - 1$ independent evolution equations and a system of 2 coupled equations. The independent evolution equations are given as:

$$\frac{d}{d\ln\mu_f^2} f_{NS;\pm,v}(x,\mu_f^2) = P_{NS;\pm,v} \otimes f_{NS;\pm,v}(x,\mu_f^2) \tag{1.28}$$

where the non-singlet PDF combinations are given as $f_{NS;\pm} = \left(f_{q_i} \pm f_{\bar{q}_i}\right) - \left(f_{q_j} \pm f_{\bar{q}_j}\right)$ and $f_{NS;v} = \sum_{i=1}^{n_f}\left(f_{q_i} - f_{\bar{q}_i}\right)$. The coupled equations are given as:

$$\frac{d}{d\ln\mu_f^2}\begin{pmatrix} f_\Sigma(x,\mu_f^2) \\ f_g(x,\mu_f^2)\end{pmatrix} = \begin{pmatrix} P_{qq} & P_{qg} \\ P_{gq} & P_{gg}\end{pmatrix} \otimes \begin{pmatrix} f_\Sigma(x,\mu_f^2) \\ f_g(x,\mu_f^2)\end{pmatrix} \tag{1.29}$$

where $f_\Sigma = \sum_{i=1}^{n_f} f_{q_i} + f_{\bar{q}_i}$ is the singlet PDF and $f_g$ is the gluon PDF. Some factors of $n_f$ are present in the splitting functions' matrix, which have been absorbed in the definition of the splitting functions, in Eq. (1.29). The evolution of PDFs is a crucial feature in the determination of the PDFs as will be discussed in Sec. 2.2.3.

known to have very high uncertainties (around 50%) and therefore are unsuitable for any precision phenomenology program.

### 1.4.3 Factorizable processes

It has been discussed above how QCD factorization allows for the separation of cross sections into a convolution of perturbative and non-perturbative objects. In this section, the processes which benefit from this factorization are listed. However, before proceeding, it is important to introduce FFs, which are analogous to the PDFs. While PDFs are space-like objects that describe the distribution of finding specified partons with a given momentum fraction $x$ inside a hadron, FFs are time-like objects that describe the distribution of specified partons fragmenting into a hadron where the hadron carries a fraction $z$ of the parton's momentum. We denote the FFs as $D_h^p(z, \mu_f^2)$, where $h$ is the final state hadron and $p$ is the parton that fragments into the hadron. With this, we can now list the processes that benefit from QCD collinear factorization.

Processes with one non-perturbative object:

- Deep inelastic scattering (DIS): $l + H \rightarrow l' + X$ where $H$ requires a PDF to be fully specified. It is given as: $f_p \otimes C_p$.
- Single inclusive annihilation (SIA): $e^+ + e^- \rightarrow h + X$ where $h$ requires a FF to be fully specified. It is given as: $C_p \otimes D_p$.

Processes with two non-perturbative objects:

- Sufficiently inclusive hadron-collider (such as LHC) processes (excluding inclusive hadron production): $H_1 + H_2 \rightarrow X$ where $H_1$ and $H_2$ require PDFs to be fully specified. It is given as: $f1_{p1} \otimes f2_{p2} \otimes \hat{\sigma}_{p1,p2}$.
- Semi-inclusive deep inelastic scattering (SIDIS): $l + H \rightarrow l' + h + X$ where $H$ requires a PDF and $h$ requires a FF to be fully specified. It is given as: $f_{p1} \otimes C_{p1,p2} \otimes D_{p2}$.

Processes with three non-perturbative objects:

- Single inclusive hadron production in hadronic collisions: $H_1 + H_2 \rightarrow h + X$ where $H_1$ and $H_2$ require PDFs and $h$ requires a FF to be fully specified. It is given as: $f1_{p1} \otimes f2_{p2} \otimes C_{p1,p2,p3} \otimes D_{p3}$.

These processes widely cover the breadth of processes used in the determination of collinear PDFs and FFs as discussed in this thesis.

### 1.4.4 Mellin space

The collinear factorization procedure relies on the convolution operator. Given the definition of the convolution operator in Eq. (1.24), and its use in evolution equations, which contain differential operators, one often lands in a situation where one needs to solve integro-differential equations. This is a complicated task, and hence, it can be useful to transform the equations from $x$ space to Mellin space. The transformation to Mellin space is governed by the Mellin transform, which is defined as:

$$\tilde{f}(N) = \int_0^1 dx\, x^{N-1} f(x), \tag{1.30}$$

where a function with a $\sim$ on top of it denotes a function in Mellin space and $N$ is the Mellin variable. The inverse Mellin transform is given as:

$$f(x) = \frac{1}{2\pi i}\int_{c-i\infty}^{c+i\infty} dN\, x^{-N}\tilde{f}(N), \tag{1.31}$$

where $c$ is a constant that is chosen such that the contour of integration is to the right of all singularities of $\tilde{f}(N)$. The Mellin transform is useful in QCD collinear factorization as it allows for the convolution operator to be transformed into a simple product operator. To see this, consider equations (1.24) and (1.30). Let us begin with taking a Mellin transform of a convolution:

$$\widetilde{(f\otimes g)}(N) = \int_0^1 dx\, x^{N-1}(f\otimes g)(x) \tag{1.32}$$

$$= \int_0^1 dx\, x^{N-1}\left[\int_x^1 \frac{dy}{y} f\left(\frac{x}{y}\right) g(y)\right] \tag{1.33}$$

$$= \int_0^1 dy\, \frac{g(y)}{y}\left[\int_0^y dx\, x^{N-1} f\left(\frac{x}{y}\right)\right] \tag{1.34}$$

Making the substitution $z = \frac{x}{y} \Rightarrow x = yz,\ dx = y\,dz$, we obtain:

$$\int_0^y dx\, x^{N-1} f\left(\frac{x}{y}\right) = \int_0^1 y\,dz\,(yz)^{N-1} f(z) \tag{1.35}$$

$$= y^N \int_0^1 dz\, z^{N-1} f(z) \tag{1.36}$$

$$= y^N \tilde{f}(N) \tag{1.37}$$

Substituting back, we find:

$$\widetilde{(f\otimes g)}(N) = \tilde{f}(N)\int_0^1 dy\, y^{N-1} g(y) \tag{1.38}$$

$$= \tilde{f}(N)\tilde{g}(N) \tag{1.39}$$

This property is particularly useful in the case of integro-differential equations which become ordinary differential equations when they undergo Mellin transformation. It is therefore widely used in computational software focused on DGLAP evolution or coefficient functions' computation.

# Chapter 2

# Crash course on the NNPDF approach

## 2.1 Introduction

In Sec. 1.4, it was shown that cross sections involving hadronic bound states can be expressed as a convolution of hard partonic cross sections and PDFs (or FFs), using QCD factorization. In this chapter, the discussion will be focused on understanding the fundamental aspects of the process of performing a fit that allows for the extraction of PDFs (or FFs). The discussion will be based on the methodology, techniques and computational framework of the NNPDF collaboration, which uses neural networks to parametrize PDFs.

The problem of PDF determination is an inverse problem, i.e. a problem where the goal is to use experimental observations to infer the underlying causal factors that produced them. This is, as it is the PDFs and the hard partonic cross sections that, together, lead to the observed cross sections (in addition to sources of theoretical and experimental errors), and our task, is to use the observed cross sections to infer and extract the PDFs.

## 2.2 Ingredients for a PDF fit

There are a few key ingredients that need to be prepared before a PDF or an FF fit can be performed. The first is all the experimental data, including the central values and the uncertainties, that needs to be put in a consistent and a unified format. The second includes the theoretical computations of the hard partonic cross sections associated to all the corresponding experimental data, again, in a proper format which will allow its use in an efficient manner in the fitting process. The third ingredient has to do with the scale dependence of the PDFs. Every data point in an experimental dataset is measured at a specific energy scale. A PDF fit is performed at an initial scale, and therefore the PDFs need to be evolved to the scale of the data points before they can be convolved with the hard partonic cross sections to produce the theoretical predictions, thereby allowing a comparison with the experimental data. In this section, all these concepts will be discussed in some detail.

### 2.2.1 Experimental data and its implementation

Experimental data that goes into a PDF fits comes from many different experiments, conducted over many years, and many different processes. Consequently, the different datasets are released in different formats. There has been some progress to standardize the delivery of the data in recent years with a lot of LHC data being released on HEPData [9], however, even then, there are significant differences such that it is not possible to directly use the files obtained from HEPData in a PDF fit. This is where an important step comes into play, where the experimental data is processed and implemented in a consistent manner. Up to the NNPDF4.0 family of PDF sets, the NNPDF collaboration had been implementing data in its framework using C++ code and stored the data in specific data files. However, recently, it has transitioned to a new data implementation standard where data is implemented using Python and stored in YAML format. A part of the PhD work in this thesis was also devoted to the designing and enhancing of the new data format. This whole redesign offers a lot of flexibility that is desirable for the whole process. One example of its usage is that at times, experimental datasets provide a few variants of uncertainties or correlations. The new format allows for dataset variants that can be used with their tag, thus allowing an efficient method to perform comparisons between the different variants. There are many such use cases that now benefit from these improvements. The actual data implementation process will not be discussed here, and an interested reader is encouraged to refer to the documentation of the NNPDF code [10].

The focus of this section will be on how the experimental datasets are made ready to be used in the fitting procedure. All cross section distributions are provided as a number of data points, or bins. Each bin is characterized by numerical ranges of the kinematical variables, and has an associated central value and a set of statistical and systematic uncertainties. The numerical ranges of the kinematical variables are the most straightforward parts of the data implementation, as they are almost always implemented as is.

The central values of the cross sections are also straightforward, as long as the uncertainties associated to the measurement are symmetric. However, if the uncertainties are asymmetric, they need to be symmetrized before they can be used. Sometimes, the experimental papers provide the symmetrization prescription, but more often than not, it is not provided. In such cases, we rely on d'Agostini's symmetrization procedure [11]. Consider a data point given as: $V^{\Delta_+}_{\Delta_-}$, where $V$ is the central value, $\Delta_-$ is the left uncertainty and $\Delta_+$ is the right uncertainty. Note that $\Delta_\pm$ contain the sign and the value, i.e. it is not just the numerical part of the uncertainty. The symmetrization procedure requires that the central value is shifted to the average of the lower and upper uncertainties, such that:

$$V_{\text{shifted}} = V + \frac{\Delta_+ + \Delta_-}{2} \qquad (2.1)$$

and the symmetrized uncertainty is defined as:

$$\Delta_{\text{sym}} = \sqrt{\left(\frac{\Delta_+ - \Delta_-}{2}\right)^2 + 2\left(\frac{\Delta_+ + \Delta_-}{2}\right)^2} \tag{2.2}$$

Besides this, there is generally not much to be done with the central values.

The uncertainties, are generally the part that require the most care. As will be discussed in Sec. 2.3, the fitting procedure requires the optimization of a $\chi^2$ function, which in its definition, contains the covariance matrix of the uncertainties. The covariance matrix is an $N_{\text{dat}} \times N_{\text{dat}}$ matrix, where $N_{\text{dat}}$ is the number of data points, and it is defined as follows:

$$C_{ij} = \sum_{n=1}^{\text{N, uncorr.}} \sigma_i^n \sigma_j^n \delta_{ij} + \sum_{n=1}^{\text{M, corr., add.}} A_{n,ij} \sigma_i^n \sigma_j^n + \sum_{n=1}^{\text{L, corr., mult.}} B_{n,ij} \sigma_i^n \sigma_j^n \tag{2.3}$$

where there are three types of uncertainties: uncorrelated, additive correlated and multiplicative correlated. $A_{n,ij}$ and $B_{n,ij}$ are correlations between the uncertainties of $i^{\text{th}}$ and $j^{\text{th}}$ data point. The distinction between additive and multiplicative uncertainties is needed as the multiplicative uncertainties lead to d'Agostini bias, as discussed in Sec. 2.3.2. If the uncertainties are provided as a list of uncorrelated, additive correlated and multiplicative correlated uncertainties, they are implemented as is, but the lack of correlations leads to treating all the correlated uncertainties as being $100\%$ correlated. On the other hand, sometimes, the uncertainties are provided in the form of a covariance matrix, or a list of uncertainties and a correlation matrix. In the latter case, the uncertainties and the correlation matrix is combined to form the covariance matrix:

$$(\text{cov})_{ij} = \sigma_i \sigma_j (\text{corr})_{ij}$$

Once the covariance matrix is available, it needs to be decomposed into 'artificial uncertainties' which are treated as additive correlated uncertainties. The reason for this is that the NNPDF framework expects a list of uncertainties (and their type) for every distribution, and it generates the covariance matrix internally. The following is the algorithm used to decompose the covariance matrix into a list of uncertainties.

Given a covariance matrix $\mathrm{C} \in \mathbb{R}^{n \times n}$ for $n$ data points, the algorithm computes an artificial uncertainty matrix $\mathrm{A} \in \mathbb{R}^{n \times n}$ as follows:

1. Reconstruction: The covariance matrix C is constructed from a flattened list of its elements.

2. Eigen-decomposition: Compute eigenvalues $\{\lambda_j\}_{j=1}^n$ and eigenvectors $\{\mathrm{v}_j\}_{j=1}^n$ of C:

$$\mathrm{C}\mathrm{v}_j = \lambda_j \mathrm{v}_j.$$

3. Positive-semidefinite check: Verify that all eigenvalues satisfy

$$\lambda_j \geq \epsilon,$$

where $\epsilon \approx -10^{-10}$ accounts for numerical precision. For covariance matrices corresponding to $m$ number of normalized observables, up to $m$ eigenvalues may be allowed, to be approximately zero without raising an error.

4. Construction of artificial uncertainty matrix:

$$A_{ij} = \sum_{\substack{k=1 \\ \lambda_k > 0}}^{n} v_{ik} \sqrt{\lambda_k}\, \delta_{jk},$$

where $\delta_{jk}$ is the Kronecker delta.

Here, the $i$-th row of A corresponds to the artificial uncertainties associated with the $i$-th data point, while the columns relate to the contributions from each eigenmode of the covariance matrix. The matrix A thus encodes the square root decomposition of the positive part of C, providing a basis for incorporating artificial uncertainties into the dataset.

These intricacies cover majority of the cases one might encounter when implementing experimental data in the NNPDF framework. With experimental data implemented, one needs to look at the next ingredient, which is the hard partonic cross sections.

### 2.2.2 Partonic cross sections and interpolation grids

As is discussed in detail in Sec. 2.3, the fitting procedure optimizes a $\chi^2$ function through gradient descent, and the definition of the $\chi^2$ function requires the theoretical predictions, which involve a convolution of the PDFs with the hard partonic cross sections. The convolution requires the computation of an integral over the partonic cross sections multiplied by the PDFs, see Eq. (1.24). The computation of this integral is a very computationally expensive task, and this combined with the fact that the $\chi^2$ optimization procedure requires the convolution to take place repeatedly, means it is simply not possible to do so, from a practical perspective. To overcome this, the hard partonic cross sections are computed and stored in an interpolation grid. This reduces the convolution integral to a multi-dimensional array multiplication, making it a few orders of magnitude faster.

The idea of interpolation concerns functions which are provided at a finite set of points. Consider a function $f$, for which a definition is not available, but rather a set of points $\{(x_i, f_i)\}_{i=1}^{N}$ is given, where $f_i = f(x_i)$. The goal of interpolation is to construct a function $g$ such that

$$g(x_i) = f_i \quad \forall i = 1, \ldots, N.$$

A well defined interpolation function $g$ can be used to compute the value of $f$ at any point $x$ in the domain of $f$, i.e. $g(x) = f(x)$. This approach is naturally suited for PDFs and FFs, which are often provided at a finite set of points, e.g. through LHAPDF grids. The possibility of using interpolation grids for PDFs has existed for quite some time, through APPLgrid [12] and fastNLO [13, 14], and more recently, through PineAPPL [15], which is a library developed and currently used by NNPDF. There exist a number of interpolation techniques, and the choice used in PineAPPL is that of Lagrange interpolation.

Consider a set of points $\{x_0, x_1, \dots, x_n\}$, where $x_a \neq x_b$ for all $a \neq b$. This set consists of the interpolation nodes. A polynomial basis $P_m(x)$ can then be constructed where the degree of the polynomial, $m \leq n$. The $j^{\text{th}}$ polynomial is defined as:

$$P_j(x) = \prod_{\substack{0 \leq m \leq n \\ m \neq j}} \frac{x - x_m}{x_j - x_m} \tag{2.4}$$

The Lagrange interpolating polynomial for the nodes through their corresponding values $\{f_0, f_1, \dots, f_n\}$ is then given as a linear combination of the polynomials:

$$L(x) = \sum_{j=0}^{n} f_j P_j(x) \tag{2.5}$$

To see how this works, consider the example of DIS where a structure function is defined as:

$$F_l = \sum_p \int_x^1 \frac{d\hat{x}}{\hat{x}} f_p\left(\frac{x}{\hat{x}}, Q\right) C_p^l(\hat{x}, Q) \tag{2.6}$$

which can be decomposed into the Lagrange interpolating polynomial basis as:

$$F_l = \sum_p \sum_{j=0}^{n} f_j^p(Q) \cdot \int_x^1 \frac{d\hat{x}}{\hat{x}} P_j\left(\frac{x}{\hat{x}}\right) C_p^l(\hat{x}, Q) \tag{2.7}$$

where $f_j^p(Q)$ is the value of the PDF $f_p$ at the interpolation node $j$ and at the energy scale $Q$. The set of computed values of

$$\int_x^1 \frac{d\hat{x}}{\hat{x}} P_j\left(\frac{x}{\hat{x}}\right) C_p^l(\hat{x}, Q)$$

for all $j$ forms an array, which is called a subgrid. The subgrid's dimensionality is based on the number of convolutions that need to be performed, so it is a 1D subgrid above, but would be a 2D subgrid for LHC processes. Once the subgrids are computed, the convolution integral is replaced by a simple array multiplication between the subgrids and the PDF values at the interpolation nodes, as shown below:

$$F_l = \sum_p \sum_{j=0}^{n} f_j^p(Q) \cdot \int_x^1 \frac{d\hat{x}}{\hat{x}} P_j\left(\frac{x}{\hat{x}}\right) C_p^l(\hat{x}, Q) = \sum_p \sum_{j=0}^{n} f_j^p(Q) \cdot S_{j,p}^l(Q) \tag{2.8}$$

where $S^l_{j,p}(Q)$ is the value of the subgrid at the interpolation node $j$ and parton $p$.

In the above example, consider the summation over the partons $p$. There needs to be a separate subgrid for each parton, or more generally, a separate subgrid for each partonic channel. Also consider that the coefficient functions $C^l_p$ define a series of functions, for each perturbative order, and hence there needs to be a separate subgrid for each perturbative order as well. Finally, any experimental data contains a number of bins, each of which corresponds to a different $x$ value, and hence there needs to be a separate subgrid for each bin as well. An interpolation grid is then simply a multi Cartesian product of bins, orders and channels, i.e. it is a 3-dimensional array of subgrids.

With these interpolation grids encoding the hard partonic cross sections, it is time to look at the next ingredient, which is the evolution kernel operators and their integration into the interpolation grids.

### 2.2.3 Evolution operators and fast interpolation grids

In the last section, it was shown how interpolation grids can be used to efficiently convolve PDFs with hard partonic cross sections. However, it was also evident that an interpolation grid involves a lot of subgrids, thus there are a lot of array multiplications that need to be performed to compute the convolution. While the interpolation grids are a common technique used by many researchers in the field, the NNPDF collaboration uses a specialized version of an interpolation grid, called an FK table (fast-kernel table). In this section, the concept and the benefits of an FK table will be discussed.

The PDFs are parametrized at a parametrization scale $Q_0$, whereas each experimental data point is measured at a different scale $Q$. To be able to consistently perform the convolution, it is then necessary to evolve the PDFs from the parametrization scale $Q_0$ to the scale $Q$ for each data point. This is done using the DGLAP evolution equations, as was discussed in Sec. 1.4.2. The DGLAP evolution equations can be cast as evolution kernel operators (EKOs), which are tensors that evolve the PDFs from one scale to another. An EKO for a dataset can be produced using a software by the same name, EKO [16]. A detailed mathematical discussion is beyond the scope of this section, and an interested reader is encouraged to refer to the relevant literature [16, 17]. For the purposes of this section, it is sufficient to know that an EKO object, which is a tensor can be contracted with the interpolation grids (or rather the subgrids inside the interpolation grids) to produce new interpolation grids that can be directly convolved with the PDFs at the parametrization scale $Q_0$ to produce the theoretical predictions at the scale $Q$. The nature of these new interpolation grids is closely related to the discussion on FK tables, but before the discussion on FK tables, it is important to have a discussion on perturbative orders.

As was discussed in Sec. 2.2.2, an interpolation grid is a 3-dimensional array of subgrids, where one of the dimensions is 'orders'. The orders in a PineAPPL subgrid are given as:

$$\alpha_s^k \alpha_{em}^l log^m\left(\frac{\mu_r^2}{Q^2}\right) log^n\left(\frac{\mu_f^2}{Q^2}\right)$$

where each combination of $k$, $l$, $m$ and $n$ corresponds to a different specific order. The separation of the couplings' perturbative orders is straightforward to understand, in that, the matrix elements of the hard partonic cross sections need to be multiplied by the coupling raised to the power consistent with the perturbative order of the matrix element. The separation of the logarithms that depend on the renormalization and factorization scales, by their powers, has to do with the dependence of the PDFs on the factorization scale, and the dependence of the couplings (and hence the matrix elements) on the renormalization scale. In a calculation that was performed at all orders in perturbation theory, the dependence on the renormalization and factorization scales is removed, however, in a calculation that is performed at a finite order, these scales lead to some scale dependence. This dependence leads to the introduction of theoretical uncertainties, and to take into account these uncertainties, one can vary the renormalization and factorization scales, that allows for the computation of a theory covariance matrix, which can be used in the process of PDF fitting, to improve the quality of the fit. This is further discussed in Sec. 2.3.1.

While a PDF fit is being performed by itself, i.e. no other SM parameters are being fitted, the other SM parameters are fixed. This means that in a regular PDF fit, the value of the couplings is fixed. This allows the possibility to factor out the PDFs from the cross sections, where by the partonic cross sections at each order are multiplied by the couplings raised to the power consistent with the perturbative order, and summed together. This allows for a simplification in the structure of the interpolation grids, where the dimension of the orders is reduced to a single point, at which all orders are combined. This is one of the key features of the FK tables, in that, the subgrids in an FK table are not separated by orders, but rather, the subgrids from the same bin and channel, at different orders, are multiplied by their respective couplings, and scale dependent logarithms, and summed together to form a single subgrid. An FK table has a fixed value of a coupling and depends on a specific choice of the renormalization and factorization scales. The variation of renormalization and factorization scales happens by generating multiple FK tables, each corresponding to a different choice of the renormalization and factorization scales. Another key feature of the FK tables is that it does not contain the simple subgrids, obtained by the convolution integral of partonic cross sections with Lagrange polynomials, but rather, these subgrids are also contracted with the EKO tensors. As was discussed in Sec. 1.4.2, the computation of the DGLAP evolution equations is simpler to perform in the evolution basis, as opposed to the flavor basis, and hence, the EKO tensors are computed in the evolution basis. To allow for the convolution of the interpolation grids with the EKOs, the channel dimension of the grid has to be transformed from the flavor basis to the evolution basis, by taking the appropriate linear combinations of the subgrids. Hence, the final FK table is a special type of interpolation grid, where the perturbative orders of couplings and scale dependent logarithms have been summed together, the channels have been transformed into the evolution basis, and the subgrids have been contracted with the EKOs. The parametrization of the PDFs also happens in the evolution basis, and hence, the FK tables provide a very efficient way to compute the theoretical predictions during the PDF fitting procedure.

## 2.3 Fitting a PDF

In the previous sections, it was shown how the experimental data and the FK tables (which combine the hard partonic cross sections and the DGLAP evolution operators) are prepared for the PDF fitting procedure. In this section, the actual process of fitting a PDF will be described.

### 2.3.1 Key concepts

The process of fitting a PDF is a regression problem, achieved by optimization, where the goal is to optimize a $\chi^2$ function, which is defined as:

$$\chi^2 = \frac{1}{N_{\text{dat}}} \sum_{ij} (T - E)_i \, C_{ij}^{-1} \, (T - E)_j \tag{2.9}$$

where $T$ is the theoretical prediction vector, $E$ is the experimental data vector, $C_{ij}$ is the covariance matrix of the uncertainties, and $N_{\text{dat}}$ is the number of data points. However, the determination of the PDFs is not simply about extracting a PDF function, but rather, about extracting a PDF function with its uncertainties, i.e. the experimental (and theoretical) uncertainties need to be properly propagated through the fitting procedure. The method used by the NNPDF collaboration to propagate the uncertainties is the Monte Carlo sampling method. It proceeds as follows:

1. An ensemble of pseudo-data is generated by considering the experimental data and their uncertainties. The uncertainties are assumed to be Gaussian, and the pseudo-data is generated by sampling from a Gaussian distribution with the uncorrelated uncertainties shifting each data point independently, and the correlated uncertainties shifting all data points together.

2. Each set of pseudo-data undergoes the fitting procedure, where a $\chi^2$ optimization is performed to extract a replica from the space of parameters of the PDFs.

3. A large number of replicas are generated, that are assumed to be independent, and follow a Gaussian distribution. The optimal number of replicas is determined by requiring that the difference between the moments (central value, variance, etc.) computed over the sampled pseudo-data and the corresponding moments of the original experimental data is less than a few percent. See Ref. [18] for details.

4. The final PDF is then obtained by taking the mean of the replicas, and the uncertainties are obtained by taking the standard deviation of the replicas.

The uncertainties go into the PDF fitting procedure through the covariance matrix, which governs the Monte Carlo sampling that generates the pseudo-data and which is also used in the definition of the $\chi^2$ function. The uncertainties are not restricted to those associated to the experimental data, but also include the theoretical uncertainties, which can originate from many sources, including the truncation of the computation of the couplings, partonic cross sections and the DGLAP evolution equations at a finite order. These uncertainties can be taken into account by varying the scales upon which the specified perturbative series depends, i.e. the renormalization and factorization scales. The variation is carried out by multiplying the scales by a factor of 2 and 1/2, and the differences between the theoretical predictions at these scales and the central value of the theoretical predictions is then used to compute a theory covariance matrix. A mathematical outlook into the theory covariance matrix formalism is beyond the scope of this thesis, but an interested reader is encouraged to refer to the relevant literature in Ref. [19-21]. An important point to note is that the experimental and theoretical covariance matrices are combined to form a total covariance matrix, which is then used in the computation of the $\chi^2$ function, and to obtain the uncertainties by which the experimental data is fluctuated to generate the pseudo-data.

### 2.3.2 The process

The NNPDF collaboration uses neural networks to parametrize the PDFs, which is shown in Fig. 2.1. This means a PDF $f_k(x, Q)$ assumes the following form:

$$x f_k(x, Q) = A_k x^{\alpha_k}(1-x)^{\beta_k}\mathrm{NN}_k(x, \theta),\ k = 1, \ldots, 8 \tag{2.10}$$

where $A_k$ is a normalization constant determined by momentum and valence sum rules for the PDF, $\alpha_k$ and $\beta_k$ are the exponents that control the behavior of the PDF in the low and high $x$ regions, respectively, and $\mathrm{NN}_k(x, \theta)$ is a neural network, where $\theta$ is the set of parameters of the neural network.

The 8 output nodes correspond to the 8 PDFs being parametrized, in the evolution basis, which are:

- $g$: $g$
- $\Sigma$: $\sum_i^{N_f} q_i^+$
- $V$: $\sum_i^{N_f} q_i^-$
- $V_3$: $u^- - d^-$
- $V_8$: $u^- + d^- - 2s^-$
- $T_3$: $u^+ - d^+$
- $T_8$: $u^+ + d^+ - 2s^+$
- $T_{15}$: $u^+ + d^+ + s^+ - 3c^+$

where $q_i^{\pm} = q_i \pm \bar{q}_i$. The choice of using evolution basis over flavor basis to parametrize the PDFs is related to the decoupling of PDFs in the evolution basis, as discussed in Sec. 1.4.2. The reason that

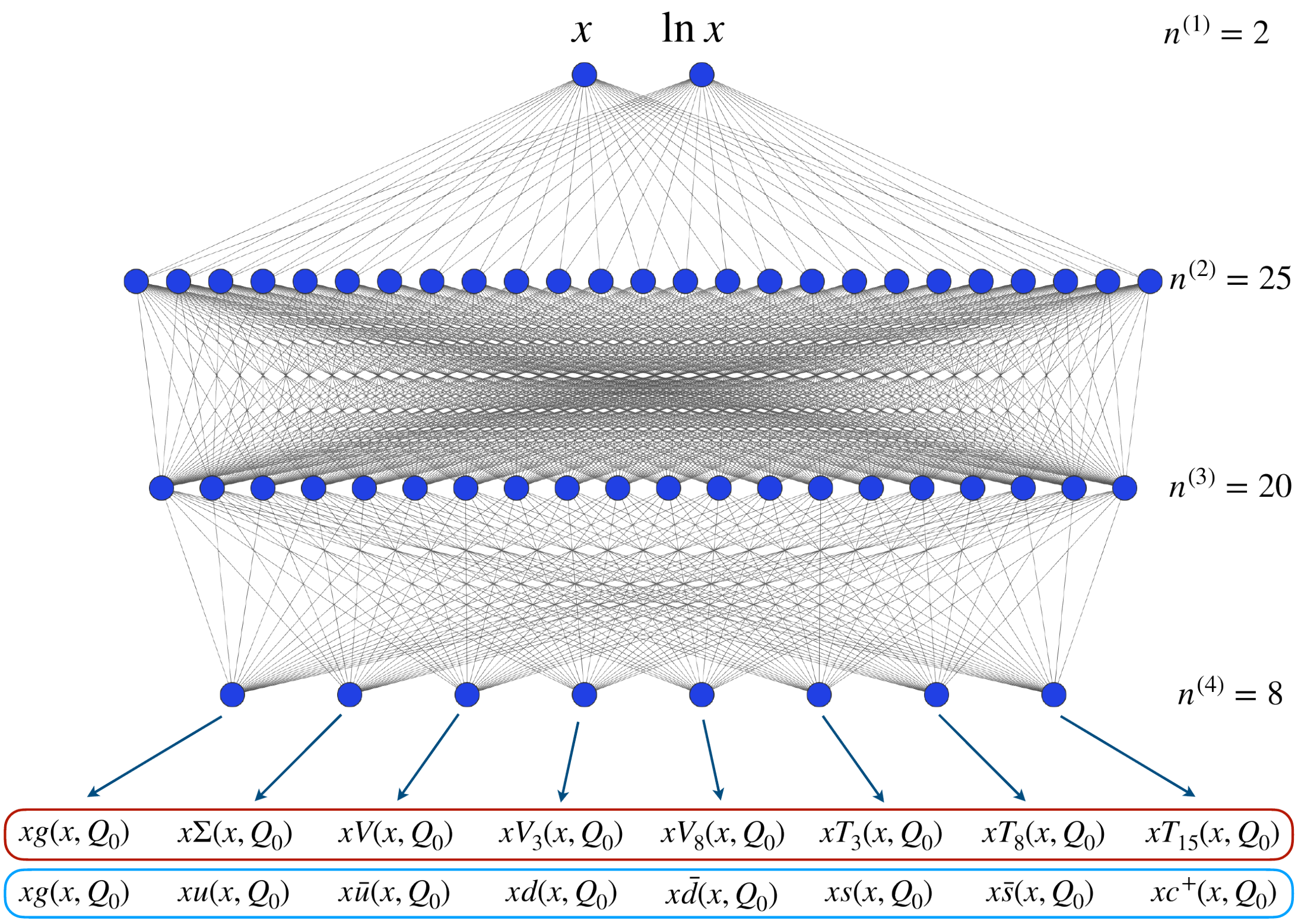

Figure 2.1: This figure shows the architecture of the neural network used to parametrize the PDFs. The output layer has 8 nodes, corresponding to the 8 PDFs in the evolution basis. Ref. [22].

8 PDFs are sufficient in the evolution basis is that the aim is to fit 8 parton distributions in the flavor basis as well, which are: $g$, $u$, $\bar{u}$, $d$, $\bar{d}$, $s$, $\bar{s}$ and $c = \bar{c}$.

To be able to parametrize the PDFs using a neural network, a crucial step is to determine the hyperparameters of the neural network, such as the number of layers, the number of nodes in each layer, the activation function, the learning rate, etc. This is done by performing hyperparameter optimization [23]. Once the structure of the neural network is determined, the fitting procedure can be performed, whereby each replica is obtained by a particular set of pseudo-data, three quarters of which are used to train the neural network, and the remaining one quarter is used to validate the neural network. However, before the training can be performed, a number of parameters need to be defined, such as the preprocessing exponents, i.e. the values of $\alpha_k$ and $\beta_k$ in (2.10), or the $T0$ covariance matrix, The $T0$ covariance matrix is needed as the multiplicative uncertainties lead to a systematic downward shift of the PDFs, known as the d'Agostini bias [24, 25], and therefore, a redefinition of the covariance matrix is needed to account for this bias. A detailed explanation on the $T0$ covariance matrix can be found in [24]. For both the preprocessing exponents and the $T0$ covariance matrix, some preliminary values are chosen, often based on previous fits, and once the fitting procedure is performed, the values obtained can be used to perform an iterated fit with the newer values. This is typically done until the values for both stabilize.

The neural network itself is defined as a feed-forward neural network [26], but with added Lagrange multipliers to the loss function definition, which act as penalty terms to ensure that the PDFs satisfy positivity and integrability constraints. An interested reader is encouraged to refer to the rel-evant sections on imposing positivity and integrability in Ref. [22].

Once the required number of replicas is generated, the central PDF is obtained by taking the mean of the replicas, and the $1\sigma$ uncertainties are obtained by taking the standard deviation of the replicas, assuming that the replicas are independent and follow a Gaussian distribution. A schematically similar procedure is followed to obtain polarized PDFs or FFs as well.

# Chapter 3

# Advancing PDFs

## 3.1 Introduction

A fully accurate determination of PDFs is a hard and a non-trivial inverse problem. This is because PDFs are infinite dimensional objects that need to be extracted from a finite set of experimental data. The experimental data is also subject to statistical and systematic uncertainties which need to be carefully propagated into the PDFs. Furthermore, the theoretical predictions pertaining to the partonic cross sections that go into the determination are truncated and therefore accurate to a given perturbative order, and this introduces theoretical uncertainties due to the missing higher orders (or incomplete higher orders). To this end, PDF determinations have to be continuously and iteratively improved.

And indeed in recent years, there has been major progress in determining the proton's PDFs [27–30], driven by three main factors: a larger input dataset, especially due to high-precision measurements from the LHC; better accuracy in theoretical calculations, now reaching aN$^3$LO in the strong coupling; and deeper study of methodological issues, particularly in how PDF uncertainties are estimated. Some groups [22, 31–33] regularly update their PDF results using broad datasets, while others focus on smaller, more specific data sets [34–36]. These different PDF sets often disagree, sometimes more than their stated uncertainties would suggest—and those uncertainties can vary widely between sets. To explore why these differences happen, many benchmark studies have been carried out over the years [37–52]. One such study [51] showed that when three different methods [32, 33, 53] were used on the same data and theory inputs, the central values of the PDFs were similar, but the uncertainties were somewhat different. These differences likely come from the use of different methods.

These differences make it harder to interpret what the PDF determinations tell us about the proton's internal structure. This issue is especially clear in the ongoing debate over whether the proton contains intrinsic charm quarks [54–56]. They also reduce the sensitivity of key LHC studies that rely on PDFs. This includes both the measurement of the fundamental SM parameters such as the strong coupling $\alpha_s(m_Z)$, the $W$-boson mass $m_W$, and the effective leptonic mixing angle $\sin^2\theta^{\ell}_{\text{eff}}$, and searches for new physics, whether through direct signals (like resonances) or indirect effects (like those described by EFTs). The first issue is seen, for example, in the strong dependence on PDFs of the high-mass forward-backward asymmetry in Drell-Yan gauge boson production [57-59]. The second is shown by how PDFs interact with possible EFT effects in measurements of high-$p_T$ top-quark pair and Drell-Yan production cross sections [60-66].

Recent LHC analyses have made it clear that the current situation is far from ideal. Consider the following three examples. First, the ATLAS determination of the strong coupling $\alpha_s(m_Z)$ using neutral-current Drell-Yan differential measurements in the transverse momentum of the $Z$ boson [67]. This is the most precise determination of $\alpha_s$ ever done by a single experiment, with a quoted uncertainty of $\delta = 9 \cdot 10^{-4}$. Of this, the uncertainty due to the PDF is estimated to be the largest part, $\delta_{\text{pdf}} = 5 \cdot 10^{-4}$, based on the MSHT20 aN$^3$LO fit [68]. However, if one defines the PDF uncertainty as the difference between the central predictions obtained using the CT18A [32] and NNPDF4.0 [22] sets, the resulting uncertainty is four times larger, $\delta_{\text{pdf}} = 2 \times 10^{-3}$. Second, the CMS measurement of the effective leptonic mixing angle $\sin^2 \theta^{\ell}_{\text{eff}}$ [69]. In this case, the PDF uncertainty is estimated as $\delta_{\text{pdf}} = 0.14\%$ using the CT18Z set [32], but the spread in central values between CT18 [32] and MSHT [33] is about five times larger, $\delta_{\text{pdf}} = 0.7\%$. Third, the updated ATLAS measurement of the $W$ boson mass at 7 TeV [70]. Here, the estimated PDF uncertainty is $\delta_{\text{pdf}} = 7.7\ (14.6)$ MeV in the transverse momentum of the lepton $p_T^{\ell}$ (transverse mass $m_T$) channels. But the difference between NNPDF4.0 and MSHT20 leads to a value that is roughly twice as large, $\delta_{\text{pdf}} = 17\ (21)$ MeV. A similar issue arises in the precise $m_W$ measurement carried out by the CMS collaboration [71]. Each of these analyses uses a different baseline PDF set. Using an alternative PDF choice can result in central values that lie outside the quoted PDF uncertainty range.

Given the importance of the PDFs and the impact they have on precision phenomenology as discussed, it is imperative to have a sufficient grasp over the methodological aspects and the choices made to perform a PDF determination. As the NNPDF collaboration moves towards its next PDF determination, a large part of my PhD has been to study the crucial inclusions, updates and changes made, that will distinguish the next determination from the current one. In Sec. 3.2, I will present on a study that aims to assess the differences that might arise when one moves from the use of K-factors to the inclusion of exact NNLO corrections in a PDF determination. This section is based on the work I presented in [72]. In Sec. 3.3, I will present on a study that aims to quantify the generalization power of the most widely used PDF sets by confronting them with new data, that was not included in their respective fitting procedures. This section is a reproduction of the work we presented in [73]. In Sec. 3.4, I will present on a study that assesses the impact of including new data from gluon-sensitive processes to understand its impact on the existing fits. This will be followed by a concise summary and outlook.

## 3.2 K-factors vs exact NNLO corrections

At the time of writing of this thesis, there is a large dedicated effort towards aN3LO PDF determinations [68, 74, 75]. However, it represents the state-of-the-art of the field. The most used PDF sets in the context of LHC phenomenology are currently at the NNLO accuracy. These include, but are not limited to, NNPDF4.0 [22], CT18 [32] and MSHT20 [33].

However, the NNLO PDF sets may not necessarily be based on the exact NNLO theoretical corrections, but rather on K-factor approximations. This is a result of the limited availability of tools for theoretical predictions. In Sec. 2.2.2 the use of interpolation grids to perform convolutions in for PDF determinations was discussed. The need for interpolation grids is the starting point of the limitation. While an MC event generator that can perform the full NNLO calculations for a given process may exist, it might not be interfaced to an interpolation grid library, or it might not even be publicly available, making it impossible to interface it with an interpolation grid library, unless the authors of the code decide to do so. In such cases, the only available pieces of information may be the theoretical predictions of cross sections at NLO and NNLO, obtained by convolving them with a particular PDF set, as provided by the code authors. The way one proceeds in this case is by computing the K-factors per datapoint which are defined as follows:

$$k = \frac{\hat{\sigma}_{\text{NNLO}} \otimes \text{PDF}_{\text{NNLO}}}{\hat{\sigma}_{\text{NLO}} \otimes \text{PDF}_{\text{NNLO}}} \tag{3.1}$$

where $\hat{\sigma}$ is the partonic cross section. Each K-factor is then applied as a multiplicative correction to the NLO cross section. This leads to NNLO predictions that are approximations, as the K-factor is a blanket correction, applied to all the partonic channels with the same value. There is however, no a priori reason to expect that all the partonic channels will receive the exact same multiplicative factor correction at the next order. In fact, they generally do not. Furthermore, there is an intrinsic dependence of the K-factor on the PDF set used. The use of K-factors is therefore a potential area where one can improve upon, by moving towards exact NNLO corrections while performing the PDF determination. The move away from K-factors to exact NNLO corrections should allow for an improvement in the accuracy and the precision of the PDF.

As we move towards the use of exact NNLO corrections in future PDF determinations, we need to be able to disentangle the impact of moving from K-factors to exact NNLO corrections from other changes in the methodology. As a case study, I consider here the top pair production process. This process has been included in the major PDF determinations by the means of NNLO K-factors. However, it is now possible to produce interpolation grids for the top pair production process at exact NNLO accuracy due to the interfacing of MATRIX [76-84] with PineAPPL [15, 85].

To perform this comparison, the first step is to produce the top pair production PineAPPL grids using MATRIX. The grids are computed using a dynamical scale choice of

$$\mu_r = \mu_f = \mu = H_T/4$$

where

$$H_T = \sqrt{m_t^2 + p_{T,t}^2} + \sqrt{m_t^2 + p_{T,\bar{t}}^2}.$$

Here $m_t$ is the mass of the top quark and $p_{T,t(\bar{t})}$ is the transverse momentum of the top quark (anti-quark). This choice is made to maximize perturbative convergence as suggested in [86]. The mass of the top quark is set to $m_t = 172.5$ GeV. The grids are produced with a Monte Carlo precision of 0.1% on the total cross section. This choice was made to strike a balance between precision and computational time. The choice of 0.1% precision on the total cross section leads to a precision that is better than 1% to 2% per individual bin, depending on the size of the individual bin and its kinematical region.

In Table 3.1, the datasets used in the comparison are listed. All of these datasets were included in NNPDF4.0 [22] and NNPDF4.0 with MHOUs [20].

| Dataset | Observable | $N_{dat}$ | Ref. |
|---|---|---|---|
| ATLAS $t\bar{t}$ 7 TeV | $\sigma_{t\bar{t}}$ | 1 | [87] |
| ATLAS $t\bar{t}$ 8 TeV | $\sigma_{t\bar{t}}$ | 1 | [87] |
| ATLAS $t\bar{t}$ 13 TeV | $\sigma_{t\bar{t}}$ | 1 | [88] |
| ATLAS $t\bar{t}$ $2\ell$ 8 TeV | $1/\sigma\, d\sigma/d\lvert y_{t\bar{t}}\rvert$ | 5 | [89] |
| ATLAS $t\bar{t}$ $\ell$+jets 8 TeV | $1/\sigma\, d\sigma/d\lvert y_t\rvert$ | 5 | [90] |
| ATLAS $t\bar{t}$ $\ell$+jets 8 TeV | $1/\sigma\, d\sigma/d\lvert y_{t\bar{t}}\rvert$ | 5 | [90] |
| CMS $t\bar{t}$ 5 TeV | $\sigma_{t\bar{t}}$ | 1 | [91] |
| CMS $t\bar{t}$ 7 TeV | $\sigma_{t\bar{t}}$ | 1 | [92] |
| CMS $t\bar{t}$ 8 TeV | $\sigma_{t\bar{t}}$ | 1 | [92] |
| CMS $t\bar{t}$ 13 TeV | $\sigma_{t\bar{t}}$ | 1 | [93] |
| CMS $t\bar{t}$ $2\ell$ 8 TeV | $1/\sigma\, d^2\sigma/dm_{t\bar{t}}d\lvert y_t\rvert$ | 16 | [94] |
| CMS $t\bar{t}$ $\ell$+jets 8 TeV | $1/\sigma\, d\sigma/dy_{t\bar{t}}$ | 10 | [95] |
| CMS $t\bar{t}$ $2\ell$ 13 TeV | $d\sigma/dy_t$ | 10 | [96] |

Table 3.1: The top pair production experimental datasets from LHC for which we perform the comparisons between their K-factors and exact NNLO corrections. The datasets are listed along with the observable considered and the number of data points in each dataset.

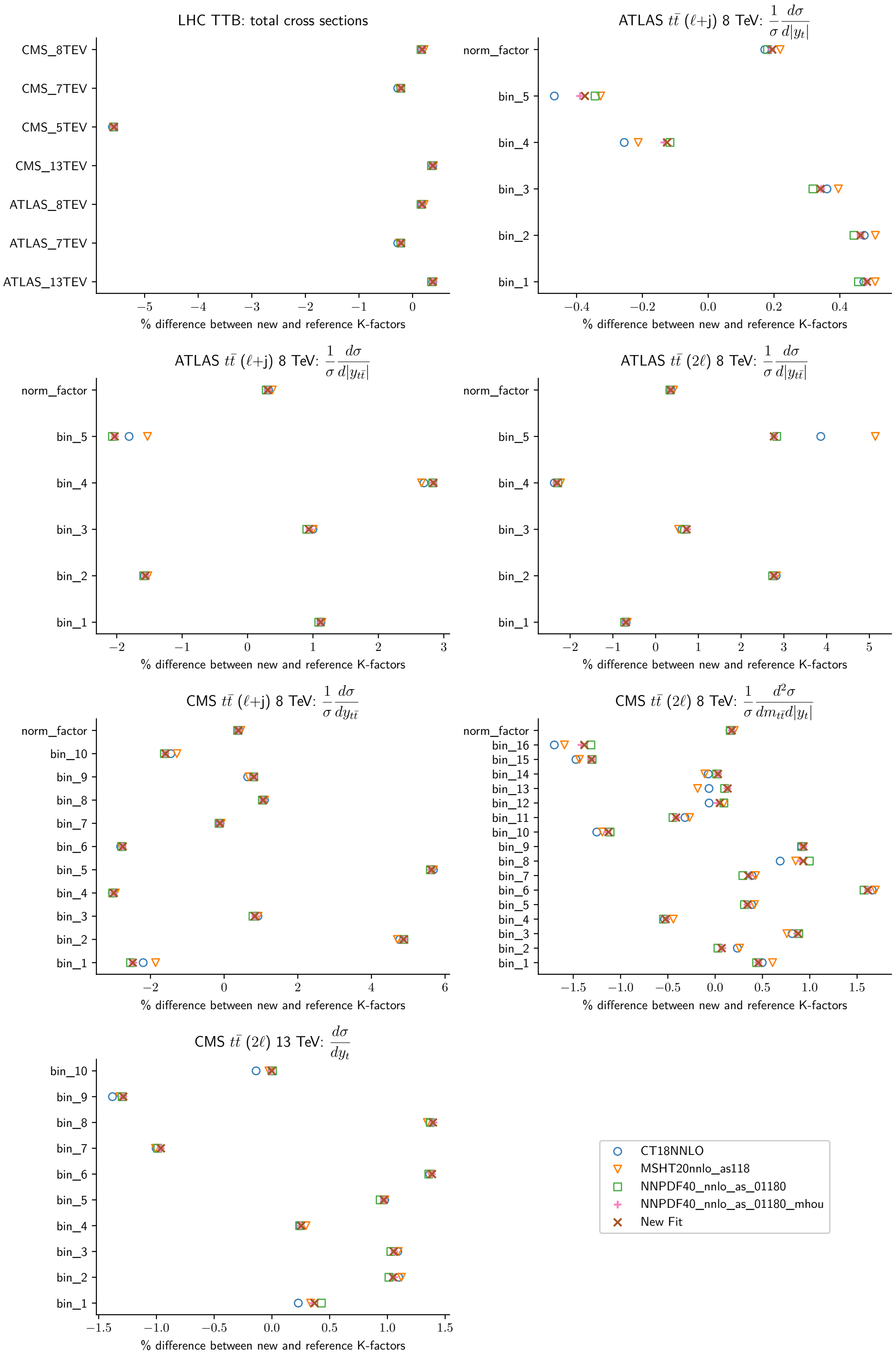

Figure 3.1: This figure shows the percentage difference between K-factors that were used in NNPDF4.0 (the reference K-factors) and K-factors obtained from pure NNLO grids convolved with CT18 [32], MSHT20 [33], NNPDF4.0 [22], NNPDF4.0 with MHOUs [20] and a New Fit (which is otherwise analogous to NNPDF4.0 with MHOUs) where the exact NNLO $t\bar{t}$ grids are used during the fitting procedure.

To perform a systematic comparison, we start with K-factors that were used in the determination of NNPDF4.0 [22] (and NNPDF4.0 with MHOUs [20]), which were obtained using NLO predictions from mg5_aMC [97], and NNLO predictions from publicly available fastNLO tables [98, 99] and top++ [100]. These act as our reference K-factors. We proceed by taking the $t\bar{t}$ PineAPPL grids and convolving them with some select PDF sets, namely CT18 [32], MSHT20 [33], NNPDF4.0 [22], NNPDF4.0 with MHOUs [20]. Performing this convolution at NNLO and NLO allows for the computation of K-factors (using Eq. 3.1), where each set of K-factors depends on the specific PDF set used to compute them. In addition, we perform a new fit where the exact NNLO $t\bar{t}$ grids are used during the fitting procedure (and all else remains same as NNPDF4.0 with MHOUs). We also compute K-factors for this new fit. In Fig. 3.1, the percentage difference between the reference K-factors and the K-factors obtained during this study are shown. The results vary from dataset to dataset with the percentage difference w.r.t. the reference K-factors, going as high as 5-6% for one bin. However, the percentage differences between the K-factors computed using the select PDFs (and the new fit) are extremely small, indicating a consistency between the K-factors obtained using different PDF sets. It should also be noted that these small percentage differences are of statistical origin, as can be seen from their fluctuation between the negative and positive percentages bin by bin, originating from the statistical uncertainties in the MC integration procedure. These can be reduced by requiring high precision from the MC generators, which comes with a computational cost. Nevertheless, this demonstrates that for the large part, K-factors are able to capture the NNLO corrections fairly well, and as we move towards the use of exact NNLO corrections for top pair production (and possibly for other processes as well) in the fitting procedure, it is reasonable to expect the impact on the PDFs to be minimal.

## 3.3 Confronting PDFs with new data

A crucial study before the process of PDF determination is understanding the new data that is planned to be included in the PDF determination. One important part of this study is to see how the existing PDFs (that do not include the new data) perform at predicting the new data. This not only allows us to understand the generalization power of the existing PDF sets, but also allows us to understand how impactful each dataset might be when included in the new PDF determination. In [73], my collaborators and I took the most widely used PDF sets and confronted them with the new data from the run 2 of the LHC for the following processes: single inclusive jet production, di-jet production, top-quark pair production and Drell-Yan gauge boson production, and the new data from HERA for the following processes: single inclusive jet production and di-jet production. This study comprehensively demonstrates the predictive power the existing PDF sets while taking into account experimental uncertainties, MHO uncertainties, PDF uncertainties and $\alpha_s(M_z)$ uncertainties. In this section, I will present the results of this study.

### 3.3.1 Introduction and motivation

In Sec. 3.1, we briefly discussed how the choice of PDFs can considerably impact precision phenomenology, for example, in the determination of SM parameters. These considerations highlight the importance of understanding the origin of the differences observed when computing theoretical predictions with different PDF sets. Complementing existing benchmark studies that tackle this question, here we investigate whether existing PDF sets can be discriminated according to their predictive power of high-precision measurements not included in their determination. We will specifically consider cross sections measured by the LHC run II for Drell-Yan gauge boson, top-quark pair, single inclusive jet and di-jet production, and by HERA for single inclusive jet and di-jet production. We will compare these experimental data to theoretical predictions computed at NNLO accuracy in perturbative QCD and quantitatively assess their mutual agreement. We will take into account all sources of theoretical uncertainty in this assessment, namely PDF, $\alpha_s$, and MHO uncertainties. We will study the dependence of this assessment on the input PDF set. This exercise is an extension of the future test introduced in [101].

The outline of this section is as follows. In Sec. 3.3.2 I present the considered LHC and HERA measurements and the computation of the corresponding theoretical predictions. In Sec. 3.3.3 I describe how we quantitatively assess the agreement between experimental data and theoretical predictions, and in particular how we account for PDF, $\alpha_s$, and MHO uncertainties in this assessment. In Sec. 3.3.4 I present a selection of representative results for each class of measurements, highlighting the relative contribution of the various sources of theoretical uncertainty in the description of the data, and commenting on features that are common to or different from various PDF sets. I summarize our findings in Sec. 3.3.5. Two appendices complement this section. Appendix A quantifies the impact of regularizing ill-conditioned experimental covariance matrices in the assessment of the data-theory comparison. Appendix B collects the complete set of results not shown in Sec. 3.3.4.

### 3.3.2 Experimental data and theoretical predictions

In this section, I present the experimental data considered in this work and the details of the corresponding theoretical computations. The data has been selected according to the following criteria.

- We consider datasets for scattering processes that provide information on PDFs of different partons (quarks, antiquarks, gluon) in a broad kinematic region of $x$ and $Q^2$. For a given process, we select the dataset based on the largest integrated luminosity available.

- We avoid datasets that are already included in PDF determinations used to compute theoretical predictions, to avoid double-counting. The only exception is the recent re-analysis of the $Z$ data at a center-of-mass energy of 8 TeV by ATLAS [67].

- We consider datasets for which the corresponding theoretical predictions can be computed at NNLO in perturbative QCD using event generators interfaced to fast interpolation grids. This avoids reliance on $K$-factors and allows one to readily evaluate predictions upon changes of input PDF set and factorization and renormalization scales.

- We only consider datasets for which the corresponding experimental information is publicly available, in particular through the HEPdata repository [9].

Taking into account these requirements, the ATLAS, CMS, LHCb, H1, and ZEUS datasets that are considered in this study are summarized in Table 3.2, classified by process type. For each dataset we indicate the experiment, the final-state channel, the measured differential distribution(s), the center-of-mass energy, the integrated luminosity, the number of data points (after kinematic cuts), and the corresponding publication reference. For the ATLAS and CMS top-quark pair production and for the CMS single inclusive jet production datasets, we list all the separate distributions provided by the corresponding analyses. In this work, we select a subset of these distributions, which we deem most representative as explained in Sec. 3.3.3.2. In the following, we discuss the main features of these datasets and describe the associated theoretical calculations.

| Process | Experiment | Final State | Observable | $\sqrt{s}$ (TeV) | $\mathcal{L}$ (fb$^{-1}$) | $n_{\rm dat}$ | Ref. |
|---|---|---|---|---|---|---|---|
| LHC $W, Z$ | ATLAS | Z $p_T$ spectrum | $\left(\frac{1}{\sigma}\right)\frac{d\sigma}{dp_T^{\ell\ell}}$ | 13 | 36.1 | 38 | [102] |
| | CMS | W incl. prod. | $\frac{d\sigma}{d\lvert\eta\rvert}$ | 13 | 35.9 | 36 | [103] |
| | LHCb | Z incl. forward prod. | $\frac{d\sigma}{dy^Z}$ | 13 | 5.1 | 18 | [104] |
| | ATLAS | Z incl. prod. | $\frac{d\sigma}{d\lvert y\rvert}$ | 8 | 20.2 | 7 | [67] |
| top-pair | ATLAS | all-hadronic | $\left(\frac{1}{\sigma}\right)\frac{d\sigma}{dm_{t\bar{t}}}$ | 13 | 36.1 | 9 | [105] |
| | | | $\left(\frac{1}{\sigma}\right)\frac{d\sigma}{d\lvert y_{t\bar{t}}\rvert}$ | 13 | 36.1 | 12 | [105] |
| | | | $\left(\frac{1}{\sigma}\right)\frac{d^2\sigma}{d\lvert y_{t\bar{t}}\rvert dm_{t\bar{t}}}$ | 13 | 36.1 | 11 | [105] |
| | ATLAS | $\ell$+jets | $\left(\frac{1}{\sigma}\right)\frac{d\sigma}{dm_{t\bar{t}}}$ | 13 | 36.1 | 9 | [106] |
| | | | $\left(\frac{1}{\sigma}\right)\frac{d\sigma}{dp_T^t}$ | 13 | 36.1 | 8 | [106] |
| | | | $\left(\frac{1}{\sigma}\right)\frac{d\sigma}{d\lvert y_t\rvert}$ | 13 | 36.1 | 5 | [106] |
| | | | $\left(\frac{1}{\sigma}\right)\frac{d\sigma}{d\lvert y_{t\bar{t}}\rvert}$ | 13 | 36.1 | 7 | [106] |
| | CMS | $\ell$+jets | $\left(\frac{1}{\sigma}\right)\frac{d\sigma}{dm_{t\bar{t}}}$ | 13 | 137 | 15 | [107] |
| | | | $\left(\frac{1}{\sigma}\right)\frac{d\sigma}{dp_T^t}$ | 13 | 137 | 16 | [107] |
| | | | $\left(\frac{1}{\sigma}\right)\frac{d\sigma}{d\lvert y_{t\bar{t}}\rvert}$ | 13 | 137 | 10 | [107] |
| | | | $\left(\frac{1}{\sigma}\right)\frac{d\sigma}{d\lvert y_t\rvert}$ | 13 | 137 | 11 | [107] |
| | | | $\left(\frac{1}{\sigma}\right)\frac{d^2\sigma}{d\lvert y_{t\bar{t}}\rvert dm_{t\bar{t}}}$ | 13 | 137 | 35 | [107] |
| LHC jets | ATLAS | incl. jet $R = 0.4$ | $\frac{d^2\sigma}{dp_T d\lvert y\rvert}$ | 13 | 3.2 | 177 | [108] |
| | CMS | incl. jets $R = 0.4$ (0.7) | $\frac{d^2\sigma}{dp_T d\lvert y\rvert}$ | 13 | 36.3 (33.5) | 78 | [109] |
| | ATLAS | di-jets $R = 0.4$ | $\frac{d^2\sigma}{dm_{jj} d\lvert y^*\rvert}$ | 13 | 3.2 | 136 | [108] |
| HERA jets | H1 | incl. jet (low $Q^2$) | $\frac{d^2\sigma}{dQ^2 dp_T}$ | 0.319 | 0.29 | 37 | [110] |
| | H1 | incl. jet (high $Q^2$) | $\frac{d^2\sigma}{dQ^2 dp_T}$ | 0.319 | 0.351 | 24 | [111] |
| | ZEUS | incl. jet | $\frac{d^2\sigma}{dQ^2 dE_T}$ | 0.300 | 0.038 | 30 | [112] |
| | ZEUS | incl. jet | $\frac{d^2\sigma}{dQ^2 dE_T}$ | 0.319 | 0.082 | 30 | [113] |
| | H1 | di-jets (low $Q^2$) | $\frac{d^2\sigma}{dQ^2 d\langle p_T\rangle}$ | 0.319 | 0.29 | 37 | [110] |
| | H1 | di-jets (high $Q^2$) | $\frac{d^2\sigma}{dQ^2 d\langle p_T\rangle}$ | 0.319 | 0.351 | 24 | [111] |
| | ZEUS | di-jets | $\frac{d^2\sigma}{dQ^2 d\langle E_T\rangle}$ | 0.319 | 0.374 | 22 | [114] |

Table 3.2: The ATLAS, CMS, LHCb, H1, and ZEUS datasets considered in this work, classified by process type. For each dataset we indicate the experiment, the final-state channel, the measured differential distribution(s), the center-of-mass energy $\sqrt{s}$, the integrated luminosity $\mathcal{L}$, the number of data points $n_{\rm dat}$ (after kinematic cuts), and the corresponding publication reference. For the CMS single inclusive jet production and for the ATLAS and CMS top-quark pair production datasets, we list the separate distributions provided by the corresponding analyses.

#### 3.3.2.1 Drell-Yan weak boson production at the LHC

Neutral- and charged-current Drell-Yan production is used to probe quark-flavor PDF separation, through rapidity distributions in the central (ATLAS and CMS) and forward (LHCb) regions [115, 116], and the gluon PDF, through transverse momentum distributions [117]. In the former case, the leading partonic channel is initiated by quarks and antiquarks; in the latter case, a non-zero $p_T$ distribution arises from the $gq(\bar{q})$ partonic initial state followed by a hard $g \to q\bar{q}$ splitting. Here we consider three LHC Run II representative measurements for each of these categories: one by ATLAS [102], one by CMS [103], and one by LHCb [118]. All these measurements correspond to a center-of-mass energy of 13 TeV. We also consider the recent re-analysis of the inclusive $Z$ boson production measurement at a center-of-mass energy of 8 TeV by ATLAS extrapolated to the full leptonic phase space [67].

The ATLAS Run II measurement [102] corresponds to an integrated luminosity of 36.1 fb$^{-1}$. It consists of the $Z$-boson production cross section, reconstructed from the combination of events resulting from electron and muon decays, differential in the transverse momentum of the dilepton pair $p_T^{\ell\ell}$. The measurement is performed in a fiducial phase space, defined by the lepton transverse momentum $p_T^{\ell} > 27$ GeV, the absolute lepton pseudorapidity $|\eta_\ell| < 2.5$, and the dilepton invariant mass $66 < m_{\ell\ell} < 116$ GeV. Cross sections are provided for both the absolute distribution and the distribution normalized to the fiducial cross section. The full breakdown of correlated systematic uncertainties is available and taken into account.

The CMS Run II measurement [103] corresponds to an integrated luminosity of 35.9 fb$^{-1}$. It consists of the $W^\pm$ boson production cross section, reconstructed from the combination of events resulting from electron and muon decays. This measurement is presented as a double differential distribution in the absolute lepton rapidity $|\eta|$, with $|\eta| < 2.4$, and in the lepton transverse momentum $p_T^{\ell}$, with $26 < p_T^{\ell} < 56$ GeV. It is available for each $W$ polarization state and averaged over polarizations. For each boson, 18 equally large bins in $|\eta|$ and a single bin in $p_T^{\ell}$ are provided. The full breakdown of correlated systematic uncertainties is available and taken into account.

The LHCb Run II measurement [104] corresponds to an integrated luminosity of 5.1 fb$^{-1}$. It consists of the $Z$-boson production cross section, reconstructed only from muon decays, in the fiducial phase space defined by the muon transverse momentum $p_T^{\mu} > 20$ GeV, the dimuon invariant mass $60 < m_{\mu\mu} < 120$ GeV, and the muon rapidity $2.0 < \eta_\mu < 4.5$. The presented cross section is differential in the rapidity of the $Z$ boson $y^Z$. The full breakdown of correlated systematic uncertainties is available and taken into account.

Finally, we consider the recent ATLAS measurement of $Z$ boson production based on the 2012 dataset at a center-of-mass energy of 8 TeV, which corresponds to an integrated luminosity of 20.2 fb$^{-1}$ [67]. The measurement is extrapolated to the full phase space of the decay electrons and muons in the dilepton rapidity range $|y| < 3.6$, and covers the $Z$ pole invariant mass region, $80 \leq m_{\ell\ell} \leq 100$ GeV. We specifically consider the cross section differential in $|y|$. The dependence on the transverse momentum of the dilepton pair is integrated over. The precision of this measurement, excluding the luminosity uncertainty, ranges from 0.2%, for $|y| \leq 2.0$, to 0.9% at more forward rapidities. This measurement is based on a re-analysis of events that were previously used in another two measurements [119, 120] from which double- and triple-differential distributions in the fiducial region for the final-state leptons were reconstructed. The distributions are differential, respectively, in the invariant mass $m_{\ell\ell}$ and absolute rapidity $|y|$ of the dilepton pair, and in $m_{\ell\ell}$, $|y|$, and the cosine of the Collins-Soper angle, $\cos\theta^*$. The covered invariant mass region extends below, across and above the $Z$ peak. The double differential measurement was included in the MSHT20 [33] and NNPDF4.0 [22] PDF fits. For this reason, the new measurement [67] does not fulfil the second selection criterion established at the beginning of this section. We make an exception for this measurement because, first, it exhibits a significant PDF dependence, and, second, it underlies the most precise determination of the strong coupling ever performed at a hadron collider, in which PDF uncertainties are the leading uncertainties. In Sec. 3.3.4.2 we will discuss the interplay between the original [119, 120] and new [67] measurements.

For all these measurements, theoretical predictions accurate to NNLO QCD are computed in the form of PineAPPL interpolation grids [15] with NNLOjet [121, 122]. The computation incorporates in particular the NNLO QCD corrections to the transverse momentum distributions of the $Z$ boson from Refs. [123, 124]. The central renormalization and factorization scales are set to

$$\mu_F = \mu_R = \sqrt{m_{\ell\ell}^2 + (p_T^{\ell\ell})^2}\,, \qquad \mu_F = \mu_R = M_V\,, \tag{3.2}$$

respectively for the $Z$ transverse momentum distribution and the gauge boson rapidity distributions (with $M_V$ the gauge boson mass, $Z$ or $W$). In the former case, we also apply a kinematic cut $p_T^{\ell\ell} >$ 30 GeV to remove the region where resummation corrections, not accounted for by our fixed-order computation, may be relevant [117, 125, 126]. Electroweak, QED, and photon-induced corrections, though known, are not considered here.

#### 3.3.2.2 Top quark pair production at the LHC

Top-quark pair production at the LHC, which is initiated by gluon-gluon scattering, primarily probes the gluon PDF at large $x$ [62, 77, 78, 86, 127–130]. In addition to their PDF sensitivity, top-quark pair cross sections also constrain the top-quark mass $m_t$ and the strong coupling $\alpha_s(m_Z)$ [131, 132]. Here we consider the ATLAS measurement reconstructed from the all-hadronic [105] and lepton+jets [106] final states, and the CMS measurement reconstructed from the lepton+jets final state [107]. We consider only parton-level measurements presented in terms of the kinematic variables of the final-state top quarks. The reason being that theoretical computations accurate to NNLO QCD for particle-level measurements [133] are not available in a numerical format suitable for this analysis. All these measurements were taken during the LHC Run II, at a center-of-mass energy of 13 TeV.

The ATLAS measurements correspond to an integrated luminosity of 36 fb$^{-1}$. Cross sections are provided, absolute and normalized to the total cross section, as single and double differential distributions in various kinematic variables. For the sake of this work, we consider a subset of them, either the normalized or the absolute differential distributions. The choice depends on the stability of the experimental covariance matrix, as we will explain in Sec. 3.3.3.2. For the all-hadronic measurement, we choose the single differential absolute (normalized) distributions in the invariant mass (absolute rapidity) of the top-quark pair, $m_{t\bar{t}}$ ($y_{t\bar{t}}$), and the double differential absolute distribution in these two variables. For the lepton+jets measurement, we choose the single differential normalized distributions in the invariant mass of the top quark pair, $m_{t\bar{t}}$, in the transverse momentum of the top quark, $p_T^t$, in the absolute rapidity of the top-quark pair, $y_{t\bar{t}}$, and in the absolute rapidity of the top-quark, $y_t$.

The CMS measurement corresponds to an integrated luminosity of 137 fb$^{-1}$, that is, all the events recorded during the LHC Run II. Absolute and normalized cross sections are provided as single and double differential distributions in various kinematic variables. We select a subset of them, specifically the single differential normalized distributions in $m_{t\bar{t}}$, $|y_{t\bar{t}}|$ , $|y_t|$, and $p_T^t$, and the double differential normalized distribution in $(m_{t\bar{t}}, |y_{t\bar{t}}|)$.

Theoretical predictions accurate to NNLO QCD are computed using MATRIX [76], which has been interfaced to PineAPPL [15]. The central factorization and renormalization scales are set to

$$\mu_R = \mu_F = H_T/2 = \sqrt{m_t^2 + \left(p_T^t\right)^2}\Big/2\,, \tag{3.3}$$

where $m_t$ and $p_T^t$ are the mass and the transverse momentum of the top quark. This choice follows the recommendation of [134]. A value of $m_t^{\rm pole} = 172.5$ GeV has been used for the top-quark pole mass, consistently with the latest PDG average [135]. These computations have been benchmarked, when possible, against FastNLO tables [98] generated with the code presented in [136]. Electroweak, QED, and photon-induced cross sections are not included.

#### 3.3.2.3 Jet production at the LHC

Collider jet production at the LHC is a traditional probe of the gluon PDF in the large-$x$ region [137–140], though it provides also information on the large-$x$ quark PDFs. Here we consider the ATLAS measurement of single inclusive jet and di-jet production [108], and the CMS measurement of single inclusive jet production [109]. Both of them were taken during the LHC Run II, at a center-of-mass energy of 13 TeV.

The ATLAS measurements correspond to an integrated luminosity of 3.2 fb$^{-1}$. Whereas this amounts to only a small fraction of the events recorded during Run II, no other unfolded measurements of single inclusive jet or di-jet production based on a larger sample have been presented by ATLAS to date. The single inclusive jet measurement is presented as a set of double differential cross sections in the jet transverse momentum $p_T$, with 100 GeV $\leq p_T \leq$ 3.5 TeV, and the jet absolute rapidity $|y|$, with $|y| < 3.0$. The di-jet measurement is presented as a set of double differential cross sections in the di-jet invariant mass $m_{jj}$, with 300 GeV $\leq m_{jj} \leq$ 9 TeV, and the half absolute rapidity separation between the two leading jets, $|y^*|$, with $|y^*| < 3.0$. single inclusive jets and di-jets are reconstructed by means of the anti-$k_T$ clustering algorithm [141] for a jet radius of $R = 0.4$. The full breakdown of correlated systematic uncertainties is available, separately for the single inclusive jet and di-jet measurements, and taken into account.

The CMS measurement corresponds to an integrated luminosity of 36.3 (33.5) fb$^{-1}$ and a jet radius of $R = 0.4$ ($R = 0.7$). We consider the measurement with $R = 0.4$. Cross sections are double differential in the jet transverse momentum $p_T$, with $97 \leq p_T \leq 3.1$ TeV, and in the jet absolute rapidity $|y|$, with $|y| < 2.0$. The experimental covariance matrix of the measurement is provided and taken into account.

For all the aforementioned measurements, theoretical predictions, accurate to NNLO QCD in the leading color approximation, were computed with the NNLOjet code [142]. The central factorization and renormalization scales where chosen as

$$\mu_F = \mu_R = p_T , \qquad \mu_F = \mu_R = m_{jj} , \tag{3.4}$$

respectively for single inclusive jets and di-jets. These predictions were released [143] as interpolation grids in the APPLfast format through the Ploughshare website [144]. For the sake of this work, these grids have been converted to the PineAPPL format [15]. We do not account for NLO electroweak corrections or photon-initiated contributions, neither for single inclusive jets nor for di-jets. Monte Carlo uncertainties due to the generation of a finite number of events are generally smaller than MHO and $\alpha_s$ uncertainties, and are thus ignored.

#### 3.3.2.4 Jet production at HERA

Jet production in DIS can probe the gluon PDF at large $x$ as well. This process was measured at HERA by the H1 and ZEUS experiments and demonstrated to constrain the gluon PDF and the strong coupling [145–147] in comparison to fits based on inclusive DIS measurements only. Here we consider four H1 measurements [110, 111, 148] and three ZEUS measurements [112–114] for single inclusive jet and di-jet cross sections, as indicated in Table 3.2.

The H1 measurements correspond to the HERA-II data-taking period with a center-of-mass energy of 319 GeV. Two pairs of single inclusive jet and di-jet measurements are available, which focus on different regions of the exchanged boson virtuality $Q^2$: a low-$Q^2$ pair, $5.5 \leq Q^2 \leq 80$ GeV$^2$, and a high-$Q^2$ pair, $150 \leq Q^2 \leq 15000$ GeV$^2$. The integrated luminosity is, respectively, 290 pb$^{-1}$ and 351 pb$^{-1}$. On top of the virtuality $Q^2$, cross sections are differential in the jet transverse momentum $p_T$ or the di-jet average momentum $\langle p_T \rangle$, respectively for the single inclusive jet and the di-jet measurements. Massless jets are identified using the $k_T$ algorithm with the $R$ parameter set to $R = 1$. Experimental correlations are available for all measurements, including for points in different single inclusive jet and di-jet bins at different $Q^2$.

The ZEUS measurements correspond to the HERA-I data-taking period, with a center-of-mass energy of 300 GeV and an integrated luminosity of 38 pb$^{-1}$, and to the HERA-II data-taking period, with a center-of-mass energy of 319 GeV and an integrated luminosity of 82 pb$^{-1}$ and 374 pb$^{-1}$. Cross sections are presented as differential distributions in the vector boson virtuality $Q^2$, with $Q^2 \geq 125$ GeV$^2$, and the jet transverse energy $E_T$ or the di-jet average transverse energy $\langle E_T \rangle$, respectively for the single inclusive jet and the di-jet measurements. Experimental correlations are available across bins within the same set, but not across bins in different datasets. Unlike inclusive DIS structure functions [149], no combination between the H1 and ZEUS results exists to date. A final ZEUS mea-surement of single inclusive jet production cross has been published recently [145], however we do not consider it because NNLO QCD corrections to matrix elements are not readily available in a format suitable for fast convolution with PDFs.

Theoretical predictions accurate to NNLO QCD were computed with the NNLOjet code [150, 151] in the zero-mass variable-flavor-number scheme. The central factorization and renormalization scales are

$$\mu = \mu_F = \mu_R = Q^2 + (p_T)^2 , \qquad \mu = \mu_F = \mu_R = Q^2 + \langle p_T \rangle^2 , \tag{3.5}$$

respectively for single inclusive jets and di-jets. Data points for which $\mu \leq 10$ GeV were excluded to ensure that the scale is larger than the $b$-quark mass. This is necessary because jets are built from massless partons. As in the case of LHC jets, theoretical predictions were released as interpolation grids through the Ploughshare website [144]. We convert these grids to the PineAPPL format [15].

### 3.3.3 Methodological framework

In this section, we describe how we quantitatively assess the agreement between the measurements presented in Sec. 3.3.2 and the corresponding theoretical predictions for different PDF sets. We first define the figure of merit that we use, and specifically explain how we take into account experimental and theoretical uncertainties in it. We then discuss how this figure of merit may become misleading if the experimental covariance matrix is ill-conditioned, and illustrate how we identify and handle such cases. We finally review the PDF sets that we consider as input for the computation of the theoretical predictions.

#### 3.3.3.1 Figure of merit

We quantify the agreement between experimental data and theoretical predictions by computing the (reduced) $\chi^2$ for each dataset

$$\chi^2 = \frac{1}{n_{\rm dat}} \sum_{i,j=1}^{n_{\rm dat}} \left( T_i^{(0)} - D_i \right) \left( {\rm cov}^{-1} \right)_{ij} \left( T_j^{(0)} - D_j \right) , \tag{3.6}$$

where $n_{\rm dat}$ is the number of data points in the considered dataset, $\{D_i\}$ are the central values of the experimental data, $\{T_i^{(0)}\}$ are the corresponding theoretical predictions, and ${\rm cov}_{ij}$ is the covariance matrix. Theoretical predictions are evaluated, for both Monte Carlo and Hessian PDF sets, as the prediction obtained with the central PDF $f^{(0)}$, $T_i^{(0)} = T_i(f^{(0)})$. Note that the $\chi^2$ in Eq. (3.6) is normalized to the number of data points. Therefore, in case of perfect agreement between data and theory, one expects $\chi^2 \sim 1$, with statistical fluctuations of the order of the standard deviation of the $\chi^2$ distribution, $\sigma_{\chi^2} = \sqrt{2/n_{\rm dat}}$.

To evaluate Eq. (3.6), one needs to compute the covariance matrix ${\rm cov}_{ij}$. In addition to experimental uncertainties, one should consider all the relevant sources of theoretical uncertainties, in particular, those associated to missing higher orders (MHO), to PDFs, and to the value of the strong coupling $\alpha_s(m_Z)$. Assuming that all of these theoretical uncertainties follow a Gaussian distribution and that they are mutually independent, they can be incorporated in the covariance matrix following the formalism developed in [20, 152]. Specifically, in this formalism the covariance matrix in Eq. (3.6) reads as

$$\mathrm{cov}_{ij} = \left(\mathrm{cov}_{\mathrm{exp}}\right)_{ij} + \left(\mathrm{cov}_{\mathrm{th}}\right)_{ij} , \tag{3.7}$$

where the theory covariance matrix is in turn the sum of a MHO, PDF, and $\alpha_s$ contributions

$$\left(\mathrm{cov}_{\mathrm{th}}\right)_{ij} = \left(\mathrm{cov}_{\mathrm{mho}}\right)_{ij} + \left(\mathrm{cov}_{\mathrm{pdf}}\right)_{ij} + \left(\mathrm{cov}_{\mathrm{as}}\right)_{ij} . \tag{3.8}$$

The experimental covariance matrix is sometimes provided together with the experimental measurements, otherwise, in most cases, we reconstruct it from knowledge of experimental uncertainties as

$$\left(\mathrm{cov}_{\mathrm{exp}}\right)_{ij} = \delta_{ij}\sigma_i^{(\mathrm{uncorr})}\sigma_j^{(\mathrm{uncorr})} + \sum_{\ell=1}^{n_{\mathrm{corr}}} \sigma_{i,\ell}^{(\mathrm{corr})}\sigma_{j,\ell}^{(\mathrm{corr})} . \tag{3.9}$$

Here $\sigma_i^{(\mathrm{uncorr})}$ is the sum in quadrature of all the uncorrelated uncertainties, and $\sigma_{i,\ell}^{(\mathrm{corr})}$ is the set of $n_{\mathrm{corr}}$ correlated uncertainties. These could be additive or multiplicative, however this distinction is not relevant here, given that Eq. (3.6) is only used to quantify the agreement between data and theory. In a fit of PDFs, this distinction is instead relevant because multiplicative uncertainties may bias the determination of the PDF central value and variance. Therefore they would require a specific treatment, by re-defining either the experimental covariance matrix with the $t_0$ prescription [24] or the figure of merit with additional nuisance parameters [153]. Note that whenever we reconstruct the experimental covariance matrix using Eq. (3.9), we implicitly assume that correlated uncertainties are 100% correlated, given that no specific correlation model is provided for the considered datasets, see Sec. 3.3.2. Decorrelation remains however possible, using the procedure summarized in Sec. 3.3.3.2, and we will actually make use of it, as discussed further below.

The contribution to the covariance matrix due to MHO is evaluated following [20, 152]. Specifically, MHO are estimated as the difference between theoretical predictions computed with fixed and varied renormalization and factorization scales, $\mu_R$ and $\mu_F$. Several prescriptions defining scale variations are possible. As is common practice in LHC analyses, we adopt the 7-point variation prescription, which gives the MHO covariance matrix

$$\left(\mathrm{cov}_{\mathrm{mho}}\right)_{ij} = \frac{1}{3}\left\{\Delta_i^{+0}\Delta_j^{+0} + \Delta_i^{-0}\Delta_j^{-0} + \Delta_i^{0+}\Delta_j^{0+} + \Delta_i^{0-}\Delta_j^{0-} + \Delta_i^{++}\Delta_j^{++} + \Delta_i^{--}\Delta_j^{--}\right\} , \tag{3.10}$$

where, for each data point $i, j$, the shifts are defined as

$$\Delta_i\left(\kappa_R, \kappa_F\right) = T_i\left(\mu_R = \kappa_R\mu_R^{(0)}, \mu_F = \kappa_R\mu_F^{(0)}\right) - T_i\left(\mu_R^{(0)}, \mu_F^{(0)}\right) , \tag{3.11}$$

with $\left(\mu_R^{(0)}, \mu_F^{(0)}\right)$ the central renormalization and factorization scales and

$$\begin{aligned} &\Delta_i^{+0} = \Delta_i\,(2,1)\ , \quad \Delta_i^{-0} = \Delta_i\,(1/2,1)\ , \quad \Delta_i^{0+} = \Delta_i\,(1,1/2)\ , \\ &\Delta_i^{0-} = \Delta_i\,(1,1/2)\ , \quad \Delta_i^{++} = \Delta_i\,(2,2)\ , \quad \Delta_i^{--} = \Delta_i\,(1/2,1/2)\ . \end{aligned} \tag{3.12}$$

The shifts in the NNLO theory predictions associated to the scale variations, Eq. (3.11), are directly evaluated from the PineAPPL grids [15]. In general, the 7-point MHO theory covariance matrix defined by Eq. (3.10) differs from the envelope prescription to estimate MHO uncertainties frequently used in LHC studies.

The contribution to the covariance matrix due to PDF uncertainties is determined, for each of the PDF sets considered (see Sec. 3.3.3.3), using the definition of covariance between the random variables $T_i$ and $T_j$

$$\left(\text{cov}_{\text{pdf}}\right)_{ij} = \mathbb{E}\left[(T_i - \mathbb{E}[T_i])(T_j - \mathbb{E}[T_j])\right]\ , \tag{3.13}$$

where $\mathbb{E}[X]$ denotes the expectation value of the random variable $X$. For Hessian PDF sets, Eq. (3.13) reads

$$\left(\text{cov}_{\text{pdf}}\right)_{ij} = \sum_{k=1}^{n_{\text{eig}}} \left(T_i^{(k)} - T_i^{(0)}\right)\left(T_j^{(k)} - T_j^{(0)}\right)\ , \tag{3.14}$$

where $T_i^{(k)} = T_i(f^{(k)})$ is the theoretical prediction computed with the PDF associated to the $k$-th eigenvalue $f^{(k)}$, and $T_i^{(0)} = T_i(f^{(0)})$ is the theoretical prediction computed with the central PDF $f^{(0)}$. We use the symmetric definition of Hessian PDF uncertainties, since we assume that PDF uncertainties are Gaussian. In case of asymmetric Hessian PDF sets, we replace the difference between $T_{i,j}^{(k)}$ and $T_{i,j}^{(0)}$ in Eq. (3.9) with the difference between predictions obtained with positive and negative eigenvectors. For Monte Carlo PDF sets, Eq. (3.13) reads

$$\left(\text{cov}_{\text{pdf}}\right)_{ij} = \frac{1}{n_{\text{rep}}} \sum_{k=1}^{n_{\text{rep}}} \left(T_i^{(k)} - \langle T_i\rangle_{\text{rep}}\right)\left(T_j^{(k)} - \langle T_j\rangle_{\text{rep}}\right)\ , \tag{3.15}$$

where $T_i^{(k)} = T_i(f^{(k)})$ is the theoretical prediction computed with the PDF associated to the $k$-th replica $f^{(k)}$, and $\langle T_i\rangle_{\text{rep}} = \frac{1}{n_{\text{rep}}}\sum_{k=1}^{n_{\text{rep}}} T_i^{(k)}$ is the average over replicas.

The contribution to the covariance matrix due to the uncertainty of the value of $\alpha_s(m_Z)$ is determined as follows. We take $\alpha_s(m_Z) = 0.118 \pm 0.001$ for all PDF sets considered, consistently with the latest PDG average [135], and we construct

$$\left(\text{cov}_{\text{as}}\right)_{ij} = \frac{1}{2}\left\{\Delta_{i,\alpha_s}^{+}\Delta_{j,\alpha_s}^{+} + \Delta_{i,\alpha_s}^{-}\Delta_{j,\alpha_s}^{-}\right\}\ , \tag{3.16}$$

where, for each data point $i, j$,

$$\begin{aligned} \Delta_{i,\alpha_s}^{+} &\equiv T_i(\alpha_s = 0.119) - T_i(\alpha_s = 0.118)\,, \\ \Delta_{i,\alpha_s}^{-} &\equiv T_i(\alpha_s = 0.118) - T_i(\alpha_s = 0.117)\,. \end{aligned} \tag{3.17}$$

The value of $\alpha_s$ in the theory predictions is varied consistently both in the matrix element and in the PDFs, a fact that is streamlined thanks to the usage of PineAPPL grids. The combination of Eq. (3.16) with Eq. (3.14) (for a Hessian set) or Eq. (3.15) (for a Monte Carlo set) reproduces the prescription of [51], according to which PDF and $\alpha_s$ uncertainties are added in quadrature.

In Sec. 3.3.4 we will quantify the agreement between experimental data and theory predictions, obtained with different PDF sets, in terms of the figure of merit given in Eq. (3.6). When accounting for all sources of experimental and theoretical uncertainties, we have

$$\chi^2_{\rm exp+th} = \frac{1}{n_{\rm dat}} \sum_{i,j=1}^{n_{\rm dat}} \left(T_i^{(0)} - D_i\right) \left(\left(\mathrm{cov}_{\rm exp} + \mathrm{cov}_{\rm mho} + \mathrm{cov}_{\rm pdf} + \mathrm{cov}_{\rm as}\right)^{-1}\right)_{ij} \left(T_j^{(0)} - D_j\right) , \tag{3.18}$$

with the individual contributions to the covariance matrix combined in quadrature. In order to understand the impact of the various sources of uncertainties entering Eq. (3.18), we will also present results for variants of this figure of merit restricted to a subset of the uncertainties, in particular

$$\chi^2_{\rm exp} = \frac{1}{n_{\rm dat}} \sum_{i,j=1}^{n_{\rm dat}} \left(T_i^{(0)} - D_i\right) \left(\left(\mathrm{cov}_{\rm exp}\right)^{-1}\right)_{ij} \left(T_j^{(0)} - D_j\right) , \tag{3.19}$$

which contains only the experimental uncertainties, and

$$\chi^2_{\rm exp+mho} = \frac{1}{n_{\rm dat}} \sum_{i,j=1}^{n_{\rm dat}} \left(T_i^{(0)} - D_i\right) \left(\left(\mathrm{cov}_{\rm exp} + \mathrm{cov}_{\rm mho}\right)^{-1}\right)_{ij} \left(T_j^{(0)} - D_j\right) , \tag{3.20}$$

defined without the contribution of the PDF and $\alpha_s$ uncertainties. In all cases, the figures of merit are presented normalized to the number of data points of each dataset considered. We emphasize that, when evaluating Eq. (3.18), PDFs enter in two different places: through the theory predictions $T_i$ and through the PDF contribution to the total covariance matrix in Eq. (3.8).

To further assess the significance of $\chi^2_{\rm exp+th}$, Eq. (3.18), as a measure of the agreement between experimental data and theoretical predictions, we will make use of two additional estimators in Sec. 3.3.4. The first estimator is the relative change in the total $\chi^2$ due to the change of input PDF set for a given dataset

$$\Delta\chi^{2(i)} = \frac{\chi^{2(i)}_{\rm exp+th} - \left\langle \chi^2_{\rm exp+th} \right\rangle_{\rm pdfs}}{\left\langle \chi^2_{\rm exp+th} \right\rangle_{\rm pdfs}} , \tag{3.21}$$

where the index $i$ runs over the $n_{\rm pdfs}$ input PDF sets considered in the analysis (see Sec. 3.3.3.3), and the average over PDF sets is evaluated as

$$\left\langle \chi^2_{\rm exp+th} \right\rangle_{\rm pdfs} = \frac{1}{n_{\rm pdfs}} \sum_{i=1}^{n_{\rm pdfs}} \chi^{2(i)}_{\rm exp+th} . \tag{3.22}$$

By construction, $\sum_i \Delta\chi^{2(i)} = 0$. This estimator gauges the relative change in the value of the $\chi^2$ for a given PDF set with respect to the average evaluated over all PDF sets considered. It therefore allows one to disentangle PDF-related effects in the $\chi^2$ from other effects.

The second estimator quantifies the difference of the $\chi^2$, computed with a given PDF, with respect to the $\chi^2$ averaged over all PDF sets in terms of the number of standard deviations of the $\chi^2$ distribution

$$\Delta n_\sigma^{(i)} = \frac{\chi^{2(i)}_{\rm exp+th} - \left\langle \chi^2_{\rm exp+th} \right\rangle_{\rm pdfs}}{\sqrt{2/n_{\rm data}}} \,. \tag{3.23}$$

This estimator allows one to compare the $\chi^2$ variation due to the choice of PDF to the expected statistical fluctuations of the $\chi^2$, and therefore check if this is significant or not. Note indeed that several of the datasets considered contain a relatively small number of data points, so that a large relative change of the $\chi^2$ in Eq. (3.21) may be simply explained by large fluctuations due to the small data sample.

#### 3.3.3.2 Stability of the experimental covariance matrix

The interpretation of the agreement of theoretical predictions with experimental data, as quantified by the value of the $\chi^2$, requires some care. As discussed in Ref. [154], an inaccurate determination of experimental uncertainty correlations, in otherwise very precise data, may result in an ill-conditioned experimental covariance matrix, which leads in turn to anomalously large values of the $\chi^2$.

A metric to measure the conditioning of an experimental covariance matrix was introduced in Ref. [154], see, in particular, Eq. (26). This was defined as the inverse of the smallest singular value of the experimental correlation matrix, and called condition number $Z$. The value $(\sqrt{2}Z)^{-1}$ was then demonstrated to be related to the amount by which experimental correlations need to be determined to ensure that the $\chi^2$ remains stable, namely that it does not vary by more than one standard deviation, $\sigma_{\chi^2} = \sqrt{2/n_{\rm dat}}$. A large value of $Z$ indicates a dataset for which small variations of the correlation model can potentially lead to large $\chi^2$ variations for unchanged data and theory and vice-versa. In Ref. [154] a reasonable threshold was defined to be $Z = 4$. This value corresponds to assuming that correlations on uncertainties of the order of a few percent, such as those that affect the LHC measurements considered in this work, be estimated with an absolute uncertainty of less than 0.18. This means that if the correlation between two bins is estimated to be 1.00, while its real value is 0.82, one can expect that the $\chi^2$ deviates from one by more than $1\sigma$ even if experimental data and theoretical predictions are perfectly consistent. Note that the smaller the uncertainty, the higher the required precision with which correlations need be known to be within a $1\sigma$ variation of the $\chi^2$. As explicitly shown in Ref. [154] as part of a toy model (see in particular Eq. (29) and Fig. 3), for a 1% (2.5%) uncertainty, correlations ought to be known with a precision of 2% (12%). The choice $Z = 4$ (the same for all experiments) should therefore be seen as a practical diagnostic choice.

In some cases, a large value of $Z$ may not imply a pathological behavior of the experimental data. A typical case is the one in which the luminosity uncertainty, which by definition is 100% correlated across all bins of a given dataset, is the largest of all uncertainties. In this case, we expect the condition number $Z$ to be large. The same would happen with any experimental uncertainty that is fully correlated across bins for specific experimental reasons. In these cases, we should compute $Z$ upon excluding these uncertainties from the reconstruction of the experimental covariance matrix. For the sake of this work, we single out only the luminosity uncertainty as $100\%$ correlated, given that we do not have complete knowledge of which other uncertainties are also undoubtedly $100\%$ correlated. We then split the experimental covariance matrix into two components, one that contains only the luminosity uncertainty, and one that contains all of the other uncertainties. We compute $Z$, which we call $Z_{\mathscr{L}}$, for the latter covariance matrix and regularize it, if necessary; we then compute the $\chi^2$ using the sum of the original luminosity covariance matrix and the regularized covariance matrix. Clearly, this procedure might decorrelate systematic uncertainties that owe to be 100% correlated, or decorrelate too much some other systematic uncertainties. Nevertheless, we consider that the procedure remains a useful diagnosis tool when information on specific decorrelation models, provided under the experimental guidance, are not available. In this respect, we shall also note that here our aim is not to characterize a dataset for inclusion (or not) in a PDF determination, but rather to comparatively assess the ability of various PDF sets to describe the data. We consider that the application of our regularization procedure does not alter our judgement on such an ability (see Appendix A).

An alternative estimator to assess the conditioning of the experimental correlation matrix, sometimes used in experimental analyses, is $\lambda_\rho$, defined as the ratio of the smallest to the largest eigenvalues of the experimental correlation matrix. A small value of $\lambda_\rho$ indicates a large spread of eigenvalues, with the directions associated to the smallest ones almost degenerate. These degeneracies are those that lead to a ill-conditioned matrix.

In Table 3.3 we display, for each dataset listed in Table 3.2 and separately for each observable, the number of data points, $n_{\rm dat}$, and the condition numbers $\lambda_\rho$, $Z$, and $Z_{\mathscr{L}}$. For normalized distributions $Z = Z_{\mathscr{L}}$ by construction. For datasets which do not provide the breakdown of systematic uncertainties but instead only the overall covariance matrix, $Z_{\mathscr{L}}$ is computed by subtracting from this covariance matrix a covariance matrix constructed only from the $100\%$-correlated luminosity uncertainty. In the case of the CMS top-quark pair distribution, this procedure is however not applied, given that the measurement is the combination of events recorded with different luminosities. We therefore leave the corresponding entry blank in Table 3.3. Whenever a dataset is presented in different variants, for example as absolute or normalized distributions or for two different values of the jet radius $R$, we indicate with a (*) the one used in Sec. 3.3.4. We select the distributions that feature the lowest value of $Z_{\mathscr{L}}$.

| Process | $n_{\text{dat}}$ | $\lambda_\rho$ | $Z$ | $Z_{\mathscr{L}}$ |
|---|---|---|---|---|
| ATLAS 13 TeV $Z$ $1/\sigma d\sigma/dp_T^{\ell\ell}$ | 38 | $1.9\times10^{-1}$ | 1.10 | 1.10 |
| CMS 13 TeV $W^+$ $d\sigma/d\|\eta\|$ | 18 | $8.3\times10^{-5}$ | 25.1 | 19.0 |
| CMS 13 TeV $W^-$ $d\sigma/d\|\eta\|$ | 18 | $8.9\times10^{-5}$ | 26.0 | 18.0 |
| LHCb 13 TeV $Z$ $d\sigma/dy^Z$ | 17 | $1.9\times10^{-3}$ | 5.92 | 2.09 |
| ATLAS 8 TeV $Z$ $d\sigma/d\|y\|$ | 7 | $3.2\times10^{-4}$ | 21.6 | 2.10 |
| ATLAS 13 TeV $t\bar{t}$ all hadr. $d\sigma/dm_{t\bar{t}}$ (*) | 9 | $2.4\times10^{-3}$ | 7.27 | 7.24 |
| ATLAS 13 TeV $t\bar{t}$ all hadr. $1/\sigma\, d\sigma/dm_{t\bar{t}}$ | 9 | $3.9\times10^{-5}$ | 64.7 | 64.7 |
| ATLAS 13 TeV $t\bar{t}$ all hadr. $d\sigma/d\|y_{t\bar{t}}\|$ | 12 | $3.3\times10^{-3}$ | 5.27 | 5.25 |
| ATLAS 13 TeV $t\bar{t}$ all hadr. $1/\sigma\, d\sigma/d\|y_{t\bar{t}}\|$ (*) | 12 | $8.9\times10^{-2}$ | 1.77 | 1.77 |
| ATLAS 13 TeV $t\bar{t}$ all hadr. $d^2\sigma/dm_{t\bar{t}}\, d\|y_{t\bar{t}}\|$ (*) | 11 | $4.4\times10^{-3}$ | 4.83 | 4.81 |
| ATLAS 13 TeV $t\bar{t}$ all hadr. $1/\sigma\, d^2\sigma/dm_{t\bar{t}}\, d\|y_{t\bar{t}}\|$ | 11 | $9.4\times10^{-5}$ | 52.1 | 52.1 |
| ATLAS $t\bar{t}$ $\ell+j$ $d\sigma/dm_{t\bar{t}}$ | 9 | $5.2\times10^{-4}$ | 16.2 | 15.9 |
| ATLAS 13 TeV $t\bar{t}$ $\ell+j$ $1/\sigma\, d\sigma/dm_{t\bar{t}}$ (*) | 9 | $3.0\times10^{-3}$ | 7.62 | 7.62 |
| ATLAS 13 TeV $t\bar{t}$ $\ell+j$ $d\sigma/dp_T^t$ | 8 | $5.8\times10^{-4}$ | 16.8 | 16.6 |
| ATLAS 13 TeV $t\bar{t}$ $\ell+j$ $1/\sigma\, d\sigma/dp_T^t$ (*) | 8 | $2.5\times10^{-3}$ | 8.46 | 8.46 |
| ATLAS 13 TeV $t\bar{t}$ $\ell+j$ $d\sigma/d\|y_t\|$ | 5 | $1.5\times10^{-3}$ | 11.7 | 11.5 |
| ATLAS 13 TeV $t\bar{t}$ $\ell+j$ $1/\sigma\, d\sigma/d\|y_t\|$ (*) | 5 | $9.6\times10^{-2}$ | 2.06 | 2.06 |
| ATLAS 13 TeV $t\bar{t}$ $\ell+j$ $d\sigma/d\|y_{t\bar{t}}\|$ | 7 | $6.2\times10^{-4}$ | 15.7 | 15.4 |
| ATLAS 13 TeV $t\bar{t}$ $\ell+j$ $1/\sigma\, d\sigma/d\|y_{t\bar{t}}\|$ (*) | 7 | $7.8\times10^{-2}$ | 2.26 | 2.26 |
| CMS 13 TeV $t\bar{t}$ $\ell+j$ $d\sigma/dm_{t\bar{t}}$ | 15 | $1.1\times10^{-2}$ | 3.90 | – |
| CMS 13 TeV $t\bar{t}$ $\ell+j$ $1/\sigma\, d\sigma/dm_{t\bar{t}}$ (*) | 15 | $3.0\times10^{-2}$ | 3.51 | 3.51 |
| CMS 13 TeV $t\bar{t}$ $\ell+j$ $d\sigma/dp_T^t$ | 16 | $7.5\times10^{-3}$ | 4.04 | – |
| CMS 13 TeV $t\bar{t}$ $\ell+j$ $1/\sigma\, d\sigma/dp_T^t$ (*) | 16 | $1.3\times10^{-1}$ | 1.78 | 1.78 |
| CMS $t\bar{t}$ $\ell+j$ $d\sigma/d\|y_t\|$ | 11 | $3.3\times10^{-3}$ | 5.75 | – |
| CMS 13 TeV $t\bar{t}$ $\ell+j$ $1/\sigma\, d\sigma/d\|y_t\|$ (*) | 11 | $2.7\times10^{-1}$ | 1.36 | 1.36 |
| CMS 13 TeV $t\bar{t}$ $\ell+j$ $d\sigma/d\|y_{t\bar{t}}\|$ | 10 | $1.2\times10^{-3}$ | 9.68 | – |
| CMS 13 TeV $t\bar{t}$ $\ell+j$ $1/\sigma\, d\sigma/d\|y_{t\bar{t}}\|$ (*) | 10 | $1.9\times10^{-1}$ | 1.53 | 1.53 |
| CMS 13 TeV $t\bar{t}$ $\ell+j$ $d^2\sigma/dm_{t\bar{t}}\, d\|y_{t\bar{t}}\|$ | 35 | $8.1\times10^{-5}$ | 22.4 | – |
| CMS 13 TeV $t\bar{t}$ $\ell+$ jj $1/\sigma\, d^2\sigma/dm_{t\bar{t}}\, d\|y_{t\bar{t}}\|$ (*) | 35 | $1.8\times10^{-4}$ | 17.2 | 17.2 |
| ATLAS 13 TeV single-inclusive jets $d^2\sigma/dp_T d\|y\|$ | 177 | $2.6\times10^{-5}$ | 16.9 | 16.2 |
| CMS 13 TeV single-inclusive jets ($R=0.4$) $d^2\sigma/dp_T d\|y\|$ (*) | 78 | $1.1\times10^{-4}$ | 13.3 | 13.1 |
| CMS 13 TeV single-inclusive jets ($R=0.7$) $d^2\sigma/dp_T d\|y\|$ | 78 | $9.0\times10^{-5}$ | 14.8 | 14.5 |
| ATLAS 13 TeV di-jets $d^2\sigma/dm_{jj}d\|y^*\|$ | 136 | $3.8\times10^{-5}$ | 16.8 | 15.6 |
| H1 single-inclusive-jets (low $Q^2$) $d^2\sigma/dQ^2dp_T$ | 48 | $7.6\times10^{-3}$ | 6.00 | 5.91 |
| H1 single-inclusive-jets (high $Q^2$) $d^2\sigma/dQ^2dp_T$ | 24 | $7.0\times10^{-3}$ | 1.46 | 1.19 |
| ZEUS single-inclusive jets (low luminosity) $d^2\sigma/dQ^2dE_T$ | 30 | $5.0\times10^{-2}$ | 1.87 | 1.82 |
| ZEUS single-inclusive jets (high luminosity) $d^2\sigma/dQ^2dE_T$ | 30 | $1.9\times10^{-2}$ | 2.56 | 2.43 |
| H1 di-jets (low $Q^2$) $d^2\sigma/dQ^2d\langle p_T\rangle$ | 48 | $9.0\times10^{-2}$ | 7.67 | 7.42 |
| H1 di-jets (high $Q^2$) $d^2\sigma/dQ^2\, d\langle p_T\rangle$ | 24 | $1.0\times10^{-1}$ | 1.60 | 1.45 |
| ZEUS di-jets $d^2\sigma/dQ^2d\langle E_T\rangle$ | 22 | $1.5\times10^{-2}$ | 2.83 | 2.72 |

Table 3.3: The number of data points, $n_{\text{dat}}$, the condition numbers $\lambda_\rho$, $Z$, and $Z_{\mathscr{L}}$ for all datasets considered, see the text for their definition. When the $Z_{\mathscr{L}}$ estimator cannot be unambiguously computed (as explained in the text) the corresponding entry is left blank. Whenever different variants or distributions exist for a dataset, we indicate with a (*) the one used in Sec. 3.3.4.

The values of the condition numbers $\lambda_\rho$ and $Z$ reported in Table 3.3 consistently indicate that the experimental correlation and covariance matrices are ill-conditioned for a subset of the analyzed datasets, according to the criterion of Refs. [22, 154] ($Z > 4$). For some of them, such as the ATLAS $d\sigma^Z/d|y_{\ell\bar{\ell}}|$ measurement at 8 TeV, and to a lesser extent for LHCb $d\sigma^Z/dy_{\ell\bar{\ell}}$, this high $Z$ value is explained by the dominance of the luminosity uncertainty: in these cases, $Z_{\mathcal{L}}$ is indeed significantly smaller than $Z$. For all the other datasets, $Z \sim Z_{\mathcal{L}}$. Relatively high values of $Z$ are found for the ATLAS and CMS single inclusive jet and di-jet datasets, a fact that was already observed in the case of the corresponding measurements at 8 TeV, for which various decorrelation models have been proposed and tested [35, 138, 154–156]. We finally observe that the value of $Z$ can fluctuate by a large amount across different differential measurements in the same dataset. For instance, the 13 TeV ATLAS $t\bar{t}$ hadronic dataset provides single differential distributions in $m_{t\bar{t}}$ and in $|y_{t\bar{t}}|$, associated to values of $Z$ respectively of 64.5 and 1.77.

The $\chi^2$ of the datasets listed in Table 3.3 will therefore need to be interpreted with care, in particular taking into account the possibility that it be spuriously high due to a misestimate of experimental correlations. To avoid this issue, in Sec. 3.3.4 we will compute the $\chi^2$ upon regularization of the experimental covariance matrix, for all the datasets with $Z_{\mathcal{L}} > 4$. We use the procedure laid out in Ref. [154]. This procedure consists in clipping the singular values of the correlation matrix to a constant, whenever these are smaller than that, while leaving the rest of the singular vectors unchanged. This way, directions that do not contribute to instability are not affected and the alteration to the original matrix is minimal. The clipping constant is chosen to be $\delta^{-1} = Z$, where the value of $Z = 4$ was determined empirically in Ref. [154]. The values of the $\chi^2$ computed with the unregularized experimental covariance matrix are collected in Appendix A.

#### 3.3.3.3 PDF sets

The computation of the theoretical predictions that enter the $\chi^2$ require a choice of PDFs as input. In this work, we consider the following PDF sets: ABMP16 [31], CT18, CT18A, and CT18Z [32], MSHT20 [33], NNPDF3.1 [157], NNPDF4.0 [22], PDF4LHC15 [44], and PDF4LHC21 [51]. These PDF sets are the most widely used by LHC experimental collaborations in their analyses. The main features of each of them are summarized as follows.

ABMP16 [31]. This PDF determination is based on DIS, Drell-Yan, single top and top-quark pair production measurements. The underlying theory calculations are based on a Fixed flavor Number (FFN) scheme, with $n_f = 3, 4, 5$. The strong coupling constant is determined alongside the PDFs yielding $\alpha_s(m_Z) = 0.1147 \pm 0.0008$ with $n_f = 5$, though a variant with a fixed value $\alpha_s(m_Z) = 0.118$ is also provided. The PDFs are parametrized at the input scale $Q_0 = 1$ GeV with a fixed functional form. The charm PDF is assumed to be purely perturbative, therefore it is generated by partonic DGLAP evolution above the charm quark mass, whose value is a parameter of the fit. Hessian symmetric PDF uncertainties are determined from variations $\Delta\chi^2 = 1$.

CT18 [32]. The CT18 family of PDF determinations is based on DIS, Drell-Yan, single inclusive jet, and top-quark pair production measurements. The underlying theory calculations are based on a General Mass Variable flavor Number (GM-VFN) scheme, specifically ACOT-$\chi$ [158-161], and use a fixed value of the strong coupling as input. Parton distributions are parametrized at the input scale $Q_0 = 1.3$ GeV, equal to the charm pole mass $m_c^{\text{pole}}$, in terms of Bernstein polynomials, the charm PDF is purely perturbative, and Hessian symmetric PDF uncertainties are determined by means of a dynamical tolerance factor $\Delta\chi^2 > 1$. The ATLAS 7 TeV $W/Z$ data [162] is not included in the default CT18 PDF set. Alternate sets are determined including this dataset (CT18A), a new scale choice for low-$x$ DIS data (CT18X), or all of the above with a slightly higher value of the charm mass (CT18Z).

MSHT20 [33]. This PDF determination is based on DIS, Drell-Yan, Drell-Yan with jet, single inclusive jet, and top-quark pair production measurements. The fit is based on the Thorne-Roberts variant of the GM-VFN scheme [163], and uses a fixed value of the strong coupling as input. Parton distributions are parametrized at the input scale $Q_0 = 1$ GeV in terms of Chebyschev polynomials, the charm PDF is purely perturbative (with charm pole mass $m_c^{\text{pole}} = 1.4$ GeV), and Hessian symmetric uncertainties are determined by means of a dynamical tolerance factor $\Delta\chi^2 > 1$.

NNPDF3.1 [53]. This PDF determination is based on DIS, Drell-Yan, Drell-Yan with jet, single inclusive jet, and top-quark pair production measurements. The fit is based on the FONLL GM-VFN scheme [164] and uses a fixed value of the strong coupling constant as input. Parton distributions are parametrized at the initial scale $Q_0 = 1.65$ GeV in terms of deep neural networks, optimized by means of a genetic algorithm. The charm PDF is fitted on the same footing as lighter quark flavors (with a charm pole mass $m_c^{\text{pole}} = 1.51$ GeV). PDF uncertainties are determined from a Monte Carlo sampling of experimental uncertainties.

NNPDF4.0 [22]. This PDF determination is based on DIS, Drell-Yan, Drell-Yan with jet, single inclusive jet and di-jet, single top and top-quark pair, and prompt photon production measurements. The fit is based on the same treatment of quark masses, running coupling, charm quark PDF, and uncertainty representation as NNPDF3.1. In comparison to NNPDF3.1, NNPDF4.0 is however characterized by several methodological differences: newer theoretical constraints, in particular on PDF positivity and integrability, are implemented; PDFs are parametrized with a single neural network, optimized by means of gradient descent; hyperparameters, such as those that define the architecture of the neural network, are determined by means of an automated scan of the space of models that selects the optimal one [22, 165]; and the methodology is closure tested [166].

PDF4LHC15 [44]. This PDF set is the Monte Carlo combination of the CT14 [167], MMHT2014 [168], and NNPDF3.0 [169] PDF sets. The combination is performed by first converting the CT14 and MMHT2014 Hessian PDF sets into Monte Carlo PDF sets by means of the algorithm developed in [45, 46]. For each of the three PDF sets 300 Monte Carlo replicas are generated, that are subsequently collated in a single set. The number of replicas is finally reduced by means of the compression algorithm developed in [50] or converted to a single Hessian set by means of the algorithm developed in [48].

PDF4LHC21 [51]. This PDF set is the Monte Carlo combination of the CT18′, MSHT20, and NNPDF3.1′ PDF sets. The CT18′ and NNPDF3.1′ PDF sets are variants of the CT18 and NNPDF3.1 PDF sets: both of these differ from the corresponding baseline sets for the values of the charm and bottom quark pole masses, which are set to values common to those used in MSHT20, $m_c^{\text{pole}} = 1.4$ GeV and $m_b^{\text{pole}} = 4.75$ GeV. The NNPDF3.1′ PDF set differs from NNPDF3.1 for a number of additional variations in the input dataset and in the details of the theoretical computations, see Sec. 2.3 in [51]. The combination is carried out as in PDF4LHC15.

In all cases, we use PDF sets accurate to NNLO with a common, fixed value of $\alpha_s(m_Z) = 0.118$. Note that NNLO corrections to hadronic processes were included in all of the aforementioned PDF sets by means of $K$-factors, whereas here we make predictions by means of exact NNLO computations. This fact is however immaterial, given the very weak dependence of $K$-factors on PDFs [72]. In the case of ABMP16, we use the set with $n_f = 5$ active flavors. For ABMP16, CT18, and MSHT20, we consider Hessian sets; for NNPDF3.1, NNPDF4.0, PDF4LHC15, and PDF4LHC21, we consider Monte Carlo sets composed of 100 replicas. In Fig. 3.2 we compare the partonic luminosities, defined by Eqs. (1-4) of [170], obtained with the ABMP16, CT18, MSHT20, NNPDF4.0, and PDF4LHC21 PDF sets. Results are displayed as a function of the invariant mass of the final state $m_X$ at a center-of-mass energy $\sqrt{s} = 13$ TeV and are normalized to PDF4LHC21. Comparison using other PDF sets can be seen in [29].

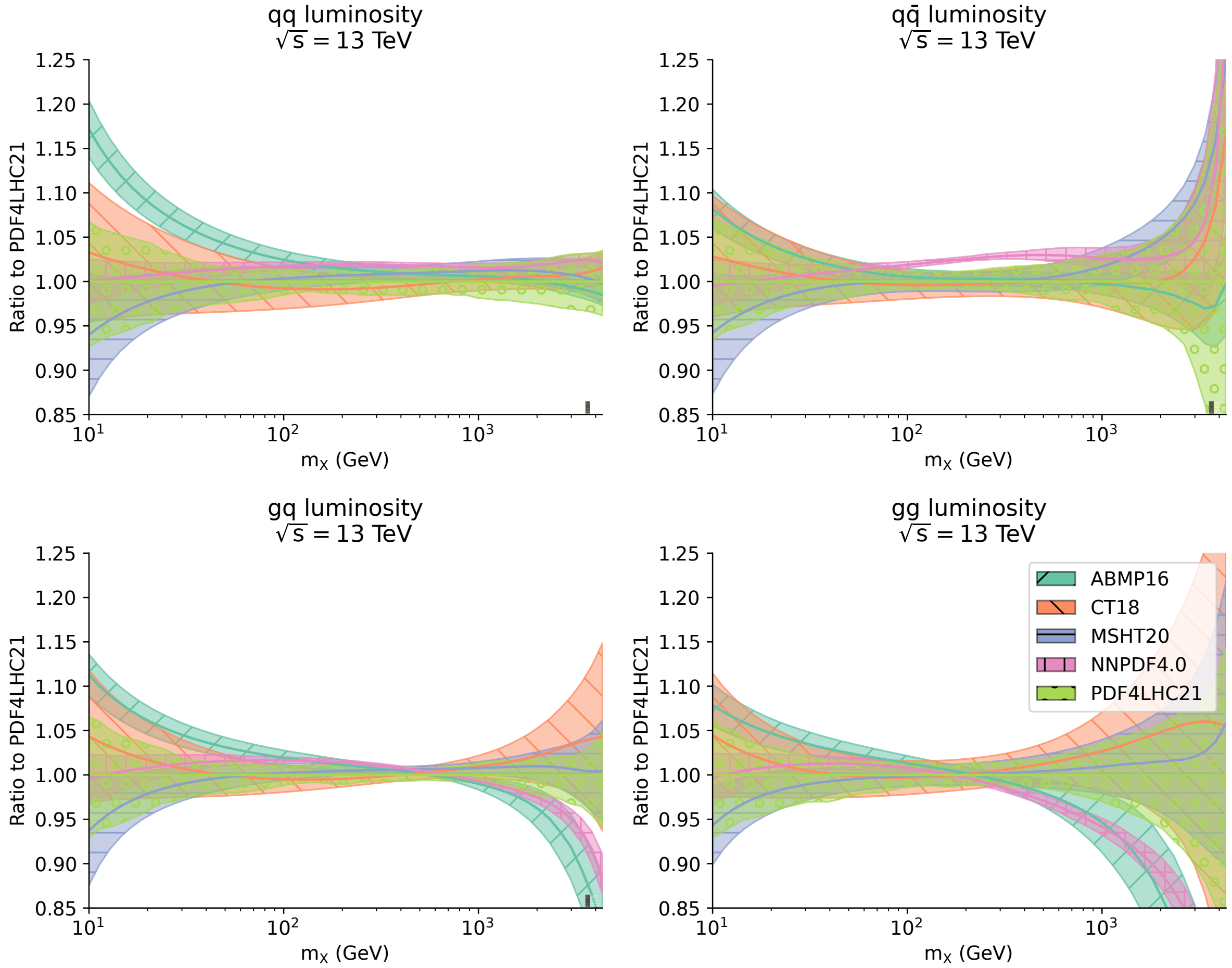

Figure 3.2: The quark-quark (top left), quark-antiquark (top right), gluon-quark (bottom left), and gluon-gluon (bottom right) partonic luminosities, Eqs. (1-4) of [170], as a function of the invariant mass of the final state $m_X$ at a center-of-mass energy $\sqrt{s} = 13$ TeV obtained with the ABMP16, CT18, MSHT20, NNPDF4.0, and PDF4LHC21 PDF sets. Results are normalized to PDF4LHC21.

We do not consider PDF sets including QED corrections [171–173], aN$^3$LO QCD corrections [68, 74] or MHOU [20], the reason being that these are not commonly used in LHC experimental analyses. This said, the computation of the $\chi^2$ does not change if one uses any of these PDF sets. We will study the phenomenological implications of QED, aN$^3$LO, and of MHOU corrections to the PDFs in the appraisal of LHC data in future work. An exception is represented by the high-precision ATLAS 8 TeV inclusive dilepton rapidity measurement [67], for which predictions based on the NNPDF4.0 QED [173], MHOU [20], and aN$^3$LO [74] PDF sets will be considered in Sec. 3.3.4.2.

### 3.3.4 Results

In this section, we quantify the agreement between the experimental data and the corresponding theoretical predictions presented in Sec. 3.3.2 according to the estimators and upon variations of the input PDF sets discussed in Sec. 3.3.3. We examine datasets for each process in turn. For each of these, we provide three complementary ways of visualizing the data-theory agreement: a table with the values of $\chi^2_{\rm exp+th}$, and $\chi^2_{\rm exp}$, Eqs. (3.18) and (3.19), evaluated with all the PDF sets summarized in Sec. 3.3.3.3[1]; a set of histograms in which the total $\chi^2_{\rm exp+th}$, Eq. (3.18), is split into the components $\chi^2_{\rm exp+mho}$, Eq. (3.20), and $\chi^2_{\rm exp+th}$, Eq. (3.19), albeit only for CT18, MSHT20, NNPDF4.0, and PDF4LHC21; and a set of data-theory comparison plots, only for NNPDF4.0 and PDF4LHC21, in which the PDF+$\alpha_s$ and MHO uncertainties are displayed separately for selected data points. For all measurements with $Z_{\mathscr{L}} > 4$ (see Table 3.3), the experimental covariance matrix is regularized as explained in Sec. 3.3.3.2. We finally provide a collective visualization of the $\Delta\chi^{2(i)}$ and $\Delta n_\sigma^{(i)}$ estimators, Eqs. (3.22) and (3.23), for the CT18, MSHT20, NNPDF4.0, and PDF4LHC21 PDF sets. The values of $\chi^2_{\rm exp+th}$ obtained without regularization of the experimental covariance matrix are given, for the subset of measurements with $Z_{\mathscr{L}} > 4$, in Appendix A. Additional histogram and data-theory comparison plots, for the subset of measurements not highlighted in this section, are given in Appendix B.

#### 3.3.4.1 Drell-Yan weak boson production measurements at 13 TeV

We start by considering the three LHC Drell-Yan weak boson production measurements at a center-of-mass energy of 13 TeV outlined in Sec. 3.3.2. The values of $\chi^2_{\rm exp}$ and $\chi^2_{\rm exp+th}$, computed with each of the PDF sets summarized in Sec. 3.3.3.3, are reported in Table 3.4. The experimental covariance matrix of the CMS dataset is regularized as explained in Sec. 3.3.3.2, see Appendix A for the unregularized values of $\chi^2_{\rm exp+th}$. The breakdown of $\chi^2_{\rm exp+th}$ into $\chi^2_{\rm exp+mho}$ and $\chi^2_{\rm exp}$ is displayed in Fig. 3.3. The data-theory comparison is displayed in Fig. 3.4. Each plot consists of three panels: the upper one displays the measured and predicted cross sections, with experimental and total (MHO and PDF+$\alpha_s$) theoretical uncertainties; the middle one displays the same cross sections normalized to the experimental central value; the lower one displays the relative PDF+$\alpha_s$ and MHO uncertainties separately. Experimental error bars correspond to the total uncorrelated uncertainty. Correlated uncertainties are included by shifting the central experimental value, by an amount determined as explained in Appendix B of [153].

[1] For ease of reference, in this table we also provide the number of data points $n_{\rm dat}$ and the standard deviation of the $\chi^2$ per point distribution $\sqrt{2/n_{\rm dat}}$

| Dataset | $n_{\rm dat}$ | $\sqrt{2/n_{\rm dat}}$ | | ABMP16 | CT18 | CT18A | CT18Z | MSHT20 | NNPDF3.1 | NNPDF4.0 | PDF4LHC15 | PDF4LHC21 |
|---|---|---|---|---|---|---|---|---|---|---|---|---|
| ATLAS 13 TeV $Z\ \frac{1}{\sigma}\frac{d\sigma}{dp_T^{\ell\ell}}$ | 38 | 0.23 | $\chi^2_{\rm exp+th}$ | 0.36 | 0.31 | 0.42 | 0.59 | 0.40 | 0.39 | 0.50 | 0.31 | 0.38 |
| | | | $\chi^2_{\rm exp}$ | 0.80 | 1.18 | 2.38 | 4.91 | 1.58 | 1.20 | 2.20 | 0.83 | 1.64 |
| CMS 13 TeV $W^+\ \frac{d\sigma}{d\|\eta\|}$ | 18 | 0.33 | $\chi^2_{\rm exp+th}$ | 1.31 | 1.20 | 1.11 | 1.06 | 1.26 | 0.85 | 0.96 | 1.15 | 0.98 |
| | | | $\chi^2_{\rm exp}$ | 1.41 | 1.67 | 1.30 | 1.31 | 1.37 | 0.97 | 1.12 | 1.38 | 1.27 |
| CMS 13 TeV $W^-\ \frac{d\sigma}{d\|\eta\|}$ | 18 | 0.33 | $\chi^2_{\rm exp+th}$ | 1.56 | 1.15 | 1.11 | 1.10 | 1.43 | 1.12 | 1.60 | 1.14 | 1.20 |
| | | | $\chi^2_{\rm exp}$ | 1.60 | 1.89 | 1.43 | 1.38 | 1.57 | 1.64 | 1.95 | 1.54 | 1.54 |
| LHCb 13 TeV $Z\ \frac{d\sigma}{dy^Z}$ | 18 | 0.33 | $\chi^2_{\rm exp+th}$ | 2.14 | 2.19 | 2.26 | 2.08 | 2.28 | 2.21 | 2.26 | 2.15 | 2.07 |
| | | | $\chi^2_{\rm exp}$ | 2.28 | 3.09 | 2.91 | 2.62 | 2.66 | 2.70 | 2.48 | 3.06 | 2.67 |

Table 3.4: The values of $\chi^2_{\rm exp+th}$, Eq. (3.18), and of $\chi^2_{\rm exp}$, Eq. (3.19), for the ATLAS, CMS, and LHCb Drell-Yan gauge boson production measurements at the LHC 13 TeV of Table 3.2, computed with each of the PDF sets summarized in Sec. 3.3.3.3. The experimental covariance matrix of the CMS dataset is regularized as explained in Sec. 3.3.3.2. The unregularized values of $\chi^2_{\rm exp}$ are collected in table A.1 of Appendix A.

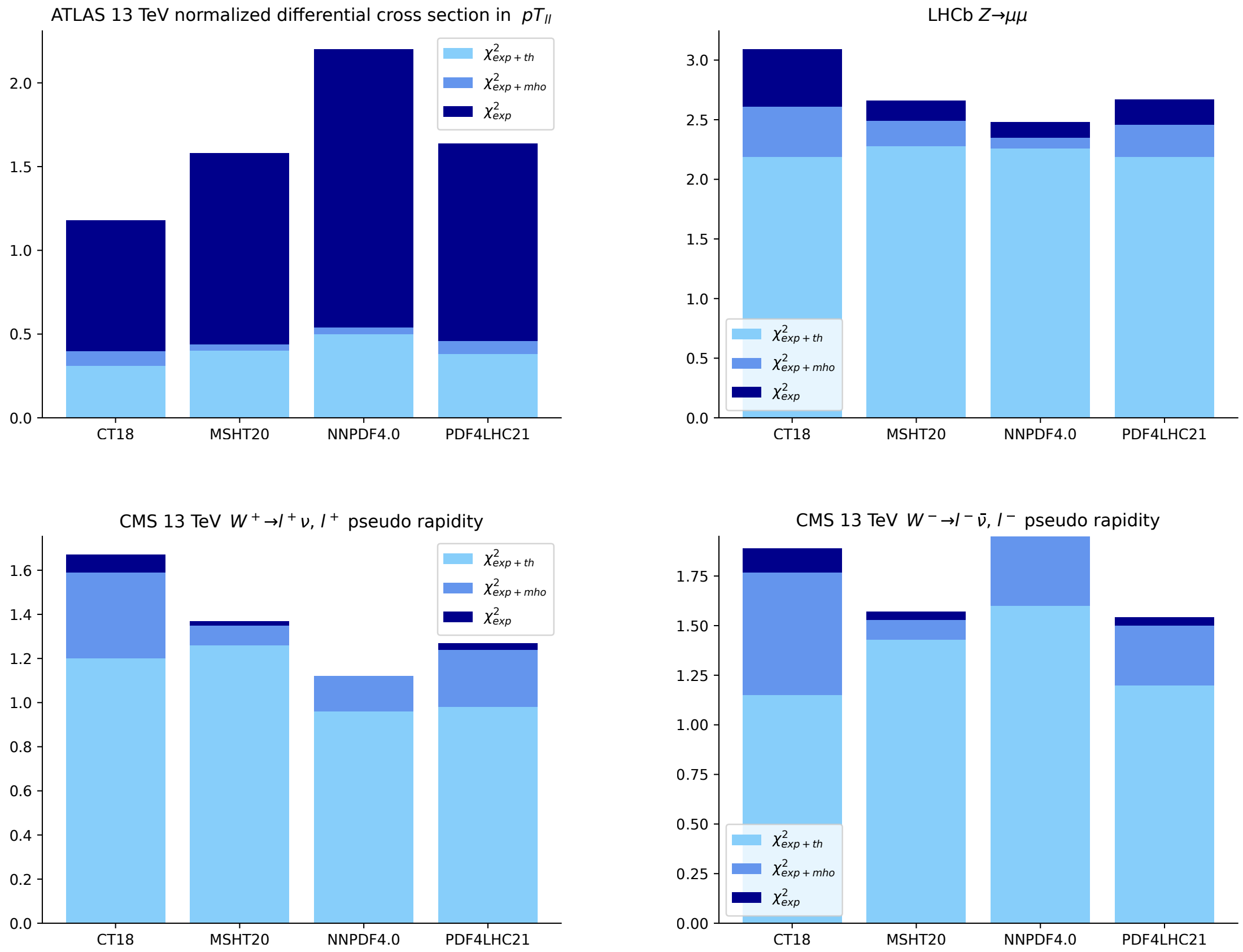

Figure 3.3: The breakdown of $\chi^2_{\text{exp+th}}$, Eq. (3.18), into $\chi^2_{\text{exp+mho}}$, Eq. (3.20), and $\chi^2_{\text{exp}}$, Eq. (3.19), for the ATLAS ($n_{\text{dat}} = 38$, $\sqrt{2/n_{\text{dat}}} = 0.23$), CMS (each set made of $n_{\text{dat}} = 18$, $\sqrt{2/n_{\text{dat}}} = 0.23$), and LHCb ($n_{\text{dat}} = 18$, $\sqrt{2/n_{\text{dat}}} = 0.33$) Drell-Yan gauge boson production measurements at the LHC 13 TeV. Note that the inclusion of MHOUs has a negligible impact on the $\chi^2$ values of the CMS 13 TeV $W$ distributions when theoretical predictions are computed with NNPDF4.0, hence the dark blue component of the corresponding histogram is hardly visible.

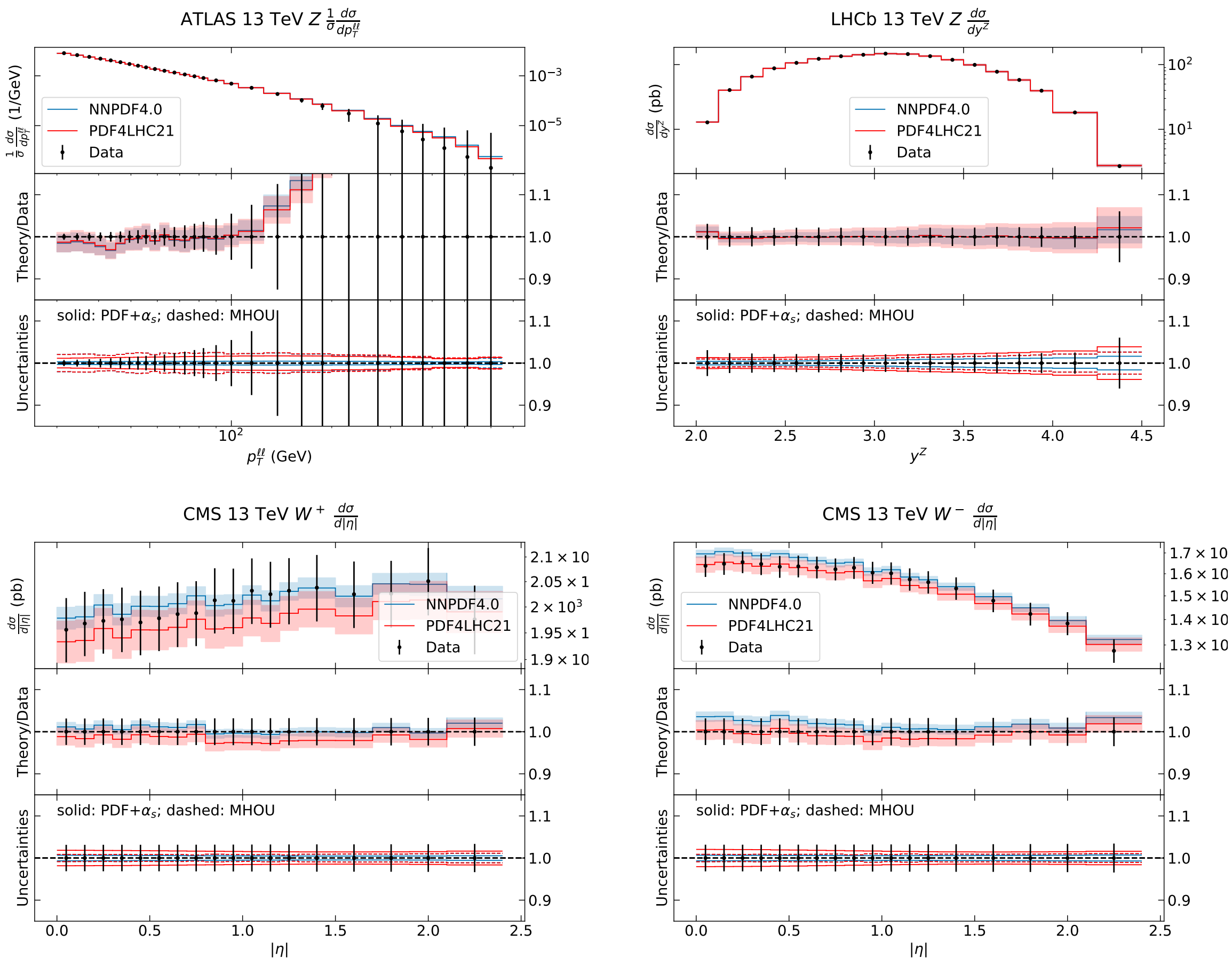

Figure 3.4: Data-theory comparison for the ATLAS, CMS, and LHCb Drell-Yan gauge boson production measurements at the LHC 13 TeV of Table 3.2. (Upper panels) The measured and predicted cross sections, with experimental and total (MHO and PDF+$\alpha_s$) theoretical uncertainties. (Middle panels) The same cross sections normalized to the experimental central value. (Lower panels) The relative PDF+$\alpha_s$ (dashed) and MHO (solid) uncertainties separately. In all panels, the experimental error bars correspond to the total uncorrelated uncertainty. Correlated uncertainties are kept into account by shifting the central experimental value as explained in Appendix B of [153].

From inspection of Table 3.4 and of Fig. 3.3, we observe that the values of $\chi^2_{\rm exp+th}$, computed with different input PDFs, are generally closer to each other than the corresponding values of $\chi^2_{\rm exp}$. This fact suggests that the inclusion of theory uncertainties is essential to assess the predictive power of a given PDF set. Moreover, the values of $\chi^2_{\rm exp+th}$ are very similar across PDF sets: this is manifest in the case of the ATLAS and LHCb datasets, and true on average for the CMS dataset. In this latter case, the PDF sets with larger values of $\chi^2_{\rm exp+th}$ on the $W^+$ dataset have the smaller values of $\chi^2_{\rm exp+th}$ on the $W^-$ dataset, and the other way around. For the CMS $W^+$ distribution, whereas the total $\chi^2_{\rm exp+th}$ remains within $1\sigma$ of the $\chi^2$ distribution per unit point for all PDF sets, the experimental $\chi^2_{\rm exp}$ is close to one only for NNPDF3.1 and NNPDF4.0. In the case of the CMS $W^-$ distribution, the ABMP16 and NNPDF4.0, and to a lesser extent the MSHT20 set do yield a somewhat worse description, in that the corresponding values of $\chi^2_{\rm exp+th}$ are almost $2\sigma$ away from the unit expectation. For the purely experimental $\chi^2_{\rm exp}$, the predictions obtained with NNPDF4.0 and CT18 are almost $3\sigma$ away from one. Despite these differences, when all uncertainties are kept into account, we cannot single out a PDF set that, overall, generalizes better than another on these datasets.

The breakdown of $\chi^2_{\text{exp+th}}$ into its theoretical components depends on the dataset and on the PDF set. The component due to MHO, gauged from the difference between $\chi^2_{\text{exp}}$ and $\chi^2_{\text{exp+mho}}$, dominates the ATLAS measurement, irrespective of the PDF set, whereas it is less prominent in the other datasets. For CMS, this is almost immaterial, irrespective of the PDF set. For LHCb, irrespective of the PDF set, this is typically as large as the component due to PDF+$\alpha_s$ uncertainties, gauged from the difference between $\chi^2_{\text{exp+mho}}$ and $\chi^2_{\text{exp+th}}$. This latter component may depend on the PDF set, being usually larger for PDF sets affected by the largest uncertainties, such as CT18 and PDF4LHC21, see Fig. 3.2. All these facts are a consequence of how the various partonic channels contribute to the cross sections of these processes. The ATLAS measurement receives its leading contribution, which is $\mathcal{O}(\alpha_s)$, from the quark-gluon partonic luminosity. The CMS and LHCb measurements receive their leading contributions, which are $\mathcal{O}(\alpha_s^0)$, from quark-antiquark partonic luminosities, yet in different regions of $x$, given that they are central and forward rapidity measurements: the former at intermediate values of $x$; the latter at large values of $x$.

The quality of the data description is generally good, being $\chi^2_{\text{exp+th}} \sim 1$, except for LHCb, for which $\chi^2_{\text{exp+th}} \sim 2$, irrespective of the PDF set. For ATLAS, $\chi^2_{\text{exp+th}} \sim 0.4$, again irrespective of the PDF set. This value is anomalously small, and may point towards the fact that MHOU are actually overestimated by scale variations. Discrepancies between data and theory that may lead to these results are seen in Fig. 3.4, where the alignment of experimental data and theoretical predictions is optimal, within their uncertainties, for all datasets. We therefore conclude that the somewhat high $\chi^2_{\text{exp+th}}$ for LHCb is due to experimental correlations, and will likely decrease once the dataset is included in a fit. Note finally that the quality of the data description of the CMS measurement would have been rather worse, at face value, had the regularization procedure described in Sec. 3.3.3.2 not been applied. The values of $\chi^2_{\text{exp+th}}$ obtained without it are reported in Appendix A. As we can see from Fig. 3.4, theoretical predictions are almost spot on experimental measurements. The otherwise very large values of the $\chi^2$ obtained without regularization are spurious, and denote an ill-conditioning of their experimental covariance matrix.

#### 3.3.4.2 The ATLAS 8 TeV inclusive $Z$ boson production measurement

We then consider the ATLAS measurement of Drell-Yan Z boson production at the LHC 8 TeV outlined in Sec. 3.3.2. The values of $\chi^2_{\text{exp}}$ and $\chi^2_{\text{exp+th}}$, computed with each of the PDF sets summarized in Sec. 3.3.3.3, are reported in Table 3.5. The breakdown of $\chi^2_{\text{exp+th}}$ into $\chi^2_{\text{exp+mho}}$ and $\chi^2_{\text{exp}}$ and the data-theory comparison are displayed in Fig. 3.5, in the same format as Figs. 3.3 and 3.4.

| Dataset | $n_{\rm dat}$ | $\sqrt{2/n_{\rm dat}}$ | | *ABMP16* | *CT18* | *CT18A* | *CT18Z* | *MSHT20* | *NNPDF3.1* | *NNPDF4.0* | *PDF4LHC15* | *PDF4LHC21* |
|---|---|---|---|---|---|---|---|---|---|---|---|---|
| ATLAS 8 TeV $Z\ \frac{d\sigma}{d\lvert y\rvert}$ | 7 | 0.53 | $\chi^2_{\rm exp+th}$ | 4.25 | 1.52 | 1.52 | 1.18 | 1.37 | 1.61 | 3.83 | 1.23 | 1.09 |
| | | | $\chi^2_{\rm exp}$ | 7.36 | 14.0 | 4.63 | 4.31 | 2.14 | 4.70 | 7.90 | 7.41 | 1.93 |

Table 3.5: Same as Table 3.4 for the ATLAS Drell-Yan gauge boson production measurements at the LHC 8 TeV [67].

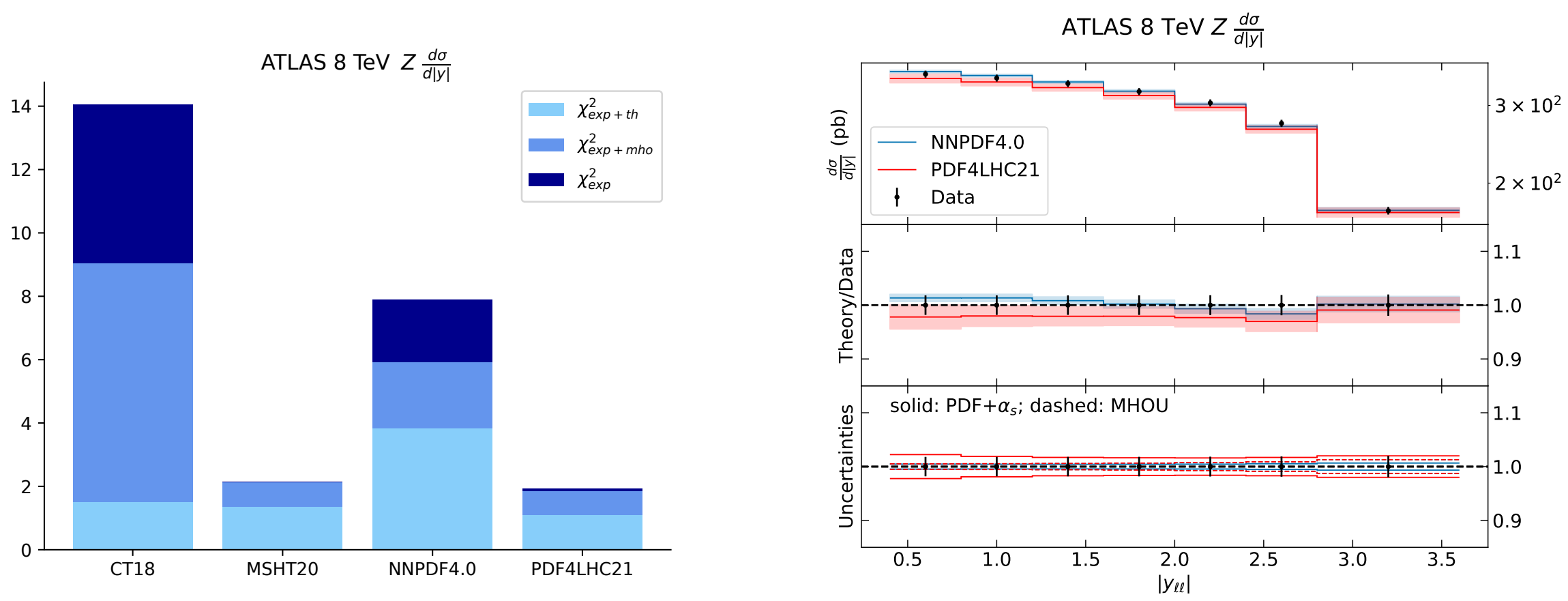

Figure 3.5: Same as Figs. 3.3 (left) and 3.4 (right) for the ATLAS Drell-Yan gauge boson production measurements at the LHC 8 TeV [67] ($n_{\rm dat} = 7$, $\sqrt{2/n_{\rm dat}} = 0.53$).

From inspection of Table 3.5 and Fig. 3.5, we observe that the values of $\chi^2_{\rm exp+th}$ decrease significantly with respect to $\chi^2_{\rm exp}$. As already remarked for the other Drell-Yan data, this fact further indicates that a careful account of theoretical uncertainties is crucial to assess the predictive power of a PDF set. For CT18 and NNPDF4.0, the MHO and PDF+$\alpha_s$ contributions to the $\chi^2$ have approximately the same size, and are relatively large. For MSHT20 and PDF4LHC21, the MHO contribution to the $\chi^2$ is essentially immaterial. This is possibly due to the fact that there is a large variance in the quality of the description of this dataset before including theoretical uncertainties in the computation of the $\chi^2$: even if all PDF sets provide an unsatisfactory description of the data, MSHT20 and PDF4LHC21 have a $\chi^2_{\rm exp}$ of order 2, whereas all of the others have a $\chi^2_{\rm exp}$ of order 5-10. Once theoretical uncertainties are included, one gets $\chi^2_{\rm exp+th}$ of the order of 1, except for ABMP16 and NNPDF4.0, for which $\chi^2_{\rm exp+th}$ is equal to 3.47 and 3.83. The discrepancy between experimental data and theoretical predictions obtained with NNPDF4.0 instead of PDF4LHC21 is visible in the right panel of Fig. 3.5. The shape of the NNPDF4.0 prediction displays a peculiar dip around a value of the dilepton rapidity of 2.7.

This is a case in which NNPDF4.0 seems to perform poorly. This fact is a little surprising because, as discussed in Sec. 3.3.4.1, an earlier version [120] of the ATLAS measurement of [67] was included in NNPDF4.0. This is a substantial difference with respect to all of the other measurements examined in this work, which, albeit corresponding to production processes already included in NNPDF4.0, are completely out of sample. For this reason, we consider that, only in this case, some more extensive investigations are needed. To this purpose, we recompute the values of $\chi^2_{\rm exp}$ and $\chi^2_{\rm exp+th}$ using the NNPDF4.0 PDF sets that include QED corrections [173], MHOUs [20], and aN$^3$LO corrections and MHOUs [74]. All these PDF sets include the ATLAS Drell-Yan $Z$ boson production measurements at 8 TeV presented in [119, 120]. Furthermore, to understand the interplay of these measurements with the new version considered here [67] (see Sec. 3.3.2 for details), we perform the following additional fits:

(a) a fit equivalent to the NNLO NNPDF4.0 baseline fit excluding the ATLAS measurement of [120];

(b) a fit equivalent to the NNLO NNPDF4.0 baseline fit in which the ATLAS measurement of [120] is replaced with that of [67];

(c) a fit equivalent to fit (b), in which the ATLAS measurement of [67] is weighted as explained in Sec. 4.2.3 of [22];

(d) a fit equivalent to fit (b), including MHOUs.

| Dataset | $n_{\rm dat}$ | $\sqrt{2/n_{\rm dat}}$ | | *NNPDF4.0* | *aN$^3$LO MHOU* | *MHOU* | *QED* | *fit (a)* | *fit (b)* | *fit (c)* | *fit (d)* |
|---|---|---|---|---|---|---|---|---|---|---|---|
| ATLAS 8 TeV $Z$ $\frac{d\sigma}{d\|y\|}$ [67] | 7 | 0.53 | $\chi^2_{\rm exp+th}$ | 3.83 | 3.32 | 3.33 | 3.93 | 3.43 | 2.24 | 0.17 | 1.95 |
| | | | $\chi^2_{\rm exp}$ | 7.90 | 8.42 | 8.38 | 8.77 | 7.24 | 3.49 | 0.18 | 3.19 |
| ATLAS 8 TeV $Z$ $\frac{d\sigma}{d\|y\|}$ [120] | 60 | 0.18 | $\chi^2_{\rm exp+th}$ | 1.08 | 1.05 | 1.01 | 1.09 | 1.08 | 1.06 | 1.02 | 1.01 |
| | | | $\chi^2_{\rm exp}$ | 1.23 | 1.18 | 1.11 | 1.25 | 1.24 | 1.28 | 1.41 | 1.17 |
| ATLAS 8 TeV $Z$ $\frac{d\sigma}{d\|y\|}$ (at $Z$-peak) [120] | 24 | 0.29 | $\chi^2_{\rm exp+th}$ | 1.30 | 1.27 | 1.27 | 1.29 | 1.31 | 1.30 | 1.28 | 1.28 |
| | | | $\chi^2_{\rm exp}$ | 1.31 | 1.28 | 1.27 | 1.30 | 1.31 | 1.31 | 1.28 | 1.31 |

Table 3.6: Same as Table 3.5 for the baseline NNPDF4.0 PDF set and for the additional NNPDF4.0-like PDF sets described in the text. Values are displayed separately for the measurement from [67], for the measurement from [120], and for the subset of the latter corresponding to the invariant mass bin of the $Z$ peak.

The values of $\chi^2_{\rm exp}$ and $\chi^2_{\rm exp+th}$ computed with the baseline NNPDF4.0 PDF set and with all the aforementioned PDF sets are collected in Table 3.6. Values are displayed for the ATLAS measurement of [67], which is included only in fits (b), (c), and (d), for the ATLAS measurement of [120], which is included in the NNPDF4.0, aN$^3$LO MHOU, MHOU, and QED fits, and for the subset of the ATLAS measurement of [120] corresponding to the invariant mass bin of the $Z$ peak. This way, the kinematic coverage is the same as in [67]. The results corresponding to the NNPDF4.0 baseline fit are the same as in Table 3.5.

From Table 3.6, we make the following conclusions. The ATLAS dataset of [120] is described fairly well by NNPDF4.0, including the bin corresponding to the $Z$-peak invariant mass, whereas the dataset of [67] is not, even when accounting for theoretical uncertainties in the computation of $\chi^2_{\rm exp+th}$. This state of affairs does not change if one considers variants of the NNPDF4.0 PDF sets including N$^3$LO corrections, MHOUs, or QED corrections. It is therefore unlikely that theoretical inaccuracy is a limitation in the description of the ATLAS measurement of [67]. The ATLAS dataset of [120] is described with comparable quality by a PDF set determined from a fit without it (fit (a)); in this case, the description of the ATLAS dataset of [67] does not improve in a significant way. If instead one tries to fit the ATLAS measurement of [67] (fit (b)), the value of $\chi^2_{\rm exp+th}$ ($\chi^2_{\rm exp}$) improves by about $2\sigma$ ($8\sigma$). At the same time, the description of the ATLAS measurement of [120] does not change in a significant way. We therefore conclude that the old and new measurements are not in tension between each other. The picture can be further improved if one repeats fit (b) with inclusion of MHOUs (fit (d)): in this case, the values of $\chi^2_{\rm exp+th}$ and $\chi^2_{\rm exp}$ reduce by about another half of a sigma. One may finally wonder whether the ATLAS measurement of [67] is in tension with other datasets included in NNPDF4.0. In this respect, fit (c) reveals that an overly good description of the dataset can be achieved if it is overweighted. The description of the ATLAS measurement of [120] is not significantly altered, in comparison to the other fits, thus confirming that the two measurements are consistent with each other. However, the global fit quality deteriorates significantly: the total $\chi^2_{\rm exp}$ per point increases from 1.16 (in the default NNPDF4.0 fit) to 1.24. Because there are about 4600 fitted data points, this corresponds to a worsening of about $4\sigma$. Inspection of individual dataset figures reveals that this is due to a deterioration in the description of several Drell-Yan measurements, which see the following increase in the value of $\chi^2_{\rm exp}$: D0 $W$ muon asymmetry production [174] from 1.91 to 4.27 ($n_{\rm dat} = 9$); ATLAS $W$, $Z$ production, 7 TeV [175] from 1.67 to 3.07 ($n_{\rm dat} = 53$); LHCb $Z$ production, 7 TeV [176] from 1.65 to 2.48 ($n_{\rm dat} = 9$); LHCb $W^\pm$ production, 7 TeV [177] from 1.97 to 4.12 ($n_{\rm dat} = 29$); and LHCb $Z$ production, 8 TeV [178] from 1.33 to 2.32 ($n_{\rm dat} = 17$). We therefore conclude that the ATLAS measurement of [67] is in tension with these other datasets.

In summary, the ATLAS measurement of [67] is consistent with its earlier version [120], but in tension with other Drell-Yan measurements included in NNPDF4.0. An acceptable description of it can be achieved if this is included in the fit and if MHOU are taken into account. The value of $\chi^2_{\rm exp+th} = 1.95$ is indeed only slightly less than $2\sigma$ away from the unit expectation. This level of disagreement does not appear to be more pathological than that observed for few other datasets included in NNPDF4.0, and is also similar to that observed for other datasets examined in the following. Note however that, for all of the other datasets examined in this work, we will look at the data-theory (dis)agreement only before inclusion in a fit. Refitting could possibly improve the overall picture, however investigations along this direction are beyond the scope of this work, as they will be part of the dataset selection involved with a future NNPDF release.

#### 3.3.4.3 Top-quark pair production measurements

We continue by discussing the LHC top-quark pair production measurements outlined in Sec. 3.3.2, see also Table 3.3. The values of $\chi^2_{\rm exp}$ and $\chi^2_{\rm exp+th}$, computed for each of the PDF sets summarized in Sec. 3.3.3.3, are reported in Table 3.7. The experimental covariance matrix is regularized as explained in Sec. 3.3.3.2 for the following datasets: the ATLAS all-hadronic absolute single differential distribution in the invariant mass of the top-quark pair and double differential distribution in the invariant mass and absolute rapidity of the top-quark pair; the ATLAS lepton+jets normalized single differential distributions in the invariant mass of the top-quark pair and in the transverse momentum of the top quark; and the CMS lepton+jets normalized double differential distribution in the invariant mass and absolute rapidity of the top-quark pair. See Appendix A for the unregularized values of $\chi^2_{\rm exp+th}$. The breakdown of $\chi^2_{\rm exp+th}$ into $\chi^2_{\rm exp+mho}$ and $\chi^2_{\rm exp}$ is displayed in Fig. 3.6, albeit only for a representative subset of distributions, specifically: the ATLAS lepton+jets normalized cross sections, single differential in the transverse momentum of the top quark, $p_T^t$, and in the invariant mass of the top-quark pair, $m_{t\bar{t}}$; the CMS lepton+jets normalized cross sections, single differential in the absolute rapidity of the top quark and of the top-quark pair, $|y_t|$ and $|y_{t\bar{t}}|$; the ATLAS all-hadronic absolute cross section, double differential in the invariant mass and absolute rapidity of the top-quark pair; and the CMS lepton+jets normalized cross section, double differential in the invariant mass and absolute rapidity of the top-quark pair. Histogram plots for the other datasets are collected in Fig. B.1 of Appendix B. The data-theory comparison is displayed in Fig. 3.7 for the same representative subset of top-quark pair measurements of Fig. 3.6. In the case of the ATLAS and CMS double differential distributions, only the bin at the lowest invariant mass is shown. Additional results are collected in Figs. B.2-B.3 of Appendix B. Note that, for normalized distributions, we consistently do not display the last bin, which is linearly dependent from the others by construction. Hence the number of data points displayed is one unit less than the number of data points reported in Table 3.3.

| Dataset | $n_{\rm dat}$ | $\sqrt{2/n_{\rm dat}}$ | | ABMP16 | CT18 | CT18A | CT18Z | MSHT20 | NNPDF3.1 | NNPDF4.0 | PDF4LHC15 | PDF4LHC21 |
|---|---|---|---|---|---|---|---|---|---|---|---|---|
| ATLAS 13 TeV $t\bar{t}$ all hadr. $\frac{d\sigma}{dm_{t\bar{t}}}$ | 9 | 0.47 | $\chi^2_{\rm exp+th}$ | 0.84 | 0.99 | 0.97 | 0.94 | 0.97 | 0.86 | 0.81 | 0.96 | 0.93 |
| | | | $\chi^2_{\rm exp}$ | 0.88 | 1.21 | 1.16 | 1.15 | 1.12 | 0.91 | 0.84 | 1.13 | 1.06 |
| ATLAS 13 TeV $t\bar{t}$ all hadr. $\frac{1}{\sigma}\frac{d\sigma}{d\lvert y_{t\bar{t}}\rvert}$ | 12 | 0.41 | $\chi^2_{\rm exp+th}$ | 0.62 | 0.78 | 0.77 | 0.85 | 0.74 | 0.64 | 0.68 | 0.73 | 0.73 |
| | | | $\chi^2_{\rm exp}$ | 0.68 | 0.85 | 0.83 | 0.95 | 0.79 | 0.67 | 0.71 | 0.82 | 0.78 |
| ATLAS 13 TeV $t\bar{t}$ all hadr. $\frac{d^2\sigma}{dm_{t\bar{t}}d\lvert y_{t\bar{t}}\rvert}$ | 11 | 0.43 | $\chi^2_{\rm exp+th}$ | 0.92 | 1.38 | 1.39 | 1.42 | 1.48 | 1.12 | 1.22 | 1.22 | 1.39 |
| | | | $\chi^2_{\rm exp}$ | 1.05 | 2.55 | 2.38 | 2.84 | 2.08 | 1.20 | 1.29 | 2.11 | 2.07 |
| ATLAS 13 TeV $t\bar{t}\,\ell+j\,\frac{1}{\sigma}\frac{d\sigma}{dm_{t\bar{t}}}$ | 9 | 0.47 | $\chi^2_{\rm exp+th}$ | 1.41 | 1.17 | 1.17 | 1.04 | 1.18 | 1.46 | 1.39 | 1.20 | 1.19 |
| | | | $\chi^2_{\rm exp}$ | 1.67 | 1.26 | 1.26 | 1.12 | 1.27 | 1.65 | 1.57 | 1.32 | 1.31 |
| ATLAS 13 TeV $t\bar{t}\,\ell+j\,\frac{1}{\sigma}\frac{d\sigma}{dp_T^t}$ | 8 | 0.50 | $\chi^2_{\rm exp+th}$ | 0.56 | 0.54 | 0.54 | 0.52 | 0.53 | 0.56 | 0.53 | 0.53 | 0.53 |
| | | | $\chi^2_{\rm exp}$ | 0.76 | 0.68 | 0.68 | 0.67 | 0.69 | 0.77 | 0.72 | 0.68 | 0.70 |
| ATLAS 13 TeV $t\bar{t}\,\ell+j\,\frac{d\sigma}{d\lvert y_t\rvert}$ | 5 | 0.63 | $\chi^2_{\rm exp+th}$ | 1.39 | 1.05 | 1.09 | 0.92 | 1.10 | 1.70 | 1.58 | 1.09 | 1.15 |
| | | | $\chi^2_{\rm exp}$ | 1.62 | 1.17 | 1.19 | 1.00 | 1.14 | 1.86 | 1.62 | 1.26 | 1.29 |
| ATLAS 13 TeV $t\bar{t}\,\ell+j\,\frac{1}{\sigma}\frac{d\sigma}{d\lvert y_{t\bar{t}}\rvert}$ | 7 | 0.53 | $\chi^2_{\rm exp+th}$ | 0.57 | 0.43 | 0.42 | 0.58 | 0.47 | 0.58 | 0.42 | 0.42 | 0.39 |
| | | | $\chi^2_{\rm exp}$ | 0.74 | 0.57 | 0.55 | 0.99 | 0.66 | 0.65 | 0.47 | 0.56 | 0.47 |
| CMS 13 TeV $t\bar{t}\,\ell+j\,\frac{1}{\sigma}\frac{d\sigma}{dm_{t\bar{t}}}$ | 15 | 0.37 | $\chi^2_{\rm exp+th}$ | 0.24 | 0.49 | 0.51 | 0.53 | 0.57 | 0.29 | 0.33 | 0.42 | 0.44 |
| | | | $\chi^2_{\rm exp}$ | 0.37 | 1.38 | 1.30 | 1.44 | 1.15 | 0.39 | 0.42 | 1.14 | 0.98 |
| CMS 13 TeV $t\bar{t}\,\ell+j\,\frac{1}{\sigma}\frac{d^2\sigma}{dm_{t\bar{t}}d\lvert y_{t\bar{t}}\rvert}$ | 35 | 0.24 | $\chi^2_{\rm exp+th}$ | 2.77 | 2.89 | 2.87 | 2.76 | 3.36 | 3.01 | 3.61 | 2.81 | 2.81 |
| | | | $\chi^2_{\rm exp}$ | 8.37 | 14.2 | 13.7 | 16.6 | 13.1 | 7.31 | 8.14 | 13.1 | 11.6 |
| CMS 13 TeV $t\bar{t}\,\ell+j\,\frac{1}{\sigma}\frac{d\sigma}{dp_T^t}$ | 16 | 0.35 | $\chi^2_{\rm exp+th}$ | 0.78 | 0.62 | 0.63 | 0.66 | 0.64 | 0.79 | 0.81 | 0.63 | 0.65 |
| | | | $\chi^2_{\rm exp}$ | 1.31 | 0.68 | 0.70 | 0.69 | 0.72 | 1.24 | 1.17 | 0.74 | 0.78 |
| CMS 13 TeV $t\bar{t}\,\ell+j\,\frac{1}{\sigma}\frac{d\sigma}{d\lvert y_t\rvert}$ | 11 | 0.43 | $\chi^2_{\rm exp+th}$ | 1.07 | 1.54 | 1.57 | 1.81 | 1.90 | 1.22 | 1.57 | 1.38 | 1.42 |
| | | | $\chi^2_{\rm exp}$ | 1.61 | 3.08 | 2.94 | 4.02 | 2.81 | 1.46 | 1.84 | 2.77 | 2.49 |
| CMS 13 TeV $t\bar{t}\,\ell+j\,\frac{1}{\sigma}\frac{d\sigma}{d\lvert y_{t\bar{t}}\rvert}$ | 10 | 0.45 | $\chi^2_{\rm exp+th}$ | 0.94 | 2.01 | 1.89 | 2.16 | 2.44 | 1.76 | 2.71 | 1.53 | 2.00 |
| | | | $\chi^2_{\rm exp}$ | 8.65 | 11.0 | 10.7 | 12.9 | 10.4 | 8.06 | 8.72 | 10.6 | 9.82 |

Table 3.7: Same as Table 3.4 for the ATLAS and CMS top-quark pair production measurements at the LHC 13 TeV.

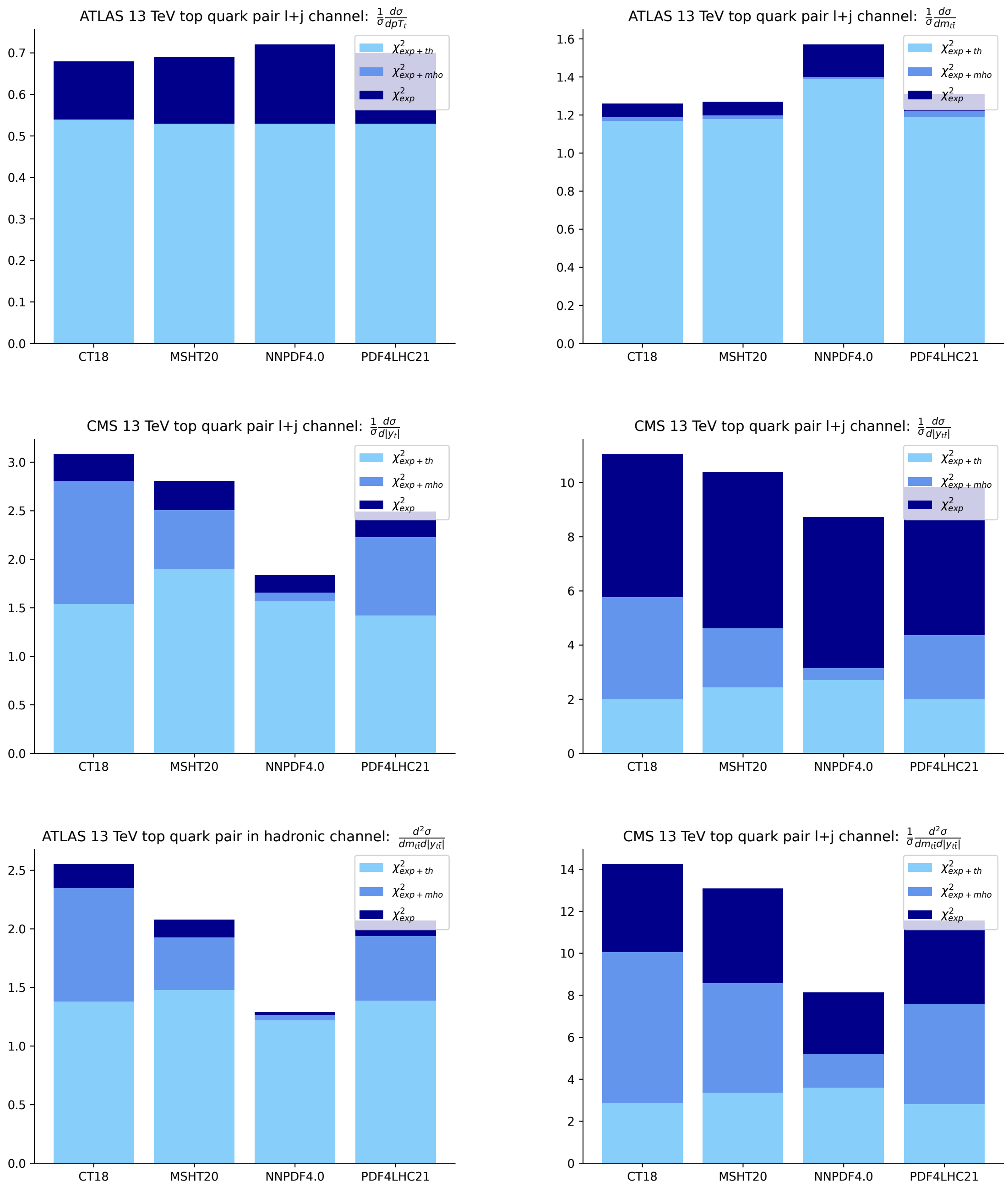

Figure 3.6: Same as Fig. 3.3 for a representative subset of top-quark pair production measurements at the LHC 13 TeV. Specifically, from top to bottom, left to right: the ATLAS lepton+jets normalized cross sections single differential in the transverse momentum of the top quark $p_T^t$ ($n_{\rm dat} = 8$, $\sqrt{2/n_{\rm dat}} = 0.50$), and in the invariant mass of the top-quark pair $m_{t\bar{t}}$ ($n_{\rm dat} = 9$, $\sqrt{2/n_{\rm dat}} = 0.47$); the CMS lepton+jets normalized cross sections, single differential in the absolute rapidity of the top quark and of the top-quark pair $|y_t|$ ($n_{\rm dat} = 11$, $\sqrt{2/n_{\rm dat}} = 0.43$) and $|y_{t\bar{t}}|$ ($n_{\rm dat} = 10$, $\sqrt{2/n_{\rm dat}} = 0.45$), the ATLAS all-hadronic absolute cross section, double differential in the invariant mass and absolute rapidity of the top-quark pair ($n_{\rm dat} = 11$, $\sqrt{2/n_{\rm dat}} = 0.43$); and the CMS lepton+jets normalized cross section, double differential in the invariant mass and absolute rapidity of the top-quark pair ($n_{\rm dat} = 35$, $\sqrt{2/n_{\rm dat}} = 0.24$). Histogram plots for the other datasets are collected in Fig.B.1 of Appendix B.

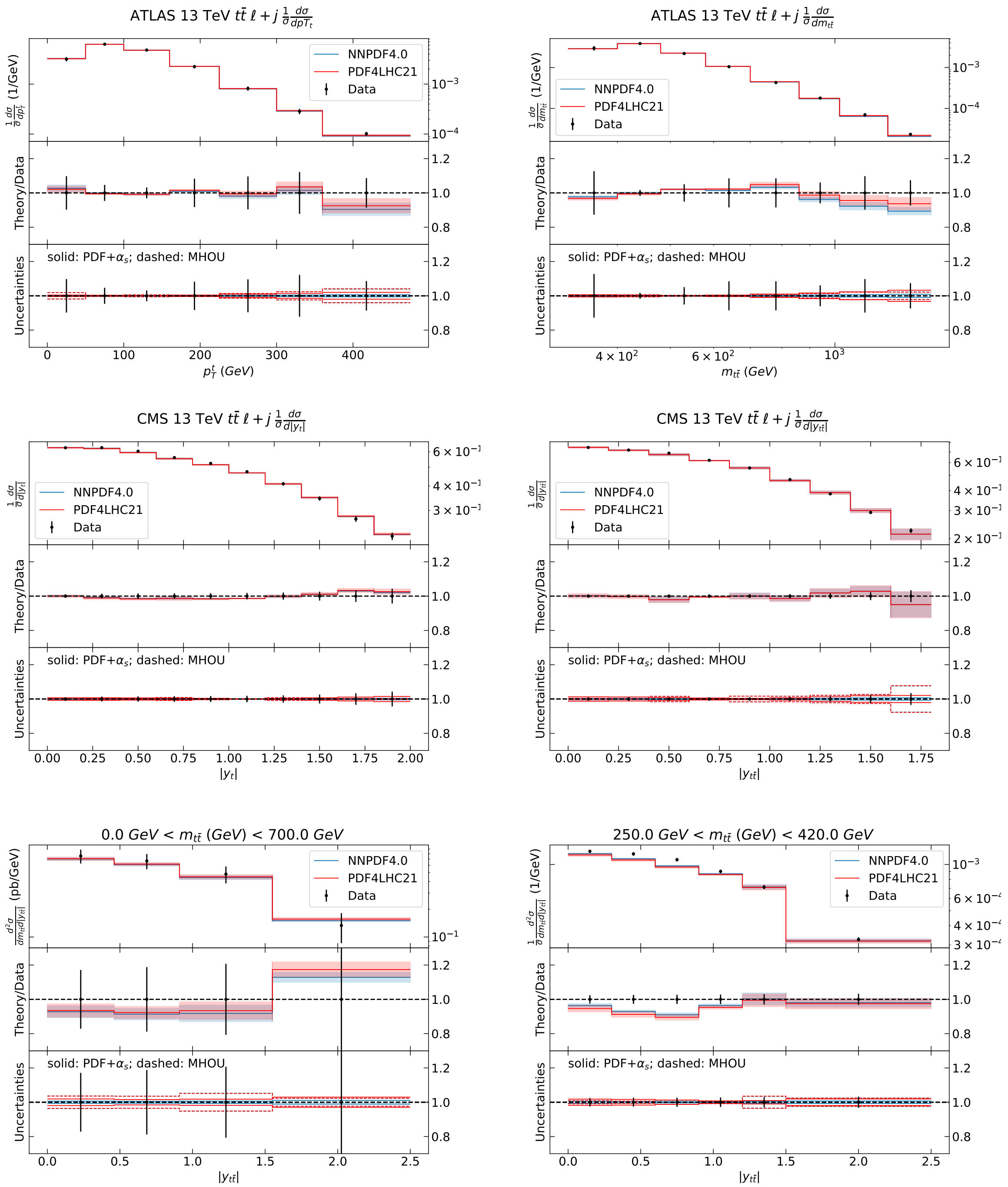

Figure 3.7: Same as Fig. 3.4 for the same representative subset of top-quark pair measurements of Fig. 3.6. In the case of the ATLAS (bottom left) and CMS (bottom right) double differential distributions, only the bin at the lowest invariant mass is shown. Additional results are collected in Figs. B.2-B.3 of Appendix B.

From inspection of Table 3.7 and of Fig. 3.6, we make considerations very similar to those made for Drell-Yan weak boson production measurements at the LHC 13 TeV. Namely, that the values of $\chi^2_{\rm exp+th}$, computed with different input PDFs, are closer to each other than the corresponding values of $\chi^2_{\rm exp}$, and that the former are generally rather similar across PDF sets. The only partial exceptions are given by the ATLAS all-hadronic double differential measurement, for which the values of $\chi^2_{\text{exp+th}}$ are away from one by slightly more than $1\sigma$ for the MSHT20 and CT18Z sets, and the ATLAS leptons+jet single differential measurement in the absolute value of the top-quark rapidity, for which the total $\chi^2_{\text{exp+th}}$ computed with NNPDF3.1 is slightly more than $1\sigma$ away from one. As far as CMS is concerned, the leptons+jet double differential distribution is poorly described by all PDF sets, with the total $\chi^2_{\text{exp+th}}$ being away from one by more than $3\sigma$ for all PDF sets. Less dramatic, but still significant, is the spread of $\chi^2_{\text{exp+th}}$ values across PDF sets for the CMS lepton+jets single differential distribution in the absolute rapidity of the top-quark and of the top-quark pair: in the former case, predictions obtained with CT18Z and MSHT20 are $2\sigma$ away from one, whereas predictions obtained with MSHT20 and NNPDF4.0 are about $3\sigma$ away from one (with NNPDF4.0 performing a little worse than MSHT20 by a quarter of a sigma).

The breakdown of $\chi^2_{\text{exp+th}}$ into its theoretical components depends on the dataset. The component due to MHO, which is relatively independent from the PDF set, prevails over the PDF+$\alpha_s$ component in the ATLAS lepton+jets distributions differential in the transverse momentum of the top quark and in the invariant mass of the top-quark pair, and in the CMS lepton+jets distribution differential in the absolute rapidity of the top-quark pair. The PDF+$\alpha_s$ component prevails in the other datasets, although it depends on the PDF set: it is generally larger for the CT18 and PDF4LHC21 PDF sets, which are affected by the largest uncertainties, see Fig. 3.2, whereas it is almost immaterial for NNPDF4.0, which has the smallest PDF uncertainties.

Overall, the quality of the data description is generally good, being $\chi^2_{\text{exp+th}} \sim 1$ for all the datasets, except in the case of the CMS normalized single differential distribution in the absolute rapidity of the top-quark pair, and double differential distribution in the absolute rapidity and invariant mass of the top-quark pair, for which $\chi^2_{\text{exp+th}} \sim 2-3$. Discrepancies between data and theory that may lead to these results are seen in Fig. 3.7, where experimental data and theoretical predictions are generally well aligned to each other, within their uncertainties, except, precisely, for the aforementioned datasets. Understanding the reason for this behavior, which is common to most PDF sets, is left to future investigations.

#### 3.3.4.4 single inclusive jet and di-jet production measurements at the LHC

We now turn to LHC single inclusive jet and di-jet production measurements outlined in Sec. 3.3.2. The values of $\chi^2_{\text{exp}}$ and $\chi^2_{\text{exp+th}}$ are reported in Table 3.8. The experimental covariance matrix is regularized as explained in Sec. 3.3.3.2 for all the datasets. Without regularization, the values of $\chi^2_{\text{exp}}$ are very poor, as they are away from one by more than 10-20$\sigma$, independently of the input PDF set. See Appendix A for the unregularized values of $\chi^2_{\text{exp+th}}$. Clearly the results that we present here do depend on regularization. However, as discussed in Appendix A, this dependence does not affect our ability to discriminate how well different PDF sets describe the data, which is the goal of this work. If, instead, we were interested to characterize the datasets for inclusion in a PDF determination (or not), we would consider other ways of decorrelating uncertainties, for instance by identifying uncertainties that, for experimental reasons, are more likely to be overcorrelated, see e.g. [156]. In this sense, the main message conveyed by the numbers in Table 3.8 is that the single inclusive jet and di-jet datasets require additional investigations on the understanding of uncertainty correlations. Only after accomplishing these investigations, which are beyond the scope of this work, one may be able to better judge how the various PDF sets comparatively generalize on them.

The breakdown of $\chi^2_{\text{exp+th}}$ into $\chi^2_{\text{exp+mho}}$ and $\chi^2_{\text{exp}}$ after regularization is displayed in Fig. 3.8. The data-theory comparison is displayed in Figs. 3.9, 3.10, and 3.11, respectively for the ATLAS and CMS single inclusive jet, and for the ATLAS di-jet measurements. In the first and second (third) cases, we plot the double differential cross section as a function of the transverse momentum of the leading jet, $p_T^j$ (the invariant mass of the two jets, $m_{jj}$), for the two outermost bins of the absolute value of the jet rapidity, $|y_j|$ (of the two-jet rapidity separation $|y^*|$). The other bins are displayed, respectively, in Figs. B.4, B.5, and B.6 of Appendix B.

| Dataset | $n_{\text{dat}}$ | $\sqrt{2/n_{\text{dat}}}$ | | *ABMP16* | *CT18* | *CT18A* | *CT18Z* | *MSHT20* | *NNPDF3.1* | *NNPDF4.0* | *PDF4LHC15* | *PDF4LHC21* |
|---|---|---|---|---|---|---|---|---|---|---|---|---|
| ATLAS 13 TeV incl. jet $\frac{d^2\sigma}{dp_T d\|y\|}$ | 177 | 0.11 | $\chi^2_{\text{exp+th}}$ | 1.85 | 1.56 | 1.64 | 1.38 | 1.67 | 1.21 | 1.51 | 1.20 | 1.25 |
| | | | $\chi^2_{\text{exp}}$ | 2.32 | 2.48 | 2.47 | 2.50 | 2.53 | 2.98 | 1.95 | 3.02 | 2.40 |
| CMS 13 TeV incl. jet $R_{0.4}$ $\frac{d^2\sigma}{dp_T d\|y\|}$ | 78 | 0.16 | $\chi^2_{\text{exp+th}}$ | 1.64 | 1.58 | 1.60 | 1.52 | 1.64 | 1.47 | 1.50 | 1.48 | 1.43 |
| | | | $\chi^2_{\text{exp}}$ | 2.05 | 2.29 | 2.25 | 2.26 | 2.23 | 2.21 | 2.02 | 2.30 | 2.18 |
| ATLAS 13 TeV di-jets $\frac{d^2\sigma}{dm_{jj} d\|y^*\|}$ | 136 | 0.12 | $\chi^2_{\text{exp+th}}$ | 1.13 | 1.08 | 1.09 | 1.05 | 1.16 | 1.09 | 1.15 | 1.01 | 0.96 |
| | | | $\chi^2_{\text{exp}}$ | 1.25 | 1.49 | 1.47 | 1.48 | 1.41 | 1.37 | 1.29 | 1.42 | 1.41 |

Table 3.8: Same as Table 3.4 for the ATLAS and CMS single inclusive and di-jet production measurements at the LHC 13 TeV.

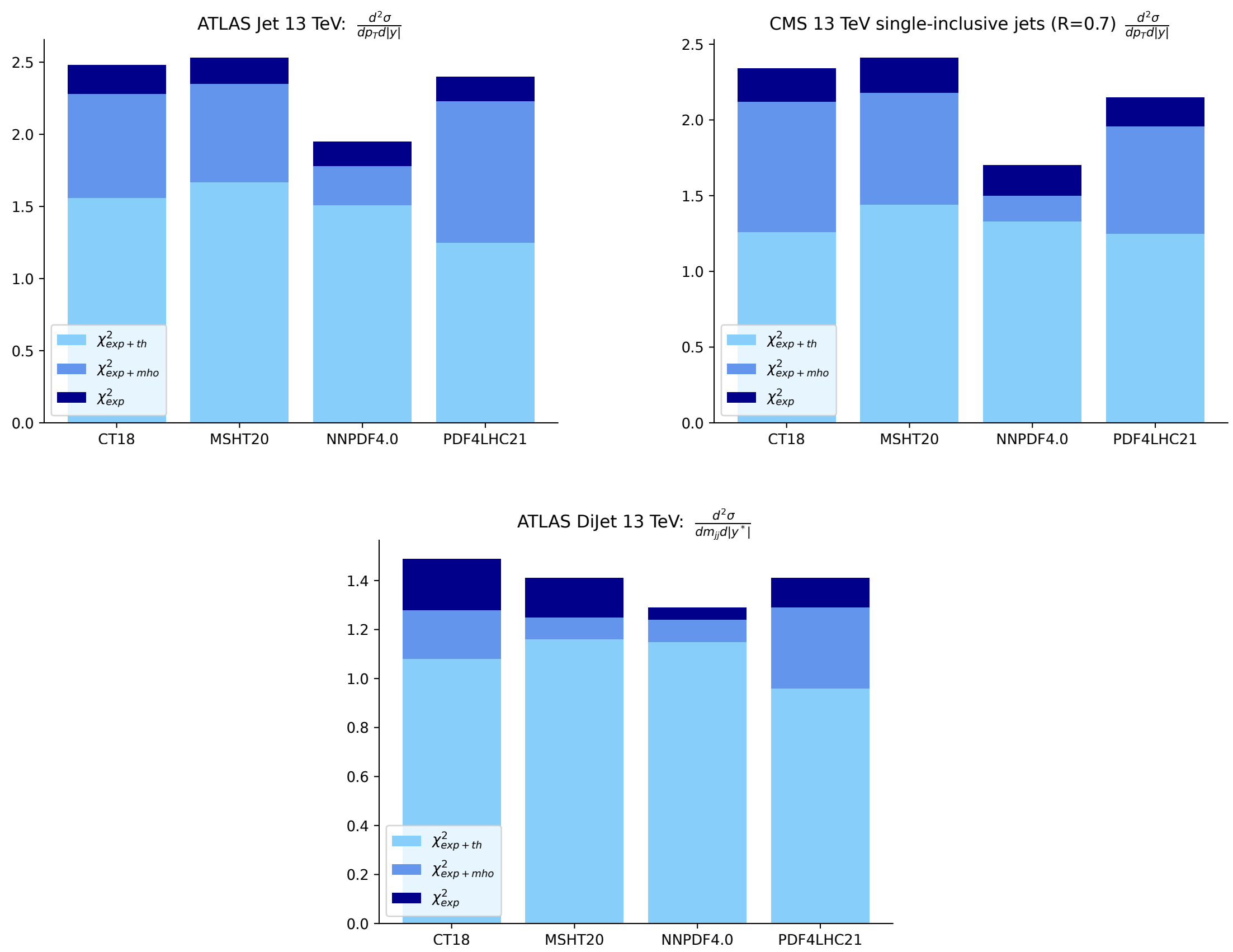

Figure 3.8: Same as Fig. 3.3 for the ATLAS and CMS single inclusive (with $n_{\rm dat} = 177$ and $n_{\rm dat} = 78$ respectively, and $\sqrt{2/n_{\rm dat}} = 0.11$ and $0.16$ correspondingly) and di-jet production measurements at the LHC 13 TeV ($n_{\rm dat} = 136$, $\sqrt{2/n_{\rm dat}} = 0.12$).

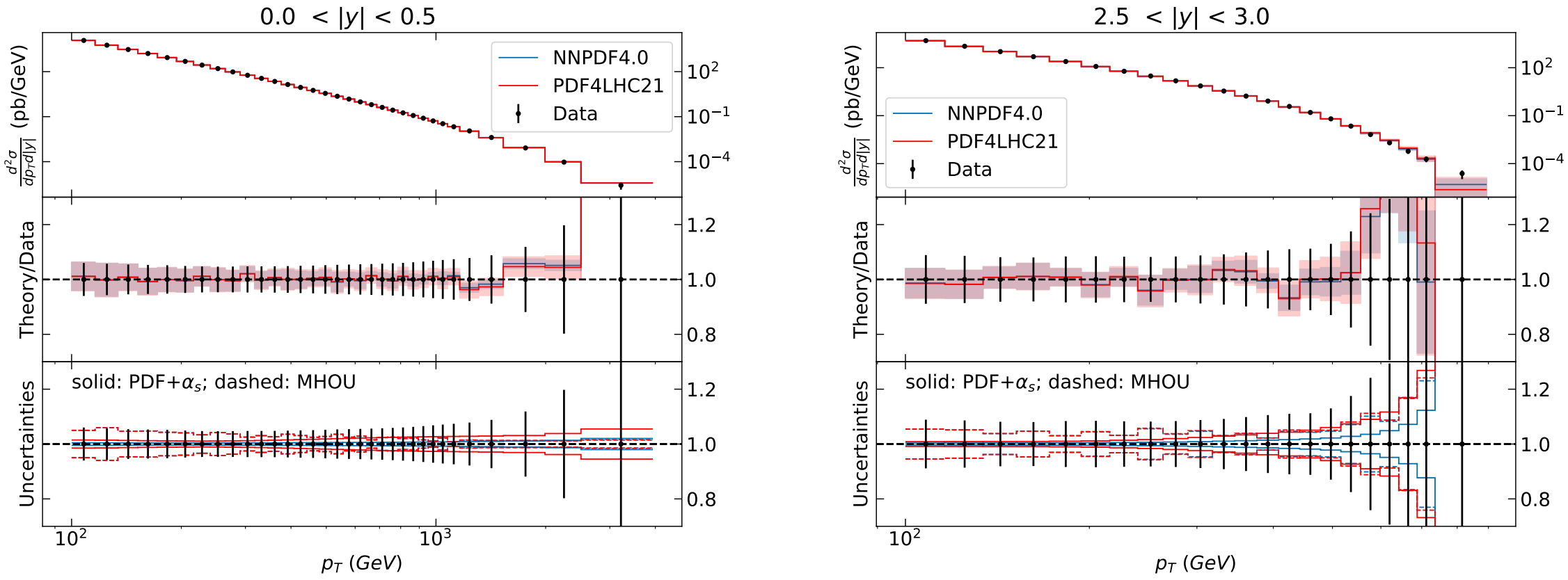

Figure 3.9: Same as Fig. 3.4 for the ATLAS single inclusive jet double differential cross section as a function of the transverse momentum of the leading jet, $p_T^j$, for the two outermost bins of the absolute value of the jet rapidity, $|y_j|$. The other bins are displayed in Fig. B.4 of Appendix B.

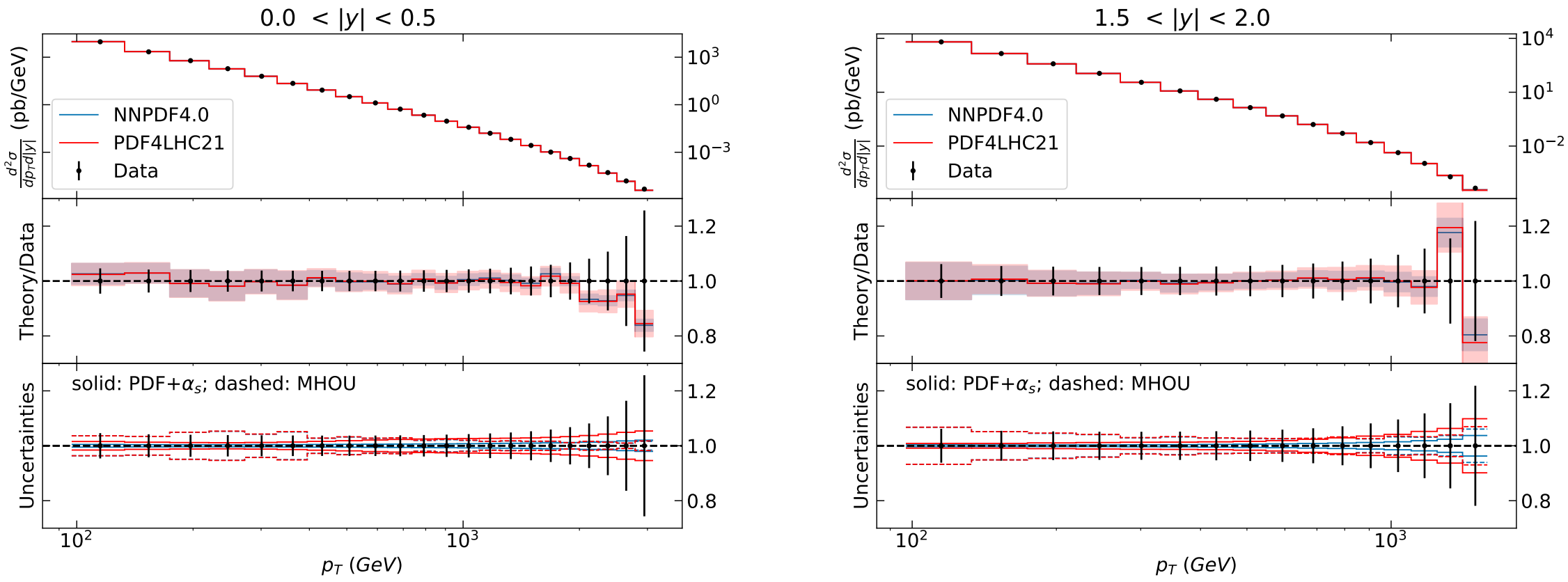

Figure 3.10: Same as Fig. 3.9 for the CMS single inclusive jet double differential cross section. The other bins are displayed in Fig. B.5 of Appendix B.

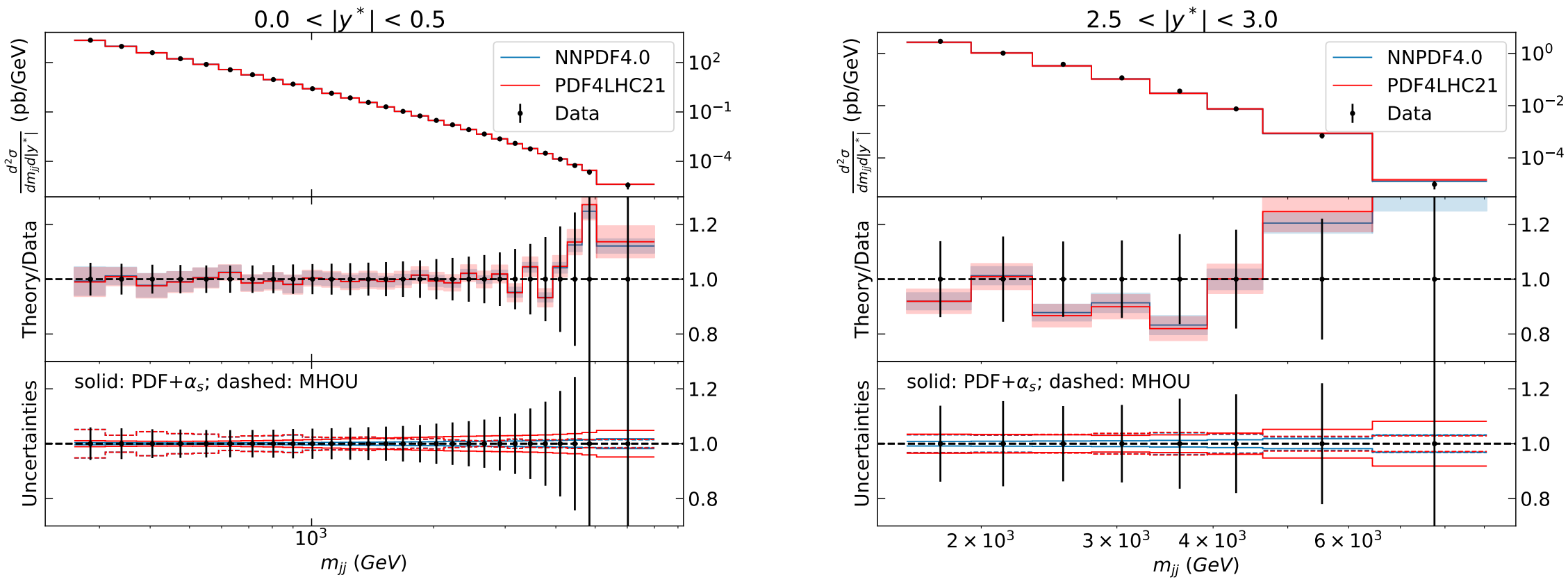

Figure 3.11: Same as Fig. 3.4 for the ATLAS di-jet double differential cross section, as a function of the invariant mass of the di-jet pair, $m_{jj}$, for the two outermost bin in the absolute rapidity separation between the two jets. The other bins are displayed in Fig. B.6 of Appendix B.

From inspection of Table 3.8 and of Fig. 3.8, very similar remarks can be drawn for the three considered datasets. First, when theory errors are not included in the computation of the $\chi^2$, the NNPDF4.0 PDF set performs better than any of the others, in the sense that the NNPDF4.0 $\chi^2_{\rm exp}$ is the closest to unity among all PDF sets, although the $\chi^2_{\rm exp}$ is still away from one by more than $10\sigma$. Some PDF sets may lead to comparatively larger values of $\chi^2_{\rm exp}$, such as APMP16, however the statistical significance of these fluctuations must be seen in units of the $\chi^2$ standard deviation, as we will further discuss in Sec. 3.3.4.6. Second, once all the theory errors are included, the values of $\chi^2_{\rm exp+th}$ become relatively close, irrespective of the input PDF set used for their computation. However, in the case of single inclusive jets, they are all a few units away from one, suggesting that once these datasets are included in the fit, they might have a significant pull on the gluon PDF. The values of $\chi^2_{\rm exp+th}$ being relatively close suggests also that, except perhaps for ABMP16, which continues to display rather large values of $\chi^2_{\rm exp+th}$ even after inclusion of theoretical uncertainties, it may be difficult to discriminate the quality of the predicting power of the various PDF sets based solely on these measurements. Third, the relatively homogeneous values of $\chi^2_{\text{exp+th}}$ occur despite the input PDF sets have very different uncertainties. For instance, PDF4LHC21 uncertainties are twice the NNPDF4.0 uncertainties, see Fig. 3.2. The breakdown of the theoretical uncertainty into its various components can be different depending on the PDF set. The MHO uncertainty remains more or less the same for all PDF sets. Conversely, the PDF+$\alpha_s$ uncertainty is the smallest for NNPDF4.0. This is consistent with the fact that NNPDF4.0 PDF uncertainties are typically the smallest among all the PDF sets considered, see Fig. 3.2. Finally, it is interesting to observe that the balance between the various components of the theoretical uncertainty depend on the kinematics. From Figs. 3.9-3.11, we see that the PDF+$\alpha_s$ (MHO) uncertainty dominates at small (large) $p_T^j$ or $m_{jj}$.

#### 3.3.4.5 single inclusive jet and di-jet production measurements at HERA

We finally discuss the HERA single inclusive jet and di-jet production measurements outlined in Sec. 3.3.4.4. The values of $\chi^2_{\text{exp}}$ and $\chi^2_{\text{exp+th}}$ are reported in Table 3.9. The experimental covariance matrix of the H1 low-$Q^2$ single inclusive jet and di-jet measurements is regularized as explained in Sec. 3.3.3.2. The unregularized values of $\chi^2_{\text{exp+th}}$ are reported in Appendix A. The breakdown of $\chi^2_{\text{exp+th}}$ into $\chi^2_{\text{exp+mho}}$ and $\chi^2_{\text{exp}}$ is displayed in Fig. 3.12, albeit only for the H1 data. The data-theory comparison is displayed in Fig. 3.13 for the highest $Q^2$ bin of the H1 single inclusive jet and di-jet differential cross sections as a function, respectively, of the transverse momentum of the leading jet and of the average transverse momentum of the jet pair. Histograms plots for the ZEUS measurements and data-theory comparison plots for the remaining H1 bins and for all of the ZEUS bins are collected in Figs. B.7-B.14 of Appendix B.

| Dataset | $n_{\rm dat}$ | $\sqrt{2/n_{\rm dat}}$ | | *ABMP16* | *CT18* | *CT18A* | *CT18Z* | *MSHT20* | *NNPDF3.1* | *NNPDF4.0* | *PDF4LHC15* | *PDF4LHC21* |
|---|---|---|---|---|---|---|---|---|---|---|---|---|
| H1 incl. jet (low $Q^2$) $\frac{d^2\sigma}{dQ^2 dp_T}$ | 37 | 0.23 | $\chi^2_{\rm exp+th}$ | 1.64 | 1.61 | 1.61 | 1.67 | 1.61 | 1.70 | 1.74 | 1.61 | 1.73 |
| | | | $\chi^2_{\rm exp}$ | 7.68 | 2.17 | 2.14 | 2.11 | 2.16 | 2.16 | 2.12 | 2.17 | 2.14 |
| H1 incl. jet (high $Q^2$) $\frac{d^2\sigma}{dQ^2 dp_T}$ | 24 | 0.29 | $\chi^2_{\rm exp+th}$ | 1.62 | 1.66 | 1.62 | 1.63 | 1.64 | 1.49 | 1.63 | 1.58 | 1.59 |
| | | | $\chi^2_{\rm exp}$ | 2.40 | 2.28 | 2.20 | 2.18 | 2.27 | 2.43 | 2.42 | 2.33 | 2.27 |
| ZEUS incl. jet (low lumi.) $\frac{d^2\sigma}{dQ^2 dE_T}$ | 30 | 0.26 | $\chi^2_{\rm exp+th}$ | 0.67 | 0.69 | 0.68 | 0.67 | 0.68 | 0.66 | 0.65 | 0.68 | 0.67 |
| | | | $\chi^2_{\rm exp}$ | 0.69 | 0.71 | 0.70 | 0.69 | 0.70 | 0.69 | 0.67 | 0.70 | 0.69 |
| ZEUS incl. jet (high lumi.) $\frac{d^2\sigma}{dQ^2 dE_T}$ | 30 | 0.26 | $\chi^2_{\rm exp+th}$ | 0.77 | 0.77 | 0.77 | 0.76 | 0.78 | 0.77 | 0.76 | 0.77 | 0.77 |
| | | | $\chi^2_{\rm exp}$ | 0.82 | 0.83 | 0.82 | 0.80 | 0.82 | 0.84 | 0.81 | 0.83 | 0.82 |
| H1 di-jets (low $Q^2$) $\frac{d^2\sigma}{dQ^2 d\langle p_T\rangle}$ | 37 | 0.23 | $\chi^2_{\rm exp+th}$ | 1.37 | 1.39 | 1.38 | 1.37 | 1.39 | 1.42 | 1.44 | 1.36 | 1.44 |
| | | | $\chi^2_{\rm exp}$ | 11.0 | 1.75 | 1.73 | 1.68 | 1.75 | 1.82 | 1.78 | 1.77 | 1.75 |
| H1 di-jets (high $Q^2$) $\frac{d^2\sigma}{dQ^2\, d\langle p_T\rangle}$ | 24 | 0.29 | $\chi^2_{\rm exp+th}$ | 2.21 | 2.03 | 2.00 | 1.95 | 2.03 | 1.84 | 2.12 | 1.94 | 1.97 |
| | | | $\chi^2_{\rm exp}$ | 2.63 | 2.47 | 2.37 | 2.32 | 2.42 | 2.65 | 2.63 | 2.51 | 2.45 |
| ZEUS di-jets $\frac{d^2\sigma}{dQ^2 d\langle E_T\rangle}$ | 22 | 0.30 | $\chi^2_{\rm exp+th}$ | 0.81 | 0.75 | 0.75 | 0.71 | 0.78 | 0.90 | 0.83 | 0.77 | 0.79 |
| | | | $\chi^2_{\rm exp}$ | 1.49 | 1.29 | 1.27 | 1.24 | 1.32 | 1.71 | 1.63 | 1.37 | 1.42 |

Table 3.9: Same as Table 3.4 for HERA single inclusive jet and di-jet data.

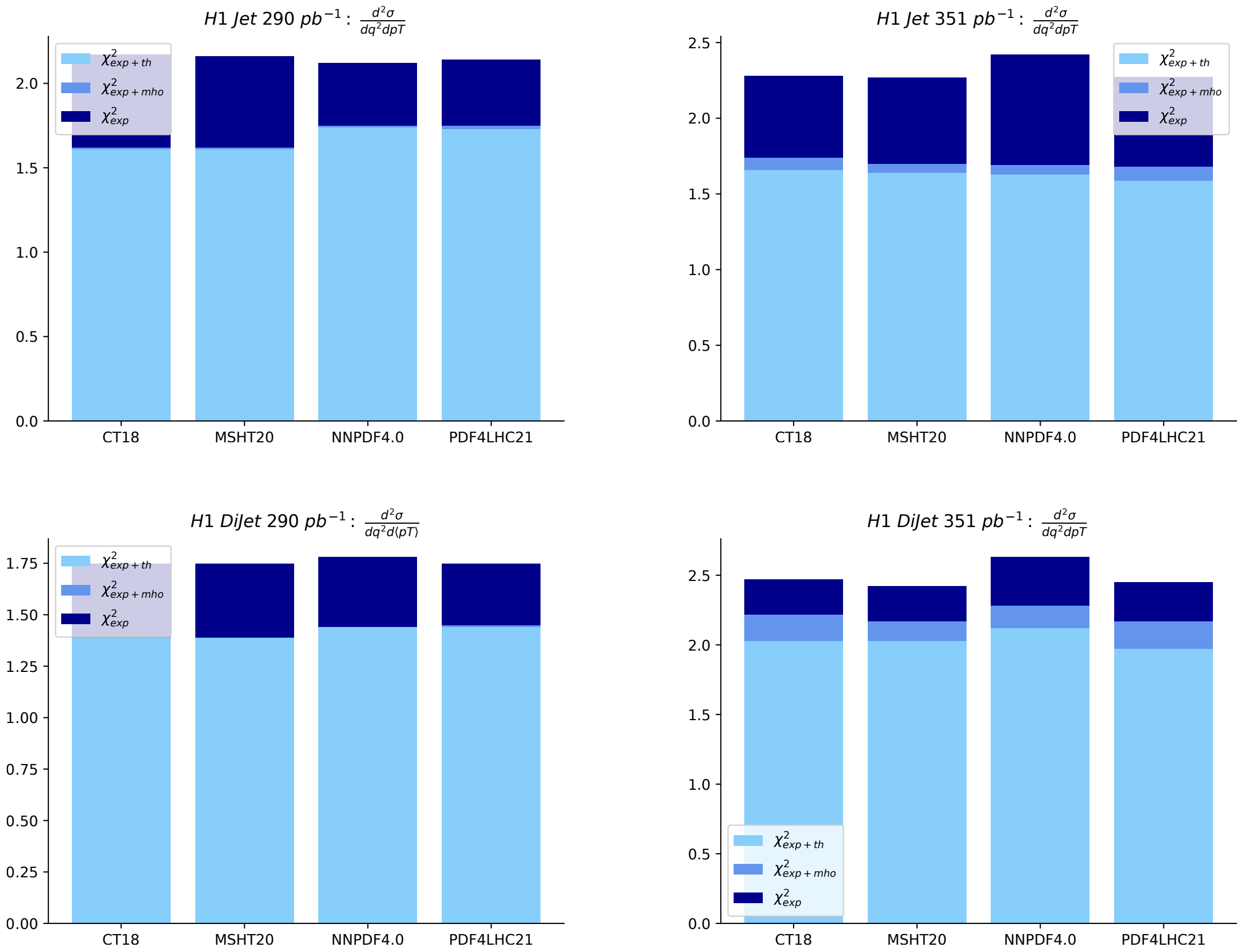

Figure 3.12: Same as Fig. 3.3 for the H1 single inclusive jet (top) and di-jet (bottom) datasets. For the measurements on the left plots $n_{\rm dat} = 37$ and $\sqrt{2/n_{\rm dat}} = 0.23$, while for the measurements on the right plots $n_{\rm dat} = 24$ and $\sqrt{2/n_{\rm dat}} = 0.29$.

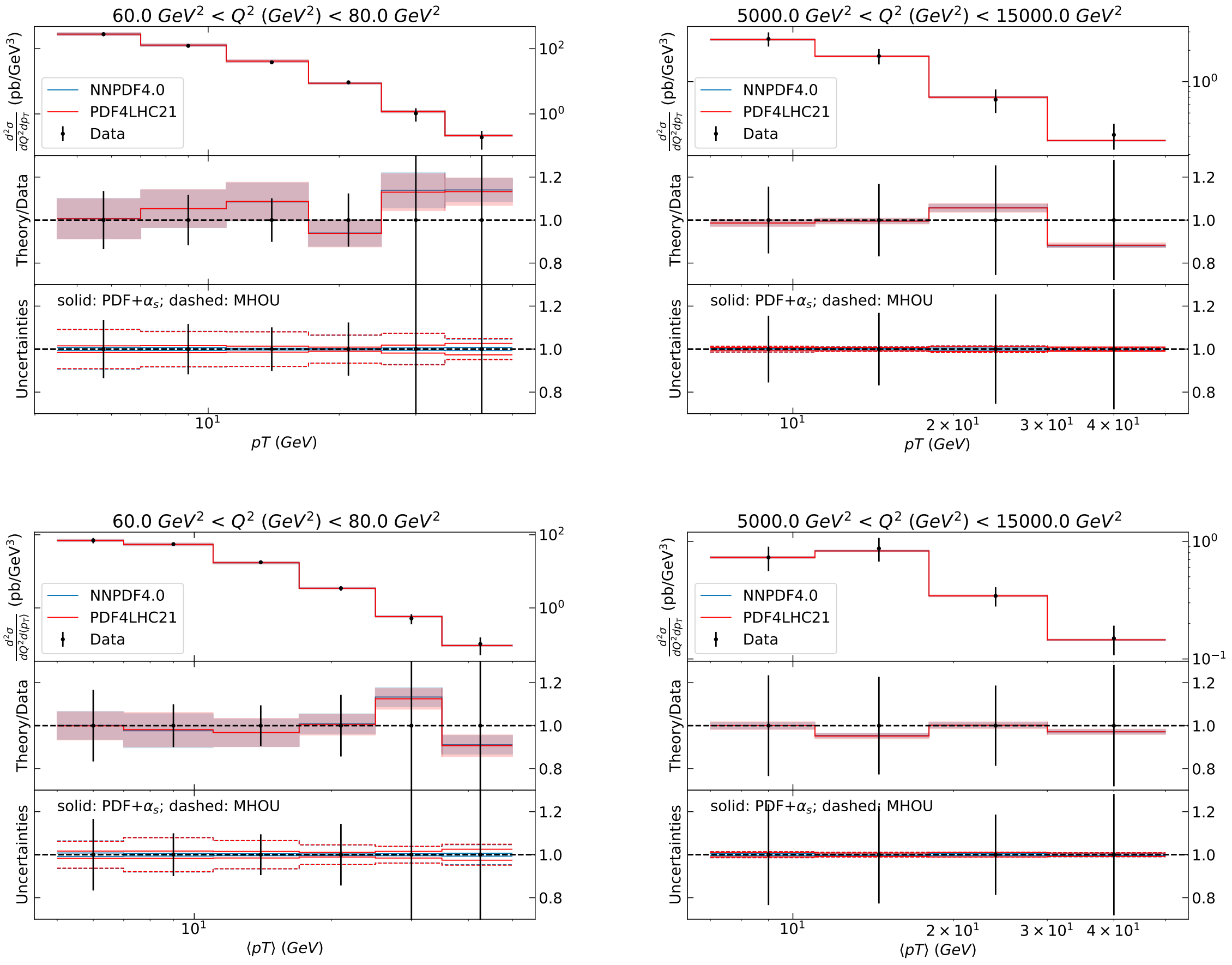

Figure 3.13: Same as Fig. 3.4 for the largest $Q^2$ bins of the H1 single inclusive jet (top) and di-jet (bottom) datasets. All the other bins are collected in Figs. B.8-B.11 of Appendix B.

From inspection of Table 3.9 and Fig. 3.12, we observe that the values of $\chi^2_{\mathrm{exp+th}}$ and of $\chi^2_{\mathrm{exp}}$ are very similar when different input PDF sets are used. All PDF sets generalize equally well on these datasets. The largest component of $\chi^2_{\mathrm{exp+th}}$ is due to MHO, in a proportion which is roughly the same across PDF sets. The PDF+$\alpha_s$ component of $\chi^2_{\mathrm{exp+th}}$ is almost immaterial (for the H1 low-$Q^2$ dataset), very small (for the H1 high-$Q^2$ single inclusive jet dataset), or as large as the MHO component (for the H1 high-$Q^2$ di-jet dataset). The quality of the data description is generally very good, with $\chi^2_{\mathrm{exp+th}} \sim 1$ for all the datasets, except for the H1 high-$Q^2$ dataset, in which case $\chi^2_{\mathrm{exp+th}} \sim 2$. Investigations into the reasons for this behavior, which is consistent throughout PDF sets, will be left to future work. For now, we remark that the agreement between experimental data and the corresponding theoretical predictions, as seen in Fig. 3.13, is generally good, except for specific bins that display larger fluctuations between the two.

#### 3.3.4.6 Combined interpretation

We now combine the results described in the previous sections to gather the overall agreement between the considered experimental data and the corresponding theoretical predictions. To this pur-pose, in Fig. 3.14, we display $\Delta\chi^{2(i)}$, the relative change in the total $\chi^2_{\text{exp+th}}$ due to the change of input PDF set with respect to the average $\chi^2_{\text{exp+th}}$ over PDF sets, see Eq. (3.21). The PDF sets considered here are ABMP16, CT18, MSHT20, NNPDF4.0, and PDF4LHC21. All the datasets listed in Table 3.2 are considered, except for the 8 TeV ATLAS Drell-Yan rapidity distribution [67]. The reason being that this dataset, extensively discussed in Sec. 3.3.4.2, is included in MSHT20 and NNPDF4.0 in the form of an earlier analysis [120], whereas all the other datasets are not included in any PDF set. Furthermore, all the other data sets are for the LHC Run II. The datasets are grouped by category: LHC Drell-Yan, LHC top-quark pair, LHC single inclusive jet and di-jet, and HERA single inclusive jet and di-jet production cross sections. The circumference corresponding to $\Delta\chi^2 = 0$ is highlighted with a solid curve. In Fig. 3.15 we display, in the same format, $\Delta n_\sigma^{(i)}$, the difference between the total $\chi^2_{\text{exp+th}}$ computed with the $i$-th PDF set and the average $\chi^2_{\text{exp+th}}$ over PDF sets, normalized to the standard deviation of the $\chi^2$ distribution, see Eq. (3.23). Figures 3.14 and 3.15 should be inspected together: the latter provides an assessment of the statistical significance of fluctuations from the average $\Delta\chi^2 = 0$ seen in the former, in units of the $\chi^2$ standard deviation. Large fluctuations may have low statistical significance if a dataset has a small number of data points and the other way around.

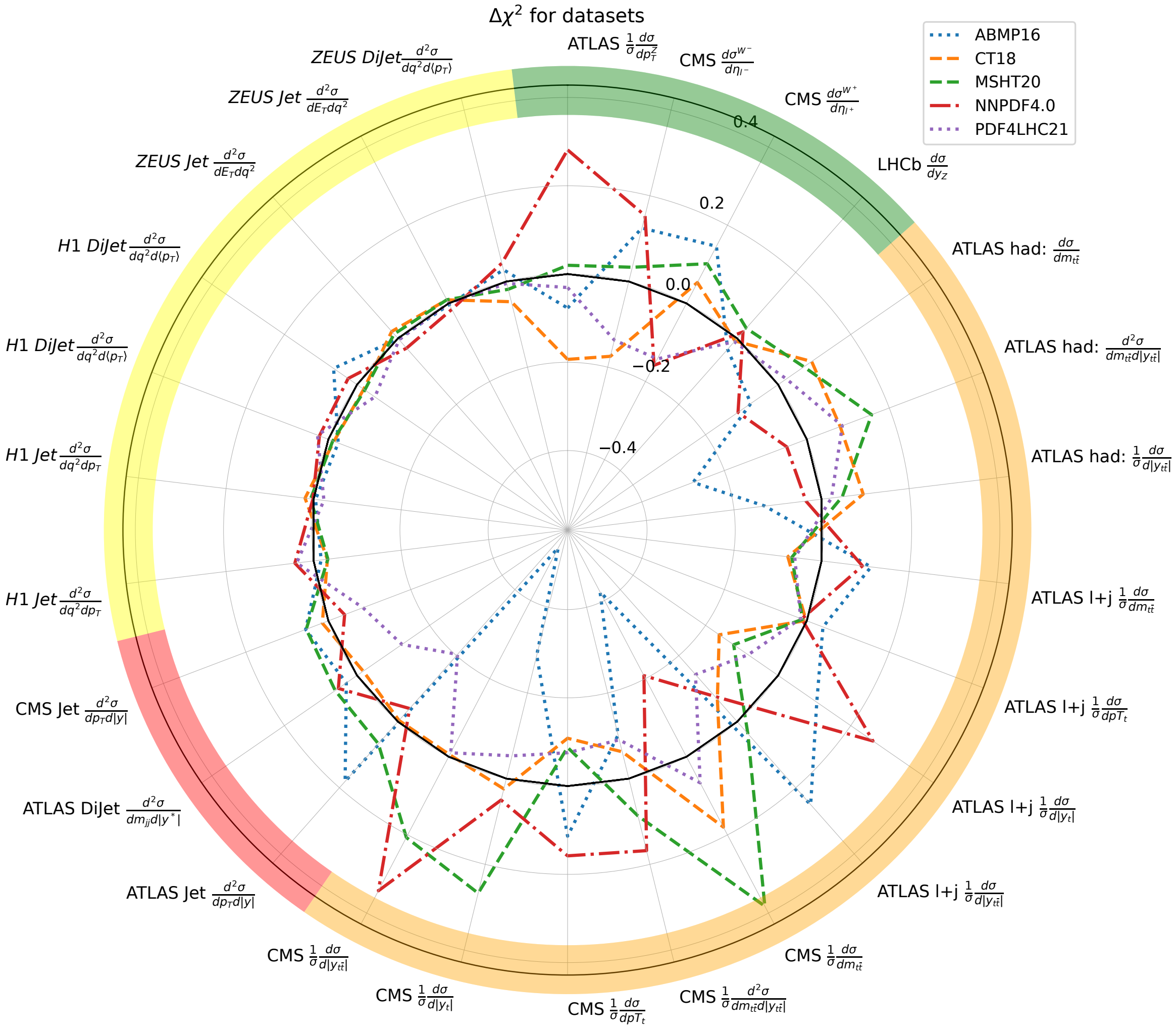

Figure 3.14: The relative change in the total $\chi^2_{\mathrm{exp+th}}$ due to the change of $i$-th input PDF set, $\Delta\chi^{2(i)}$, with respect to the average $\chi^2_{\mathrm{exp+th}}$ over PDF sets, see Eq. (3.21). The PDF sets considered here are ABMP16, CT18, MSHT20, NNPDF4.0, and PDF4LHC21. The datasets are grouped by category: LHC Drell-Yan, LHC top-quark pair, LHC single inclusive jet and di-jet, and HERA single inclusive jet and di-jet production cross sections. The circumference corresponding to $\Delta\chi^2 = 0$ is highlighted with a solid curve.

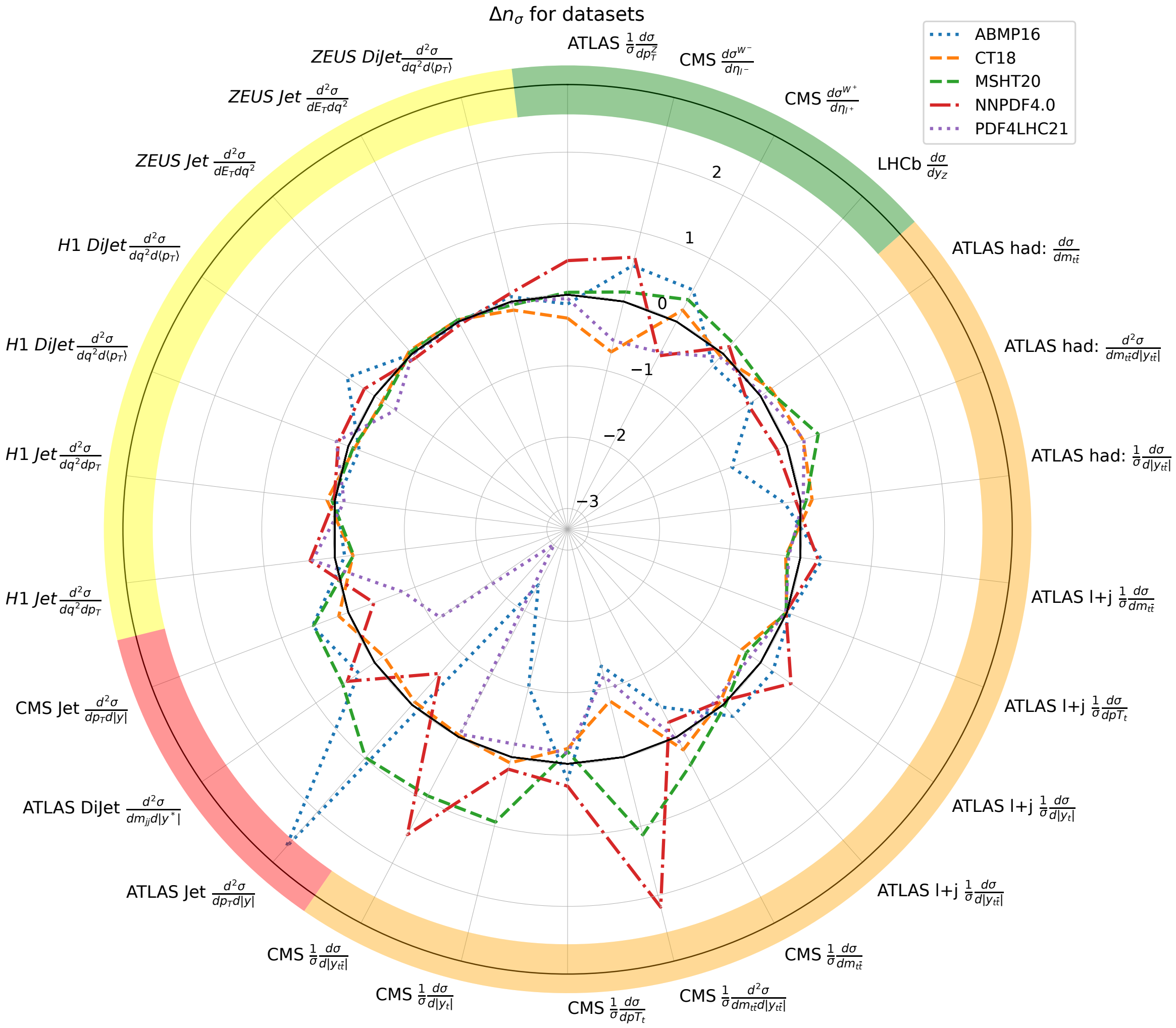

Figure 3.15: Same as Fig. 3.14 now for $\Delta n_\sigma$, see Eq. (3.23).

On the basis of Figs. 3.14 and 3.15, we conclude that the various classes of datasets are described to a different level of accuracy. However, whereas the value of $\Delta\chi^2$ displays sizeable fluctuations depending on the input PDF set, especially in the top-quark pair and jet sectors, we realize that discrepancies with respect to the average over PDF sets is almost always within $\Delta n_\sigma = 1$. The most relevant excess occurs with the ABMP16 PDF set in the case of the ATLAS and CMS single inclusive jet measurements, and with the NNPDF4.0 PDF set in the case of the CMS double differential and rapidity-differential top-quark pair measurements. In these cases, the excess is between one and three sigma. This fact may be explained by assuming that these measurements disfavor the softer (harder) large-$x$ gluon of ABMP16 (NNPDF4.0). We also note an anomalous deficiency, close to $|\Delta n_\sigma| = 3$, with the PDF4LHC21 PDF set in the case of the ATLAS single inclusive jet measurements, and with the ABMP16 PDF set in the case of the CMS rapidity-differential top-quark pair measurement. We therefore conclude that, whereas HERA jet and LHC Drell-Yan measurements may not be able to discriminate between PDF sets, LHC jet and top-quark pair measurements may help put stronger constraints on PDFs, especially those datasets for which the largest fluctuations among different PDF sets are observed in terms of $\Delta n_\sigma$.

As for the general trend displayed by individual PDF sets, on top of the aforementioned dataset-specific considerations, we remark two interesting facts. First, the NNPDF4.0 PDF set, despite displaying the smallest uncertainties among all the PDF sets considered in this work (see Fig. 3.2), provides a description of the data which is overall not worse than the others (with the aforementioned few exceptions). We therefore conclude that theoretical predictions obtained with NNPDF4.0 are overall as accurate as those obtained with the other PDF sets, despite their smaller PDF uncertainties, once experimental, MHO, and $\alpha_s$ uncertainties are taken into account, and the regularization of the covariance matrix is applied when needed. This is true on average on the examined datasets: as discussed, there are cases in which NNPDF4.0 performs better than other PDF sets, others in which it performs worse than these, and others in which the experimental precision of the datasets cannot discriminate among different PDF sets. Second, the PDF4LHC21 PDF set generally displays the value of $\Delta\chi^2$ and $\Delta n_\sigma$ closest to zero among all the PDF sets considered in this work. This fact is unsurprising, given that PDF4LHC21 is the unweighted average of the CT18, MSHT20 and NNPDF3.1 PDF sets. Deviations from the mean $\Delta\chi^2 = 0$ and $\Delta n_\sigma = 0$, obtained with individual PDF sets, cancel out by construction. In this sense, PDF4LHC21 is a conservative PDF set, as already illustrated in [51], although it remains the least precise.

### 3.3.5 Summary and Outlook

In this section we have compared theoretical predictions, computed at NNLO accuracy in perturbative QCD using different input PDF sets, with a wide array of experimental measurements, typically not yet included in PDF determination. Specifically, we have considered differential cross sections measured at the LHC, for Drell-Yan gauge boson, top-quark pair, single inclusive jet and di-jet production, and at HERA, for single inclusive jet and di-jet production. We have considered the most widely used PDF sets in LHC experimental analyses, namely, ABMP16, CT18 (and its variants), MSHT20, NNPDF3.1, NNPDF4.0, PDF4LHC15, and PDF4LHC21. We have accounted for all the relevant sources of experimental and theoretical uncertainties, in particular due to PDFs, $\alpha_s$, and MHOUs.

The aim of this work has been twofold. First, to test the predictive power of different PDF sets, by assessing the goodness with which they describe the datasets not included in their determination. Second, to quantify the various sources of uncertainty that enter theoretical predictions, specifically PDF, $\alpha_s$, and MHO uncertainties. These two objectives are becoming increasingly relevant given the ever higher precision of LHC experiments to determine SM parameters, such as the strong coupling $\alpha_s(m_Z)$, the $W$-boson mass $m_W$, and the effective lepton mixing angle $\sin^2\theta^\ell_{\rm eff}$. This precision is now comparable to, if not better than, that obtained at LEP. This outstanding result requires a careful estimate of all of the sources of uncertainties that accompany it, in particular the PDF uncertainty, which is often dominant in LHC measurements.

In this work we have considered the data-theory agreement between predictions obtained with a broad range of input PDF sets and experimental data, but we have not included any of the examined datasets in a PDF determination. Only the simultaneous inclusion of a given subset of statistically independent datasets considered in this work in future PDF determinations will allow us to determine how the resulting PDFs can adapt to the data, and possibly improve the description of the data, as well as the precision of PDFs.

The main outcome of our investigations is summarized in the overview plots presented in Sec. 3.3.4.6. We have found that the ABMP16, CT18, MSHT20, NNPDF4.0, and PDF4LHC21 PDF sets provide a comparable description of all of the datasets considered in this work on average, once all sources of theoretical uncertainty are taken into account. We have therefore concluded that all PDF sets have a similar predictive power and generalize similarly well to unseen data. Incorporating PDF, $\alpha_s$ and MHO uncertainties is crucial to reach this conclusion. These outcomes may seem counter-intuitive given that individual PDF sets differ amongst each other for their central values and uncertainties, by an amount that is not always encompassed by the latter. Within this general picture, the NNPDF4.0 and PDF4LHC21 sets represent opposite cases. On the one hand, the NNPDF4.0 set has by far the smallest uncertainties of all PDF sets, hence it is the most precise. In spite of this fact, it describes the examined data, on average, as well as the other PDF sets. On the other hand, the PDF4LHC21 set has some of the largest uncertainties of all PDF sets, hence it is the least precise. This is by construction, given that it is the combination of three different PDF sets. However, the fact that it describes the data as well as the other PDF sets means that it does not need to be as accurate as these.

The only exception to this overall trend is represented by the ATLAS 8 TeV inclusive measurement of the $Z$ rapidity distributions extrapolated to the full phase space, which underlies the recent $\alpha_s(m_Z)$ extraction from the companion measurement differential in the transverse momentum of the $Z$ boson. In this case we have found that, despite the excellent agreement of NNPDF4.0 theoretical predictions with the central values of the experimental data, the peculiar slope in rapidity, combined with the dominance of the luminosity normalization uncertainty, leads to a poor $\chi^2$. The $\chi^2$ is instead better for other PDF sets because of their larger PDF uncertainty. This may therefore be a case in which the accuracy of the NNPDF4.0 set does not match its very high precision. We have investigated whether this is truly the case. We have found that using variants of the NNPDF4.0 set that incorporate MHOU and/or aN$^3$LO corrections improves the $\chi^2$ only marginally. We have also observed that the NNPDF4.0 set and any of its variants provide an excellent description of the earlier ATLAS measurement of the $Z$ boson rapidity distributions based on the same collider events. We have finally determined that the $\chi^2$ of this dataset can be lowered only if it is included, and overweighted, in a fit, at the price of a slight deterioration in the description of the other datasets. We have therefore concluded that there are residual tensions between this dataset and the other datasets in NNPDF4.0.

Two important by-products of this work have been the computation of fast interpolation grids, accurate to NNLO, and the implementation of the experimental information, in the NNPDF format, for all the considered datasets. These facts will allow us to streamline their inclusion in future NNPDF releases. The fast interpolation framework has two major advantages. First, we will abandon the use of $K$-factors to account for NNLO corrections in partonic matrix elements. Second, we will be able to readily vary the renormalization and factorization scales in the computation of theoretical predictions and determine MHOU.

To conclude, as the LHC experiments finalize the Run II legacy measurements, start to release datasets based on the Run III luminosity, and prepare for the HL-LHC era, our analysis demonstrates the importance of testing the predictive power of PDFs on a broad set of high-precision measurements with state-of-the-art theoretical predictions, which must crucially include all possible sources of theoretical uncertainty. The methodology laid out in this section can be applied to any upcoming and future LHC measurements that may eventually provide a clear guidance concerning which PDF sets are preferred by the experimental data.

## 3.4 Towards next-generation PDFs

Once the studies evaluating the generalization power of the existing PDFs with the new data are completed, the next step is to determine the specific new datasets that are optimal for inclusion in PDF determination, and assess the impact they would have on the existing PDFs. Their impact could manifest in multiple ways, such as a better constraint on a particular flavor, or a non-negligible deviation of a particular flavor from the previous PDFs. This study on the determination of the new datasets to be used, and their impact should be carried out in a systematic and incremental manner, to ensure that differences arising from different sources are not entangled, but rather their origins are well understood. With this aim, the study presented in this section was carried out with a particular focus on datasets that are sensitive to the gluon PDF.

### 3.4.1 Introduction

A method for improving the accuracy of PDFs is by including theoretical uncertainties in the fitting procedure. In Ref. [20], a new variant of NNPDF4.0 was released, that included MHO uncertainties. In this work, we build on the NNPDF4.0 with MHOUs by inclusion of a subset of new data, to which, the gluon PDF is particularly sensitive. In particular, we include new top pair, single-inclusive jet, and dijet production datasets from the LHC Run 2, as well as single-inclusive jet and dijet production datasets from HERA in our PDF determination. These datasets have been discussed in previous subsections 3.3.2.2, 3.3.2.3 and 3.3.2.4. Whereas the LHC Run 2 datasets are truly being incorporated for the first time in this work, the HERA jets datasets were included during the NNPDF4.0 determination by means of reweighting. In this study, the HERA jets datasets are also explicitly included in the fit.

To perform this determination, our baseline fit, however, is not the NNPDF4.0 with MHOUs, but rather an improved version of it. The improvement comes in the form of having the theoretical predictions for all the top pair datasets be included at exact NNLO accuracy, as opposed to theoretical predictions at NLO accuracy with NNLO K-factors. Besides these differences, the baseline fit is produced using the exact NNPDF4.0 methodology. With a baseline defined, we are able to study the effect individual distributions have on the PDF determination, and we select the datasets that maximize the consistency of the fit and impact on the gluon PDF.

This section is organized as follows: in Sec. 3.4.2, I discuss the datasets considered. In Sec. 3.4.3, I begin with a discussion that shows how I perform the determination of impact and consistency for individual distributions, and then list the distributions selected for the final fit in Sec. 3.4.3.1. This is followed by comparing the NewFit with our baseline fit in Sec. 3.4.3.2, and comparing the NewFit with select PDF sets from other collaborations in Sec. 3.4.3.3. In Sec. 3.4.4, I study the impact of the NewFit on luminosities and cross sections. I conclude by summarizing our results in Sec. 3.4.5.

## 3.4.2 Experimental and theoretical input

### 3.4.2.1 Data sets

The experimental datasets we consider include top pair production at the LHC, and the inclusive jet and dijet production at the LHC and HERA. As highlighted in the introduction of this section, these datasets have been thoroughly discussed in the context of the study in the previous sections, and therefore they will not be repeated here. In Table 3.10, the distributions are however, relisted, to include the exact kinematical coverage for each of them.

| Dataset | Observable | Kin$_1$ | Kin$_2$ |
|---|---|---|---|
| ATLAS $t\bar{t}$ hadr. 13 TeV | $(1/\sigma)\, d\sigma/dm_{t\bar{t}}$ | $325\, GeV \leq m_{t\bar{t}} \leq 3000\, GeV$ | - |
| | $(1/\sigma)\, d\sigma/d\lvert y_{t\bar{t}}\rvert$ | $0 \leq \lvert y_{t\bar{t}}\rvert \leq 2.4$ | - |
| | $(1/\sigma)\, d^2\sigma/dm_{t\bar{t}}\, d\lvert y_{t\bar{t}}\rvert$ | $0 \leq \lvert y_{t\bar{t}}\rvert \leq 2.5$ | $0\, GeV \leq m_{t\bar{t}} \leq 3000\, GeV$ |
| ATLAS $t\bar{t}$ $\ell$+jets 13 TeV | $(1/\sigma)\, d\sigma/dm_{t\bar{t}}$ | $325\, GeV \leq m_{t\bar{t}} \leq 2000\, GeV$ | - |
| | $(1/\sigma)\, d\sigma/dp_{T,t}$ | $0\, GeV \leq p_{T,t} \leq 1000\, GeV$ | - |
| | $(1/\sigma)\, d\sigma/d\lvert y_t\rvert$ | $0 \leq \lvert y_t\rvert \leq 2.5$ | - |
| | $(1/\sigma)\, d\sigma/d\lvert y_{t\bar{t}}\rvert$ | $0 \leq \lvert y_{t\bar{t}}\rvert \leq 2.5$ | - |
| | $(1/\sigma)\, d^2\sigma/dm_{t\bar{t}}\, dp_{T,t}$ | $325\, GeV \leq m_{t\bar{t}} \leq 2000\, GeV$ | $0\, GeV \leq p_{T,t} \leq 1000\, GeV$ |
| CMS $t\bar{t}$ $\ell$+jets 13 TeV | $(1/\sigma)\, d\sigma/dm_{t\bar{t}}$ | $250\, GeV \leq m_{t\bar{t}} \leq 3500\, GeV$ | - |
| | $(1/\sigma)\, d\sigma/dp_{T,t}$ | $0\, GeV \leq p_{T,t} \leq 1500\, GeV$ | - |
| | $(1/\sigma)\, d\sigma/d\lvert y_t\rvert$ | $0 \leq \lvert y_t\rvert \leq 2.5$ | - |
| | $(1/\sigma)\, d\sigma/d\lvert y_{t\bar{t}}\rvert$ | $0 \leq \lvert y_{t\bar{t}}\rvert \leq 2.4$ | - |
| | $(1/\sigma)\, d^2\sigma/dm_{t\bar{t}}\, d\lvert y_{t\bar{t}}\rvert$ | $0 \leq \lvert y_{t\bar{t}}\rvert \leq 2.2$ | $250\, GeV \leq m_{t\bar{t}} \leq 3500 GeV$ |
| ATLAS incl. jet (R=0.4) 13 TeV | $d^2\sigma/dp_T\, d\lvert y\rvert$ | $0 \leq \lvert y\rvert \leq 3$ | $100\, GeV \leq p_T \leq 894\, GeV$ |
| ATLAS dijet (R=0.4) 13 TeV | $d^2\sigma/dm_{jj}\, d\lvert y^*\rvert$ | $0 \leq \lvert y^*\rvert \leq 3$ | $260\, GeV \leq m_{jj} \leq 9066\, GeV$ |
| CMS incl. jet (R=0.4) 13 TeV | $d^2\sigma/dp_T\, d\lvert y\rvert$ | $0 \leq \lvert y\rvert \leq 2$ | $97\, GeV \leq p_T \leq 1588 GeV$ |
| CMS incl. jet (R=0.7) 13 TeV | $d^2\sigma/dp_T\, d\lvert y\rvert$ | $0 \leq \lvert y\rvert \leq 2$ | $97\, GeV \leq p_T \leq 1588 GeV$ |
| H1 incl. jet 319 GeV (290 pb$^{-1}$) | $d^2\sigma/dq^2\, dp_T$ | $4.5\, GeV \leq p_T \leq 50\, GeV$ | $5.5\, GeV^2 \leq q^2 \leq 80\, GeV^2$ |
| H1 dijet 319 GeV (290 pb$^{-1}$) | $d^2\sigma/dq^2\, d\langle p_T\rangle$ | $5\, GeV \leq \langle p_T\rangle \leq 50\, GeV$ | $5.5\, GeV^2 \leq q^2 \leq 80\, GeV^2$ |
| H1 incl. jet 319 GeV (351 pb$^{-1}$) | $d^2\sigma/dq^2\, dp_T$ | $7\, GeV \leq p_T \leq 50\, GeV$ | $150\, GeV^2 \leq q^2$ |
| H1 dijet 319 GeV (351 pb$^{-1}$) | $d^2\sigma/dq^2\, d\langle p_T\rangle$ | $7\, GeV \leq \langle p_T\rangle \leq 50\, GeV$ | $150\, GeV^2 \leq q^2$ |
| ZEUS incl. jet 300 GeV (38.6 pb$^{-1}$) | $d^2\sigma/dE_T\, dq^2$ | $8\, GeV \leq E_T \leq 100\, GeV$ | $125\, GeV^2 \leq q^2$ |
| ZEUS incl. jet 319 GeV (82 pb$^{-1}$) | $d^2\sigma/dE_T\, dq^2$ | $8\, GeV \leq E_T \leq 100\, GeV$ | $125\, GeV^2 \leq q^2$ |
| ZEUS dijet 319 GeV (374 pb$^{-1}$) | $d^2\sigma/d\langle E_{T,B}\rangle\, dq^2$ | $8\, GeV \leq \langle E_{T,B}\rangle \leq 60\, GeV$ | $125\, GeV^2 \leq q^2$ |

Table 3.10: This table lists the datasets considered for inclusion in our NewFit with details on the kinematical coverage of each of the distribution considered.

#### 3.4.2.2 Theoretical predictions

The theoretical predictions for the hard partonic cross sections, are generated and stored in the form of interpolation grids such that they are independent of the PDFs. As these datasets are exactly as those covered in Sec. 3.3, the theoretical predictions are the same as those described in Sec. 3.3.2.2, 3.3.2.3 and 3.3.2.4.

### 3.4.3 Impact on the gluon PDF

#### 3.4.3.1 Assessment of the data sets

To meaningfully assess the impact of the new data, we need to perform a comparison in a manner which minimizes the differences arising from other ingredients that go into the process of production of the theoretical predictions. This means defining a baseline which is like NNPDF4.0 (with MHOUs), but varies in aspects where progress has been made since the release of NNPDF4.0 (with MHOUs). In particular, our new baseline varies from NNPDF4.0 through the inclusion of theoretical uncertainties corresponding to missing higher orders, in the covariance matrix. It varies from both, NNPDF4.0 and NNPDF4.0 with MHOUs by moving away from K-factor approximation and using exact QCD NNLO corrections for top pair production at the LHC. These updated theoretical predictions for the old (i.e. ones which were already included in NNPDF4.0 (with MHOUs)) top pair production datasets are computed in the exact same manner, as those for the new top pair production datasets, as described in section 3.4.2.2. We emphasize that we do not expect significant differences between the new baseline used in this study and NNPDF4.0 with MHOUs, due to top pair production K-factors generally being able to capture the NNLO effects fairly well as shown in Sec. 3.2. Another important aspect to note about the new baseline is that it is consistent with NNPDF4.0 (with MHOUs) in terms of the use of the old CMS measurement of $t\bar{t}\ \ell$+jets at 13 TeV with a luminosity of 35.8 $\mathrm{fb}^{-1}$.

To perform a global fit, we intend to include as many distributions as possible while avoiding double counting and therefore when experimental correlations are not provided for the different distributions amongst a given dataset, we have to choose a particular distribution. For this, we aim to select the distribution that maximizes consistency. This selection is based on 3 figures of merit: the stability of the covariance matrix, the dataset $\chi^2$ and the global $\chi^2$. The stability of the covariance matrix is characterized by the condition metric, $Z$, of the covariance matrix, as prescribed in Ref. [154]. To compute the dataset $\chi^2$ and the global $\chi^2$, we take all the distributions in our baseline plus the single distribution being considered and perform a new fit with 100 replicas each, and from this dataset-specific fit, we obtain the dataset $\chi^2$ and the global $\chi^2$ from the central replica. As such, we perform 37 PDF fits in total, 1 for each dataset being considered. In Table 3.11, the Z values and the $\chi^2$ values are presented for each dataset. The choice of distributions to be selected is then performed by opting for those that simultaneously have low $Z$ and $\chi^2$ as compared to other distributions in the dataset. This may not necessarily mean choosing one with the lowest value. Consider for example, the CMS top pair production in $\ell + j$ channel as shown in Table 3.11, where the distribution chosen does not have the lowest $Z$ or the lowest $\chi^2$ but rather has both on the lower end as compared to the other distributions amongst the entire dataset.

| Distribution | $N_{dat}$ | Z | dataset $\chi^2$ | global $\chi^2$ | $n_\sigma$ |
|---|---|---|---|---|---|
| ATLAS $t\bar{t}$ hadr. 13 TeV: $d\sigma/dm_{t\bar{t}}$ | 9 | 7.27 | 0.92 | 1.12 | -0.17 |
| ATLAS $t\bar{t}$ hadr. 13 TeV: $1/\sigma\, d\sigma/dm_{t\bar{t}}$ | 9 | 64.73 | 1.38 | 1.13 | 0.81 |
| ATLAS $t\bar{t}$ hadr. 13 TeV: $d\sigma/d\|y_{t\bar{t}}\|$ | 12 | 5.27 | 0.75 | 1.13 | -0.61 |
| ATLAS $t\bar{t}$ hadr. 13 TeV: $1/\sigma\, d\sigma/d\|y_{t\bar{t}}\|$ | 12 | 1.77 | 0.73 | 1.13 | -0.66 |
| ATLAS $t\bar{t}$ hadr. 13 TeV: $d^2\sigma/dm_{t\bar{t}}\, d\|y_{t\bar{t}}\|$ | 11 | 4.83 | 1.54 | 1.13 | 1.27 |
| ATLAS $t\bar{t}$ hadr. 13 TeV: $1/\sigma\, d^2\sigma/dm_{t\bar{t}}\, d\|y_{t\bar{t}}\|$ | 11 | 52.14 | 1.64 | 1.13 | 1.50 |
| ATLAS $t\bar{t}$ $\ell$+jets 13 TeV: $d\sigma/dm_{t\bar{t}}$ | 9 | 16.19 | 0.92 | 1.13 | -0.17 |
| ATLAS $t\bar{t}$ $\ell$+jets 13 TeV: $1/\sigma\, d\sigma/dm_{t\bar{t}}$ | 9 | 7.62 | 1.32 | 1.13 | 0.68 |
| ATLAS $t\bar{t}$ $\ell$+jets 13 TeV: $d\sigma/dp_{T,t}$ | 8 | 16.80 | 0.86 | 1.13 | -0.28 |
| ATLAS $t\bar{t}$ $\ell$+jets 13 TeV: $1/\sigma\, d\sigma/dp_{T,t}$ | 8 | 8.46 | 0.77 | 1.13 | -0.46 |
| ATLAS $t\bar{t}$ $\ell$+jets 13 TeV: $d\sigma/d\|y_t\|$ | 5 | 11.71 | 1.39 | 1.13 | 0.62 |
| ATLAS $t\bar{t}$ $\ell$+jets 13 TeV: $1/\sigma\, d\sigma/d\|y_t\|$ | 5 | 2.06 | 1.39 | 1.13 | 0.62 |
| ATLAS $t\bar{t}$ $\ell$+jets 13 TeV: $d\sigma/d\|y_{t\bar{t}}\|$ | 7 | 15.69 | 0.35 | 1.13 | -1.22 |
| ATLAS $t\bar{t}$ $\ell$+jets 13 TeV: $1/\sigma\, d\sigma/d\|y_{t\bar{t}}\|$ | 7 | 2.26 | 0.38 | 1.13 | -1.16 |
| ATLAS $t\bar{t}$ $\ell$+jets 13 TeV: $\frac{d^2\sigma}{dm_{t\bar{t}}dp_{T,t}}$ | 15 | 9.85 | 2.34 | 1.14 | 3.67 |
| ATLAS $t\bar{t}$ $\ell$+jets 13 TeV: $\frac{1}{\sigma}\frac{d^2\sigma}{dm_{t\bar{t}}dp_{T,t}}$ | 15 | 5.91 | 1.83 | 1.14 | 2.27 |
| CMS $t\bar{t}$ $\ell$+jets 13 TeV: $d\sigma/dm_{t\bar{t}}$ | 15 | 3.90 | 0.55 | 1.13 | -1.23 |
| CMS $t\bar{t}$ $\ell$+jets 13 TeV: $1/\sigma\, d\sigma/dm_{t\bar{t}}$ | 15 | 3.51 | 0.43 | 1.13 | -1.56 |
| CMS $t\bar{t}$ $\ell$+jets 13 TeV: $d\sigma/dp_{T,t}$ | 16 | 4.04 | 0.84 | 1.13 | -0.45 |
| CMS $t\bar{t}$ $\ell$+jets 13 TeV: $1/\sigma\, d\sigma/dp_{T,t}$ | 16 | 1.78 | 0.87 | 1.13 | -0.37 |
| CMS $t\bar{t}$ $\ell$+jets 13 TeV: $d\sigma/d\|y_t\|$ | 11 | 5.75 | 2.04 | 1.14 | 2.44 |
| CMS $t\bar{t}$ $\ell$+jets 13 TeV: $1/\sigma\, d\sigma/d\|y_t\|$ | 11 | 1.36 | 1.97 | 1.13 | 2.27 |
| CMS $t\bar{t}$ $\ell$+jets 13 TeV: $d\sigma/d\|y_{t\bar{t}}\|$ | 10 | 9.68 | 3.36 | 1.14 | 5.28 |
| CMS $t\bar{t}$ $\ell$+jets 13 TeV: $1/\sigma\, d\sigma/d\|y_{t\bar{t}}\|$ | 10 | 1.53 | 3.36 | 1.13 | 5.28 |
| CMS $t\bar{t}$ $\ell$+jets 13 TeV: $d^2\sigma/dm_{t\bar{t}}\, d\|y_{t\bar{t}}\|$ | 35 | 22.41 | 5.07 | 1.19 | 17.03 |
| CMS $t\bar{t}$ $\ell$+jets 13 TeV: $1/\sigma\, d^2\sigma/dm_{t\bar{t}}\, d\|y_{t\bar{t}}\|$ | 35 | 17.23 | 4.78 | 1.16 | 15.81 |
| ATLAS incl. jet (R=0.4) 13 TeV: $d^2\sigma/dp_T\, d\|y\|$ | 177 | 16.87 | 2.85 | 1.23 | 17.40 |
| ATLAS dijet (R=0.4) 13 TeV: $d^2\sigma/dm_{jj}\, d\|y^*\|$ | 136 | 16.80 | 1.76 | 1.17 | 6.27 |
| CMS incl. jet (R=0.4) 13 TeV: $d^2\sigma/dp_T\, d\|y\|$ | 78 | 13.30 | 2.73 | 1.18 | 10.80 |
| CMS incl. jet (R=0.7) 13 TeV: $d^2\sigma/dp_T\, d\|y\|$ | 78 | 14.81 | 2.40 | 1.16 | 8.74 |
| H1 incl. jet 319 GeV (290 pb$^{-1}$): $d^2\sigma/dq^2\, dp_T$ | 48 | 5.99 | 2.65 | 1.14 | 8.08 |
| H1 dijet 319 GeV (290 pb$^{-1}$): $d^2\sigma/dq^2\, d\langle p_T\rangle$ | 48 | 7.67 | 2.84 | 1.15 | 9.01 |
| H1 incl. jet 319 GeV (351 pb$^{-1}$): $d^2\sigma/dq^2\, dp_T$ | 24 | 1.46 | 1.33 | 1.13 | 1.14 |
| H1 dijet 319 GeV (351 pb$^{-1}$): $d^2\sigma/dq^2\, d\langle p_T\rangle$ | 24 | 1.57 | 1.33 | 1.14 | 1.14 |
| ZEUS incl. jet 300 GeV (38.6 pb$^{-1}$): $d^2\sigma/dE_T\, dq^2$ | 30 | 1.87 | 0.65 | 1.13 | -1.36 |
| ZEUS incl. jet 319 GeV (82 pb$^{-1}$): $d^2\sigma/dE_T\, dq^2$ | 30 | 2.56 | 0.77 | 1.13 | -0.89 |
| ZEUS dijet 319 GeV (374 pb$^{-1}$): $d^2\sigma/d\langle E_{T,B}\rangle\, dq^2$ | 22 | 2.83 | 0.89 | 1.13 | -0.36 |

Table 3.11: This table lists all the distributions considered for the NewFit together with the number of data points, the condition metric $Z$, the dataset $\chi^2$, the global $\chi^2$ and the number of standard deviations $n_\sigma$ from the global $\chi^2$ of the dataset $\chi^2$. The green rows indicate the distributions that were selected for the NewFit based on the criteria described in Sec. 3.4.3.1.

#### 3.4.3.2 New fit and its comparison with the baseline fit

With the baseline fit fully defined and the method for including new datasets in the new fit fully explained, I proceed with performing a new fit (referred to as 'NewFit' in this section henceforth), which includes old datasets from the baseline + new datasets selected in this study (as explained in Sec. 3.4.3.1 and shown in Table 3.11). Both, the baseline fit and the NewFit are produced with 100 replicas each. With this, we proceed with the comparison between the NewFit, and our baseline fit.

One of the measures to compare the differences between two PDFs of the same flavor is by looking at their distances (defined in Ref. [179]). In the left plot in figure 3.16, the distances between the PDFs of the NewFit and the baseline fit are shown. For most of the PDFs, the distance is less than or equal to $\sim 5$, which indicates an agreement of within $0.5\ \sigma$. The second highest distance is for the charm and the anti-charm PDFs which leads to a standard deviation of $\sim 1.3\ \sigma$ at its maximum. The distance between the gluon PDFs is the largest whereby it reaches values of standard deviations as high as $\sim 1.8$ $\sigma$ and $\sim 2.0\ \sigma$. In the right plot in figure 3.16, the distances between the variance of the two fits are shown. This shows the differences between the uncertainties of the PDFs of the two fits. The variance distances are always much less than $\sim 4$, indicating the size of the uncertainties of the two fits are in agreement within $0.4\ \sigma$. From this observation, we can deduce that the differences in PDFs which are shown below, are mainly due to the central values and not the uncertainties.

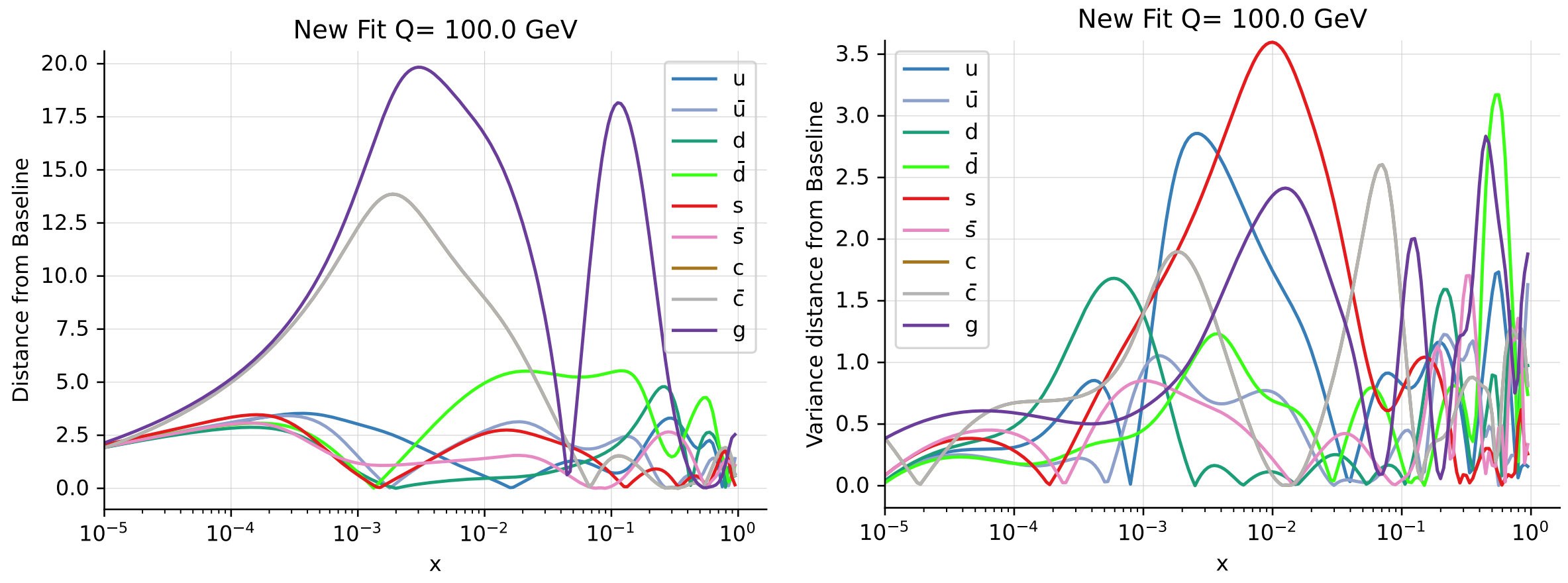

Figure 3.16: The left diagram shows the distances between the PDFs of different flavors of the NewFit and the Baseline Fit, and the right diagram show the distances between the uncertainties of the PDFs of different flavors of the NewFit and the Baseline Fit.

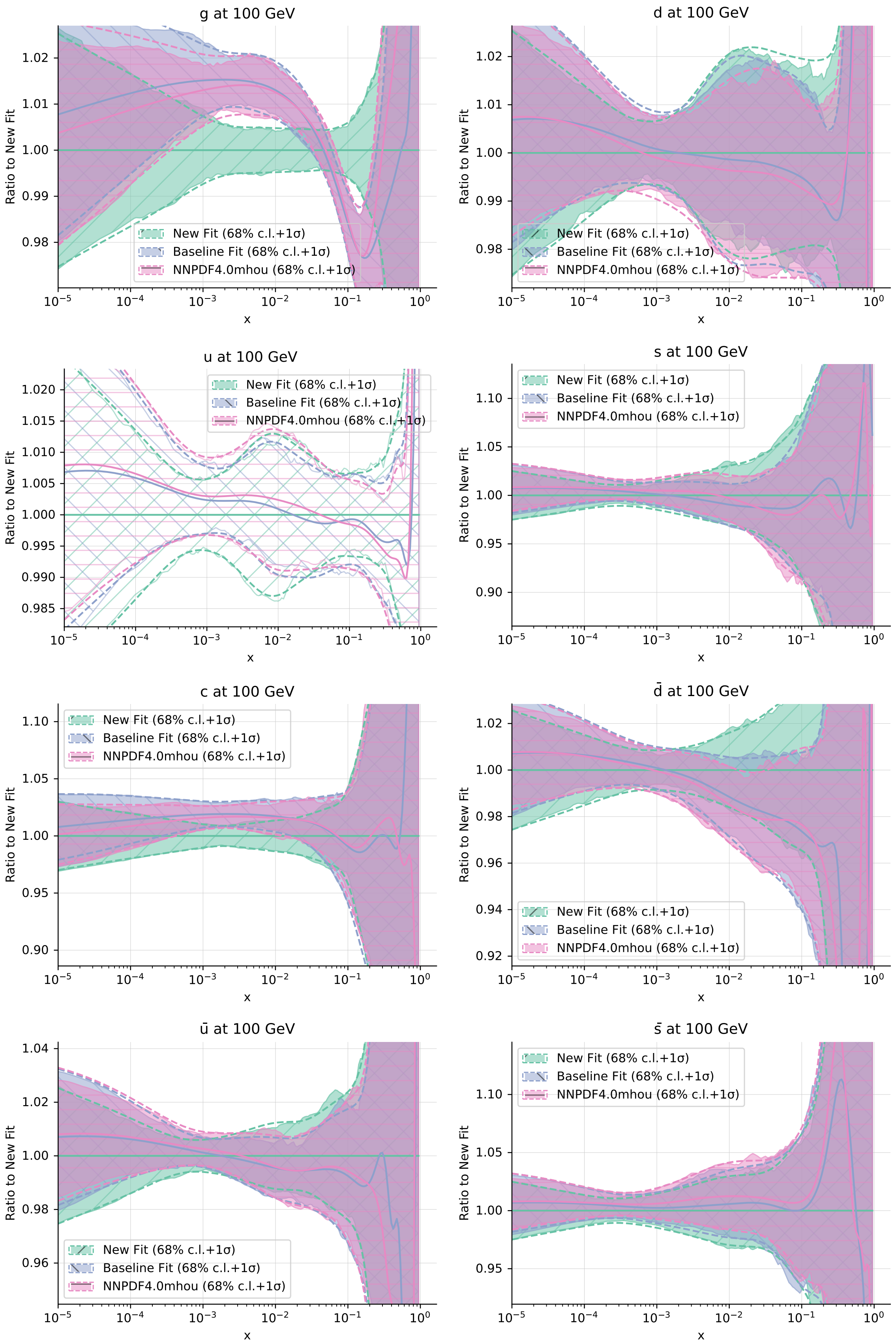

Figure 3.17: A comparison between the PDFs of each flavor from the NewFit, the Baseline Fit and NNPDF4.0mhou.

In figure 3.17, the explicit comparison between PDFs of different flavors is shown. As expected from figure 3.16, the largest differences are observed in the gluon PDFs, specifically in the x-range of $10^{-3}$ to $10^{-2}$. The plot shows that the gluon PDF of the NewFit has an uncertainty of $\sim$ 0.5%, whereas the difference between the central value of the gluon PDF of the NewFit and the uncertainty band of the baseline fit is $\sim$ 1%, indicating a 2 $\sigma$ difference, consistent with what was observed in figure 3.16. The difference between the central values of the gluon PDFs of both the fits is $\sim$ 3% indicating a 3 $\sigma$ difference. For all the other flavors, the uncertainty bands of the two fits overlap in all x regions (except for a very small range for the charm and anti-charm PDFs around x $\sim 2 \times 10^{-3}$), showing a good agreement between the two fits for these flavors.

#### 3.4.3.3 Comparison with other PDF sets

In this section, a comparison between the NewFit and other PDF sets, namely CT18 [32] and MSHT20 [33] is presented. In figure 3.18, the PDF comparison plots for the individual flavors is shown. As there was full compatibility between the NewFit and the baseline fit (except for the gluon PDF), one would expect the PDF comparisons to be analogous to the comparison between NNPDF4.0, CT18 and MSHT20. This is indeed the case for all the quark PDFs, see Fig. 5.6 of [22] where the comparison between NNPDF4.0, CT18 and MSHT20 is shown. The plot of the gluon PDF comparison is the most interesting. In Fig 5.6 of [22], there are a number of x regions where there is incompatibility between NNPDF4.0 and CT18, and NNPDF4.0 and MSHT20. In the case of the NewFit, we observe compatibility with CT18 in all x regions, whereas with MSHT20, we observe compatibility in all x regions except for a very small x-region ($x \sim 10^{-1}$).

### 3.4.4 Impact on phenomenology

#### 3.4.4.1 Luminosities

Any differences in the PDFs between the NewFit and the baseline fit would translate into differences in the luminosities involving the specific partonic channels. Figure 3.19 shows how the luminosities differ between the NewFit and the baseline fit. As expected, the quark-quark and the quark-anti-quark luminosities are compatible within their uncertainties between the two fits. We observe differences in the gluon-gluon and gluon-quark luminosities. In particular, the gluon-gluon luminosity is reduced by more than $1\sigma$ in the $\sim$ 50-150 GeV region. The maximum reduction happens around the 90 GeV region, however the most consequential reduction happens around the Higgs mass region, where the central value of the luminosity of the NewFit is lower than the central value of the baseline fit by $2 - 2.5\%$ or a little more than $2.5\sigma$ difference. This translates to a reduction in the Higgs cross section in the gluon-gluon channel. As gluon-gluon fusion is the dominant production mechanism for the Higgs boson, this result has noteworthy implications for precision Higgs physics at the LHC.

As for gluon-quark luminosity, there are differences present between the two fits. In the $\sim$ 100-250 GeV region, the central value of the luminosity of the NewFit is lower than the central value of the luminosity of the baseline fit by a little less than $3\sigma$. Furthermore, the luminosity of the NewFit is higher by as much as $2.5-3\sigma$ in the TeV region, with potential implications for new physics searches. For completeness, we also show the comparison between the luminosities of the NewFit, CT18 and MSHT20 in figure 3.20.

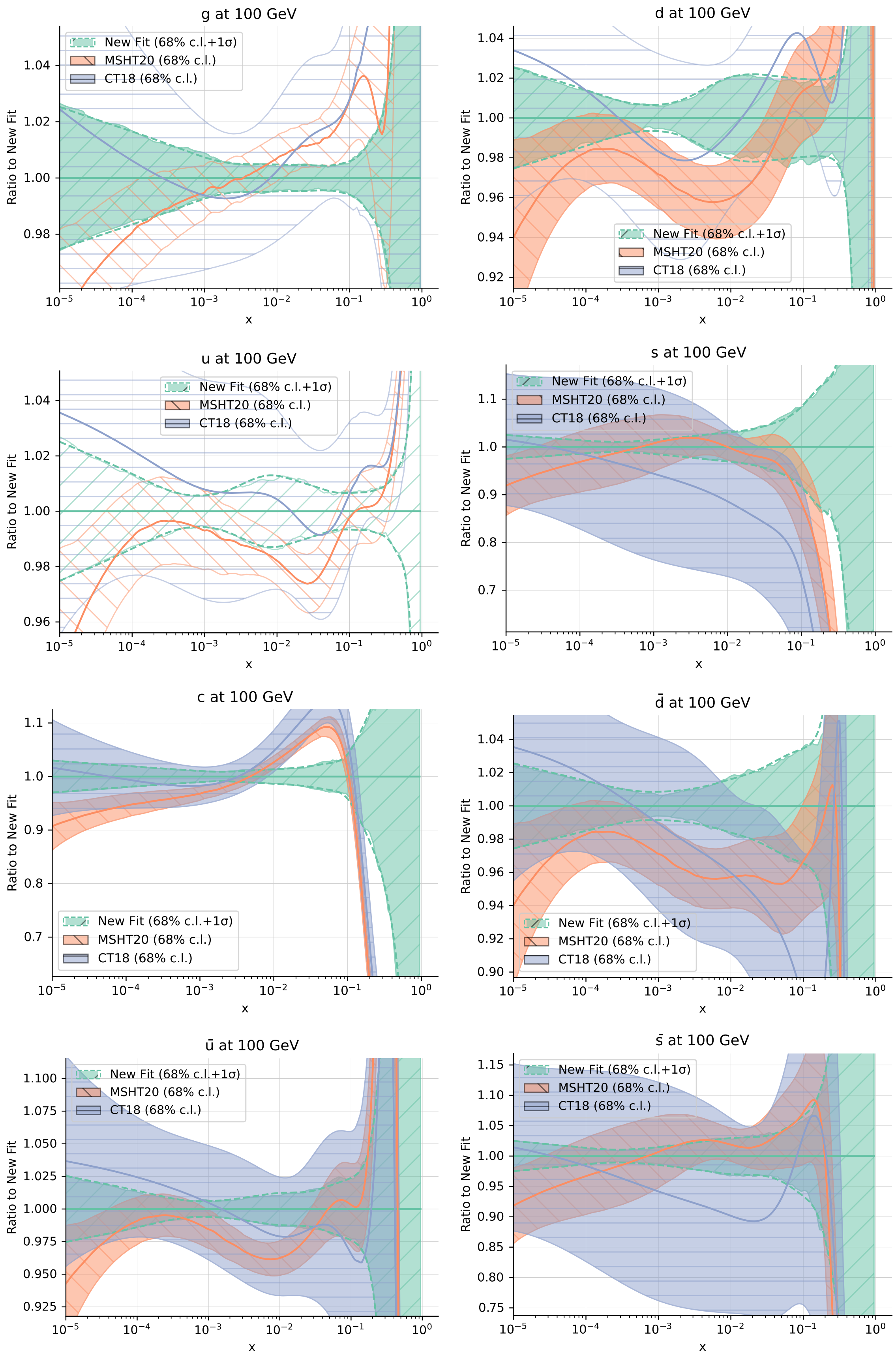

Figure 3.18: A comparison between the PDFs of each flavor from the NewFit, CT18 and MSHT20. (It should be noted that CT18 and MSHT20 do not parametrize charm, while NNPDF does, which is also what is done in the NewFit.)

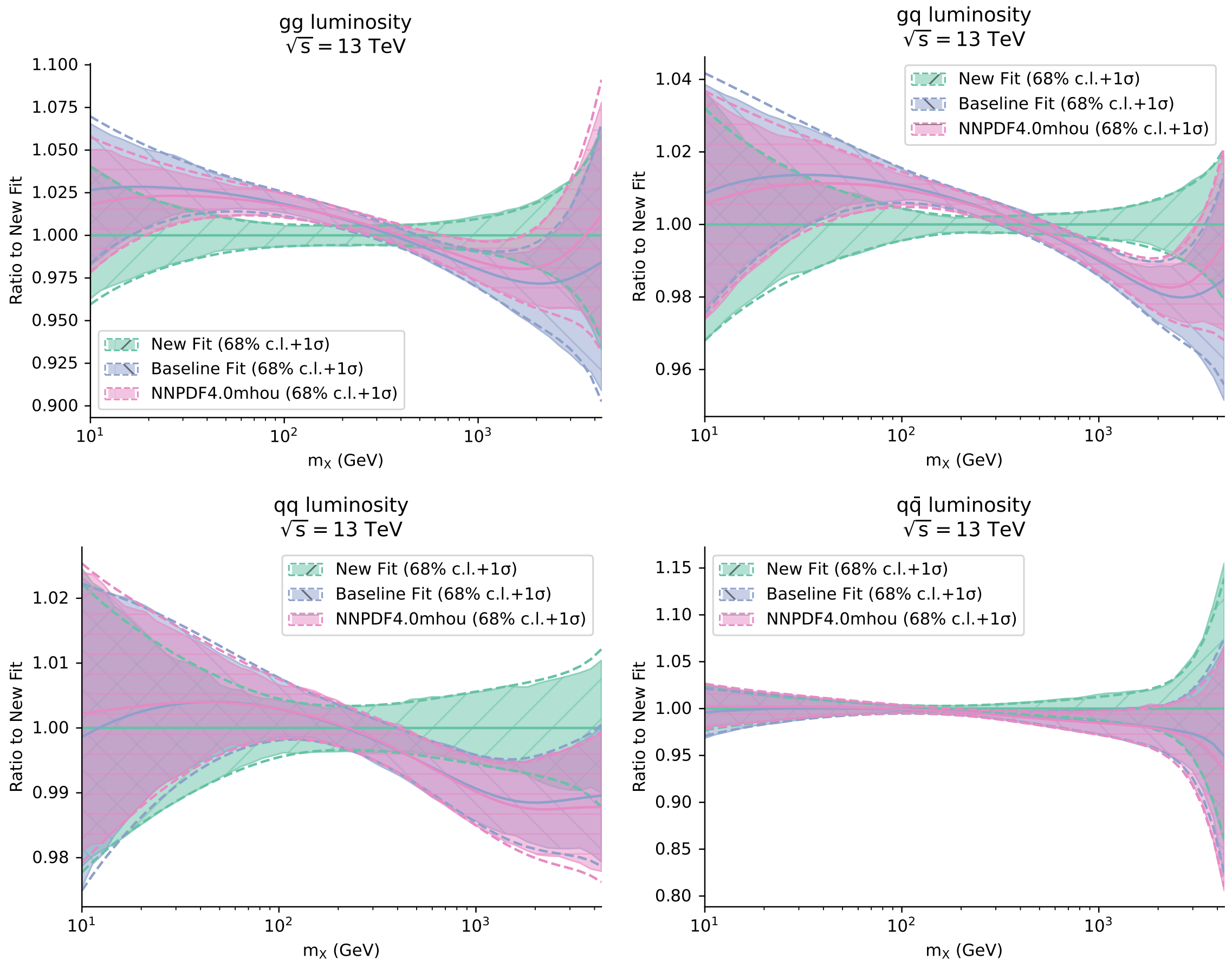

Figure 3.19: A comparison between the luminosities of the NewFit, the Baseline fit and NNPDF4.0mhou. The luminosities are shown for the gluon-gluon channel (top left), gluon-quark channel (top right), quark-quark channel (bottom left) and quark-anti-quark channel (bottom right).

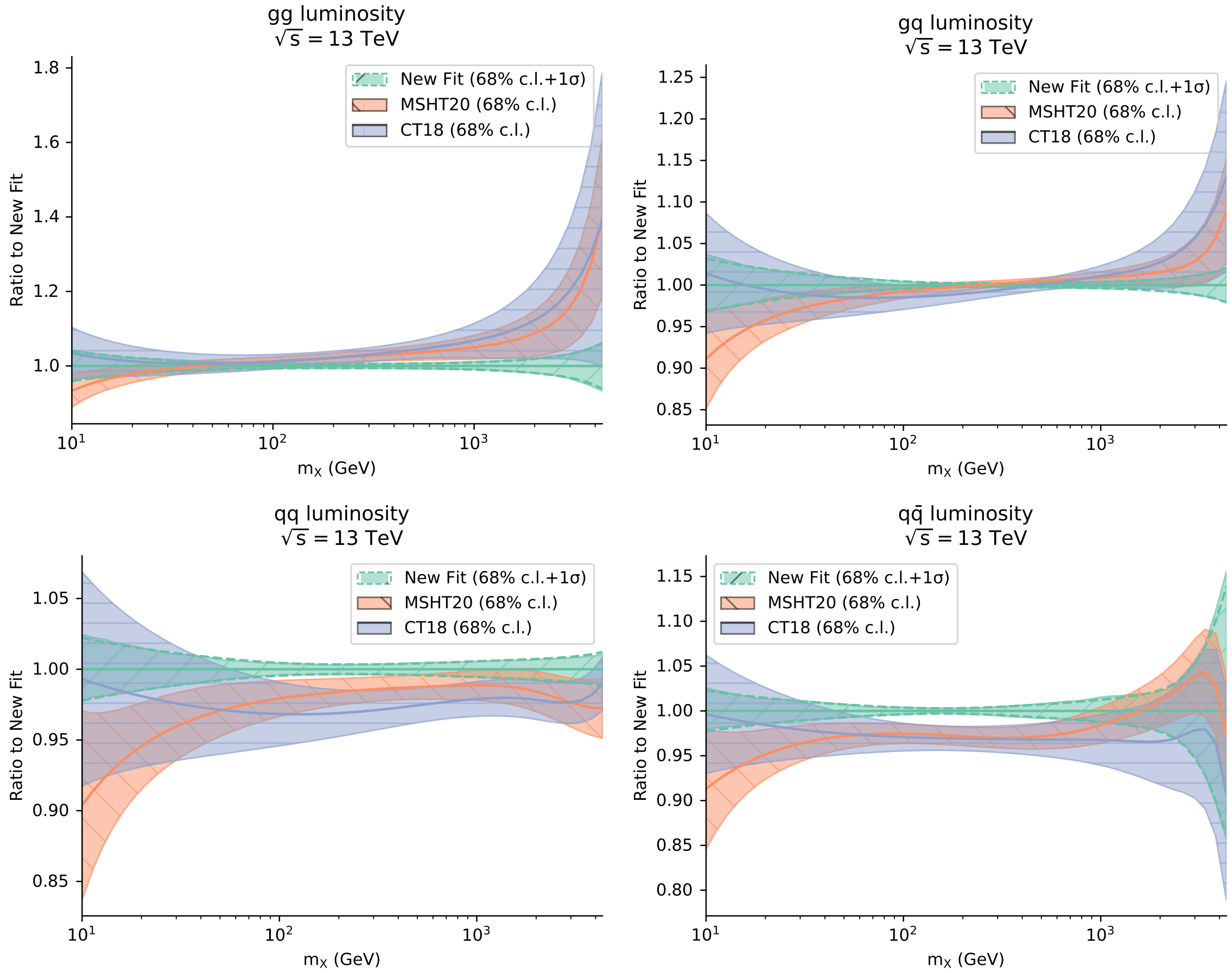

Figure 3.20: A comparison between the luminosities of the NewFit, CT18 and MSHT20. The luminosities are shown for the gluon-gluon channel (top left), gluon-quark channel (top right), quark-quark channel (bottom left) and quark-anti-quark channel (bottom right).

#### 3.4.4.2 Cross sections

Having seen the comparisons for the PDFs and the luminosities corresponding to the NewFit, the baseline fit, CT18 and MSHT20, we now turn our attention to the integrated cross sections. The computation of these cross section is performed in the exact same manner as was done in Sec. 9.2 of [22], and therefore we refer to that reference for the technical details of the computation. In figures 3.21 and 3.22, our predictions for the total LHC cross-sections at 14 TeV are shown. For Drell Yan production, in the first three plots of Fig. 3.21, we see that the NewFit and the baseline fit are right on top of each other, with the only difference being a slight reduction in the uncertainty of the NewFit. For gauge boson pair production, in the last three plots of Fig. 3.21, we see compatibility between the results of the NewFit and the baseline fit, within their uncertainty bands. In Fig. 3.22, we show the total cross sections for top pair production, and for Higgs production in various channels, and the only cross section that is incompatible within the $1\sigma$ band between the NewFit and the baseline fit is the gluon fusion channel for the Higgs production. The NewFit leads a cross section of 32.5 pb whereas the baseline fit leads to a cross section of $\sim$ 33.2 pb. This translates to the NewFit giving a cross section that is $\sim$ 2.5% lower than the baseline fit. In figure 3.19, we saw that the gluon PDF of the NewFit was lower than the baseline by 1.5% in the specified x-region, and the square of this number should translate to the reduction in the cross section. This is also consistent with the luminosity plot in figure 3.19. With this, we clearly see that the inclusion of the new data in the NewFit has a significant impact on the gluon PDF which has important phenomenological implications for the Higgs physics at the LHC.

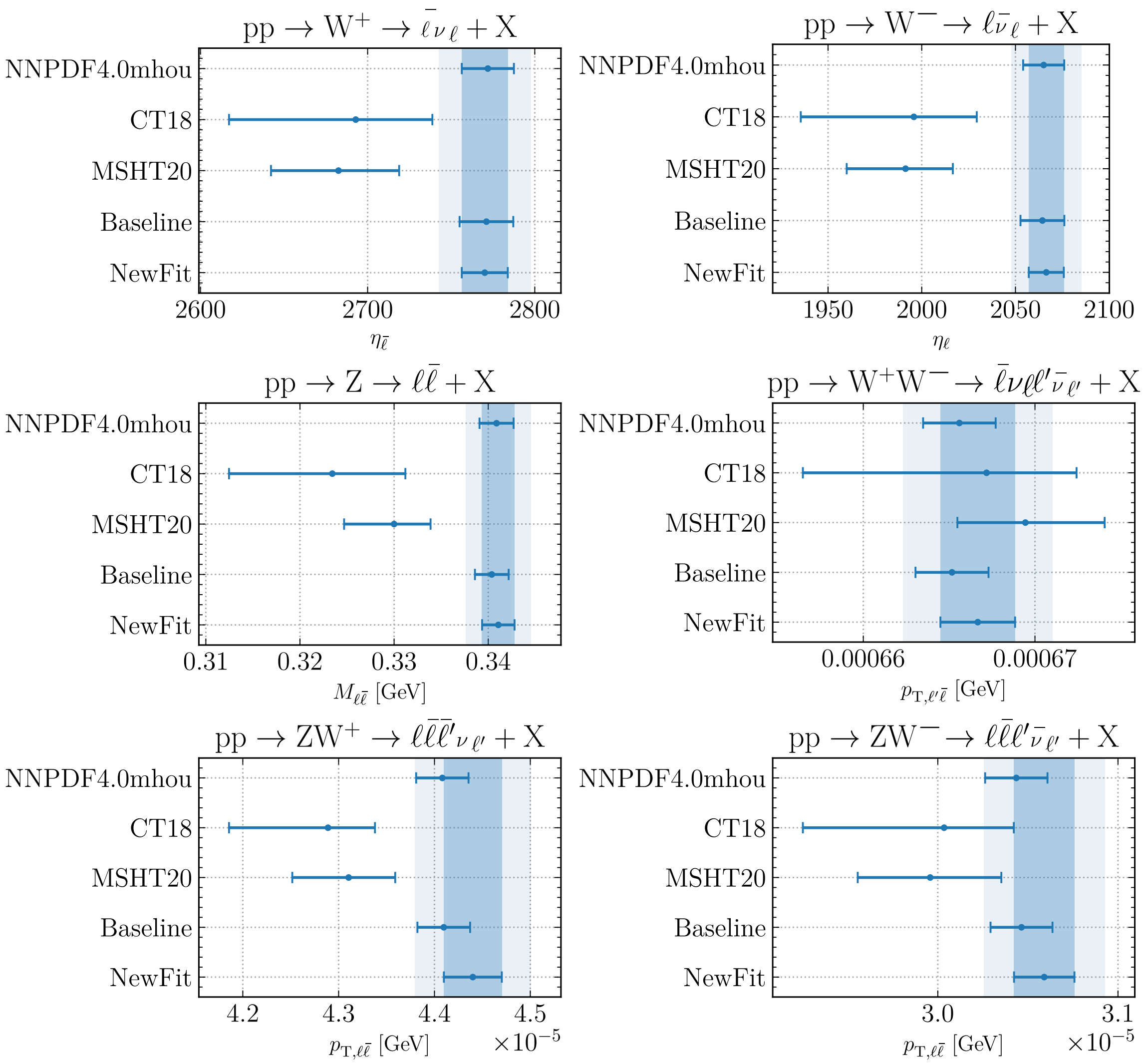

Figure 3.21: Total LHC cross-sections at 14 TeV for Drell Yan production (first 3) and gauge boson pair production (last 3) obtained with NewFit, Baseline Fit, CT18, MSHT20 and NNPDF4.0mhou with $\alpha_s(m_Z) = 0.118$. The dark and light bands represent $1\sigma$ and $2\sigma$ uncertainty bands respectively, for the NewFit.

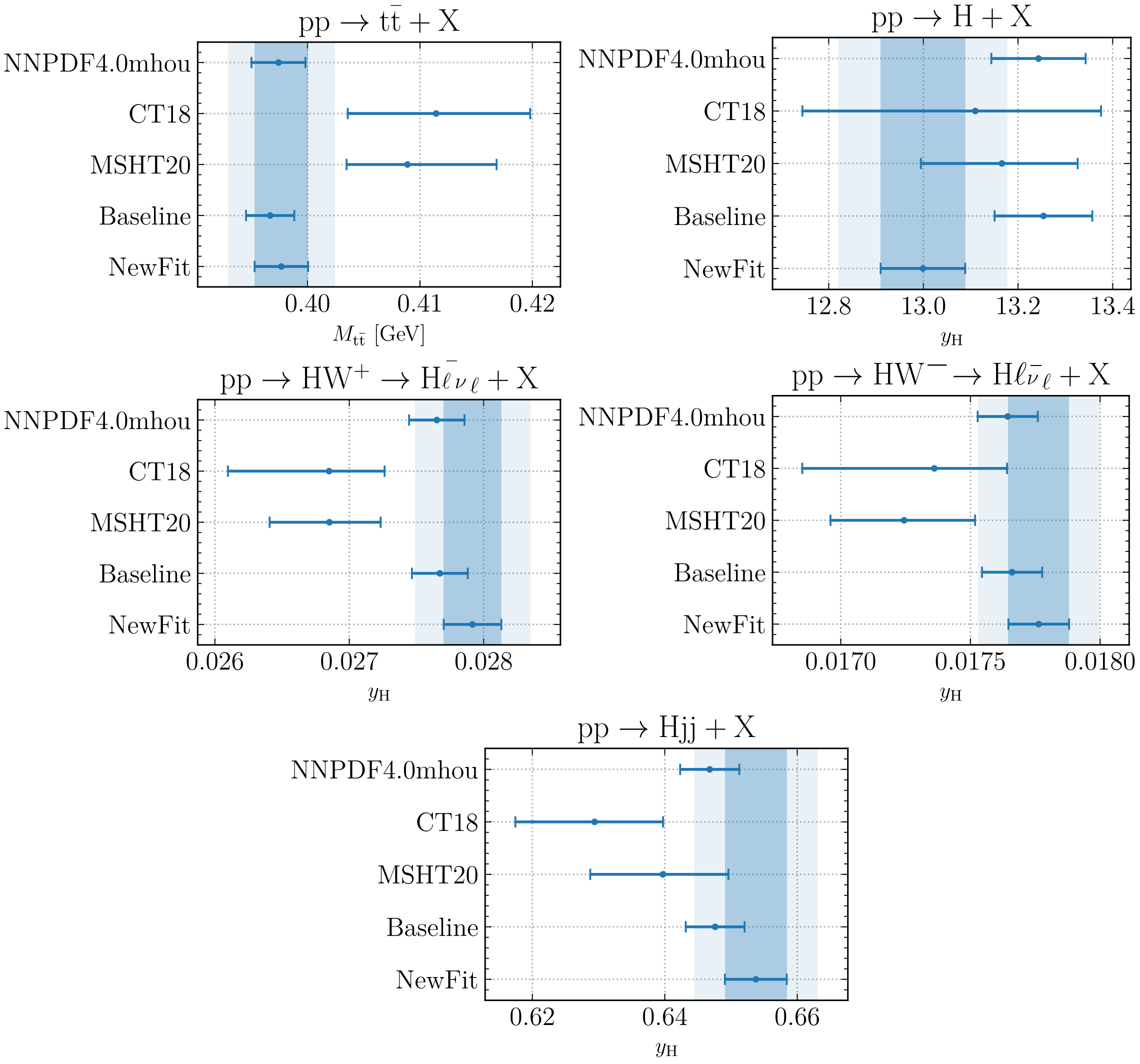

Figure 3.22: Same as Fig. 3.21 but for top pair production, and for Higgs production in various channels: gluon fusion, associated production with a W boson, and vector boson fusion.

### 3.4.5 Summary and outlook

In this section, we set out to assess the impact of new and high precision LHC and HERA data corresponding to processes that are sensitive to the gluon PDF, in particular, single-inclusive jet and dijet production at the LHC Run II and HERA, and top pair production at the LHC Run II. To start, we performed a baseline fit that was NNPDF4.0 with MHOUs-like, but included the exact NNLO QCD corrections for the old top pair production datasets.

We then systematically assessed the impact and consistency of the individual distributions by including the specified distribution, one at a time, to our baseline fit. Through this, we were able to select the distributions that maximized the consistency of the fit. This allowed us to perform our PDF determination, the NewFit, which includes the new high-precision data. We found that the NewFit is largely consistent with the baseline fit, for all the flavors of PDFs, except for the gluon PDF. The gluon PDF of the NewFit is lower than the baseline fit in the x-region of $10^{-3}$ to $10^{-2}$. At the same time, the gluon PDF of the NewFit achieves a better agreement with CT18 and MSHT20 as compared to the NNPDF4.0.

The reduction of the gluon PDF in the x-region of $10^{-3}$ to $10^{-2}$ translates to a reduction in the gluon-gluon luminosity in the mass range that includes the Higgs boson mass. Consequently, the computation of the total cross section for the Higgs boson production in the gluon fusion channel using the NewFit shows a reduction of $\sim 2.15\%$ as compared to the baseline fit. This is an important result as the gluon fusion channel is the dominant production channel for the Higgs boson and leads to important phenomenological implications for precision Higgs physics at the LHC.

## 3.5 Summary and future outlook

The process of producing a new and state-of-the-art PDF set is a complex and lengthy task requiring lots of incremental improvements. In this chapter, I have discussed the projects I carried out towards the endeavor of producing the next-gen PDF set by the NNPDF collaboration. This includes

- assessing the impact of moving from NNLO K-factors to pure NNLO corrections, by means of comparing K-factors produced using various different PDF sets,
- assessing the compatibility and generalization power of the most widely used PDF sets by confronting them with high-precision data from the Run II of the LHC, and from the jets data from HERA which has typically not been used in PDF determinations and,
- assessing the impact on the gluon PDF when new high-precision data that is particularly sensitive to the gluon PDF is included in the fit, such as top pair production data from the LHC and jets data from LHC and HERA.

These projects are complemented by the plethora of other projects and tasks that have been carried out by my colleagues in the NNPDF collaboration. A significant amount of work still remains that we are currently working on which together will allow for the release of a new and improved PDF set in the near future.

# Chapter 4

# Advancing FFs

## 4.1 Introduction

Fragmentation functions are an important part of our understanding of non-perturbative QCD. They encode the long distance dynamics amongst quarks and gluons that leads to their hadronization. As such, they provide us with the insights into the formation of hadrons as well as into the hadronic structure, as they are a key object used in the description of some processes that are used to study the internal structure of nucleons. Just like PDFs, they need to be determined by performing a global fit using experimental data. A very key aim of this thesis has been to develop the machinery that would allow the current state-of-the-art NNPDF methodology to be used to determine fragmentation functions for light hadrons, such as pions, kaons, and protons. This would allow for the release of NNFF2.0, a much more advanced and robust successor to NNFF1.0 [180], which was released in 2017. It will be more advanced and robust due to multiple aspects such as the inclusion of SIDIS data in the fitting procedure at NNLO accuracy, inclusion of theoretical uncertainties from scale variations in the covariance matrix, and the use of hyperoptimization to set the hyperparameters of the neural network. While this chapter has a very similar name to that of chapter 3 which focuses on PDFs, the goals of this chapter are different. Unlike the PDF focused studies, which were based on existing NNPDF software to move towards a next generation determination, the FF focused studies are aimed at building from scratch the capability to perform FF determination. This is done by extending existing tools in some cases, while developing new tools in others. In Sec. 4.2, I will present on the extension of EKO, a tool that solves DGLAP evolution equations, to support time-like evolution. In Sec. 4.3, I will present on the extension of PineAPPL, a library for interpolation grids, to support multiple convolutions of PDFs and FFs. In Sec. 4.4, I will present on the development of a new software, vhf, that allows for the computation of theoretical predictions for SIA and SIDIS. I will discuss some underlying details of the code and show a proof of concept of the code's ability to produce theoretical predictions by presenting data-theory comparisons.

## 4.2 Evolution kernel operators

In Sec. 1.4.2, it was explained that PDFs and FFs are objects that are parametrized at an initial scale and can then be evolved to different energy scales using the DGLAP evolution equations. In Sec. 2.2.3, it was explained that the fitting procedure requires the use of evolution kernel operators to generate FK tables such that the partonic cross sections (at an arbitrary given scale) can be seamlessly convolved with FFs extracted at the parametrization scale. To this end, EKO [16] was developed as part of the NNPDF framework, to generate the evolution kernel operators that are combined with interpolation grids to produce the FK tables.

The DGLAP evolution equations rely on splitting functions. The splitting functions used in the evolution of PDFs are space-like splitting functions, while the splitting functions used in the evolution of FFs are time-like splitting functions. EKO only contained the implementation of space-like splitting functions, and hence was not able to generate the evolution kernel operators for FFs. It was therefore necessary to extend EKO by implementing the time-like splitting functions and allowing for the possibility to generate the evolution kernel operators for FFs. EKO is designed to be a code that solves the evolution equations in Mellin space (as is discussed below) but provides the solution in momentum space. The theoretical concepts underlying the working of EKO have already been discussed in various sections before, for example, DGLAP evolution in Sec. 1.4.2, Mellin space in Sec. 1.4.4, and Lagrange interpolation in Sec. 2.2.2. Therefore, in this section, the only focus is on the equations that are actually implemented as part of this extension of EKO's functionality. It should be noted that splitting functions in Mellin space are called anomalous dimensions, and they are related to the splitting functions in momentum space by the following relation:

$$\gamma_{ij}(N, \alpha_s) = -\mathscr{M}(P_{ij}(z, \alpha_s))(N) \tag{4.1}$$

where $\gamma_{ij}(N, \alpha_s)$ represent the anomalous dimensions in Mellin space, and $\mathscr{M}$ is the Mellin transform, as defined in Eq. (1.30). Note the extra minus sign in the above equation, which is a convention that is used EKO.

The time-like splitting functions have been computed up to NNLO in existing literature [4, 181–186]. At LO, the time-like splitting functions are identical to the space-like splitting functions, but with a slight caveat in their usage in the evolution equations. As opposed to the coupled evolutions equations in Eq. (1.29), the time-like coupled evolution equations are given by:

$$\frac{d}{d\ln\mu_f^2}\begin{pmatrix} D_\Sigma(x,\mu_f^2) \\ D_g(x,\mu_f^2) \end{pmatrix} = \begin{pmatrix} P_{qq} & P_{gq} \\ P_{qg} & P_{gg} \end{pmatrix} \otimes \begin{pmatrix} D_\Sigma(x,\mu_f^2) \\ D_g(x,\mu_f^2) \end{pmatrix} \tag{4.2}$$

i.e. the off-diagonal terms are swapped. The $n_f$ factors of the elements in the splitting function matrix have been absorbed in the definition of the splitting functions for simplicity. The independent evolution equations are the same in the time-like case as in the space-like case in Eq. (1.28). At LO, there are 4 splitting functions in total:

$$P_{qq}^{(0)}(z), \quad P_{gq}^{(0)}(z), \quad P_{qg}^{(0)}(z), \quad P_{gg}^{(0)}(z)$$

At NLO, there are 6 splitting functions:

$$P_{+}^{(1)}(z), \quad P_{-}^{(1)}(z), \quad P_{ps}^{(1)}(z), \quad P_{gq}^{(1)}(z), \quad P_{qg}^{(1)}(z), \quad P_{gg}^{(1)}(z)$$

At NNLO, there are 7 splitting functions:

$$P_{+}^{(2)}(z), \quad P_{-}^{(2)}(z), \quad P_{v}^{(2)}(z), \quad P_{ps}^{(2)}(z), \quad P_{gq}^{(2)}(z), \quad P_{qg}^{(2)}(z), \quad P_{gg}^{(2)}(z).$$

The four splitting functions in the Eq. (4.2) are the ones in the singlet sector. The remaining, $P_{\pm,v}$ are in the non-singlet sector, and $P_{qq}^{(1/2)}(z) = P_{+}^{(1/2)}(z) + P_{ps}^{(1/2)}(z)$. The non-singlet splitting functions are convolved with their non-singlet PDF counterparts, which were defined in Sec. 1.4.2. The implementation of these time-like coefficient functions in EKO was complemented with a benchmark against the expressions present in MELA [187], which is another DGLAP evolution library that solves the evolution equations in Mellin space.

In a variable flavor scheme, when quark mass thresholds are crossed, the distributions behave differently above and below the threshold. To allow for these regimes to smoothly transition, matching conditions are used. The discussion on matching conditions is beyond the scope of this section, but it is worth noting that time-like matching conditions are known up to NLO [188], and these conditions were also been fully implemented in EKO. In recent times, NNLO matching conditions have also been partially computed [189], however, these are not yet implemented in EKO.

## 4.3 Interpolation grids

In Sec. 2.2.2, the need for using interpolation grids to enable efficient convolution was discussed. It was also mentioned that a library called PineAPPL was developed for this purpose by the NNPDF collaboration. While PineAPPL is a robust and versatile tool, it was originally designed with LHC processes in mind, and therefore focused on the convolution of interpolation grids with 2 PDFs. This design limited its applicability and prevented its use in processes involving fragmentation functions or a different number of convolutions, such as $p+p \to \pi + X$, which involves 3 non-perturbative objects.

Another limitation concerned the treatment of perturbative orders in PineAPPL. As described in Sec. 2.2.3, the 'orders' dimension of a grid accounts for the powers of $\alpha_s$, $\alpha_{em}$, $\log\left(\frac{\mu_r^2}{Q^2}\right)$, and $\log\left(\frac{\mu_f^2}{Q^2}\right)$. However, the term "factorization scale" in this context actually refers to the scale associated with the initial-state hadrons, i.e., PDFs. In processes involving final-state hadrons, there is an additional factorization scale known as the fragmentation scale, that originates from the factorization procedure and upon which fragmentation functions have a scale dependence.

To extend PineAPPL for use in FF fits, it was necessary to support an arbitrary number of convolutions involving various types of non-perturbative objects such as PDFs, polarized PDFs, and FFs, and to introduce the fragmentation scale. For this purpose, a new version of PineAPPL, v1.0.0 was developed and released. This version includes a redesigned data structure for the grids and all the aforementioned functionalities. Additionally, significant improvements were made to PineAPPL's interfaces in Python, C, and Fortran. This section provides a brief overview of the new features.

Even before its latest release, PineAPPL offered several distinguishing features relative to other fast interpolation libraries, most notably, the ability to include higher order EW corrections, and perform scale variation on the factorization and renormalization scales. It supported the consistent inclusion of electroweak corrections alongside QCD corrections. The grid format encoded the perturbative order and partonic sub-channel decomposition explicitly, enabling detailed inspection of different contributions. Scale dependence was also embedded, allowing variations to be performed a posteriori. Furthermore, a comprehensive user interface, including a command-line tool and language bindings, was provided to support various grid operations.

In this section, some of the new developments introduced in PineAPPL v1.0.0 are highlighted. Sec. 4.3.1 focuses on the extended capability to support an arbitrary number of hadronic states, while Sec. 4.3.2 discusses the updated internal data structure that enables this functionality.

### 4.3.1 Support for an arbitrary number of hadronic states

The new version of PineAPPL supports an arbitrary number of both initial- and final-state hadrons. It allows for flexible configurations of hadronic states, including polarized and unpolarized states, as well as time-like and space-like kinematics. To illustrate this, consider a generic example of hadron production in hadronic collisions:

$$N_1(P_a) + N_2(P_b) \to h_1(P_c) + X.$$

Assuming, for simplicity, that all hadrons are unpolarized, the factorized cross section takes the form:

$$\begin{aligned}\frac{d\sigma}{d\mathcal{K}}\left(\mathcal{K}, \mu_i, Q\right) = \sum_{a,b,c} \int_{x_1}^{1} \frac{d\hat{x}_1}{\hat{x}_1} \int_{x_2}^{1} \frac{d\hat{x}_2}{\hat{x}_2} \int_{z_1}^{1} \frac{d\hat{z}_1}{\hat{z}_1} \int_{Q^2_{\text{min}}}^{Q^2_{\text{max}}} dQ^2 \\ f_a^{N_1}\left(\frac{d\hat{x}_1}{\hat{x}_1}, \frac{\mu_F^2}{Q^2}\right) f_b^{N_2}\left(\frac{d\hat{x}_2}{\hat{x}_2}, \frac{\mu_F^2}{Q^2}\right) D_c^{h_1}\left(\frac{d\hat{z}_1}{\hat{z}_1}, \frac{\mu_A^2}{Q^2}\right) \\ \frac{d\sigma_{ab\to c}}{d\mathcal{K}}\left(\hat{x}_1, \hat{x}_2, \hat{z}_1, \frac{\mu_R^2}{Q^2}, \frac{\mu_F^2}{Q^2}, \frac{\mu_A^2}{Q^2}\right)\end{aligned} \tag{4.3}$$

where $i = R, F, A$, corresponds to the renormalization, factorization and fragmentation scales respectively. $Q$ corresponds to the physical energy scale of the process, such as that of the energy of the mediating gauge boson, or the transverse momentum of a specified particle. This expression is differential in the kinematic variable $\mathcal{K}$ and depends on the renormalization ($\mu_R$), factorization ($\mu_F$), and fragmentation ($\mu_A$) scales. The PDFs and FFs are denoted by $f_i^H$ and $D_i^h$, respectively. The partonic contribution, which is the focus of PineAPPL, can be systematically expanded as:

$$\frac{d\sigma_{ab\to cd}}{d\mathscr{K}}\left(\hat{x}_1, \hat{x}_2, \hat{z}_1, \frac{\mu_R^2}{Q^2}, \frac{\mu_F^2}{Q^2}, \frac{\mu_A^2}{Q^2}\right) = \sum_{k,l,m,n,p} \alpha_s^k\left(\mu_R^2\right)$$
$$\alpha_{em}^l \log^m\left(\frac{\mu_R^2}{Q^2}\right)\log^n\left(\frac{\mu_F^2}{Q^2}\right)\log^p\left(\frac{\mu_A^2}{Q^2}\right)$$
$$\times W_{ab\to c}^{(k,\ell,m,n,p)}\left(\hat{x}_1, \hat{x}_2, \hat{z}_1, Q^2, \mathscr{K}\right), \tag{4.4}$$

where $W$ is the quantity encoded in PineAPPL's internal grid format. $k$, $l$, $m$, $n$ and $p$ denote the powers of $\alpha_s$, $\alpha_{em}$, $\log\left(\frac{\mu_R^2}{Q^2}\right)$, $\log\left(\frac{\mu_F^2}{Q^2}\right)$ and $\log\left(\frac{\mu_A^2}{Q^2}\right)$ respectively, and also serve as indices of the 'orders' dimension in a PineAPPL grid. Further details on the theoretical formalism can be found in [15].

### 4.3.2 New data structure and grid representation

To enable this expanded functionality, the internal grid structure of PineAPPL has undergone major changes. In particular, the library now supports grids of arbitrary dimensionality via a custom sparse array implementation called `PackedArray`, which is optimized for the sparsity patterns common in practical applications.

The `PackedArray` structure incorporates two main design principles. First, regardless of dimensionality, the $n$-dimensional grid is stored as a linearized one-dimensional array, with multi-dimensional indices mapped to 1D indices for efficient access. Second, because many entries are zero due to phase-space constraints [12], only non-zero values are stored. Furthermore, non-zero elements that are close together are grouped, minimizing storage overhead. Each group stores only its position and size.

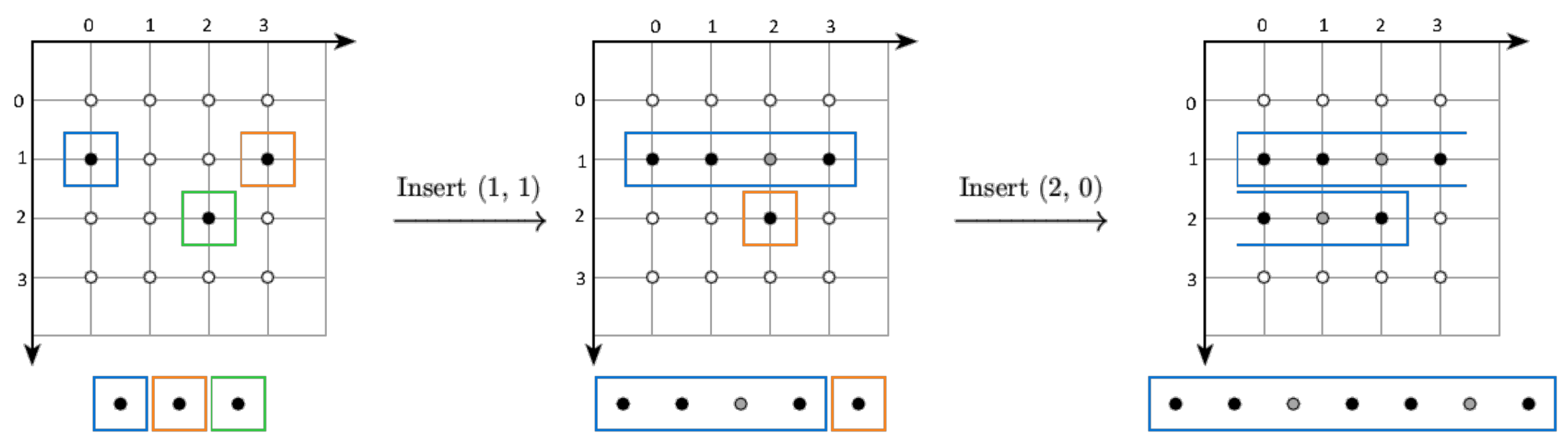

Figure 4.1: Visualization of filling the PackedArray, here in the 2D 4x4 case. It shows how the conceptual array (upper row) and the actual stored data (lower row) change when filling two elements at $(1,1)$ and $(2,0)$. Non-zero elements are indicated in black, explicitly stored zeros in gray, and implicit (non-stored) zeros in white. Elements that are grouped together are surrounded by a colored rectangle.

This grouping strategy is illustrated in Fig. 4.1. Initially, three non-zero elements (black) are present, each in separate groups (colored rectangles). Adding a new element at $(1, 1)$ causes it to merge with the group at $(1, 0)$, and even indirectly with $(1, 3)$ by storing an intermediate zero (gray). This approach is used when it is more efficient to store an explicit zero than to create multiple groups—typically the case when both `usize` and `f64` use the same number of bits. Finally, adding an element at $(2, 0)$ shows that group adjacency is determined by linear indexing, allowing groups to span across conceptual rows.

This newly introduced data structure not only supports arbitrary convolutions but also offers a more compact and efficient memory representation for high-dimensional sparse grids.

### 4.3.3 Interpolation precision

An important aspect of a library that is used for the production and usage of interpolation grids is its ability to perform interpolation with a precision that is good enough so as to not affect the theoretical predictions too much. More precisely, it is expected that the interpolation error should be equal to or lower than that of the error of the MC integrator that is used to produce the theory predictions. To illustrate this aspect of PineAPPL, let us look at some particular results from [190], where we presented the NNPDFpol2.0 PDF set. This was the first research work to utilize the new version of PineAPPL. One of the datasets used in this study corresponds to longitudinal single spin asymmetry for W$\pm$ boson production in polarized proton-proton collisions. In this measurement, one of the initial state protons in polarized whereas the other one is unpolarized and hence the new version of PineAPPL becomes necessary to use this dataset. The predictions for this process are produced using a modified version of MCFM [191]. We further modified this code to interface it with PineAPPL to be able to produce interpolation grids. We then performed a comparison between the results obtained directly from the code and the results obtained using the PineAPPL grids. Fig. 4.2 shows that the difference between the values obtained from MCFM and from the PineAPPL grid is less than the MC uncertainty of MCFM. In fact, it is always less than $0.025\%$. This result demonstrates that the usage of PineAPPL does not have any detrimental effect on the precision of the theoretical predictions.

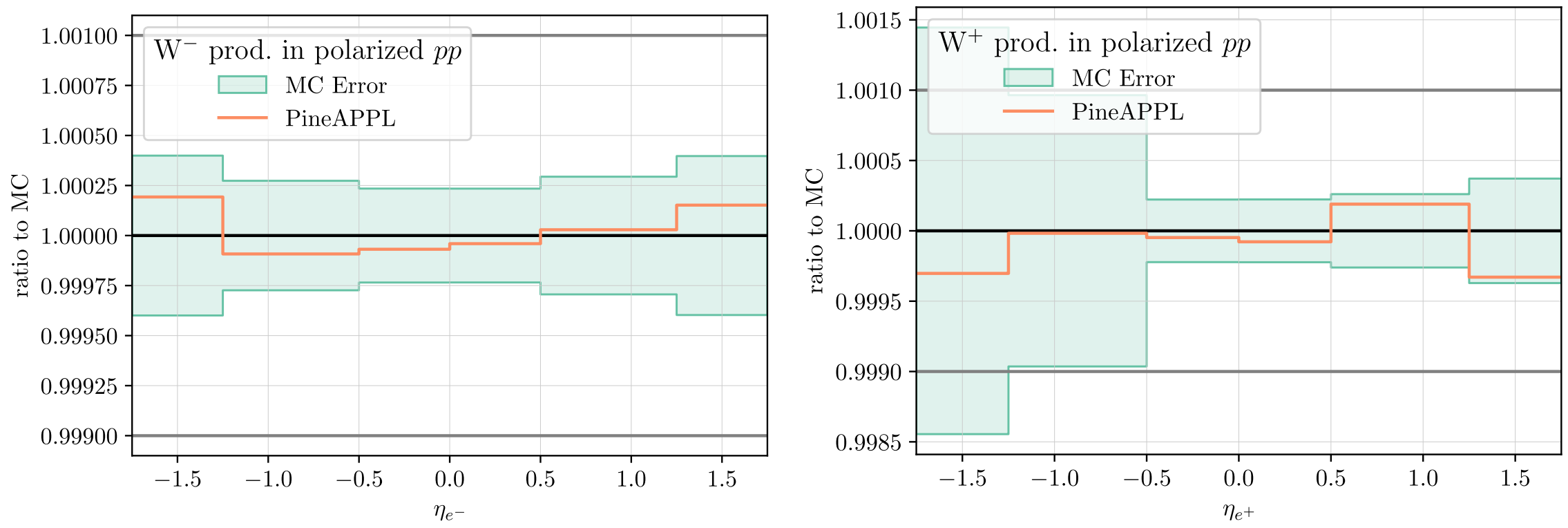

Figure 4.2: Predictions of the longitudinal single spin asymmetry for $W^-$ (left) and $W^+$ (right) boson production in polarized proton-proton collisions obtained from the code of [191] or from convolution of PDFs with PineAPPL grids generated from a modified version of the same code. The results are normalized to the former. The band corresponds to Monte Carlo uncertainties due to numerical integration performed with the code of [191].

## 4.4 A factory of virtual hadrons

This section discusses the development of a software package called vhf, acronym for Virtual Hadron Factory, which has been one of the most important and thorough projects carried out as part of this thesis. While the work discussed in the previous sections of this chapter concerned further development and extension of existing software, vhf has been single-handedly developed from scratch. The goal of vhf is to be able to compute theoretical predictions, both numerical and in the form of interpolation grids, for processes involved in the determination of the FFs. vhf has been designed not only to be able to produce theoretical predictions for processes associated to fragmentation functions, but do so while being maximally user-friendly. The codebase has been structured to be modular, allowing for easy extension and addition of new features in the future. Large part of the codebase is written in Python, while the computationally intensive parts are written in C++. This section forms the first announcement for vhf, but it will be supplemented by a dedicated paper in the future that includes more technical details of the code, and robust examples demonstrating its usage. In this section, I will begin by discussing the physics motivations, aspects and requirements of vhf, followed by qualitatively explaining the inner working of the codebase, and finally a provide demonstration of the results produced by vhf.

### 4.4.1 Physics behind vhf

The determination of FFs can utilize experimental data from three processes:

- Single inclusive annihilation (SIA): $l_1 + l_2 \to h + X$,
- Semi-inclusive deep inelastic scattering (SIDIS): $l + N \to l' + h + X$ and,
- Single inclusive hadron production in hadronic collisions: $N_1 + N_2 \to h + X$,

where $l$ denote leptons, $N$ denote nucleons, and $h$ denote final-state hadrons and $X$ denotes the rest of the final-state particles. The importance of using different processes in the fitting procedure has to do with the fact that cross sections for different processes receive their leading contributions from partonic cross sections involving different partons and hence the inclusion of a variety of processes is crucial to fit FFs for all of the partons. The computation of the theoretical predictions for these processes is carried out by using the relevant coefficient functions. Given that analytical computation of coefficient functions is a complicated task, historically, the limitations of the inclusion of different processes in the FF fits has come from the lack of availability of the coefficient functions at a given perturbative order. It should be noted that the discussion here corresponds solely to zero-mass coefficient functions.

To understand the above limitations, let us consider some existing FF sets. I will limit the focus to neural network based FF sets, which have been produced by the NNPDF collaboration and the MAP collaboration. NNPDF has released two FF sets: NNFF1.0 [180] and NNFF1.0h [192]. MAP has released one FF set: MAPFF1.0 [193, 194]. NNFF1.0 (released in 2017) is an NNLO FF set, which only includes SIA data. This is because at the time of its release, SIA was the only process for which coefficient functions had been computed to NNLO accuracy, and hence NNLO FF sets could only include SIA datasets. NNFF1.0h (released in 2018) is an NLO FF set, which includes data from SIA and hadronic collisions. It had to be released at NLO because the coefficient functions for single inclusive hadron production in hadronic collisions were only available at NLO accuracy at that time. MAPFF1.0 (released in 2021) is an NNLO FF set, which includes data from SIA and SIDIS, however, at the time of its release, the coefficient functions for SIDIS were only available at aNNLO accuracy. Since the release of these FF sets, there have been significant developments in the theoretical computation of coefficient functions. It is therefore worthwhile to first review the chronological development of the computation of the coefficient functions, thus explaining their implementation in vhf.

For SIA, the NLO coefficient functions were first computed in Ref. [195] in 1979, and the NNLO coefficient functions were computed in Ref. [196, 197] in 1996. In 2025, the $N^3LO$ coefficient functions have also been computed in Ref. [198]. At present, the $N^3LO$ coefficient functions represent the state-of-the-art of the field and can not be used in a determination of FFs because the splitting functions at $N^3LO$ are not yet available. Once these splitting functions are computed, the $N^3LO$ coefficient functions will be of significant importance in the anticipation of the planned FCC-ee. At present, coefficient functions have been implemented up to NNLO in vhf, and the $N^3LO$ coefficient functions are planned to be implemented in the future. SIA coefficient functions take into account full electroweak corrections (pertaining to the mediating gauge boson).

For fragmentation functions from SIDIS, there are two types of processes that contribute to unpolarized FFs: unpolarized SIDIS and polarized SIDIS, whereby the (un)polarized prefix of SIDIS refers to the polarization of the initial-state nucleon. For simplicity, I will restrict the discussion to unpolarized SIDIS, however, the polarized SIDIS coefficient functions are fully implemented in vhf. The unpolarized SIDIS coefficient functions at NLO were first computed in Ref. [195, 199] in 1979. The aNNLO coefficient functions were computed in Ref. [200] in 2021 and the NNLO coefficient functions were computed in Ref. [201, 202] in 2023 and 2024 by two groups independently. All of these coefficient functions take into account process that are mediated by photons only, i.e. only electromagnetic corrections are included. While this remains sufficient for the experimental data that is currently available (based on the energy scales of the datasets), future experimental data, such as that from the EIC, may definitely benefit from the inclusion of electroweak corrections. In June of 2025, and a mere couple of weeks before the writing of this section, the NNLO SIDIS coefficient functions with full electroweak corrections (pertaining to the mediating gauge boson) were computed in Ref. [203]. At present, only the electromagnetic SIDIS coefficient functions up to NNLO have been implemented in vhf, but the electroweak SIDIS coefficient functions are planned to be implemented in the near future.

For single inclusive hadron production in hadronic collisions, as was the case with SIDIS, the initial-state nucleons can be either unpolarized or polarized, and the discussion here is limited to the unpolarized case. The NLO coefficient functions were first computed in Ref. [204] in 1988. A few months before the writing of this section, the NNLO coefficient functions were computed in Ref. [205] in the March of 2025. With this development, it is finally possible to perform a full exact NNLO global FF determination. As to the best of my knowledge and at the time of the writing of this thesis, the coefficient functions for this process at both NLO and NNLO are not available in any public codebase or in a digital format from where they can be readily adapted for implementation. For NNLO in particular, the coefficient functions are not available publicly at all. The implementation of this process remains limited to some private codebases. As such this process is currently not implemented in vhf, and there are no available pathways to do so in the near future, however, such an implementation is hoped to be done in the future, making vhf a one stop shop for all the theoretical predictions for FFs. As such, for the remaining part of this section, the focus will solely be limited to SIA and SIDIS.

With a clear understanding of the availability of analytical expressions for the processes of interest in an FF determination, it is now instructive to discuss the exact form of the expressions corresponding to these processes, to build a foundation upon which their implementation in vhf will be discussed. The following discussion is not intended to be exhaustive, but rather a concise summary of the relevant expressions.

For SIA, involving an electron and a positron, the LO interaction is given as:

$$e^- + e^+ \to q + \bar{q}$$

where the quark or the antiquark may fragment into a hadron that is detected. At NLO, the interaction is given as:

$$e^- + e^+ \to q + \bar{q} + g$$

where the quark, antiquark, or the gluon may fragment into a hadron that is detected. At NNLO, the interactions are given as:

$$e^- + e^+ \to q + \bar{q} + g + g,$$

$$e^- + e^+ \to q + \bar{q} + q + \bar{q}$$

where the quark, antiquark, or the gluon may fragment into a hadron that is detected. The cross section differential in $z$ for SIA is given by:

$$\frac{d\sigma^h}{dz} = \frac{4\pi\alpha_{em}^2(Q)}{Q^2} F_2^h(z, Q) \tag{4.5}$$

where $F_2^h$ is the fragmentation structure function for the hadron $h$. $F_2^h$ is a sum of the longitudinal and transverse contributions, i.e. $F_2^h = F_L^h + F_T^h$. The decomposition of $F_i^h$ with $i = L, T$ in terms of the coefficient functions and the FFs is generally done by considering the singlet coefficient function $C_S$ and the non-singlet coefficient function $C_{NS}$ in the literature, where $C_S = C_{NS} + C_{PS}$ where $C_{PS}$ is the pure-singlet coefficient function. However, given the focus of this section is on actual implementation,

I will display the decomposition as one would practically implement it (and as is done in vhf). The fragmentation structure function, up to NNLO, is given as follows:

$$
\begin{aligned}
F_2^h = &\ \alpha_s^2 \left( \sum_q^{n_f} \hat{e}_q^2 \, C_{2,\mathrm{NS}}^{(2)} \otimes D_q^h + \langle \hat{e}_q^2 \rangle \sum_q^{n_f} C_{2,\mathrm{PS}}^{(2)} \otimes D_q^h + \sum_q^{n_f} \hat{e}_q^2 \, C_{2,g}^{(2)} \otimes D_g^h \right) \\
&+ \alpha_s \left( \sum_q^{n_f} \hat{e}_q^2 \, C_{2,\mathrm{q}}^{(1)} \otimes D_q^h + \sum_q^{n_f} \hat{e}_q^2 \, C_{2,g}^{(1)} \otimes D_g^h \right) \\
&+ \sum_q^{n_f} \hat{e}_q^2 \, C_{2,\mathrm{q}}^{(0)} \otimes D_q^h
\end{aligned} \tag{4.6}
$$

where $\hat{e}_q$ is the electroweak charge of the quark $q$ and $\langle \hat{e}_q^2 \rangle$ is the average of the square of the electroweak charge over the active quark flavors. (0), (1) and (2) denote the QCD perturbative order of the coefficient functions.

The SIA experimental dataset does not always necessarily come as the cross section differential in $z$. Sometimes, it comes as a multiplicity where the cross section is normalized to the total SIA cross section:

$$
\frac{1}{\sigma_{\mathrm{tot}}} \frac{d\sigma^h}{dz}.
$$

The total SIA cross section is given as:

$$
\sigma_{\mathrm{tot}} = \frac{4\pi\alpha_{em}^2(Q)}{3Q^2} \cdot N_c \left( \sum_q^{n_f} \hat{e}_q^2(Q) \right) \left( 1 + \frac{\alpha_s}{\pi} + r_1 \left( \frac{\alpha_s}{\pi} \right)^2 r_2 \left( \frac{\alpha_s}{\pi} \right)^3 \right) + \mathcal{O}\left(\alpha_s^4\right) \tag{4.7}
$$

where $N_c$ is the number of colors, and

$$
r_1 = 1.9857 - 0.1153 n_f,
$$

$$
r_2 = -6.6368 - 1.2001 n_f - 0.0052 n_f^2 - 1.2395 \frac{\left( \sum_q^{n_f} \hat{e}_q \right)^2}{3 \sum_q^{n_f} \hat{e}_q^2},
$$

as computed in Ref. [206]. The normalization of the differential cross section to the total cross section is carried out at the exact same perturbative order as the differential cross section.

The other complication that may need to be dealt with is that the experimental data may not be differential in $z$, but rather differential in $x_p$, $p_h$ or $\xi$. The relation between these variables and $z$ is

given as:

$$z\left(p_h\right) = 2\left(\frac{m_h^2 + p_h^2}{s}\right)^{1/2} \tag{4.8}$$

$$z\left(x_p\right) = x_p\left(1 + \frac{4}{x_p^2}\frac{m_h^2}{s}\right)^{1/2} \tag{4.9}$$

$$z\left(\xi\right) = e^{-\xi}\left(1 + 4e^{2\xi}\frac{m_h^2}{s}\right)^{1/2} \tag{4.10}$$

where $m_h$ is the mass of the hadron. As the coefficient functions are a function of $z$, the computation of the differential cross section requires the evaluation of the coefficient functions in $z$ and then the differential cross section is modified by a Jacobian factor. With this, we now move on to the discussion of SIDIS.

For SIDIS, involving an electron and a proton, the LO interaction is given as:

$$e + p \rightarrow e' + q/\bar{q} + X$$

where the quark $q$ or the antiquark $\bar{q}$ may fragment into a hadron that is detected. At NLO, the interaction is given as:

$$e + p \rightarrow e' + q/\bar{q} + g + X$$

where the quark, antiquark, or the gluon may fragment into a hadron that is detected. At NNLO, the interactions are given as:

$$e + p \rightarrow e' + q/\bar{q} + g + g + X,$$

$$e + p \rightarrow e' + q/\bar{q} + q + \bar{q} + X$$

where the quark, antiquark, or the gluon may fragment into a hadron that is detected. The SIDIS cross section is differential in $x$, $y/Q^2$ and $z$, where $x$, $y$ and $Q^2$ are the DIS variables defined in Sec. 1.4.1. As such, the expression for the differential cross sections takes two forms, and here I will present the one that is differential in $x$, $y$ and $z$, and an interested reader can find the other form in Ref. [194].

$$\frac{d^3\sigma^h}{\mathrm{dx\,dy\,dz}} = \frac{4\pi\alpha_{em}^2}{Q^2}\left[\frac{1+(1-y)^2}{2y}\mathcal{F}_T^h\left(x,z,Q\right) + \frac{1-y}{y}\mathcal{F}_L^h\left(x,z,Q\right)\right]. \tag{4.11}$$

The decomposition of the structure functions at NNLO requires 14 coefficient functions, which are presented in Ref. [202]. Just as in the case of SIA, the SIDIS experimental data can also come as a multiplicity, however, in this case, the denominator is the DIS cross section, as explained in Ref. [194]. In such a case, one needs to be able to produce theoretical predictions for SIDIS and DIS cross sections, to have the theoretical predictions for the multiplicity. The computation of the DIS cross section is not currently implemented in vhf, and is planned for the near future to allow for seamless computation of multiplicities, however, this does not represent a bottleneck as tools already exist which can compute theoretical predictions that can work harmoniously with vhf, such as Yadism [207] that can compute the DIS cross sections.

With an overview of the processes complete, one other aspect remains to be discussed. It has been shown that coefficient functions have to be convolved with the FFs (and PDFs) to produce theoretical predictions which requires a convolution integral. However, the coefficient functions may contain terms involving a dirac delta function, or a plus distribution, which has a singularity at $x = 1$ or $z = 1$ upon integration. In practice, the plus distribution terms almost always take a form as follows:

$$\left[\frac{\log^n(1-x)}{1-x}\right]_+$$

where $n$ is a non-negative integer. With such terms, a generic coefficient function involving one convolution (such as SIA or DIS) can be written as:

$$C(x) = C^r(x) + C^s(x) + C^l(x)$$

where $C^r$ is the regular term, $C^s$ is the singular term (arising from the plus distribution) and $C^l$ is the local term (arising from the delta function). The convolution integral is then given as:

$$f \otimes C = \int_x^1 \frac{d\hat{x}}{\hat{x}} f(x/\hat{x})\, C^r(\hat{x}) + \int_x^1 \left(\frac{f(x/\hat{x})}{\hat{x}} - f(x)\right) C^s(\hat{x}) + f(x)C^l(x). \tag{4.12}$$

In case of two convolutions, such as in SIDIS, such a decomposition leads to 9 terms in the convolution integral. This concludes the discussion of the physics behind vhf.

### 4.4.2 Implementation of physics in vhf

In this section, I will present an overview that demonstrates through an example, how the physics requirements are achieved by looking at the algorithm in a pedagogical manner. It should be noted that it is not the intention of this section to provide a technical overview of the codebase, or a manual demonstrating the usage of vhf, both of which will be done in a technical paper in the near future that will mark the public release of vhf.

In this section, I will demonstrate for a single bin, order and channel, the computation from a coefficient function, to an interpolation (sub)grid, to a theoretical prediction. Let us consider the longitudinal coefficient function for gluon at NLO for SIA, which only has a regular term:

$$C_L^g(z) = 2C_F\frac{1-z}{z}.$$

The first step is to define the nodes of the interpolation grid, upon which the weights will be computed.

The computation of nodes is done by following the algorithm discussed in the section on Grid representations of Ref. [15]. Let us set the number of nodes to be 50, starting from 0.01 to 1. The nodes are then:

$$[0.0100000000, 0.0120303976, 0.0144452520, 0.0173061951, 0.0206805986, \\ 0.0246406871, 0.0292621579, \ldots, 1.0000000000]$$

To each node, there is an associated Lagrange interpolating polynomial, which is discussed in Sec. 2.2.2. The slight addition made to the process of computation of the polynomials is that the polynomials are defined in a piecewise manner and we take the logarithm of every node point while computing the polynomials. This ensures that the interpolation does not deteriorate due to the fact that the nodes span a few orders of magnitude. With the polynomials defined, the convolution between the coefficient function and the interpolating polynomials is performed to produce the subgrid, whereby the first weight is the convolution between the coefficient function and the first polynomial, the second weight is the convolution between the coefficient function and the second polynomial, and so on. Let us evaluate it a value of $z = 0.2$ and $Q = 91.2$GeV. We obtain the following subgrid:

$$\begin{aligned}[0.00000e+00,\ & 0.00000e+00,\ 0.00000e+00,\ 0.00000e+00,\ 0.00000e+00, \\ 0.00000e+00,\ & 0.00000e+00,\ 0.00000e+00,\ 0.00000e+00,\ 0.00000e+00, \\ 0.00000e+00,\ & 0.00000e+00,\ 0.00000e+00,\ 0.00000e+00,\ 0.00000e+00, \\ 0.00000e+00,\ & 0.00000e+00,\ 0.00000e+00,\ 0.00000e+00,\ -2.58800e-04, \\ 1.75350e-03,\ & 3.56110e-02,\ 8.62775e-02,\ 1.30533e-01,\ 1.72890e-01, \\ 2.12756e-01,\ & 2.50206e-01,\ 2.85344e-01,\ 3.18289e-01,\ 3.49165e-01, \\ 3.78099e-01,\ & 4.05215e-01,\ 4.30637e-01,\ 4.54482e-01,\ 4.76862e-01, \\ 4.97881e-01,\ & 5.17640e-01,\ 5.36225e-01,\ 5.53730e-01,\ 5.70225e-01, \\ 5.85790e-01,\ & 6.00485e-01,\ 6.14375e-01,\ 6.27520e-01,\ 6.39970e-01, \\ 6.45440e-01,\ & 7.03265e-01,\ 5.59265e-01,\ 8.72820e-01,\ 2.36722e-01]\end{aligned}$$

Subgrids like these are stored inside a PineAPPL grid, and from there on, PineAPPL handles the rest of the convolution, however, for the purposes of demonstrating the process, let us continue up to the point where the theoretical prediction is computed. The next step is to sample the FF (that is to be used in the convolution) at the same nodes as the subgrid. For this, let us use the 'NNFF10_PIp_nlo' FF set, and sample the gluon FF. The following is the list of values of the gluon FF at the same nodes

as the subgrid:

$$[2.80597e+02,\ 2.09282e+02,\ 1.58080e+02,\ 1.21531e+02,\ 9.55307e+01,\ 7.68741e+01,\ 6.31020e+01,\ 5.24183e+01,\ 4.35019e+01,\ 3.54133e+01,\ 2.79600e+01,\ 2.13043e+01,\ 1.55890e+01,\ 1.07369e+01,\ 6.85678e+00,\ 4.02983e+00,\ 2.28254e+00,\ 1.39809e+00,\ 9.55279e-01,\ 7.18170e-01,\ 5.70952e-01,\ 4.80266e-01,\ 4.22036e-01,\ 4.03946e-01,\ 3.94195e-01,\ 3.29884e-01,\ 2.43681e-01,\ 1.75913e-01,\ 1.23161e-01,\ 8.62362e-02,\ 6.13119e-02,\ 4.11278e-02,\ 3.11541e-02,\ 2.36665e-02,\ 1.51833e-02,\ 1.06790e-02,\ 7.46530e-03,\ 5.37030e-03,\ 3.90540e-03,\ 2.83480e-03,\ 2.04560e-03,\ 1.46050e-03,\ 9.96500e-04,\ 6.27000e-04,\ 3.15500e-04,\ 5.84000e-05,\ -2.34000e-05,\ 1.98000e-05,\ 4.10000e-06,\ 0.00000e+00]$$

To perform the convolution now requires a simple dot product between the two arrays, which gives us the value: 0.5162918. Performing a direct convolution for the same value of $z = 0.2$ and $Q = 91.2$GeV, i.e. while using convolutions integrals and without the use of interpolation yields the value: 0.51661045, which equates to an interpolation error that is below permille level. The use of 50 nodes was chosen to be able to include the arrays in the text in a manageable manner, however, in practice, the number of nodes is set to 100 or more for the FF convolution, which improves the agreement between the actual result and the interpolated result by one or two orders of magnitude. This example has left out some important aspects of the computation, for example, the multiplication with the appropriate charges (as shown in Eq. (4.6)), and multiplication with prefactors such as $4\pi\alpha_{em}^2/Q^2$, which are all done in the actual implementation, at the subgrid level. The subgrids for all the channels and orders are computed and summed up to produce the theoretical prediction for a given individual bin.

### 4.4.3 Proof of concept

In this subsection, I will present data theory comparison plots for a select SIA and a select SIDIS dataset for which the theoretical predictions have been computed using vhf. The sole aim of this section is to demonstrate the working of vhf. It is not intended to be a comprehensive phenomenological study for the considered datasets.

For SIA, I will present the comparison corresponding to the experimental dataset from BELLE [208] at the center of mass energy of 10.52 GeV. This dataset provides measurements for the observable $\frac{d\sigma^{h\pm}}{dz}$, i.e. cross section differential with respect to $z$. The results are shown for the charged pion distribution $e^- + e^+ \to \pi^\pm + X$. The theoretical predictions have been computed at NLO, using NLO coefficient functions and the NLO FF set from NNFF1.0 PIsum [180]. The comparison plot is shown in Fig. 4.3. There appears to be a systematic offset between the theoretical predictions and the experimental values. This is consistent with what was shown in Fig. 5.1 of the NNFF1.0 paper [180].

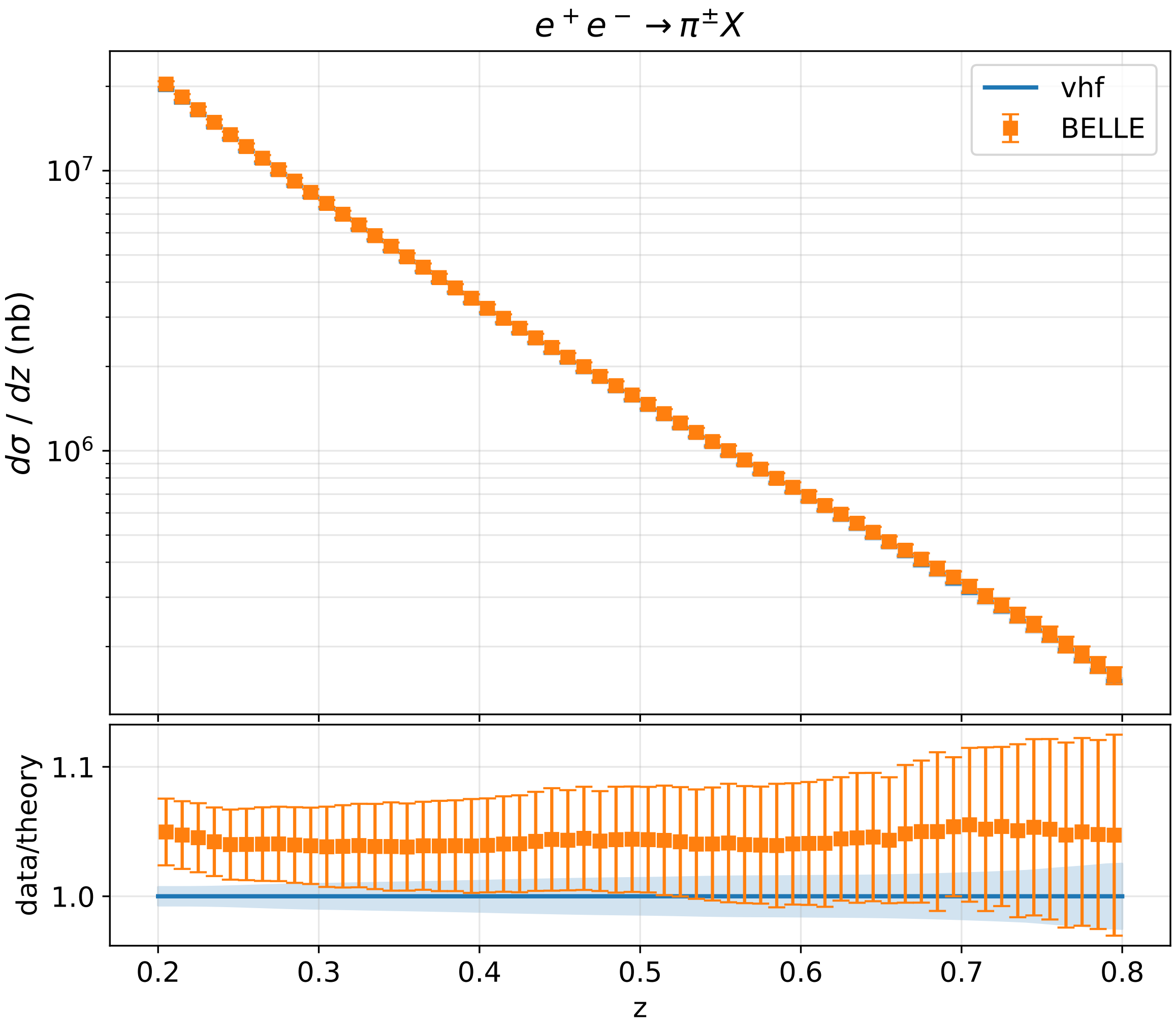

Figure 4.3: Data theory comparison for differential cross section with respect to $z$ for $\pi^\pm$ production in SIA by the BELLE collaboration at 10.52 GeV at NLO. The error bands in the theoretical predictions account for the 1 $\sigma$ uncertainty of the FF.

For SIDIS, I will present results for a polarized SIDIS observable. This is done so that it can be demonstrated that vhf is able to produce theoretical predictions for both, polarized and unpolarized SIDIS observables. In particular, I will present the data theory comparison corresponding to the experimental dataset from the HERMES experiment [209] where a 27.6 GeV electron (or positron) beam is scattered off a longitudinally polarized nucleon. The dataset, provides measurements for polarized proton and deuteron targets, and $\pi^+$, $\pi^-$, $K^+$, $K^-$, $p$, and $\bar{p}$ final-state hadrons. For simplicity, I will present the results corresponding to a proton target and $\pi^+$ final-state hadron. The measurement is of the form $A_1$ which is a ratio between the $g_1$ and $F_1$ structure functions, (see ref. [210] for more details), and is binned in $x$ (with 3 bins) and $z$ (with 7 bins). For simplicity, the results are shown for the central $x$ bin. The theoretical predictions have been computed at NNLO using the NNLO SIDIS coefficient functions (with NNLO polarized coefficient functions used for the numerator and NNLO unpolarized coefficient functions used for the denominator), the NNLO PDF set from NNPDF4.0 mhou [20], the NNLO polarized PDF set from NNPDFpol2.0 mhou [190] and the NNLO FF set from MAPFF1.0 PIp [193, 194]. The comparison plot is shown in Fig. 4.4.

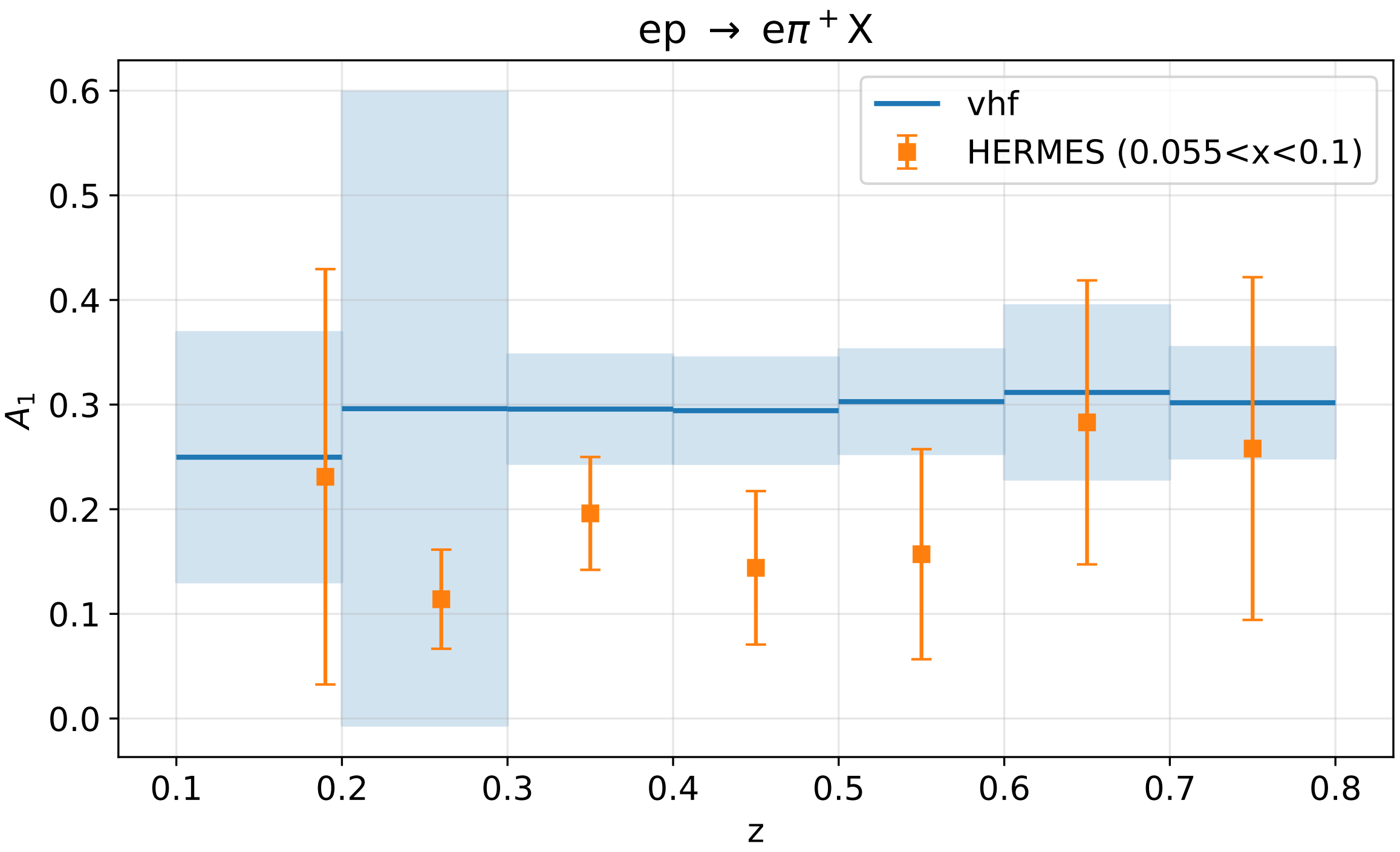

Figure 4.4: Data theory comparison for $A_1$ for $\pi^+$ production in SIDIS by the HERMES experiment at NNLO. The error bands in the theoretical predictions account for the $1\,\sigma$ uncertainties of the PDFs, polarized PDFs and FFs, added in quadrature.

## 4.5 Summary and future outlook

In this chapter, I have discussed the extension of the NNPDF framework such that it allows for the performing of global FF fits using the NNPDF methodology. These developments will allow for the release of the next-gen NNFF set. My work has focused on three main aspects:

- to extend EKO to allow for the solving of time-like DGLAP evolution equations such that the FFs can be evolved from the parametrization energy scale to the energy scale of the individual experimental datapoints,
- to extend PineAPPL to allow for the production of interpolation grids that support an arbitrary number of convolutions, support convolutions with polarized PDFs and FFs, and extend the support of scale dependence to fragmentation scale such that theoretical uncertainties can be computed by means of scale variation for FF specific processes too, and
- to develop vhf, a software designed to produce theoretical predictions for SIA and SIDIS, thus allowing for the production of PineAPPL interpolation grids for these processes that will be used in the FF determination.

With these developments completed as part of this thesis, we will now move forward with our planned studies concerning FFs. These will include phenomenological studies followed by the actual global FF determination.

# Conclusion

The ongoing (upcoming) physics program at the LHC (HL-LHC), that is dedicated to precision physics and BSM physics, and the upcoming physics program at the EIC, that will be dedicated to understanding the internal hadronic structure, require continuous progress to be made to achieve their respective aims. This is not just limited to better detectors and experimental procedures, but also includes theoretical and phenomenological aspects. Collinear PDFs, which are currently one of the dominant sources of uncertainty in the extraction of a number of SM parameters, represent one of the bottlenecks. To this end, it is crucial to work towards the determination of next generation PDFs, that are built on solid and robust methodology, and are increasingly more accurate and precise. Besides the use of PDFs for precision collider phenomenology, they are also an important piece in the puzzle, that is our understanding of hadronic substructure, which remains difficult to access due to the non-perturbativity of QCD in the energy scales associated to confinement. For this latter purpose, they are also complemented by FFs. While PDFs provide an inner look at the hadronic substructure when the hadrons are broken (i.e. are in the initial state), FFs provide an inner look at the hadronic structure when the hadrons are made (i.e. are in the final state).

The goals of this thesis have been to work towards the next generation PDFs and FFs from the NNPDF collaboration. The work towards PDFs has been focused on phenomenological studies that try to understand the impact of various aspects that will be a part of the new determination. These studies will allow for a clear understanding of the similarities and differences we will observe in the new determination with respect to the current PDFs. The work towards FFs has been focused on extending the NNPDF framework to be able to perform FF determination, and thus lies at the intersections of physics and software development, with a very large focus on writing code that works efficiently to achieve the required goals.

In this thesis, the first chapter was to review the fundamental concepts of particle physics and QCD that act as the foundation upon which collider phenomenology is carried. The second chapter was to review the NNPDF methodology and framework, which forms the basis for the determination of the PDFs and the FFs that we plan to carry out. In chapter 3, I presented the studies carried out for PDFs. The first study focused on assessing the impact of moving from the use of NNLO K-factors to exact NNLO corrections. In this study, we saw that the K-factors are relatively good at capturing the effects of the NNLO corrections and therefore the move towards the use of exact NNLO corrections should not have significant impact. The second study focused on assessing the compatibility of the most widely used current PDF sets with new high precision data, to see their generalization power, which also sheds some light on which datasets would (and would not) have significant impact upon their inclusion in the fitting procedure. This study was carried out such that major sources of theoretical uncertainties, such as from MHOUs, $\alpha_s$ and PDFs were considered. In this study, we saw that the most widely employed sets of PDFs seem to, on average, perform equally well when it comes to producing theoretical predictions for new data. This obviously comes with the caveat that different PDF sets have different precision, which might indicate that PDFs with higher precision may be more accurate, thanks to their methodology. The third study focused on assessing the impact on gluon PDF when new datasets that correspond to gluon sensitive processes are included in the fits explicitly. The inclusion of the new datasets was carried out in a systematic manner whereby the aim was to maximize consistency and impact. In this study, we saw that the new datasets have non-trivial amount of effect on the gluon PDF, leading to statistically significant differences in the gluon-gluon luminosity in the energy region associated to the Higgs mass, thereby indicating its potential effect on precision Higgs phenomenology. In chapter 4, I presented the work carried out for FFs. The first task was towards extending EKO to support time-like evolution, which is needed to evolve FFs from one energy scale to another, and is a crucial part of parametrizing the FFs at an initial scale. The second task focused on extending PineAPPL to make it structurally much more flexible to be able to support collinear distributions beyond unpolarized PDFs, such as polarized PDFs and FFs and to also support additional energy scales, such as the fragmentation scale. The third task was to develop from scratch a computational software that can produce theoretical predictions for SIA and SIDIS, which are vital for an FF determination.

With these developments stated, let me briefly discuss the future outlook. The studies towards various aspects of the next generation PDFs continue to be carried out by my NNPDF collaborators and I, as we inch towards running the actual PDF fits. On the FF front, with the development of vhf, we can produce theoretical predictions for all the experimental datasets of interest within the context of SIA and SIDIS. The next step involves the implementation of data in the NNPDF framework, as was explained in Sec. 2.2.1. This will follow with some preliminary phenomenological studies comparing the theoretical predictions using the existing FF (and PDF for SIDIS) sets with experimental data. This would be followed by a hyperoptimization study to determine the optimal hyperparemeters of the neural network that will parametrize the FFs. Once the neural network is designed, the fits will finally be performed.

# Appendix A

# Impact of the experimental covariance matrix regularization

As discussed in sec. 3.3.3.2, the experimental covariance matrix of all the datasets with $Z_{\mathscr{L}} > 4$ is regularized by means of the procedure laid out in [154]. The values of $\chi^2_{\rm exp}$ and $\chi^2_{\rm exp+th}$ reported in sec. 3.3.4 are computed accordingly. In this appendix, I recompute the values of $\chi^2_{\rm exp+th}$ with the original, unregularized experimental covariance matrix. These values, called $\chi^2_{\rm exp+th,orig}$, are compared to the $\chi^2_{\rm exp+th}$ values of sec. 3.3.4 in Table A.1. I also substantiate what the largest changes in covariances and correlations are upon applying our regularization procedure. To this purpose, I report in Table A.2, for each of the regularized data sets, the maximum relative difference of the variances $\Delta\sigma_r$ (in percent) and the maximum absolute difference of the correlation $|\Delta\rho|$ computed between the corresponding unregularized and regularized matrices.

From Table A.1, we see that the effect of regularization on the $\chi^2$ depends on the dataset. For some of these, the effect is huge, e.g. for the CMS Drell-Yan measurement or for the ATLAS and CMS single-inclusive jet and di-jet measurements. Specifically, it amounts to a reduction of the $\chi^2$ of more than $20\sigma$ for the first and of about $7\sigma$ for the latter. For others, the effect is small, e.g. for the ATLAS and CMS top-quark pair measurements or for the H1 single-inclusive jet measurements. This is unsurprising, given that the first datasets have the largest value of $Z_{\mathscr{L}}$ among all the datasets selected in Table 3.3. One may think that the regularization procedure is significantly modifying uncertainties and correlations. However, as we can see from Table A.2, changes are relatively mild: the largest relative change of uncertainties is of order $5\%$, whereas the largest absolute change of correlation is of order $0.06$. The first of these changes results in an effective increase of the diagonal elements of the covariance matrix. The increase is however moderate: typical LHC uncertainties are around a few percent, so the actual increase in uncertainty is of the order of a few permille. The second of these changes results in an effective decrease of correlations by about $6\%$. We consider this decrease to also be moderate, and at the same time we appreciate that it may be experimentally very challenging to quantify correlations with a precision of $6\%$.

It may seem counter-intuitive that a relatively small change in the covariance matrix leads to a variation of several standard deviations in the $\chi^2$. We refer the reader to [154] for a mathematical demonstration of this fact. Here we shall note that the degree of knowledge of experimental correlations related to a $1\sigma$ variation of the $\chi^2$ depends on the size of the uncertainties. The smaller the uncertainty, the higher the required degree of precision. Roughly speaking, as one can see from Fig. 3 of [154], for a $1\%$ ($2.5\%$) uncertainty, correlations ought to be known with a precision of $2\%$ ($12\%$) in order to be within a variation of the $\chi^2$ of one standard deviation. It is therefore unsurprising that the largest improvements in the value of the $\chi^2$ occur for the most precise datasets, which are affected by percent-level (if not sub-percent-level) uncertainties. That being said, we reiterate the fact that the regularization procedure applies to the covariance matrix as a whole, hence it does not discriminate across different uncertainties that could have a different physical meaning. If we assessed a dataset for inclusion in a PDF determination, understanding which uncertainties are responsible for the bad conditioning of the covariance matrix would be mandatory. However, we consider that all of these observations do not affect our ability to comparatively judge the performance of different PDF sets at describing the data. In our view, the regularization procedure does not alter the relative pattern of $\chi^2$ among different PDF sets and datasets. Therefore, the conclusions of sec. 3.3.4 continue to hold.

| Dataset | | ABMP16 | CT18 | CT18A | CT18Z | MSHT20 | NNPDF3.1 | NNPDF4.0 | PDF4LHC15 | PDF4LHC21 |
|---|---|---|---|---|---|---|---|---|---|---|
| CMS 13 TeV $W^+$ $\frac{d\sigma}{d\lvert\eta\rvert}$ | $\chi^2_{\text{exp+th}}$ | 1.31 | 1.20 | 1.11 | 1.05 | 1.26 | 0.85 | 0.96 | 1.15 | 0.98 |
| | $\chi^2_{\text{exp+th,orig}}$ | 14.6 | 10.9 | 10.8 | 10.8 | 12.2 | 10.2 | 11.2 | 10.7 | 10.5 |
| CMS 13 TeV $W^-$ $\frac{d\sigma}{d\lvert\eta\rvert}$ | $\chi^2_{\text{exp+th}}$ | 1.56 | 1.15 | 1.11 | 1.10 | 1.43 | 1.12 | 1.60 | 1.14 | 1.20 |
| | $\chi^2_{\text{exp+th,orig}}$ | 13.9 | 8.22 | 8.32 | 8.51 | 10.4 | 8.26 | 11.5 | 8.40 | 8.59 |
| ATLAS 13 TeV $t\bar{t}$ all hadr. $\frac{d\sigma}{dm_{t\bar{t}}}$ | $\chi^2_{\text{exp+th}}$ | 0.84 | 0.99 | 0.97 | 0.94 | 0.97 | 0.86 | 0.81 | 0.96 | 0.93 |
| | $\chi^2_{\text{exp+th,orig}}$ | 0.92 | 1.07 | 1.05 | 1.02 | 1.05 | 0.94 | 0.87 | 1.04 | 1.01 |
| ATLAS 13 TeV $t\bar{t}$ all hadr. $\frac{d^2\sigma}{dm_{t\bar{t}}d\lvert y_{t\bar{t}}\rvert}$ | $\chi^2_{\text{exp+th}}$ | 0.93 | 1.38 | 1.39 | 1.42 | 1.48 | 1.12 | 1.22 | 1.22 | 1.39 |
| | $\chi^2_{\text{exp+th,orig}}$ | 0.97 | 1.44 | 1.44 | 1.46 | 1.53 | 1.17 | 1.28 | 1.26 | 1.45 |
| ATLAS 13 TeV $t\bar{t}\,\ell+j$ $\frac{1}{\sigma}\frac{d\sigma}{dm_{t\bar{t}}}$ | $\chi^2_{\text{exp+th}}$ | 1.41 | 1.17 | 1.17 | 1.04 | 1.18 | 1.46 | 1.39 | 1.20 | 1.19 |
| | $\chi^2_{\text{exp+th,orig}}$ | 1.45 | 1.24 | 1.23 | 1.10 | 1.24 | 1.48 | 1.41 | 1.26 | 1.24 |
| ATLAS 13 TeV $t\bar{t}\,\ell+j$ $\frac{1}{\sigma}\frac{d\sigma}{dp_T^t}$ | $\chi^2_{\text{exp+th}}$ | 0.56 | 0.54 | 0.54 | 0.52 | 0.53 | 0.56 | 0.53 | 0.53 | 0.53 |
| | $\chi^2_{\text{exp+th,orig}}$ | 0.80 | 0.83 | 0.83 | 0.81 | 0.82 | 0.81 | 0.78 | 0.81 | 0.81 |
| CMS 13 TeV $t\bar{t}\,\ell+j$ $\frac{1}{\sigma}\frac{d^2\sigma}{dm_{t\bar{t}}d\lvert y_{t\bar{t}}\rvert}$ | $\chi^2_{\text{exp+th}}$ | 2.77 | 2.89 | 2.87 | 2.76 | 3.36 | 3.01 | 3.61 | 2.81 | 2.81 |
| | $\chi^2_{\text{exp+th,orig}}$ | 3.16 | 3.29 | 3.27 | 3.27 | 3.80 | 3.47 | 4.19 | 3.21 | 3.26 |
| ATLAS 13 TeV single-inclusive jets $\frac{d^2\sigma}{dp_T d\lvert y\rvert}$ | $\chi^2_{\text{exp+th}}$ | 1.85 | 1.56 | 1.64 | 1.38 | 1.67 | 1.21 | 1.51 | 1.20 | 1.25 |
| | $\chi^2_{\text{exp+th,orig}}$ | 3.25 | 2.77 | 2.90 | 2.49 | 2.86 | 2.17 | 2.80 | 2.28 | 2.38 |
| CMS 13 TeV single-inclusive jets ($R=0.4$) $\frac{d^2\sigma}{dp_T d\lvert y\rvert}$ | $\chi^2_{\text{exp+th}}$ | 1.64 | 1.58 | 1.60 | 1.52 | 1.64 | 1.47 | 1.50 | 1.48 | 1.43 |
| | $\chi^2_{\text{exp+th,orig}}$ | 3.04 | 2.68 | 2.69 | 2.63 | 2.94 | 2.48 | 2.70 | 2.51 | 2.62 |
| ATLAS 13 TeV di-jets $\frac{d^2\sigma}{dm_{jj}d\lvert y^*\rvert}$ | $\chi^2_{\text{exp+th}}$ | 1.13 | 1.08 | 1.09 | 1.05 | 1.16 | 1.09 | 1.15 | 1.01 | 0.96 |
| | $\chi^2_{\text{exp+th,orig}}$ | 1.73 | 1.56 | 1.58 | 1.52 | 1.72 | 1.53 | 1.70 | 1.46 | 1.41 |
| H1 single-inclusive-jets (low $Q^2$) $\frac{d^2\sigma}{dQ^2 dp_T}$ | $\chi^2_{\text{exp+th}}$ | 1.64 | 1.61 | 1.61 | 1.67 | 1.61 | 1.70 | 1.74 | 1.61 | 1.73 |
| | $\chi^2_{\text{exp+th,orig}}$ | 2.20 | 2.19 | 2.19 | 2.27 | 2.20 | 2.30 | 2.34 | 2.17 | 2.33 |
| H1 di-jets (low $Q^2$) $\frac{d^2\sigma}{dQ^2 d\langle p_T\rangle}$ | $\chi^2_{\text{exp+th}}$ | 1.37 | 1.39 | 1.38 | 1.37 | 1.39 | 1.42 | 1.44 | 1.36 | 1.44 |
| | $\chi^2_{\text{exp+th,orig}}$ | 2.28 | 2.29 | 2.29 | 2.21 | 2.31 | 2.32 | 2.34 | 2.27 | 2.34 |

Table A.1: A comparison of the values of $\chi^2_{\text{exp+th}}$, computed in sec. 3.3.4 by regularizing the experimental covariance matrix with the procedure of [154], to the corresponding values $\chi^2_{\text{exp+th,orig}}$, computed with the original, unregularized covariance matrix.

| | CMS 13 TeV $W^+$ $\frac{d\sigma}{d\lvert\eta\rvert}$ | CMS 13 TeV $W^-$ $\frac{d\sigma}{d\lvert\eta\rvert}$ | ATLAS 13 TeV $t\bar{t}$ all hadr. $\frac{d\sigma}{dm_{t\bar{t}}}$ | ATLAS 13 TeV $t\bar{t}$ all hadr. $\frac{d^2\sigma}{dm_{t\bar{t}}d\lvert y_{t\bar{t}}\rvert}$ | ATLAS 13 TeV $t\bar{t}\ \ell + j$ $\frac{1}{\sigma}\frac{d\sigma}{dm_{t\bar{t}}}$ | ATLAS 13 TeV $t\bar{t}\ \ell + j$ $\frac{1}{\sigma}\frac{d\sigma}{dp_T^t}$ | CMS 13 TeV $t\bar{t}\ \ell + j$ $\frac{1}{\sigma}\frac{d^2\sigma}{dm_{t\bar{t}}d\lvert y_{t\bar{t}}\rvert}$ | ATLAS 13 TeV single-inclusive jets $\frac{d^2\sigma}{dp_T d\lvert y\rvert}$ | CMS 13 TeV single-inclusive jets ($R = 0.4$) $\frac{d^2\sigma}{dp_T d\lvert y\rvert}$ | ATLAS 13 TeV di-jets $\frac{d^2\sigma}{dm_{jj}d\lvert y^*\rvert}$ | H1 single-inclusive-jets (low $Q^2$) $\frac{d^2\sigma}{dQ^2 dp_T}$ | H1 di-jets (low $Q^2$) $\frac{d^2\sigma}{dQ^2 d\langle p_T\rangle}$ |
|---|---|---|---|---|---|---|---|---|---|---|---|---|
| $\Delta\sigma_r$ | 5.48 | 5.45 | 2.54 | 1.37 | 3.06 | 2.71 | 4.07 | 5.54 | 5.41 | 4.91 | 2.76 | 3.73 |
| $\lvert\Delta\rho\rvert$ | 0.06 | 0.06 | 0.03 | 0.02 | 0.04 | 0.04 | 0.06 | 0.06 | 0.06 | 0.06 | 0.03 | 0.04 |

Table A.2: The maximum relative difference of the variances $\Delta\sigma_r$ (in percent) and the maximum absolute difference of the correlation $|\Delta\rho|$ computed between the corresponding unregularized and regularized matrices for the regularized datasets outlined in Table A.1.

# Appendix B

# Additional results from the study in Sec. 3.3

In this appendix, I collect additional results, complementing those presented in sec. 3.3.4, for the breakdown of $\chi^2_{\rm exp+th}$ into its $\chi^2_{\rm exp+mho}$ and $\chi^2_{\rm exp}$ components, and for the data-theory comparisons. The additional results refer to the following categories of measurements.

Top-quark pair production. Figure B.1 displays the breakdown of $\chi^2_{\rm exp+th}$ into its $\chi^2_{\rm exp+mho}$ and $\chi^2_{\rm exp}$ components for the datasets not displayed in Fig. 3.6, namely: the ATLAS all-hadronic absolute differential distribution in the invariant mass of the top-quark pair; the ATLAS all-hadronic normalized differential distribution in the absolute rapidity of the top-quark pair; the ATLAS lepton+jets normalized differential distributions in the absolute rapidity of the top quark and of the top-quark pair; and the CMS lepton+jets normalized differential distributions in the transverse momentum of the top quark and of the invariant mass of the top-quark pair. Figure B.2 displays the data-theory comparison for the top-quark pair single-differential distributions not displayed in Fig. 3.7, namely: the ATLAS all-hadronic normalized distribution differential in the absolute value of the top-quark pair rapidity; the ATLAS all-hadronic absolute distribution differential in the invariant mass of the top-quark pair; the ATLAS lepton+jets normalized distributions differential in the absolute rapidity of the top quark and of the top-quark pair; and the CMS lepton+jets normalized distributions differential in the transverse momentum of the top quark and in the invariant mass of the top-quark pair. Figure 3.4 displays the data-theory comparison for the top-quark pair bins of the ATLAS and CMS double-differential distributions not displayed in Fig. 3.7.

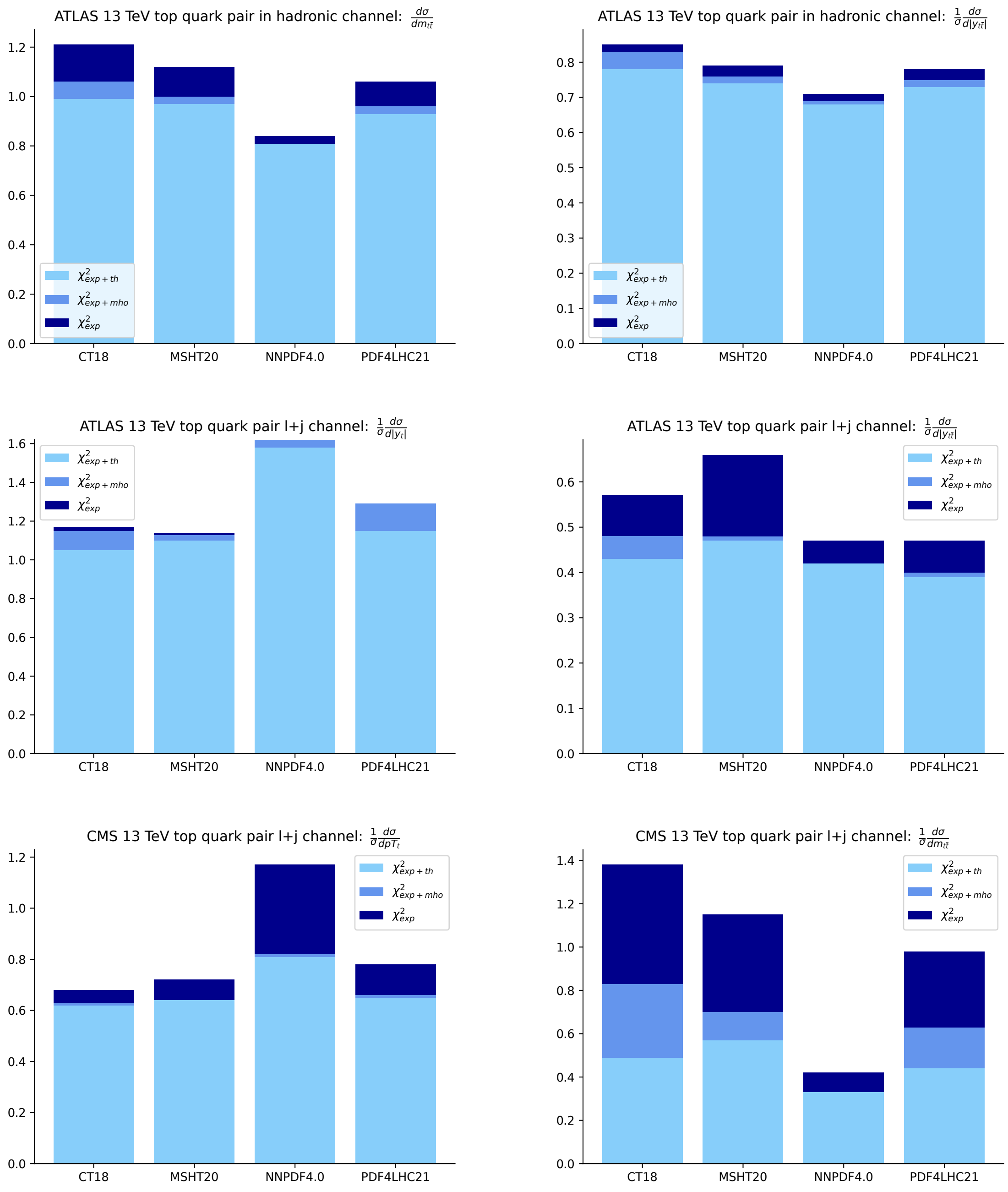

Figure B.1: Same as Fig. 3.6 for the ATLAS and CMS datasets not shown there. From top to bottom, left to right: the ATLAS all-hadronic absolute differential distribution in the invariant mass of the top-quark pair ($n_{\rm dat} = 9$, $\sqrt{2/n_{\rm dat}} = 0.47$); the ATLAS all-hadronic normalized differential distribution in the absolute rapidity of the top-quark pair ($n_{\rm dat} = 12$, $\sqrt{2/n_{\rm dat}} = 0.41$); the ATLAS lepton+jets normalized differential distributions in the absolute rapidity of the top quark ($n_{\rm dat} = 5$, $\sqrt{2/n_{\rm dat}} = 0.63$) and of the top-quark pair ($n_{\rm dat} = 7$, $\sqrt{2/n_{\rm dat}} = 0.53$); and the CMS lepton+jets normalized differential distributions in the transverse momentum of the top quark ($n_{\rm dat} = 16$, $\sqrt{2/n_{\rm dat}} = 0.35$) and of the invariant mass of the top-quark pair ($n_{\rm dat} = 15$, $\sqrt{2/n_{\rm dat}} = 0.37$).

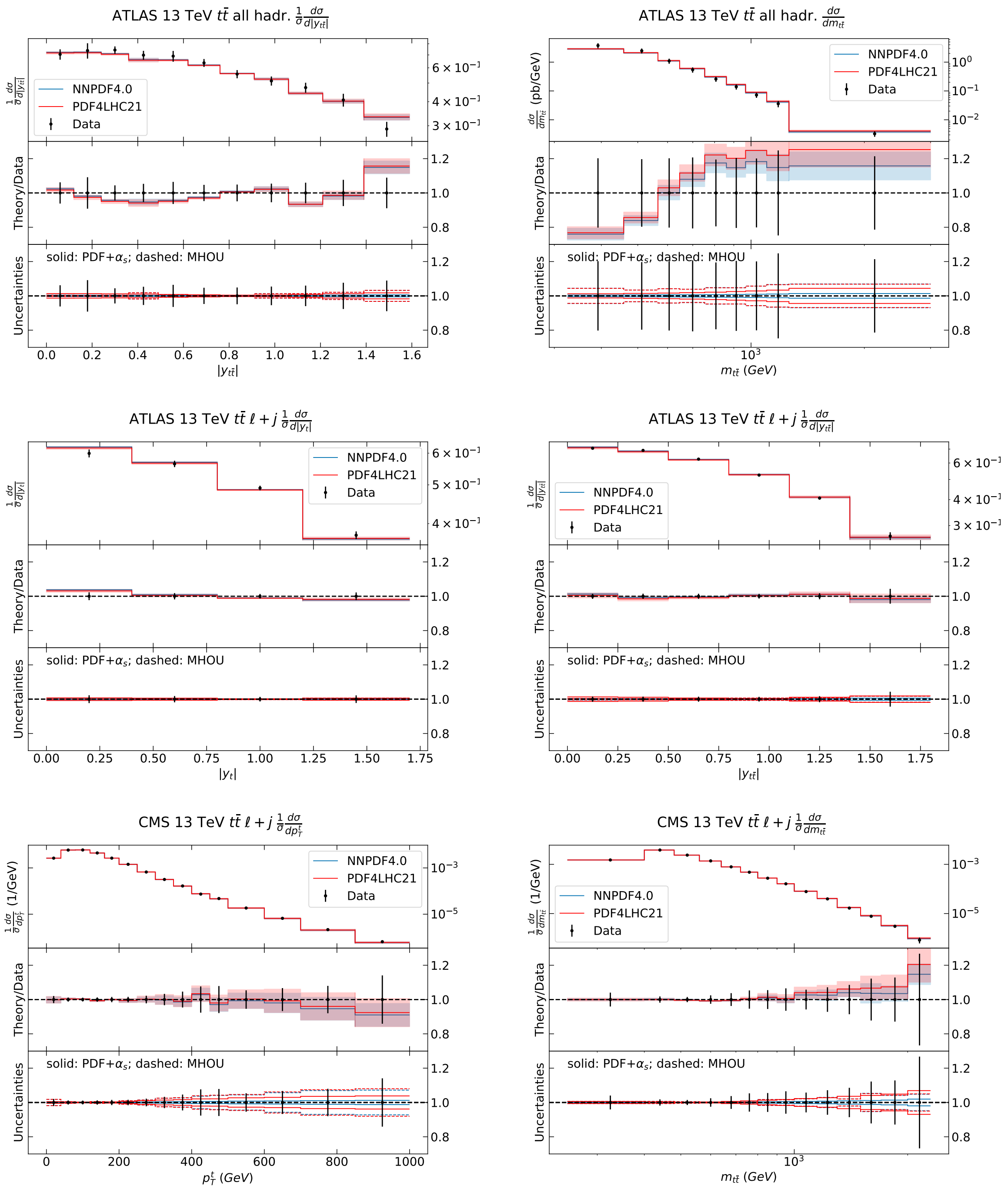

Figure B.2: Same as Fig. 3.4 for the top-quark pair single-differential distributions not displayed in Fig. 3.7, namely: the ATLAS all-hadronic normalized distribution differential in the absolute value of the top-quark pair rapidity; the ATLAS all-hadronic absolute distribution differential in the invariant mass of the top-quark pair; the ATLAS lepton+jets normalized distributions differential in the absolute rapidity of the top quark and of the top-quark pair; and the CMS lepton+jets normalized distributions differential in the transverse momentum of the top quark and in the invariant mass of the top-quark pair.

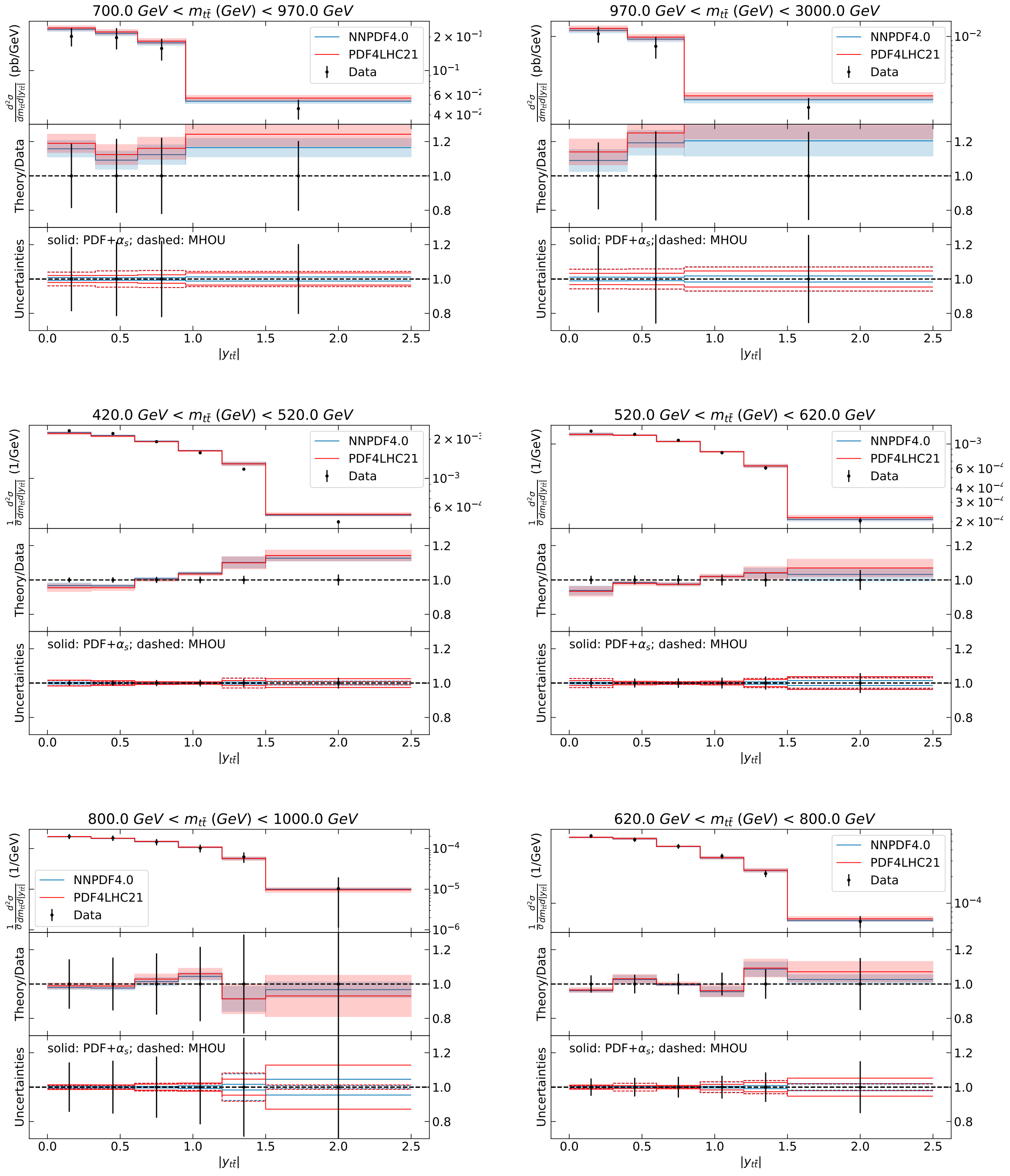

Figure B.3: Same as Fig. 3.4 for the top-quark pair bins of the double-differential distributions not displayed in Fig. 3.7: the first row corresponds to the ATLAS measurement in the all-hadronic final state; the second and third rows correspond to the CMS measurement in the lepton+jets final state.

Single-inclusive jet and di-jet production at the LHC. Figures B.4, B.5, and B.6 display the data-theory comparison for the remaining rapidity bins not shown in Figs. 3.9, 3.10, and 3.11, respectively. Figure B.4 corresponds to the ATLAS single-inclusive jet measurement; Fig. B.5 corresponds to the CMS single-inclusive jet measurement; and Fig. B.5 corresponds to the ATLAS di-jet measurement.

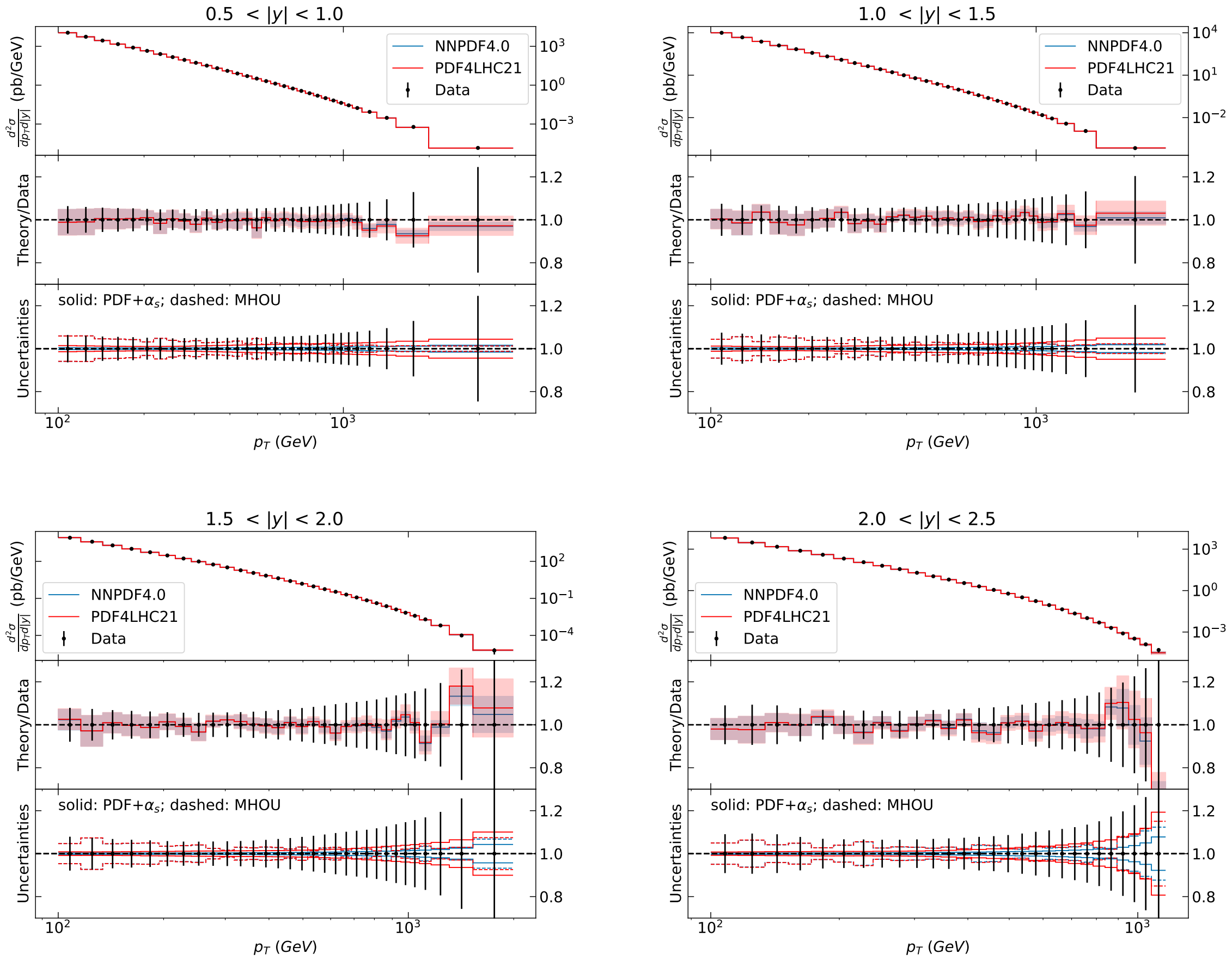

Figure B.4: Same as Fig. 3.9 for the intermediate rapidity bins.

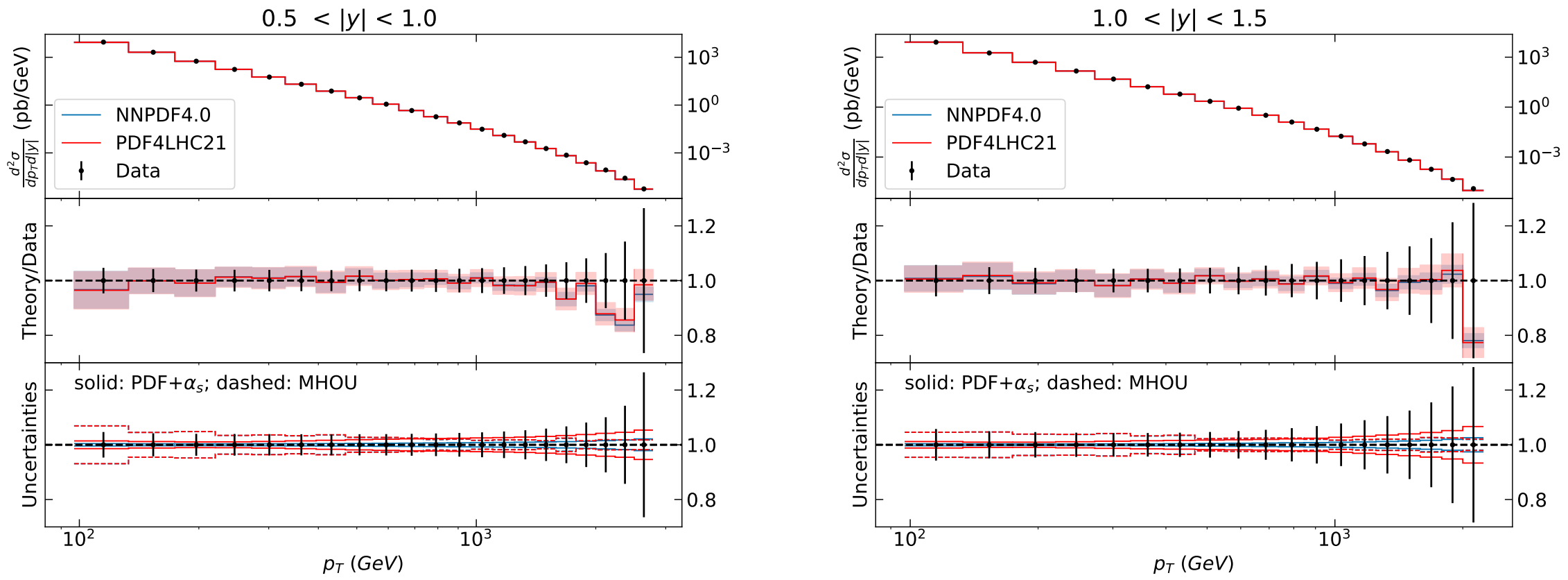

Figure B.5: Same as Fig. 3.10 for the intermediate rapidity bins.

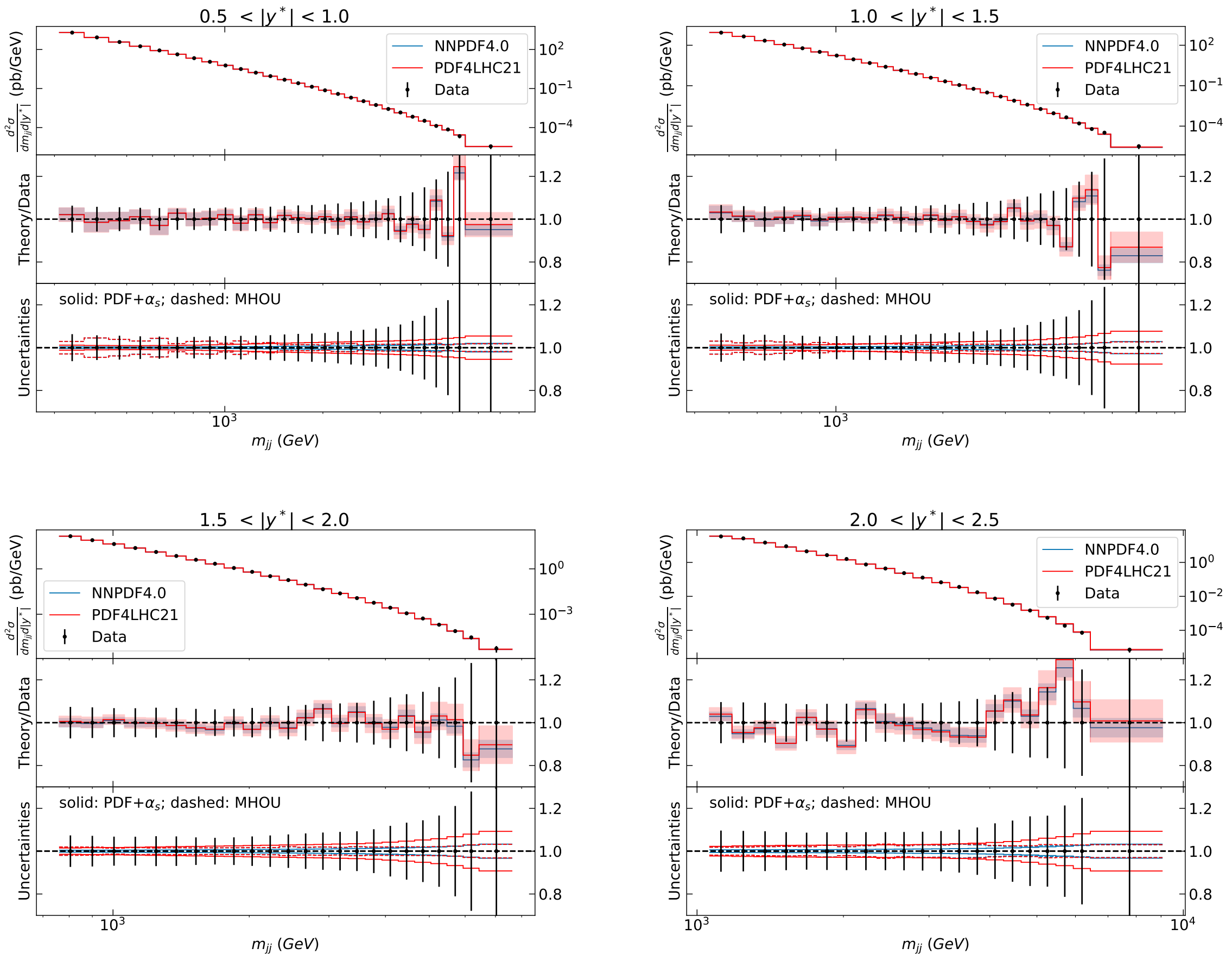

Figure B.6: Same as Fig. 3.11 for the intermediate rapidity bins.

**Single-inclusive jet and di-jet production at HERA.** Figure B.7 displays the breakdown of $\chi^2_{\rm exp+th}$ into its $\chi^2_{\rm exp+mho}$ and $\chi^2_{\rm exp}$ components for the ZEUS single-inclusive jet and di-jet measurements outlined in Table 3.3. Figures B.8-B.11 display the data-theory comparison for the H1 $Q^2$ bins not displayed in Fig. 3.13, respectively, for the low-$Q^2$ single-inclusive jet and di-jet measurements, and for the high-$Q^2$ single-inclusive jet and di-jet measurements. Figures B.12-B.14 display the data-theory comparison, respectively, for the ZEUS low-luminosity single-inclusive jet, high-luminosity single-inclusive jet, and di-jet measurements.

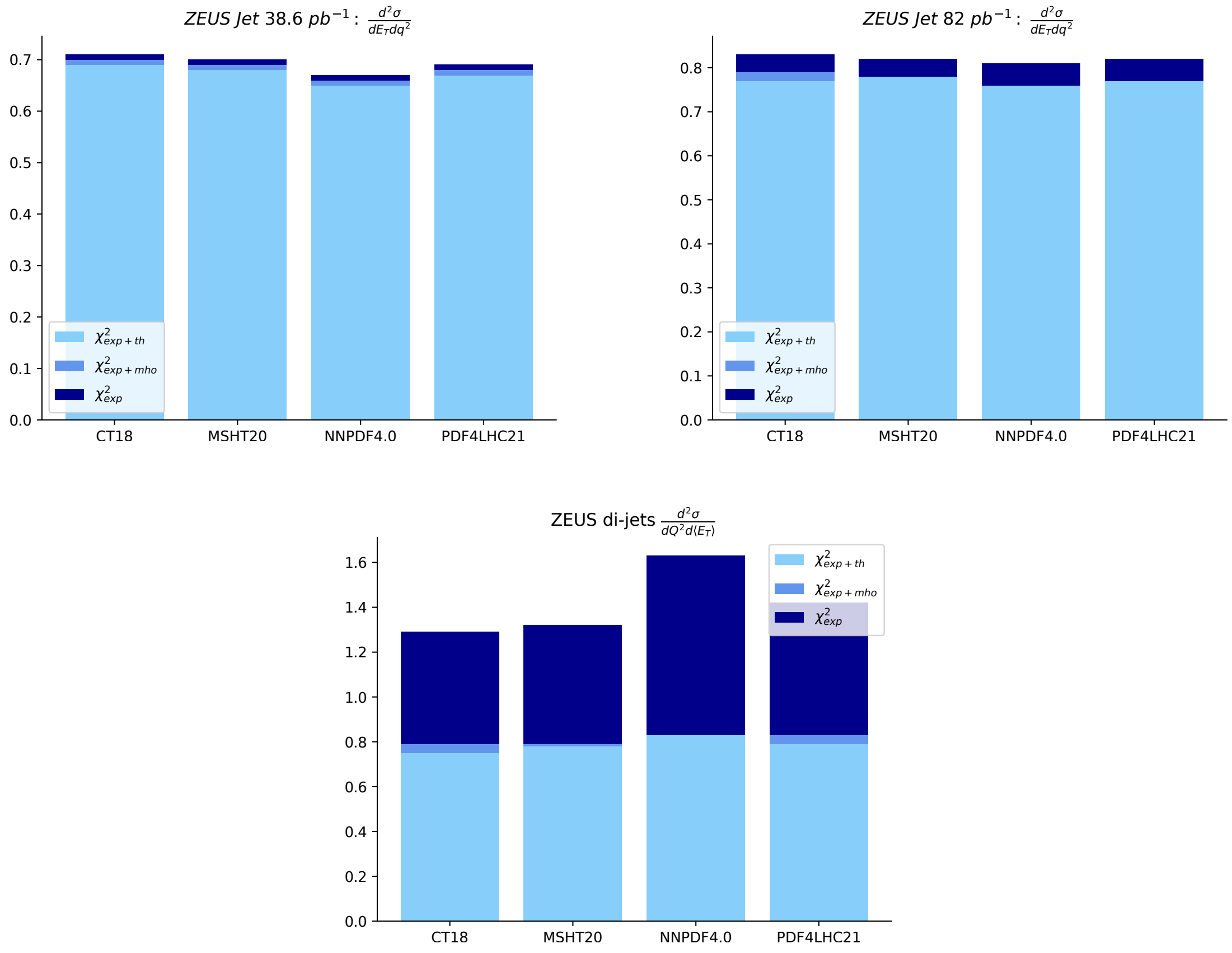

Figure B.7: Same as Fig. 3.3 for the ZEUS single-inclusive jet dataset comprising $n_{\text{dat}} = 30$ data points (and $\sqrt{2/n_{\text{dat}}} = 0.26$) (top) and di-jet (bottom) dataset comprising $n_{\text{dat}} = 22$ data points (and $\sqrt{2/n_{\text{dat}}} = 0.30$).

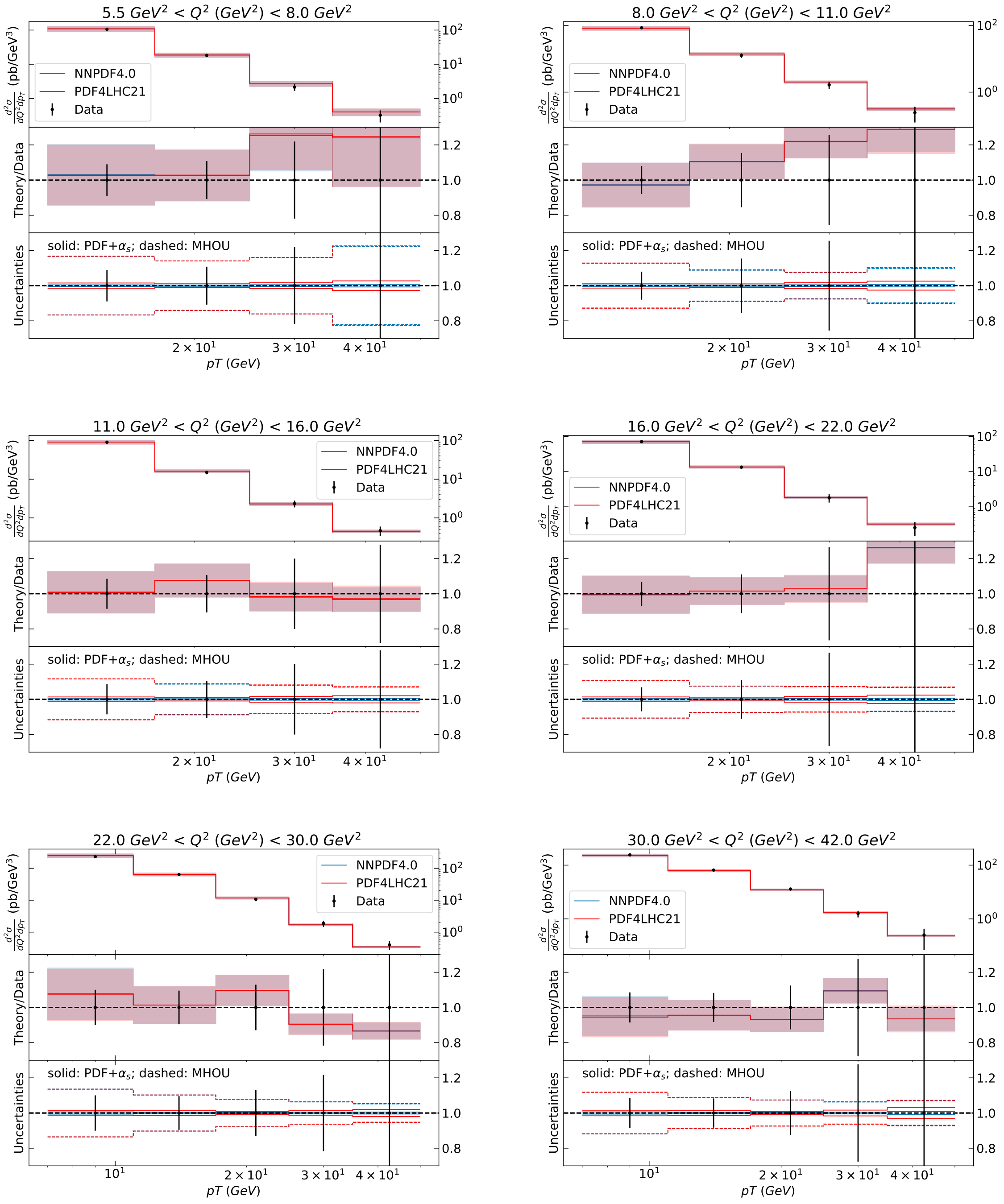

Figure B.8: Same as Fig. 3.13 for the bins of the H1 low-$Q^2$ single-inclusive jet measurement not displayed in Fig. 3.13.

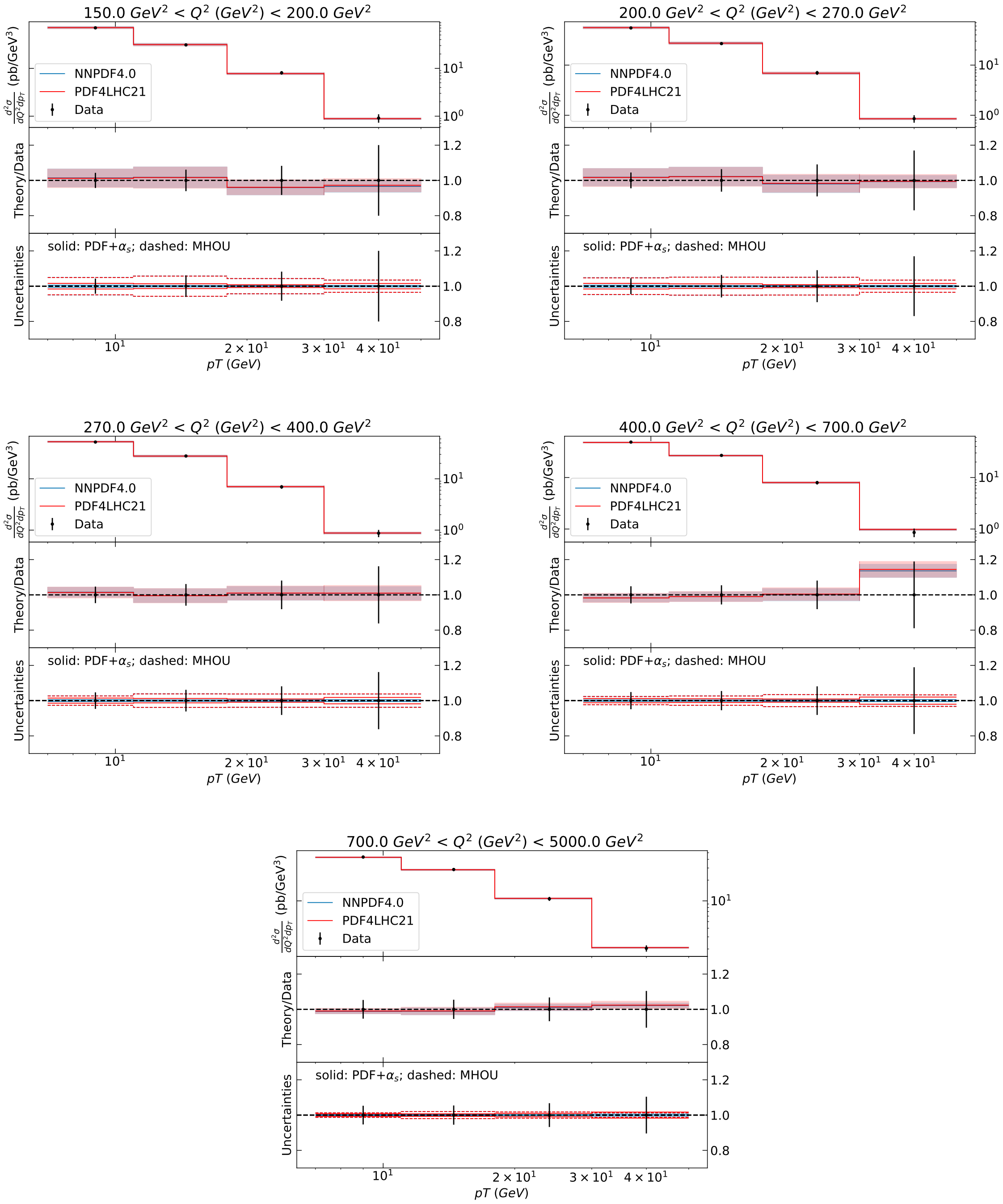

Figure B.9: Same as Fig. 3.13 for the bins of the H1 high-$Q^2$ single-inclusive jet measurement not displayed in Fig. 3.13.

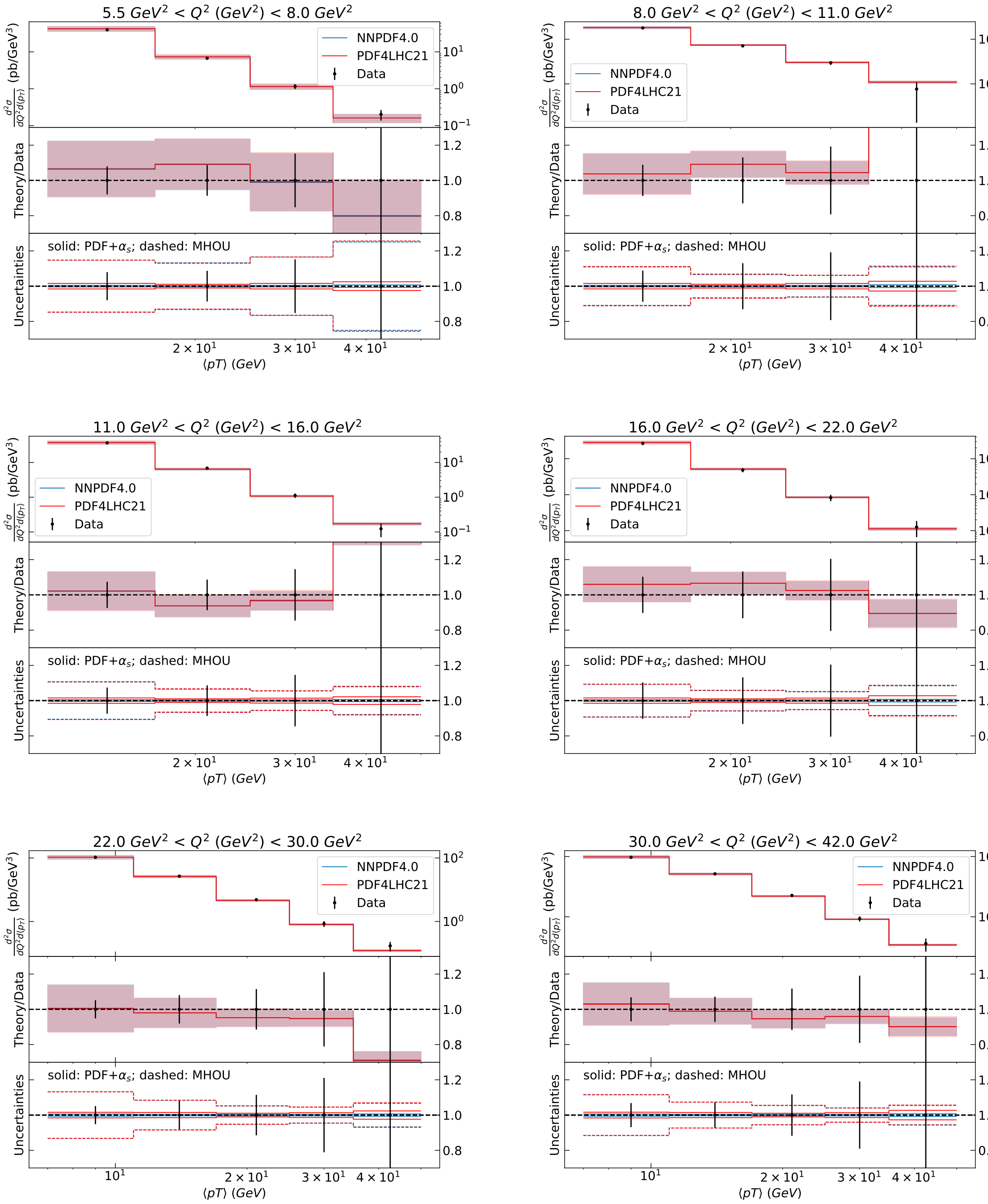

Figure B.10: Same as Fig. 3.13 for the bins of the H1 low-$Q^2$ di-jet measurement not displayed in Fig. 3.13.

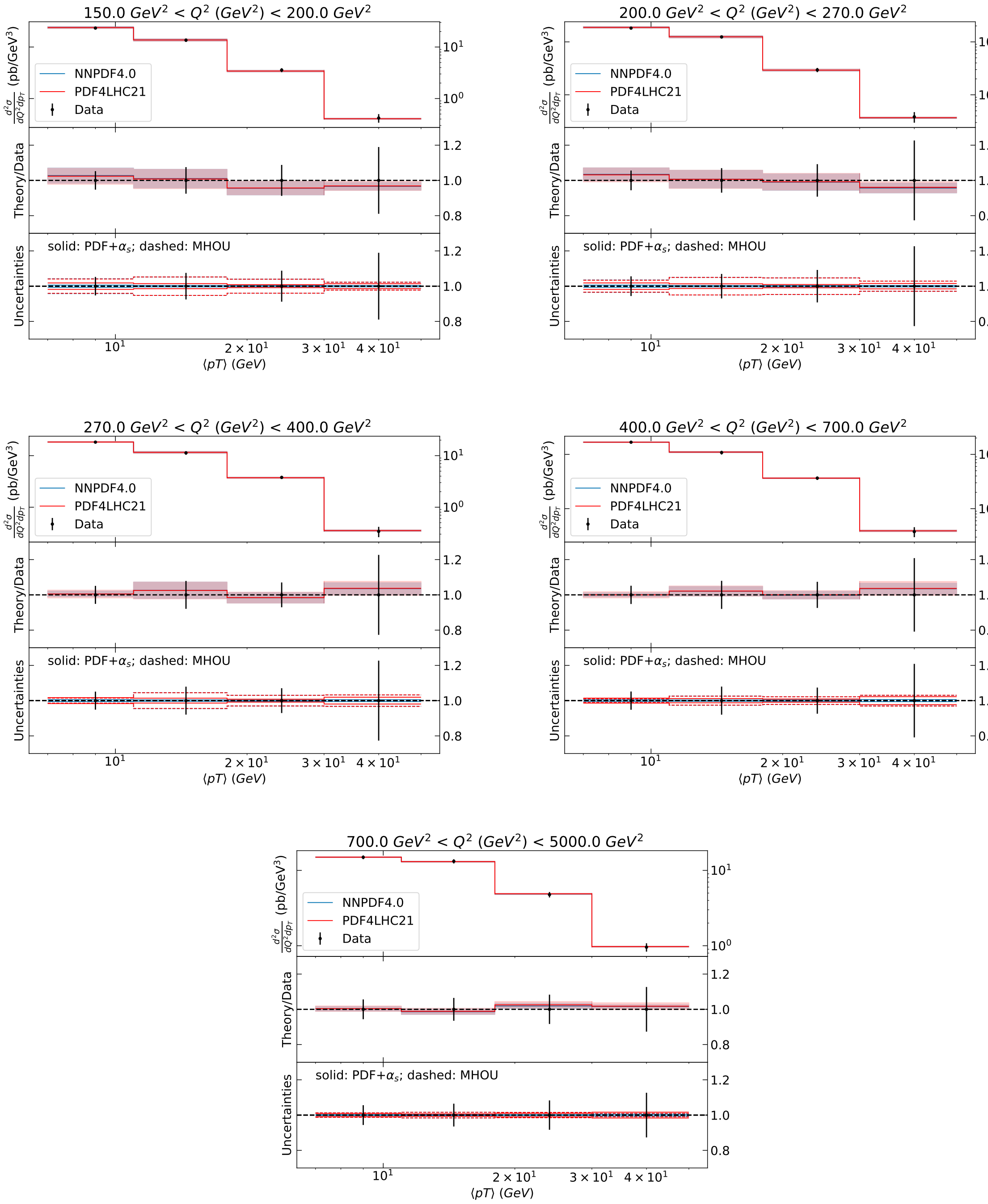

Figure B.11: Same as Fig. 3.13 for the bins of the H1 high-$Q^2$ di-jet measurement not displayed in Fig. 3.13.

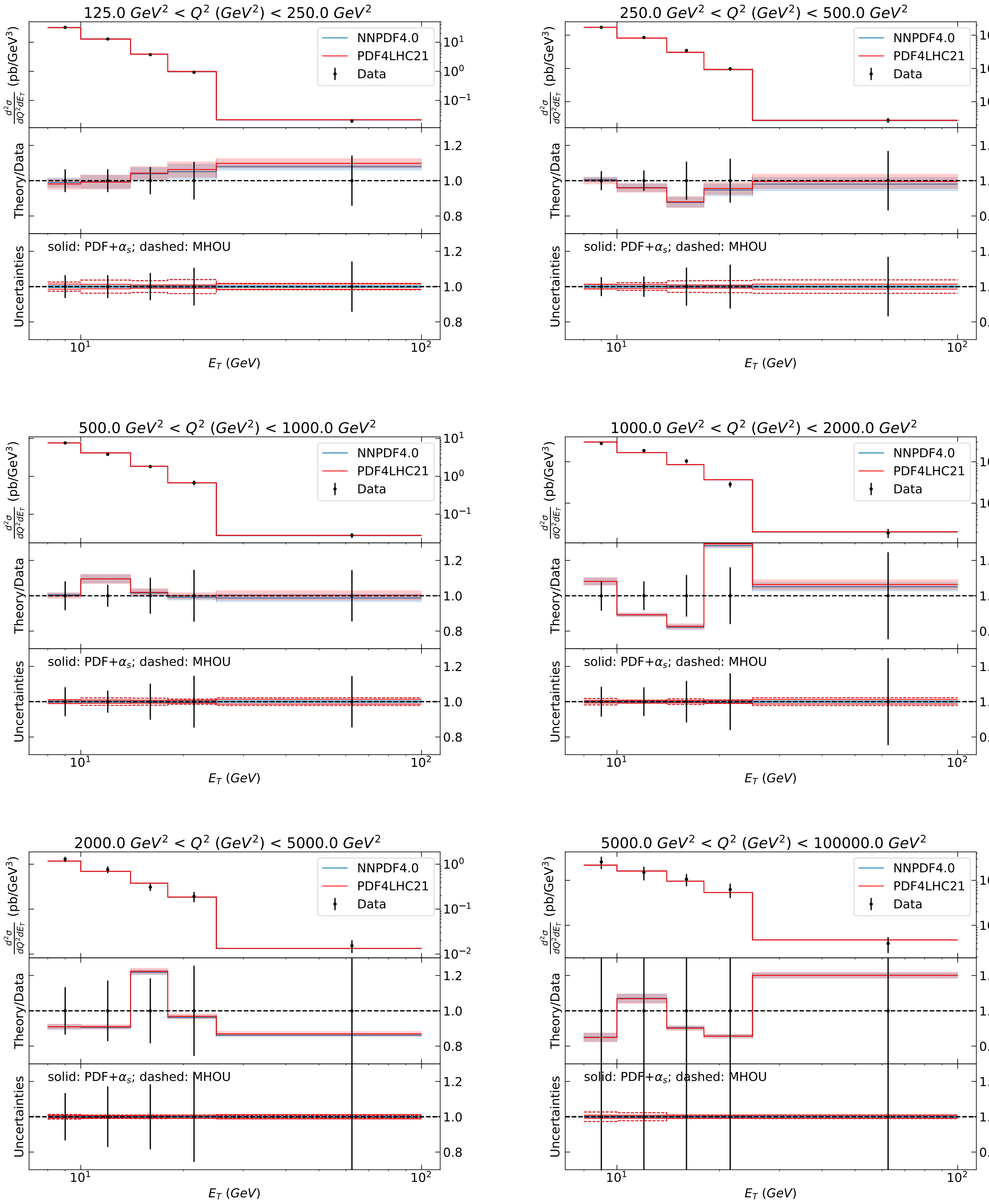

Figure B.12: Same as Fig. 3.4 for the ZEUS low-luminosity single-inclusive jet production measurement.

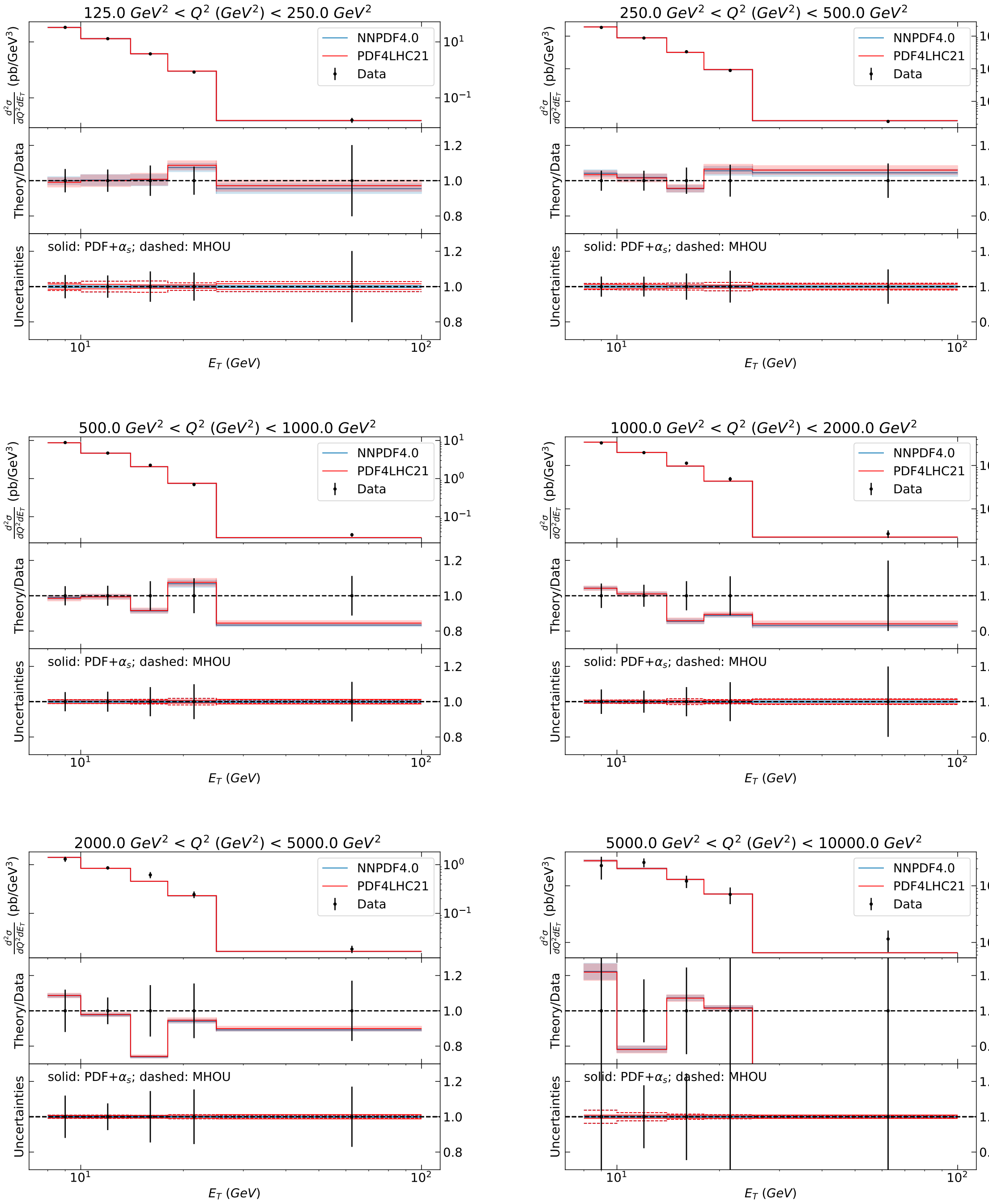

Figure B.13: Same as Fig. 3.4 for the ZEUS high-luminosity single-inclusive jet production measurement.

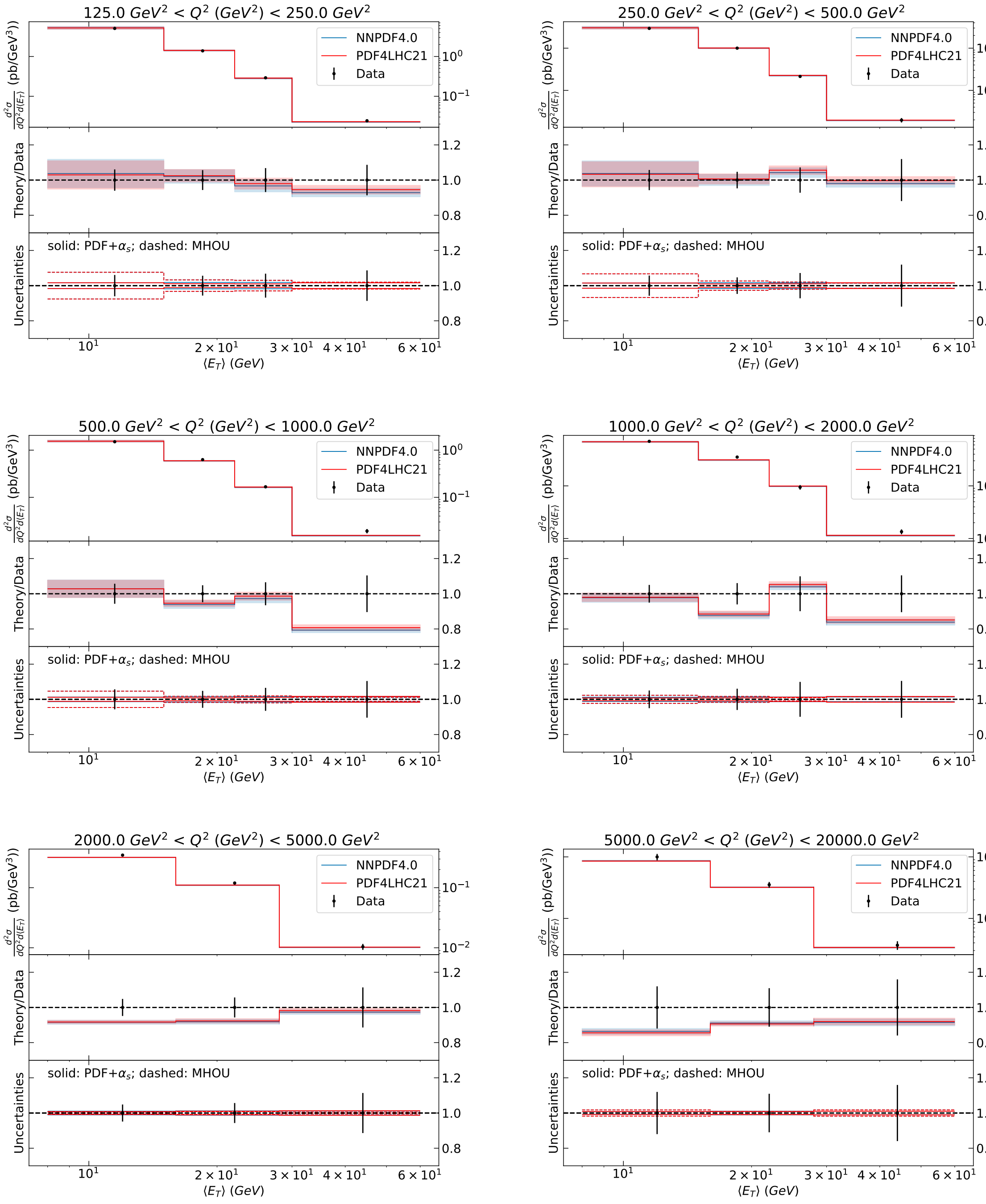

Figure B.14: Same as Fig. 3.4 for the ZEUS di-jet production measurement.